\documentclass[twocolumn]{aastex62}

\newcommand{\JOURNAL}{}

\usepackage{amsmath}
\usepackage{mathtools}

\renewcommand{\la}{\ensuremath{\left\langle}}  
\newcommand{\ra}{\ensuremath{\right\rangle}}

\newcommand{\lv}{\ensuremath{\left\lvert}}   
\newcommand{\rv}{\ensuremath{\right\rvert}}

\newcommand{\ls}{\ensuremath{\left[}}        
\newcommand{\rs}{\ensuremath{\right]}}

\newcommand{\lp}{\ensuremath{\left(}}        
\newcommand{\rp}{\ensuremath{\right)}}

\renewcommand{\vec}[1]{{\boldsymbol{\mathbf{#1}}}}
\newcommand{\mat}[1]{{\boldsymbol{\mathbf{#1}}}}

\newcommand{\hconj}{\ensuremath{\dagger}}

\newcommand{\phat}{\hat{p}}

\newcommand{\va}{\vec{a}}

\newcommand{\vd}{\vec{d}}
\newcommand{\vk}{\vec{k}}
\newcommand{\vn}{\vec{n}}
\newcommand{\vp}{\vec{p}}
\renewcommand{\vr}{\vec{r}}
\newcommand{\vs}{\vec{s}}

\newcommand{\vu}{\vec{u}}
\newcommand{\vv}{\vec{v}}

\newcommand{\vx}{\vec{x}}
\newcommand{\vq}{\vec{q}}
\newcommand{\vy}{\vec{y}}
\newcommand{\vb}{\vec{b}}

\newcommand{\vbhat}{\hat{\vec{b}}}

\newcommand{\vehat}{\hat{\vec{e}}}

\newcommand{\vphat}{\hat{\vec{p}}}
\newcommand{\vrhat}{\hat{\vec{r}}}

\newcommand{\vxhat}{\hat{\vec{x}}}
\newcommand{\vyhat}{\hat{\vec{y}}}
\newcommand{\vzhat}{\hat{\vec{z}}}

\newcommand{\vA}{\vec{A}}
\newcommand{\vE}{\vec{E}}

\newcommand{\veps}{\vec{\varepsilon}}
\newcommand{\vtheta}{\vec{\theta}}

\newcommand{\mB}{\mat{B}}
\newcommand{\mC}{\mat{C}}
\newcommand{\mE}{\mat{E}}
\newcommand{\mF}{\mat{F}}
\newcommand{\mG}{\mat{G}}
\newcommand{\mI}{\mat{I}}

\newcommand{\mM}{\mat{M}}
\newcommand{\mN}{\mat{N}}
\newcommand{\mP}{\mat{P}}
\newcommand{\mQ}{\mat{Q}}

\newcommand{\mS}{\mat{S}}
\newcommand{\mT}{\mat{T}}

\newcommand{\mV}{\mat{V}}
\newcommand{\mW}{\mat{W}}

\newcommand{\mNh}{\mat{N}^{\scriptscriptstyle -1/2}}

\DeclareMathOperator{\sinc}{sinc}

\DeclareMathOperator{\Var}{Var}
\DeclareMathOperator{\Cov}{Cov}

\DeclareMathOperator{\argmax}{argmax}

\newcommand{\appropto}{\mathrel{\vcenter{
  \offinterlineskip\halign{\hfil$##$\cr
    \propto\cr\noalign{\kern2pt}\sim\cr\noalign{\kern-2pt}}}}}

\newcommand{\calG}{\mathcal{G}}

\newcommand{\calP}{\mathcal{P}}

\newcommand{\eps}{\varepsilon}

\usepackage{hyperref}
\usepackage{xspace}
\usepackage{graphicx}

\ifdefined\JOURNAL
    \usepackage{natbib}
\else

    \let\textcite\relax
    \let\citet\relax
    \let\citep\relax
    \let\citealt\relax
    \let\citealp\relax

    \expandafter\let\csname ver@natbib.sty\endcsname\relax

    \usepackage[style=authoryear-comp,natbib=true,
                useprefix=true, giveninits=true,
                maxbibnames=5, minbibnames=3,
                isbn=false,safeinputenc]{biblatex}
    \AtEveryBibitem{\clearfield{title}}   
    \DeclareFieldFormat*{parens}{#1\addperiod\addspace}
    \renewbibmacro{in:}{}

\fi

\usepackage{savesym}
\savesymbol{tablenum}
\usepackage{siunitx}
\restoresymbol{SIX}{tablenum}
\DeclareSIUnit\degr{deg}
\DeclareSIUnit\radian{rad}

\usepackage{nomencl}

\definecolor{ucol}{rgb}{.6,0,0}
\definecolor{webgreen}{rgb}{0,.5,0}
\definecolor{halfgray}{gray}{0.55}

\usepackage{aas_macros}

\usepackage[capitalize]{cleveref}

\newcommand{\secref}[1]{Section~\ref{#1}}

\newcommand{\tcm}{21$\,$cm\xspace}  
\newcommand{\Tspin}{\ensuremath{T_s}}
\newcommand{\Tcmb}{\ensuremath{T_\gamma}}
\newcommand{\nuHI}{\ensuremath{\nu_{\scriptscriptstyle 21}}}
\newcommand{\tauHI}{\ensuremath{\tau_{\scriptscriptstyle 21}}}
\newcommand{\sHI}{\ensuremath{{\scriptscriptstyle HI}}}

\newcommand{\OmegaHI}{\ensuremath{\Omega_\sHI}}

\makenomenclature

\begin{document}

\title{Data Analysis for Precision \tcm Cosmology}
\correspondingauthor{Adrian Liu}
\email{acliu@physics.mcgill.ca}
\author[0000-0001-6876-0928]{Adrian Liu}
\affil{Department of Physics and McGill Space Institute, McGill University, Montreal, QC, Canada H3A 2T8}

\author[0000-0002-4543-4588]{J. Richard Shaw}
\affil{Department of Physics and Astronomy, University of British Columbia, Vancouver, BC, Canada V6T 1Z1}
\affil{National Research Council Canada, Herzberg Research Centre for Astronomy and Astrophysics, Dominion Radio Astrophysical Observatory, P.O. Box 248, Penticton, British Columbia, Canada V2A 6J9}

\begin{abstract}
The redshifted \tcm line is an emerging tool in cosmology, in principle permitting three-dimensional surveys of our Universe that reach unprecedentedly large volumes, previously inaccessible length scales, and hitherto unexplored epochs of our cosmic timeline. Large radio telescopes have been constructed for this purpose, and in recent years there has been considerable progress in transforming \tcm cosmology from a field of considerable theoretical promise to one of observational reality. Increasingly, practitioners in the field are coming to the realization that the success of observational \tcm cosmology will hinge on software algorithms and analysis pipelines just as much as it does on careful hardware design and telescope construction. This review provides a pedagogical introduction to state-of-the-art ideas in \tcm data analysis, covering a wide variety of steps in a typical analysis pipeline, from calibration to foreground subtraction to map making to power spectrum estimation to parameter estimation.
\end{abstract}

%
%
%

\section{Introduction}

The last few decades have seen significant progress in observing large portions of our Universe using galaxy surveys, Cosmic Microwave Background (CMB) measurements, Lyman alpha forest studies, gravitational lensing, galaxy cluster counts, and now even gravitational wave interferometers. However, in many ways, our Universe remains considerably unexplored. For example, the CMB and galaxy surveys probe the early and the late universe, respectively, but intermediate portions of our cosmic timeline have not yet been systematically surveyed. As a result, there exists a gap in direct observations between $\sim 400,000$ and $1.5\,\textrm{Gyr}$ after the Big Bang (corresponding to a missing redshift range between $z \sim 1,100$ to $z\sim 3$). To be fair, some information does exist at these intermediate redshifts, but this tends to take the form of individual astronomical objects of interest (or in the case of Lyman alpha forest studies, just a small handful of sight lines that extend to the relevant redshifts). Moreover, even at redshifts that are nominally covered by surveys, the surveys may be incomplete in a variety of ways. For example, they may not cover the entire sky because of the location of one's survey instrument. They may also lack the resolution to capture details on the finest scales, or the sensitivity to capture information from anything but the brightest objects. In general, there is considerable information content locked in our cosmos that has yet to be harnessed.

In the next few decades, \tcm cosmology has the potential to significantly enhance our understanding of our Universe. The basic idea is to take advantage of the \tcm-wavelength spectral line that arises from the hyperfine ``spin flip" transition of neutral hydrogen. By searching for absorption or emission of the redshifted \tcm line, one has a way to observe neutral hydrogen, which can in turn be used as a tracer of matter or as an indirect probe of other properties of our Universe such as its ionization state or temperature. The \tcm line of hydrogen is a ``forbidden" transition, and thus the likelihood of an individual atom making the transition is low. However, compensating for this is the sheer abundance of hydrogen in our Universe, which results in an observable signal. In addition, the weakness of the transition means that the transition is optically thin, i.e., \tcm photons are unlikely to be absorbed/emitted more than once as they travel to our telescopes. This enables a fully three-dimensional mapping of the spectral line, with the redshift providing line of sight distance information. The result will be a dataset with unprecedented reach in volume.

Using the \tcm line not only enables larger survey volumes, but also pushes state-of-the-art limits in redshift, luminosity, and scale. Extremely high redshifts are in principle accessible because one only requires neutral hydrogen, which exists prior to the formation of the first luminous objects. Moreover, even after luminous objects are formed, only the brightest ones will be seen by traditional observations. Observations in \tcm cosmology typically do not resolve such individual bright objects, instead averaging over many objects that fall inside single, relatively large, pixels. The reward of this approach, however, is that one is probing the integrated emission over the entire luminosity function, including photons from objects that would be too dim to detect individually. Without the need to resolve individual objects, instrument specifications tend to be driven by length scales dictated by cosmology. While these length scales are therefore generally large, it should be noted that the \tcm line also enables one to access scales that are much finer than most other cosmological probes. Much of this is the result of the relative ease in which spectral resolution (and therefore line of sight distance resolution) can be obtained in radio astronomy.

Motivated by the aforementioned opportunities, a large number of instruments have been constructed to make a first measurement of the cosmological \tcm signal. The last few years have seen considerable progress towards this goal: detections of the \tcm line have been made in cross correlation with galaxy surveys at $z< 1$, and there has been a tentative claim of a detection of the sky-averaged \tcm signal at $z \sim 17$. In addition, instruments seeking to measure spatial structures in the \tcm line across a wide range of redshift are now online and in many cases, have begun setting upper limits on the signal. Much of the emphasis in the field has therefore shifted over to issues of precision data analysis. In many ways, \tcm instruments are as much \emph{software telescopes} as they are hardware telescopes---it is becoming increasingly clear that analysis pipelines capable of meeting exquisite requirements in calibration, data volume, and control of systematics are just as important as careful hardware design.

This paper is a response to the recent emphasis on data analysis. Our goal is to provide a pedagogical introduction to analysis ideas and problems that are currently on the minds of analysts working in \tcm cosmology. While we do provide some overview of the theory (Section \ref{sec:Science}) and instrumentation (Sections \ref{sec:BasicObs}, \ref{sec:ObsStatus}, and \ref{sec:InterferometryBasics}), these summaries are intended to serve only as background for our more thorough discussions of analysis. As such, this paper is complementary to the many excellent previous reviews on \tcm cosmology, such as \citet{FurlanettoReview,MoralesWyitheReview2010,PritchardLoeb2012review,aviSteveBook}.

The rest of this paper is organized as follows. Section \ref{sec:Science} briefly reviews the theoretical foundations of the predicted \tcm signal. Section \ref{sec:BasicObs} then provides a quick introduction to \tcm cosmology from an observational standpoint, qualitatively discussing issues such as instrument design optimization. This is followed by a more technical discussion of interferometry in Section \ref{sec:InterferometryBasics} after a survey of the current status of the field in Section \ref{sec:ObsStatus}. We provide an overview of many systematic effects in Section \ref{sec:InstrumentSystematics}, but leave the chief technical challenge of \tcm cosmology---that of foreground contaminants---to a dedicated description in Section \ref{sec:fgs}. After a brief mid-paper summary in Section \ref{sec:Interlude}, state-of-the-art approaches to dealing with the aforementioned challenges are then described in subsequent sections, with Section \ref{sec:Calib} discussing calibration, Section \ref{sec:mapmaking} discussing map making, Section \ref{sec:pspecestimation} discussing power spectrum estimation, and Section \ref{sec:FgMitigation} discussing foreground mitigation. Section \ref{sec:MCMC} then moves onto parameter extraction and fitting data to theoretical models within the context of a power spectrum measurements. Other types of measurements are described in Section \ref{sec:BeyondPspec}. In Section \ref{sec:GlobalSig} we provide an overview of sky-averaged \emph{global signal} measurements that have generated much excitement lately because of a claimed positive detection. We summarize our conclusions in Section \ref{sec:Conc}.

\begin{table*}
\caption{\label{tab:Definitions}Glossary of important mathematical quantities.  The ``context" column gives Eq. references, typically either their defining Eq. or their first appearance in the text.}
\begin{ruledtabular}
\begin{tabular}{l p{12.0cm}l p{2cm}l p{2cm}}

Quantity & Meaning/Definition &  Context \\
\hline
$\nu_{21}$ & Rest frequency of \tcm line & Sec. \ref{sec:21cmBasics} \\
$T_s$ & Spin temperature of the \tcm line & Eq. \eqref{eq:SpinTemperature}\\
$T_\gamma$ & Temperature of the Cosmic Microwave Background & Sec. \ref{sec:21cmBasics} \\
$T_b$ or $T$ & Brightness temperature of the \tcm line & Sec. \ref{sec:21cmBasics} \\
$\overline{T}_b$ & Globally averaged \tcm brightness temperature &  Eq. \eqref{eq:GlobalSigDef}, Sec. \ref{sec:GlobalSig} \\
$H_0$ & Hubble parameter today & --- \\
$D_c$ & Line of sight comoving distance & Eq. \eqref{eq:LOScomovingdistance} \\
$z$ & Redshift & Eq. \eqref{eq:RedshiftDef} \\
$x_\textrm{H}$ & Neutral hydrogen fraction & Sec. \ref{sec:21cmBasics} \\
$\OmegaHI$ & Normalized neutral hydrogen density & Sec.\ref{sec:LowzHI} \\
 $b_\sHI$ & Neutral hydrogen bias & Sec. \ref{sec:LowzHI} \\
 $\mathbf{b}$ & Baseline vector & Sec. \ref{sec:dish_or_inter} \\
 $\mathbf{u}$ & Normalized baseline vector/Fourier dual coordinate to angle in flat-sky approximation  & Eqs. \eqref{eq:Itilde}, \eqref{eq:Itilde_inverse}, \eqref{eq:udef} \\
 $A_p$ & Power primary beam & Eq. \eqref{eq:BasicVis}, Footnote \ref{foot:Ap} \\
 $V$ & Interferometric visibility, cosmological survey volume, or Stokes V (depending on context) & Eqs. \eqref{eq:BasicVis}, \eqref{eq:SurveyVolume}, Sec. \eqref{sec:SkyEmissionStokes} \\
 $Q, U$ & Stokes Q and U parameters & Sec. \ref{sec:SkyEmissionStokes}\\
 $P(\mathbf{k})$ & Rectilinear power spectrum & Eq. \eqref{eq:PspecDef} \\
 $\phi$ & Survey volume function, frequency tapering function, or bispectrum phase (depending on context)& Eqs. \eqref{eq:Ttildeobs}, \eqref{eq:BispectrumPhase} \\
 $\xi(\mathbf{x})$ & Correlation function & Eq. \eqref{eq:CorrFct} \\
 $\Delta^2 (k) $ & ``Dimensionless" power spectrum/variance per logarithmic $k$ bin & Eq. \eqref{eq:DeltaSqDefinition} \\
 $a_{\ell m}$ & Spherical harmonic expansion coefficient & Eq. \eqref{eq:SphericalHarmonicExpansion} \\
 $Y_{\ell m}(\hat{\vr})$ & Spherical harmonic basis functions & Eq. \eqref{eq:SphericalHarmonicExpansion} \\
 $C_\ell$ & Angular power spectrum & Eqs. \eqref{eq:AngPspec}, \eqref{eq:CellGuess} \\
$\mathbf{k}_\perp$ & Fourier wavevector perpendicular to the line of sight & Eq. \eqref{eq:cylind_fourier_def} \\
$k_\parallel$ & Fourier wavenumber parallel to the line of sight & Eq. \eqref{eq:cylind_fourier_def} \\
$\widetilde{V}$ & Delay transformed visibility & Eq. \eqref{eq:DelayTransformDef} \\
$\Omega_p$ & Integral of power beam & Eq. \eqref{eq:BeamAreaDefs} \\
$\Omega_{pp}$ & Integral of square of power beam & Eq. \eqref{eq:BeamAreaDefs} \\
$T_\textrm{sys}$ & System temperature & Eq. \eqref{eq:uveta_noise} \\
$\mathbf{E} (\mathbf{x}, t) $ & Electric field measured at position $\mathbf{x}$ at time $t$ & Eq. \eqref{eq:dvEdef} \\
$\varepsilon(\hat{\vr}, \nu)$ & Contribution of electric field from direction $\hat{\vr}$ and frequency $\nu$ & Eq. \eqref{eq:dvEdef} \\
$\mathbf{A}$ & Electric field/voltage beam of an antenna & Eq. \eqref{eq:dvEdef} \\
$g$ & Instrumental gain & Eq. \eqref{eq:CalibEqn} \\
$\mathbf{B}$ & Instrument transfer matrix & Eq. \eqref{eq:discrete_vis} \\
$\mathbf{N}$ & Noise covariance matrix of data & Eq. \eqref{eq:mapmaking_likelihood} \\
$\mathbf{S}$ & Signal covariance matrix of data & Sec. \ref{sec:LinearMapMakers} \\
$\mathbf{C}$ & Total covariance matrix of data & Eq. \eqref{eq:CovDef} \\
$\mathbf{E}^\alpha$ & Quadratic estimator matrix for $\alpha$th power spectrum bandpower & Eq. \eqref{eq:generic_phat} \\
$ \mathbf{Q}^\alpha$ & Response of the covariance matrix to the $\alpha$th power spectrum bandpower & Eq. \eqref{eq:QmatrixdCdp} \\
$ \mathbf{W}$ & Window function matrix & Eqs. \eqref{eq:Wab}, \eqref{eq:phatWp}, \eqref{eq:WMF} \\
$V_{\alpha \beta}$ & Error covariance between $\alpha$th and $\beta$th power spectrum bandpowers & Eqs. \eqref{eq:Vgeneral} and \eqref{eq:VCECE} \\ 
$\hat{q}^\alpha$ & Unnormalized power spectrum bandpower estimates & Eq. \eqref{eq:qalpha} \\
$\hat{p}^\alpha$ & Normalized power spectrum bandpower estimates & Eqs. \eqref{eq:generic_phat} and \eqref{eq:pmq} \\
$\mathbf{M}$ & Power spectrum normalization matrix & Eq. \eqref{eq:pmq} \\
$\mathbf{F}$ & Fisher information matrix & Eq. \eqref{eq:Fisher}
\end{tabular}
\end{ruledtabular}
\end{table*}

\section{Science with the \tcm line}
\label{sec:Science}
\subsection{21cm line basics}
\label{sec:21cmBasics}
The \tcm line is the hyperfine transition of atomic hydrogen. The parallel alignment of the electron and proton spins is a slightly higher energy state than the anti-parallel alignment. As an atom transitions from one state to the other, it emits (or absorbs) a photon of \tcm wavelength. To study this line we use a quantity called the spin-temperature $\Tspin$ that describes the relative occupancy of the two spin states
\nomenclature{\Tspin}{The spin-temperature of an ensemble of hydrogen atoms defined from the relative occupancy of the two spin states.}%
\begin{equation}
\label{eq:SpinTemperature}
    \frac{n_1}{n_0} = 3 \: \exp{\biggl(- \frac{h \, \nuHI}{k_b \, \Tspin}\biggr)} \; ,
\end{equation}
where the factor of three comes from the relative degeneracy of the states, $n_1$ is the number of atoms in the excited hyperfine state, $n_0$ is the number in the ground hyperfine state, $h$ is Planck's constant, $k_b$ is Boltzmann's constant, and $\nuHI \approx 1420.406\,\textrm{MHz}$ is the rest frequency of the \tcm line. The physics of the spin temperature throughout the history of the Universe is complex; we will cover the relevant parts below. The key thing to note is that we observe the \emph{contrast} between the Cosmic Microwave Background (CMB) and the spin temperature. Where the spin temperature is higher than the CMB temperature, $\Tcmb$, we are emitting photons and see an excess above the CMB temperature; when it is lower than $\Tcmb$ the photons are absorbed from the CMB and we see a deficit compared to what we expect.\footnote{In this paper, we focus exclusively on absorption or emission of the \tcm line relative to the CMB. However, it may also be possible to detect \tcm absorption from a bright, high-redshift radio source \citep{ChrisRadioLoud2002,FurlanettoRadioLoud2002,XuRadioLoud2009,neosporin,CiardiRadioLoud2013,EwallWiceRadioLoud2014,CiardiRadioLoud2015,Ciardi2015RadioLoudSKA,Smelin2016}.}
\nomenclature{\Tcmb}{The temperature of the Cosmic Microwave Background at a given redshift.}

The brightness temperature $T_b$ of the \tcm line is given by
\begin{equation}
\label{eq:Tb_background}
   T_b(\vrhat, \nu) = \left[ 1 - e^{-\tauHI (\vrhat, z) } \right] \frac{T_s (\vrhat, z) - T_\gamma (z) }{1 + z},
\end{equation}
where $\vrhat$ is a unit vector that is centred on the observer and points in the direction of observation on the sky, and the observational frequency $\nu$ is related to the redshift $z$ via the standard relation:
\begin{equation}
\label{eq:RedshiftDef}
1+z = \frac{\nu_{21}}{\nu},
\end{equation}
We have additionally defined $\tauHI(z)$ as the optical depth across the \tcm line at redshift $z$, which is given by
\begin{equation}
   \tauHI(\vrhat, z) = \frac{3 \hbar c^3 A_{10} }{16 k_b \nu_{_{21}}^2 } \frac{ x_\textrm{HI} n_\textrm{H}}{(1+z)(dv_\parallel / dr_\parallel) T_s} ,
\end{equation}
where $dv_\parallel / dr_\parallel$ is the gradient of the proper velocity $v_\parallel$ along the line of sight distance $r_\parallel$, $x_\textrm{HI}$ is the fraction of hydrogen atoms that are neutral, $n_H$ is their number density, $\hbar$ is the reduced Planck's constant, $c$ is the speed of light, and $A_{10} = 2.85 \times 10^{-15}\,\textrm{s}^{-1}$ is the spontaneous emission coefficient of the \tcm line. This optical depth is typically $\lesssim 4 \%$ at all regions of interest \citep{Lewis2007}. We may therefore Taylor expand our expression for $T_b$ to arrive at
\begin{equation}
\label{eq:Tb_theory}
T_b(\vrhat, \nu) = \left(\frac{3 \hbar c^3 A_{10} }{16 k_b \nu_{_{21}}^2 } \right) \left[\frac{ x_\textrm{HI} n_\textrm{H}}{(1 + z)^2 (dv_\parallel / dr_\parallel) }  \right] \left( 1-\frac{T_\gamma  }{T_s} \right)
\end{equation}
One thing to note is that in the high spin temperature limit the observed brightness temperature is independent of the spin temperature itself. This can be understood from the fact that at high temperatures all spin microstates are equally occupied and thus the observed brightness depends only on the rate of spontaneous emission rate from the high energy state ($A_{10}$).

Observationally, $T_b$ can be probed in two ways. One is to measure the \emph{global signal}, where $T_b$ is averaged over all angles on the sky to produce a single averaged spectrum $\overline{T}_b$:
\begin{equation}
\label{eq:GlobalSigDef}
\overline{T}_b (\nu) = \int d\Omega T_b(\vrhat, \nu).
\end{equation}
Another is to try to measure the spatial fluctuations in the full \tcm brightness temperature field, either by attempting to reconstruct the $T_b(\vrhat, \nu)$ or by quantifying the statistical properties of the fluctuations. For most of this paper, we will be discussing the measurement of spatial fluctuations, but we will return to the global signal in Section \ref{sec:GlobalSig}.

Regardless of the type of observation, one sees from Equation \eqref{eq:Tb_theory} that the \tcm brightness temperature contains a rich variety of effects. For example, at certain redshifts $T_s$ is strongly coupled to the baryonic gas temperature (see sections below). This makes $T_b$ a high-redshift thermometer (albeit a rather indirect one). It is also a probe of the ionization state of hydrogen, via its dependence on $x_\textrm{HI}$. We therefore see that $T_b$ is likely to be an excellent probe of many high-redshift astrophysical processes. In addition, it is sensitive to cosmology, since the distribution of hydrogen (entering via the factor of $n_H$) is driven by the large scale cosmological distribution of matter, as is the $dv_\parallel / dr_\parallel$ velocity term, since it includes peculiar velocities in addition to the Hubble flow. Of course, with all of the aforementioned effects contributing to $T_b$, it may be difficult to cleanly probe any of them. Fortunately, different phenomena tend to dominate at different redshifts, and in what follows we provide a quick qualitative description of this.

\begin{figure*}[t]
\centering
\includegraphics[width=1.00\textwidth,trim={0cm 0cm 0cm 0cm},clip]{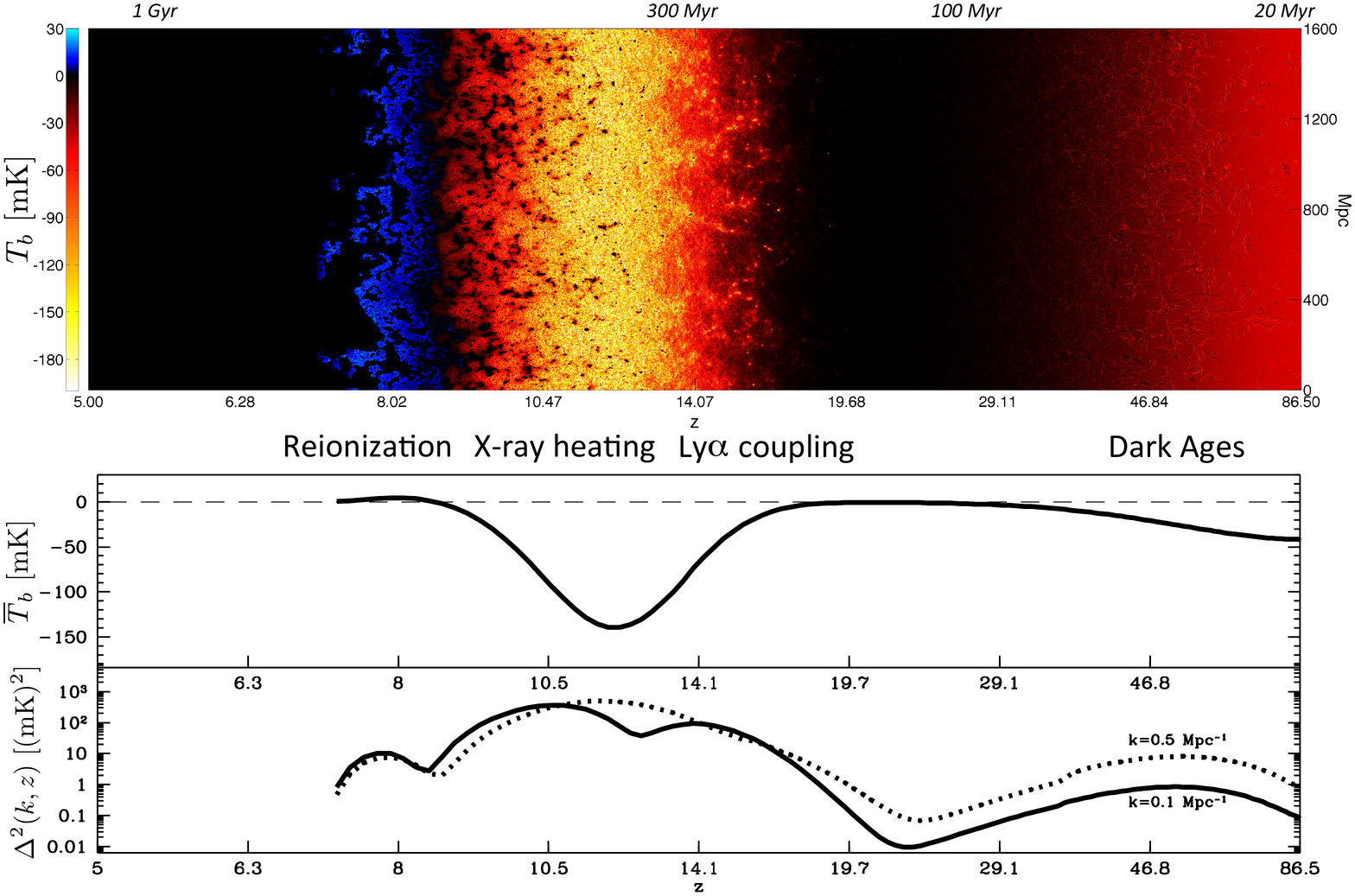}
\caption{A graphical representation of a simulated \tcm signal from $z \sim 90$ (during the Dark Ages) to the $z \sim 7$ (the end of reionization in this particular simulation). The top panel shows the \tcm brightness temperature contrast with the CMB, displayed as a two-dimensional slice of a three-dimensional volume. Because the line-of-sight distance for the \tcm line is obtained via the redshift, the horizontal axis also doubles as a redshift/evolution axis (i.e., the top panel is a picture of a \emph{light cone}, which is what a real telescope would observe). The middle panel shows the predicted \emph{global signal} of the \tcm line, where the spatial information in the vertical direction of the top panel has been averaged over to produce one average brightness temperature per redshift. The bottom panel shows the power spectrum, expressed as $\Delta^2 (k) \equiv k^3 P(k) / 2 \pi^2$, as a function of redshift. This can be thought of as a measure of the variance of spatial fluctuations as a function of spatial wavenumber $k$ (see Equation \ref{eq:DeltaSqDefinition}), with the solid line showing $k = 0.1\,\textrm{Mpc}^{-1}$ and the dotted line showing $k=0.5\,\textrm{Mpc}^{-1}$). These simulations were generated from the \texttt{21cmFAST} semi-analytic code (\citealt{Mesinger2011}; \url{https://github.com/andreimesinger/21cmFAST}), and are publicly available at \url{http://homepage.sns.it/mesinger/EOS.html} as part of the Evolution of \tcm Structure project \citep{Mesinger2016EOS}.}
\label{fig:EOS}
\end{figure*}

\subsection{Ultra high redshift: the Dark Ages}

After recombination, though the photons and baryons no longer act as a single fluid, scatterings between photons and residual electrons (and collisions with the rest of the gas) mean that the baryon temperature is coupled to the photon temperature until $z \sim 300$ when these electron-photon scatterings become rare. At this point, the gas starts to cool faster than the photons $\propto (1 + z)^2$, and because the collisional coupling between the gas temperature and the spin temperature remains strong, Equation \eqref{eq:Tb_theory} predicts a net \tcm absorption signal. All of this takes place during the \emph{Dark Ages}, a pristine cosmological era where the first luminous sources have yet to form \citep{Scott1990}, rendering the \tcm line a direct probe of density fluctuations (at least to the extent that hydrogen traces the overall matter distribution). This era stops being observationally accessible at $z \sim 30$, when collisions between neutral hydrogen atoms become rare enough that the spin degrees of freedom are no longer coupled to the kinetic degrees of freedom, and the spin temperature rises until it equilibrates with the CMB temperature. This is illustrated in the middle panel of Figure \ref{fig:EOS}, where we show the global \tcm signal. With the CMB temperature and the spin temperature in equilibrium, the brightness temperature contrast goes to zero and there is no \tcm signal to observe.

The Dark Ages are of tremendous cosmological interest. The epoch provides access to a huge number of Fourier modes of the matter density fields \citep{Loeb2004}. These fluctuations are graphically depicted in the red high redshift regions of the top panel of Figure \ref{fig:EOS}. They are a particularly clean probe of the matter field since this is prior to the formation of the first luminous sources, thus obviating the need for the modelling of complicated astrophysics. The theoretical modelling effort is in fact even simpler than, say, that needed for galaxy surveys, since at these high redshifts, matter fluctuations are in the linear perturbative regime even at very fine scales. This is important because the spatial fluctuations exist to extremely small scales, as they are not Silk damped and therefore persist down to the Jeans scale \citep{Tegmark2009}. The result is a vast number of modes that can be in principle be used to probe fundamental cosmology. The long lever arm in scale, from the largest structures to the smallest structures, enables constraints on the running of the matter power spectrum's spectral index \citep{Mao2008}. This would represent an incisive probe of the inflationary paradigm. Future measurements may also be able to detect features in the primordial power spectrum \citep{Chen2016Features}, or to detect primordial non-Gaussianity \citep{Munoz2015nonGaussianity}. In addition, the \tcm line can be used to probe the existence of relic gravitational waves from inflation, either through their direct effects on large scale structure \citep{Masui2010} or through their lensing effects \citep{Book2012}. Tests of statistical isotropy and homogeneity may also be possible \citep{Shiraishi2016}. Small-scale measurements enable measurements of the neutrino mass \citep{Mao2008} and constraints on the existence of warm dark matter \citep{Loeb2004}. Finally, the cleanliness of measurements during the Dark Ages provide a good platform for detecting exotic phenomena beyond the standard models of particle physics and cosmology (see references in \citealt{Furlanetto2019fundamentalphysics}).

Measurements during this era are extremely challenging. Not only do the extremely long wavelengths ($\lambda \approx 7$ to $70\,\textrm{m}$) make simply building an instrument with the required resolution and sensitivity extremely difficult, but as we will discuss later, the foregrounds are extremely bright (Sections \ref{sec:fgs} and \ref{sec:FgMitigation}). Additionally, the ionosphere (Section \ref{sec:Ionosphere}) is opaque at frequencies lower than a few MHz, and can cause distortions even at higher frequencies.

\subsection{High redshift: Cosmic Dawn and the Epoch of Reionization}

As the first luminous objects begin to form, the \tcm line ceases to directly trace the dark matter distribution. Radiation from these objects affects the spin temperature and ionization state of the intergalactic medium (IGM) in a spatially and temporally non-trivial way. This is reflected in the spatial fluctuations and redshift evolution of the \tcm line, which is no longer governed by the relatively simple physics of cosmological matter perturbations alone. This is both a defect and an opportunity. It is a defect because the complicated astrophysics of the era (often loosely\footnote{There is unfortunately no consistent definition of Cosmic Dawn that is agreed upon in the literature. Some authors use it to refer to a period that began when the first stars formed, and ended with the formation of larger galaxies. Others define the end of Cosmic Dawn to be when the first galaxies began to systematically reionize the intergalactic medium. Yet others consider Cosmic Dawn to be a broad term that encompasses the entire period from the formation of the first stars to the end of reionization.} referred to as ``Cosmic Dawn") means that it is hard to use the \tcm line as a clean, model-independent probe of fundamental cosmology.\footnote{One exception to this may come from the effect of velocity-induced acoustic oscillations \citep{Dalal2010,Fialkov2012VAOs,Munoz2019VAOs1,McQuinn2012VAOs}, which have the potential to serve as clean standard rulers at high redshifts, thus enabling precision measurements of the Hubble expansion rate \citep{Munoz2019VAOs2}.} On the other hand, this very complication makes the \tcm line a tremendously promising tool for understanding the nature of the first luminous objects.

\begin{figure}[t]
\centering
\includegraphics[width=0.45\textwidth,trim={3cm 0.75cm 2cm 1cm},clip]{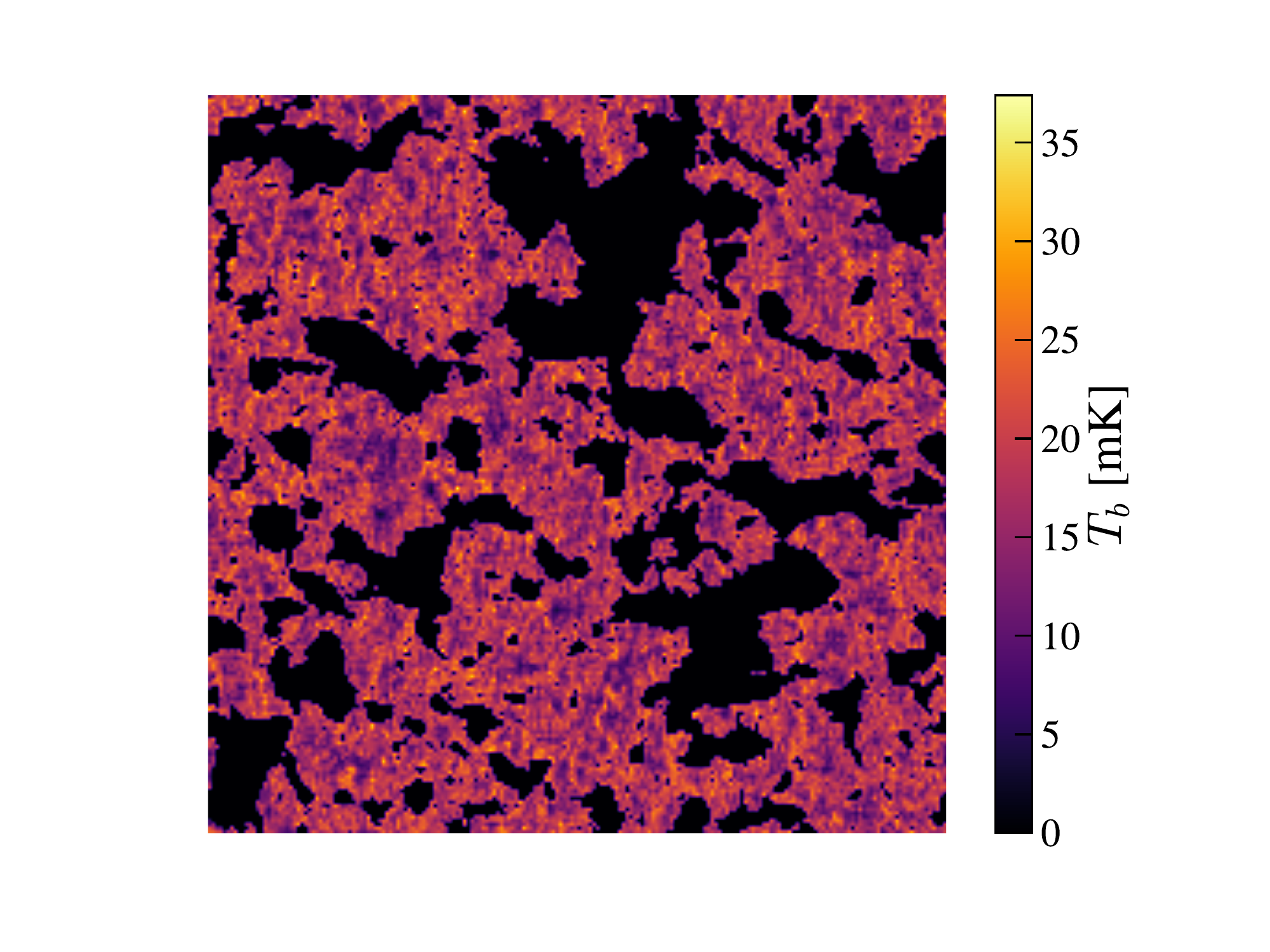}
\caption{An example \tcm brightness temperature field at $z=8.1$, generated using the \texttt{21cmFAST} semi-analytic code \citep{Mesinger2011}. With the particular astrophysical parameters and models used to generate this simulation, $z=8.1$ corresponds to a volume-weighted neutral fraction of $\sim 0.55$. Ionized regions are shown in black, with zero brightness temperature since there is no \tcm transition when there is no neutral hydrogen present. The spatial fluctuations shown here encode a wealth of information about both astrophysics and cosmology, which \tcm experiments aim to extract.}
\label{fig:21cmFAST}
\end{figure}

Figure \ref{fig:EOS} illustrates a series of three epochs following the Dark Ages:
\begin{enumerate}
\item Cosmic Dawn begins with a period of \textbf{Lyman alpha coupling}, which runs from $z \sim 20$ to $z \sim 12$ for the model shown in Figure \ref{fig:21cmFAST}. As the first stars are formed, they produce significant amounts of Ly$\alpha$ flux. This causes the Wouthuysen-Field effect, whereby Ly$\alpha$ photon absorption promotes an electron in a neutral hydrogen atom from an $n=1$ state to an $n=2$ state, only to be followed by a decay to a \emph{different} hyperfine state when the electron returns to the $n=1$ state. This enables Ly$\alpha$ photons to induce \tcm spin-flip transitions, and because of the large cross-section of Ly$\alpha$ scattering, this causes the spin temperature $T_s$ to be coupled to the gas temperature. As was the case during the Dark Ages, the gas temperature is cooler than $\Tcmb$ in this epoch, which in turn means that the spin temperature $T_s$ must also be cooler than $\Tcmb$. The result is an absorption signal that contains information about both the density field and the Ly$\alpha$ flux. As star formation continues, the Ly$\alpha$ background eventually becomes sufficiently strong for the coupling between Ly$\alpha$ photons and gas kinetics to be extremely efficient everywhere. Fluctuations from the Ly$\alpha$ background then become negligible.
\item With continued star formation (and eventually galaxy assembly), we enter a period of \textbf{X-ray heating}. High-energy photons from the first luminous objects have a heating effect on the IGM, as these photons cause photo-ionizations of HI and HeI, with the resulting photoelectrons colliding with other particles. This can result in heating, further ionizations, or atomic excitations. The heating contribution raises the gas temperature to be above $\Tcmb$, which makes the \tcm signal go from absorption to emission. Because the astrophysical sources that heat the IGM are spatially clustered, this results in spatial fluctuations in the \tcm signal. Eventually, however, the entire IGM is heated and the fluctuations sourced by X-ray heating disappear.
\item Cosmic Dawn ends with the \textbf{Epoch of Reionization (EoR)}, when sustained star formation provides enough ionizing photons to systematically ionize the IGM. Reionization does not proceed uniformly because the objects responsible for producing the ionizing photons (likely galaxies) are clustered. This is illustrated in Figure \ref{fig:21cmFAST}, where ionized bubbles form and grow around regions of high matter density because it is those regions that contain the most galaxies (and therefore produce the most ionizing photons). Since the ionized regions contain almost no neutral hydrogen,\footnote{The ``ionized" regions shown in Figure \ref{fig:21cmFAST} are not entirely ionized. At smaller scales  (which are unresolved in the simulation), there exist galaxies and neutral gas clouds that are sufficiently dense to be self-shielded from ionization \citep{Sobacchi2014,Watkinson2015selfshielding}.} the \tcm brightness temperature is close to zero there. The intricate pattern of ionized versus neutral regions gives rise to strong spatial fluctuations in the \tcm line. These fluctuations persist until the ionized bubbles are sufficiently large and numerous that they overlap, and reionization of the entire IGM is complete.
\end{enumerate}

From the bottom panel of Figure \ref{fig:EOS} (which shows the fluctuation power as a function of redshift; see Section \ref{sec:ImageVsStat} for rigorous definitions) we see that each of the three epochs shows a rise-and-fall pattern: the fluctuations increase when a source of inhomogeneity (Ly$\alpha$ coupling, X-ray heating, or ionization) becomes relevant, peak when (loosely speaking) roughly half of the volume is affected, and fall once the relevant process has uniformly affected our Universe. This provides a set of distinctive signatures to search for in observational data \citep{Lidz2008RiseAndFall,Christian2013}. However, it is important to stress that even if our qualitative description of Cosmic Dawn is accurate, the dearth of direct IGM observations at $z> 6$ mean that theoretical models remain fairly unconstrained, and there still remains a wide landscape of possible \tcm predictions. For instance, the precise timing of the three Cosmic Dawn epochs can be easily varied, to the extent that plausible models can be constructed where some of the peaks in fluctuation power merge. As another example, some models find that if $z < 8$ observations of galaxies can be extrapolated to higher redshifts, the \tcm signal may in fact never go into emission, remaining in absorption through the reionization process \citep{Mirocha2017}. Uncertainties in feedback prescriptions add to a wide variety of possible theoretical predictions \citep{Fialkov2013feedback}. It is also unclear as to what the dominant sources of reionization are. While a majority of recent models in the literature assume that galaxies are primarily responsible for reionization (see, e.g., \citealt{Robertson2010}), it would be fair to say that there is still considerable room for at least some contribution from quasars \citep{Madau2015}. Given that there are even more uncertainties in our understanding of Cosmic Dawn than just the few that we have listed here, current theoretical models should in general be treated as helpful guides rather than ironclad standard models.

Fortunately, direct observations of the \tcm line at the redshifts of Cosmic Dawn should significantly enhance our understanding. For example, the typical sizes of ionized bubbles will place constraints on the nature of sources responsible for reionization \citep{McQuinn2007}. A single \tcm measurement of the midpoint of reionization (performed, perhaps, by locating the peak of ionization fluctuation power) would inform the entire timeline of reionization when combined with CMB and Ly$\alpha$ forest data \citep{Pritchard2010BayesianCombo}. Detailed studies of \tcm fluctuations over a wide variety of spatial scales and redshift will enable constraints on the high-redshift universe, shedding light on details such as the minimum mass for halos to host active star forming galaxies, or the X-ray luminosity of galaxies, among other parameters \citep{Pober2013,Mesinger2014,Fialkov2017Xrays,Greig2017,Kern2017,Park2019}. In addition to other possible constraints that we have not discussed above, direct \tcm measurements will also play the important role of allowing models of Cosmic Dawn to be tested, rather than assumed.

\subsection{Low redshift: Post-Reionization}
\label{sec:LowzHI}
At low redshift, after the end of reionization, the neutral fraction has been driven down to $x_\sHI \sim 2\%$ \citep{Villaescusa-Navarro2018,Planck_params2015}. The gas that remains is in dense regions that have managed to self-shield against the ionizing background. Using simulations as a guide we expect the remaining HI to be found within a range of halo masses $\sim 10^{10}$ to $10^{13}\,M_\odot$. At the lowest mass end halos are not dense enough to effectively self shield, and at the high mass end where we are considering clusters, tidal stripping and ram pressure remove gas from within the constituent galaxies \citep{Villaescusa-Navarro2018}. Although the HI that remains can be either part of the cold ($T \lesssim \SI{100}{\kelvin}$) or warm ($T \gtrsim \SI{5000}{\kelvin}$) neutral medium, both temperatures are warmer than the CMB temperature within this redshift range and we expect to see the \tcm line in emission.

This redshift range is the most probed observationally. At very low redshift HIPASS ($z < 0.03$, \citealp{HIPASS}) and ALFALFA ($z < 0.06$, \citealp{2018MNRAS.477....2J}) have detected individual extra galactic objects through their \tcm emission allowing constraints on the mass function and abundance of low redshift neutral hydrogen finding that $\OmegaHI \sim (3.9 \pm 0.6)\times 10^{-4}$ \citep{2018MNRAS.477....2J}. At higher redshifts cross correlations with optical tracers have allowed detection of \tcm emission. This was initially performed with the Green Bank Telescope combined with DEEP2 and later WiggleZ which estimated the hydrogen abundance (times the bias factor $b_\sHI$) at $z \sim 0.8$ to be $\OmegaHI \, b_\sHI = 6.2^{+2.3}_{-1.5} \times 10^{-4}$ \citep{Switzer2013}.

The measurements described above are all direct measurements of \tcm emission but at higher redshifts we can use measurements of Damped Lyman Alpha systems to constrain the HI abundance up to $z = 4.9$, finding $\OmegaHI = 9.8^{+2.0}_{-1.8} \times 10^{-4}$ at $z = 4.9$ \citep{2015MNRAS.452..217C}. Together this means that unlike other epochs there is a well defined target for the amplitude of the signal.


As we expect the HI to be hosted in generally quite low mass halos, and the fraction not bound within haloes to be small (10\% at $z=5$, smaller at low redshift) the neutral hydrogen is an excellent tracer of the total mass distribution \citep{Villaescusa-Navarro2018}. This makes low redshift \tcm observations an excellent probe for cosmology, allowing us to pursue the kinds of science that are currently done by galaxy redshift surveys, but within the radio band. However, unlike galaxy redshift surveys we do not need to resolve individual galaxies (which is required only to determine their redshift), and can instead map the unresolved emission of all HI at each frequency of observation, which we can map directly to a redshift. The angular resolution is chosen to match scales of scientific interest, and each observed voxel may contain hundreds to thousands of individual galaxies. This is the idea of Intensity Mapping \citep{2004MNRAS.355.1339B,Peterson:2006,Chang2008,2008MNRAS.383..606W}, hereafter referred to as IM.


Observations within this epoch are mapping the large-scale structure of the Universe, giving them the ability to do much of the same science as optical galaxy redshift surveys. However, as they can naturally have wide fields of view and observe all redshifts in their band simultaneously, IM experiments are more readily able to quickly survey very large volumes than optical galaxy redshift surveys. And as they do not need to resolve individual galaxies, they can more easily push to high redshifts, through the redshift desert ($z\sim 1$--$3$, where optical spectroscopy is challenging) and beyond ($3 \lesssim z \lesssim 6$).

As IM is a very recent technique that is still being developed, the initial target within this era are to measure Baryon Acoustic Oscillations (BAO). These are the remnants of primordial sound waves that leave a distinct signature in the correlation structure of matter, and they can be used as a standard ruler, albeit a statistical one, to measure the expansion history of the Universe \citep{1998ApJ...504L..57E,Seo2003}. The BAO signature is quite distinct and thus more robust to systematics, making it an ideal first science target for \tcm IM experiments. As a bonus, one can push to ranges in redshifts not currently probed by existing optical surveys \citep{Peterson:2006,Chang2008,2008MNRAS.383.1195W}. By constraining the expansion history of our Universe we hope to be able to infer the properties of Dark Energy, in particular its equation of state $w(z)$, which can yield clues to a microphysical explanation. Such constraints on the equation of state could potentially shed light on recent tensions in measurements of the Hubble parameter \citep{Knox2019HubbleHunter}, particularly in the context of proposed explanations that have non-trivial time evolution of dark energy at $z \gtrsim 1$ (e.g., \citealt{DiValentino2016,DiValentino2017,Keeley2019}).

As IM is able to survey very large volumes of the Universe with precise radial distances (unlike photometric redshift surveys) it is ideal for discovering small statistical effects where measuring a large number of modes is essential. One target is looking for features in the power spectrum which might tell us about inflationary physics \citep{Chluba2018}. Another is looking for signatures of new physics by searching for non-Gaussianity in \tcm data \citep{2017PhRvD..96f3525L}.

The expansion history information that IM can obtain from BAO, combined with broadband measurements of the power spectrum shape, plus future Cosmic Microwave Background measurements gives a potent combination for probing the contents of the Universe. In particular the number of relativistic degrees of freedom and the sum of the neutrino masses \citep{2016JCAP...02..008O,2018JCAP...05..004O} can be constrained to $\lesssim \SI{20}{\milli\eV}$ by combining IM with other probes.

On extremely large scales there are general relativistic corrections to the standard observables \citep{2011PhRvD..84d3516C}, that if observed could be a stringent confirmation of the current cosmological model. To measure these, one needs to map large volumes of the Universe to build up enough samples of these scales (for which IM is ideally suited), and to combine with other probes on similar scales (e.g., photometric surveys like the Large Synoptic Survey Telescope) in order to remove sample variance effects \citep{2015PhRvD..92f3525A}.

Although one only directly measures the inhomogeneity in the distribution of HI with \tcm IM, by looking at the distortions in the observed three-dimensional field caused by gravitational lensing, one can in principle infer the \emph{total matter} distribution within the volume \citep{Foreman2018}. As the lensing displacements are typically small, upcoming \tcm experiments may only be able to see this effect in cross correlation with photometric redshift surveys, but it will be extremely powerful for following generations of instruments.

Finally, although most of the science targets we have outlined above are cosmological, there is tremendous astrophysical interest in the nature of HI at low redshift. IM by itself gives us direct access to $\Omega_\sHI(z)$, the total amount of neutral hydrogen across redshift, and by using cross-correlations against optical tracers we can start to obtain information about which galaxies host the HI \citep{2017MNRAS.470.3220W}.

\section{Observational 21cm cosmology}
\label{sec:BasicObs}

%

The broad scientific possibilities outlined in Section \ref{sec:Science} have led to a new generation of instruments that have been custom designed for $21\,\textrm{cm}$ cosmology. These instruments vary in their detailed specifications, but generally:
\begin{itemize}
\item Have high sensitivity. The cosmological $21\,\textrm{cm}$ signal is faint across all redshifts, and thus high sensitivity is essential. This drives telescope designers towards instruments with large collecting area, coupled with observational plans that call for $1000\,\textrm{hrs}$ or more of total integration time.
\item Are broadband. A key feature of $21\,\textrm{cm}$ cosmology is the ability to map our Universe in three dimensions, using redshift information to distinguish between emission from different radial distances. Doing so requires broadband instruments whose receiving elements are efficient over a broad frequency range, as well as backend electronics that are capable of simultaneously processing data over large bandwidths. Note that broad frequency coverage is helpful not only for achieving the ultimate scientific goals of $21\,\textrm{cm}$ cosmology, but also for achieving better calibration and diagnosis of systematics.
\item Are sensitive to the right scales. Spatial fluctuations in the cosmological $21\,\textrm{cm}$ signal are expected to be scientifically interesting over specific ranges of length scales. For instance, for post-reionization observations the baryon acoustic oscillation scale is of particular interest, whereas for reionization observations the typical scale of ionized bubbles would be a more appropriate scale to target. This affects the physical extent of telescopes, although as we discuss later in this section, the reasoning is subtle given the ability to probe radial fluctuations by using spectral information.
\item Are stable in time. The faintness of the cosmological signal, coupled with the dominating influence of various contaminants (see Section \ref{sec:fgs}) means that $21\,\textrm{cm}$ cosmology is synonymous with high dynamic range observations. To ensure that strong contaminants do not overwhelm a faint signal, it is crucial that one be able to minimize systematic instrumental effects. A stable instrument helps to suppress the introduction of systematics in the first place.
\end{itemize}

\subsection{Single dish telescopes or interferometers?}
\label{sec:dish_or_inter}
\begin{figure}[t]
\centering
\includegraphics[width=0.45\textwidth]{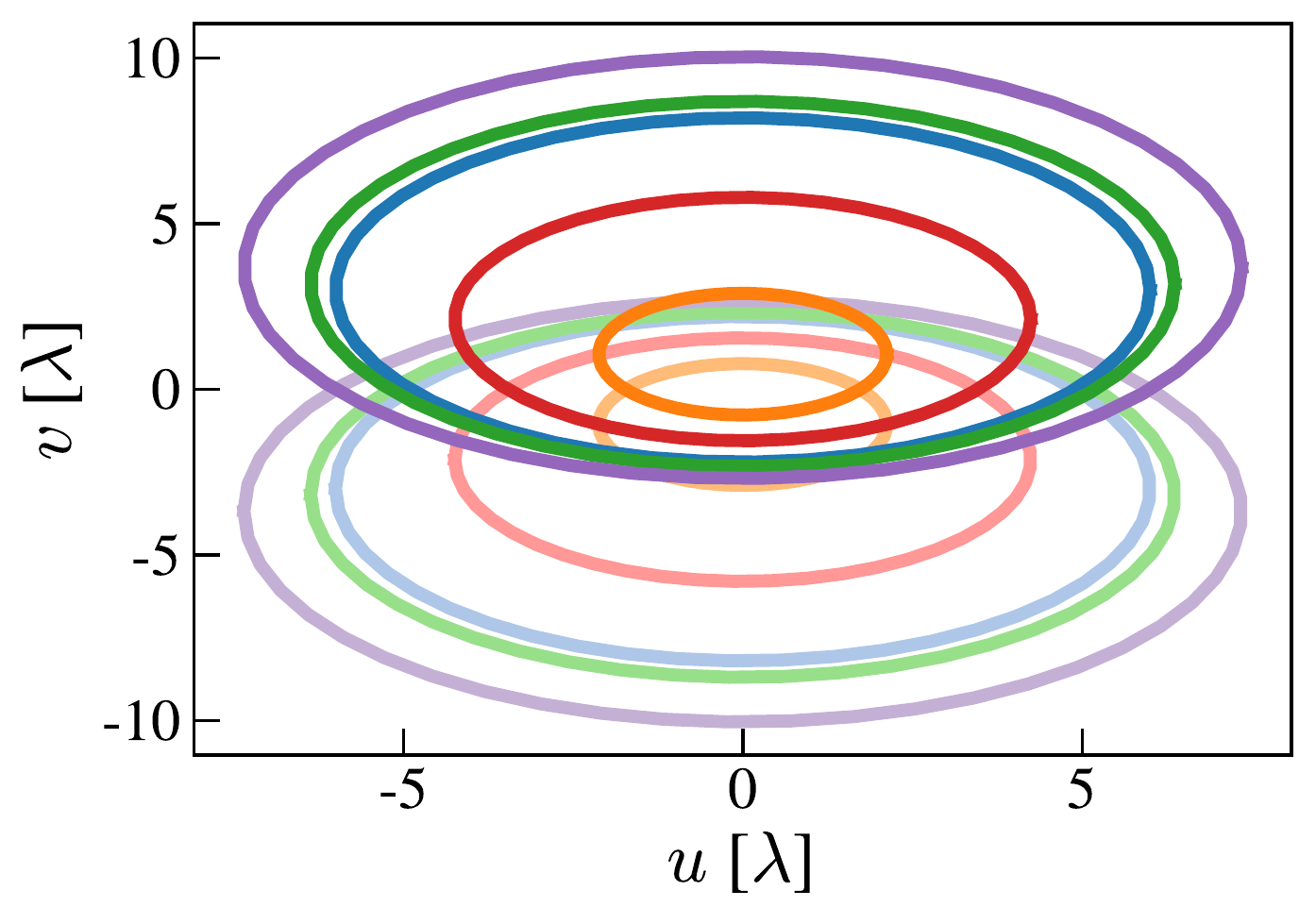}
\caption{An illustration of rotation synthesis, showing the movement of baselines through the $uv$ plane over a full 24-hour day of observations for a hypothetical interferometer comprised of $3 \times 2$ antennas arranged on a regular square grid. As the Earth rotates, the baselines sample different $uv$ modes, although of course in practice it is difficult to get $24$ hours of observations. Because a baseline $\vb$ can equally well be considered a baseline $-\vb$ (simply by reversing the order of one's antenna pairings), every sampled $(u,v)$ point is accompanied by its conjugate $(-u,-v)$ point. The conjugate points do not provide independent information, however, because the sky temperature must be a real-valued quantity. Note that we have expressed our $uv$ coordinates in units of $\lambda$, as is conventional.}
\label{fig:uv_tracks}
\end{figure}

With the aforementioned principles in mind, we examine the trade offs of various types of radio instruments that have been used (or proposed) for $21\,\textrm{cm}$ cosmology. One such trade off is the choice between a single dish telescope or an interferometer.\footnote{Global signal experiments typically use single elements that are not dishes. In this section we are primarily concerned with experiments attempting to map the spatial fluctuations of the $21\,\textrm{cm}$ signal, leaving global signal experiments to Section \ref{sec:GlobalSig}.} In its simplest form, the former consists of a monolithic single dish that focuses radio waves into a \emph{feed}, and the sky is mapped out one pixel at a time. The resolution $\theta_\textrm{dish}$ of the resulting map is then roughly determined by the characteristic size $D$ of the dish, where
\begin{equation}
\theta_\textrm{dish} = \frac{\lambda}{D},
\end{equation}
and $\lambda$ is the wavelength of observation. In contrast, an interferometer spreads out the total collecting area of a telescope into multiple receiving elements, such as multiple small dishes or dipole antennas.\footnote{For simplicity, we will refer to each receiving element of an interferometer as an antenna for the rest of this paper. Unless otherwise stated, our discussions will apply equally well to arrays of dishes.} Signals from \emph{pairs} of antennas are then multiplied together and averaged over a short period of time. This is the basic correlation operation, which is described in more detail in Section \ref{sec:interferometrypractice}. The correlated data (termed the \emph{visibility}) from each antenna pair of the interferometer roughly probes a Fourier mode that corresponds to the characteristic angular scale
\begin{equation}
\theta_\textrm{int} = \frac{\lambda}{b},
\end{equation}
where $b$ the distance between the two antennas that are being correlated. It is given the symbol $b$ because a pair of the antennas are known as a \emph{baseline}. More precisely, suppose one is considering a small enough patch of the sky that the flat-sky approximation holds. We may then define flat-sky coordinates $\boldsymbol \theta \equiv (\theta_x, \theta_y)$, and $\mathbf{u}$ is as its Fourier dual under a Fourier convention where
\begin{equation}
\label{eq:Itilde}
\widetilde{T} (\mathbf{u}) \equiv \int_{-\infty}^\infty \! d^2 \theta \, T(\boldsymbol \theta) e^{-i 2\pi \vu \cdot \boldsymbol \theta} ,
\end{equation}
and
\begin{equation}
\label{eq:Itilde_inverse}
T(\boldsymbol \theta) = \int_{-\infty}^\infty \! d^2 u \, \widetilde{T} (\mathbf{u})e^{i 2\pi \vu \cdot \boldsymbol \theta},
\end{equation}
with $T(\boldsymbol \theta)$ being the brightness temperature of the sky. A baseline defined by two antennas that are physically separated by the vector $\mathbf{b}$ can then be shown (as we do in Section \ref{sec:InterferometryBasics}) to probe $\widetilde{T}$ at
\begin{equation}
\label{eq:udef}
\mathbf{u} = \frac{\mathbf{b}}{\lambda}.
\end{equation}
To be even more quantitatively precise, under the flat-sky approximation the visibility $V$ measured by a baseline $\vb$ is given by\footnote{In this paper, we adopt an unusual convention. Typically, instead of the brightness temperature $T(\nu)$ inside the integral in Equation \ref{eq:BasicVis}, one has the intensity $I(\nu)$. We have implicitly assumed that one has already normalized the raw data using the Rayleigh-Jeans Law, i.e., $I(\nu) = 2 \nu^2 k_b T(\nu) / c^2$, where $k_b$ is Boltzmann's constant.}
\begin{equation}
\label{eq:BasicVis}
V(\vb, \nu) = \int \! d^2 \theta\, T(\boldsymbol \theta, \nu) A_p(\boldsymbol \theta, \nu) e^{-i 2\pi \vb \cdot \boldsymbol \theta / \lambda},
\end{equation}
where $A_p (\boldsymbol \theta) $ is known as the \emph{primary beam}\footnote{\label{foot:Ap}To be more precise, $A_p$ is the \emph{power} primary beam. In the literature it is not uncommon to see this denoted as $A$. However, in this paper we reserve $A$ for the electric field beam defined in Section \ref{sec:InterferometryBasics}. } and accounts for the fact that the antennas of an interferometer do not have equal sensitivity to all parts of the sky. With this extra factor (along with widefield, non-flat sky effects not shown here), one sees from comparing Equations \eqref{eq:Itilde} and \eqref{eq:BasicVis} that a baseline $\vb$ does not precisely probe the Fourier mode given by $\mathbf{u} = \vb / \lambda = \vb \nu / c$. In particular, the convolution theorem allows us to write
\begin{equation}
\label{eq:uv_convolution}
V(\vb, \nu) = \int \! d^2 u\, \widetilde{T} (\mathbf{u}, \nu) \widetilde{A}_p(\mathbf{b} \nu / c- \mathbf{u}, \nu),
\end{equation}
which shows that rather than precisely probing $\mathbf{u} = \vb \nu / c$, the visibility probes a smeared out footprint around that mode. However, if the primary beam is relatively broad, then this footprint is relatively localized, and our basic intuition remains accurate. Under this approximation, each baseline probes a particular Fourier mode, and considering all the baselines of an interferometer (i.e., all possible antenna pairings) then provides information about multiple Fourier modes of the sky. These Fourier modes can then be Fourier transformed back into the image domain to provide a map of the sky. The more unique baselines there are in an interferometer array, the closer our map will be to a true image of the sky, and the longer the baselines of an array, the higher the resolution of our map. Note that as the Earth rotates, baseline vectors rotate relative to the sky, and thus they rotate through the $uv$ plane (i.e., a two-dimensional plane with Cartesian axes given by the two components of $\mathbf{u}$). This is known as \emph{rotation synthesis}, and enables an interferometer to compile a more complete Fourier-space view of the sky, as illustrated in Figure \ref{fig:uv_tracks}. Of course, this view of the sky is still not perfect as there remain unsampled $uv$ modes. It is related to the true sky by a point spread function known as the \emph{synthesized beam}, which is given by the Fourier transform of the final $uv$ distribution (where a $1$ is placed in every pixelized cell on the $uv$ plane---a $uv$ \emph{cell}---that is sampled) that was used to make the map.

Although this picture of an interferometer's baselines tracing out tracks on the $uv$ plane is a useful one, it has some important limitations. To understand some of these limitations, let us distinguish between two types of interferometers: \emph{tracking} telescopes and \emph{drift-scan} telescopes. A tracking telescope is one where antennas are steered to track a particular field in the sky, such that the primary beam is also centred on a particular coordinate. In addition, each baseline is multiplied by a time-dependent phase factor so that the complex exponential term in the visibility has zero phase at the centre of the chosen field. (In practice, this phase centring is sometimes down in software rather than by the electronic hardware of the telescope, since it can be done after the fact). In contrast, a drift-scan telescope is one where the antennas are fixed and always point to the same location on the sky relative to the Earth (i.e., the same altitude and azimuth).\footnote{Of course, a tracking telescope can always be operated in a drift-scan mode, simply by not repointing the telescope. In the discussion that follows, we continue to use the term ``drift-scan telescope", but the reader should bear in mind that drift scanning should really be thought of as an observing mode available to all telescopes.}This point is often, but not always, the local zenith. Rather than concentrating all of one's integration time on a single field, the telescope maps out different parts of the sky as the Earth rotates, surveying a stripe on the celestial sphere.

The picture presented in Figure \ref{fig:uv_tracks} is most suitable for tracking telescopes. Since tracking telescopes focus observational time on a particular field, the coordinates needed to define the $(\theta_x, \theta_y)$ coordinate system (and by extension the $uv$ plane) can be fixed to this single field. For a drift-scan telescope, on the other hand, the field is constantly changing, and therefore there is not a single $uv$ plane that can be conveniently defined. In particular, once the Earth has rotated through an angle larger than roughly the width of the primary beam, one is essentially looking at a completely different patch of the sky. There are several ways to deal with this. One way is to define multiple $uv$ planes for different patches of the sky, with interferometers accumulating $uv$ samplings on one plane for short integration times before jumping to a different plane \citep{Cornwell1992,TegmarkZaldarriaga2010,Dillon2015Mapmaking}. Another solution is to express the sky in a three-dimensional Fourier basis, where interferometric baselines are sampling a three-dimensional $uvw$ space rather than a $uv$ plane. However, in such a formalism the interferometric visibility does not map as easily to a Fourier transform of the sky. A third solution is to use the $m$-mode formalism \citep{Shaw:2014}. A description of the $m$-mode formalism is provided in Section \ref{sec:mapmaking}. But in the current context of rotation synthesis, the essential idea is that because drift-scan telescopes map out a stripe on the sky, we may describe the entire stripe using cylindrical coordinates, effectively unrolling the stripe into a periodic Cartesian grid. A new $uv$ space can then be defined as the Fourier dual to this periodic grid. This is worked out explicitly in Appendix \ref{sec:DriftScans}, where one finds that for a drift-scan telescope, baselines do \emph{not} trace out $uv$ tracks. Despite this subtlety, the basic picture that we have presented---that at least instantaneously, an interferometric visibility probes a Fourier mode of the sky---remains a useful bit of intuition.

Having established some basic intuition for interferometry, we now return to the question of whether one should build a single dish telescope or an interferometer. The answer depends on one's application. For example, if one requires high angular resolution, it is typically advantageous to build an interferometer. Because large dishes are difficult to support mechanically, a single dish telescope cannot be made arbitrarily large. There are thus practical limitations to the resolution of single dishes. With interferometers this is much less of an issue, as antenna elements can be placed extremely far apart.

On the other hand, many interferometers have their antennas laid out in a way that results in holes in their sampling of the $uv$ plane. This results in complicated synthesized beams. A single dish does not suffer from the problem of gaps in the $uv$ plane. While the limited angular resolution of a single dish means that maps created from single dish observations do not contain information beyond some ``radius" from the origin of the $uv$ plane, the modes interior to this cutoff correspond to angular scales coarser than the angular resolution of the telescope, which can be sampled with an appropriate observational strategy that points the telescope at various parts of the sky. With a fully sampled $uv$ plane (again, up to some cutoff), the point spread function can thus be in principle inverted in one's derived maps. This can make it easier to deal with some of the systematics that we detail in Section \ref{sec:wedge} associated with contaminant emission in our data (such as the \emph{foreground} emission from our own Galaxy and other astronomical sources; see Section \ref{sec:fgs}).

In contrast to foreground systematics, noise systematics can be more difficult to deal with in single dish experiments. With a single dish observation, the instrument effectively squares electric field measurements to obtain measurements of power. Doing so results in a noise bias, since mean-zero noise in the pre-squaring data acquires a positive mean after squaring. This noise bias must be exquisitely modelled and removed. In contrast, with an interferometer one is correlating (i.e., multiplying) data from two different antennas to form a baseline, and if the noise in the two antennas are independent (which is frequently a good assumption), then the noise contribution in the result will still be mean zero. The lone exception to this is the baseline $\mathbf{b}=0$, corresponding to the correlation of an antenna with itself. For this zero-length baseline, the correlation is a squaring of the data, and one is essentially treating the antenna as a single dish telescope. The noise bias is therefore present if one chooses to use zero-length baselines. For this reason, interferometric observations often throw out the $\mathbf{b}=0$ correlation information, and thus interferometric images have zero mean, since the $\mathbf{u}=0$ mode is missing.

In the past, both types of telescopes have been used for $21\,\textrm{cm}$ cosmology, and a detailed
quantitative comparison between the two can be found in \citet{Tegmark2009}. Notable single-dish efforts in the past have included observations using the Parkes telescope and the Green Bank Telescope (see Section \ref{sec:ObsStatus}). These have resulted in detections of post-reionization \tcm fluctuations in cross-correlation with optical galaxy surveys \citep{Chang:2010,Masui:2013,Anderson:2018}. However, recent efforts have tended to favour interferometry, given the flexibility to configure an array to have the right amounts of sensitivity on precisely the right scales. Exactly what these requirements are, however, depends on the type of observations that one is attempting.

\subsection{From imaging to statistical measurements}
\label{sec:ImageVsStat}
Broadly speaking, there are two types of observations that one can target when mapping diffuse structures. The first is where the end goal is an image. Thinking about imaging in Fourier space, the process requires a telescope to have sufficient sensitivity that \emph{each} Fourier mode is measured with signal-to-noise ratio of order unity or greater. In essence, the requirement for imaging is that the amplitude and phase of each mode is reconstructed with high fidelity, enabling Fourier modes to be combined in a way that yields a position space image that resembles reality. Images are perhaps the ultimate data product of a $21\,\textrm{cm}$ cosmology experiment, since they capture the full details of the cosmological signal, including for example any non-Gaussian signatures that are present.

Without the sensitivity for imaging, one must resort to statistical measurements. Here, one takes advantage of the symmetries of the problem to combine individual modes together, thereby increasing signal to noise. For example, one might use the fact that our Universe is rotationally invariant (i.e., statistically isotropic). This means that different Fourier modes with wave vectors of the same magnitude but different orientation should---up to statistical fluctuations---carry the same information and be combinable in some way.

A prime example of a statistical measurement would be that of the \emph{power spectrum}. Measuring the power spectrum is the chief focus of most current-generation instruments, and thus much of this paper is devoted to discussing the power spectrum. Theoretical models of the spatial fluctuations of the $21\,\textrm{cm}$ signal typically make predictions for the power spectrum, providing a well-defined statistic for testing models. To define the power spectrum, we first define a three-dimensional Fourier transform of the sky:
\begin{equation}
\label{eq:cosmo_forward}
\widetilde{T} (\mathbf{k}) \equiv \int_{-\infty}^\infty d^3 r e^{-i \mathbf{k} \cdot \mathbf{r}} T(\vr),
\end{equation}
where $\vr$ is a comoving position vector, and $\vk$ a comoving wavevector.\footnote{At this point, we have (confusingly) introduced two different Fourier conventions. Unfortunately, both are standard: the convention with factors of $2\pi$ in the complex exponentials (Equations \ref{eq:Itilde} and \ref{eq:Itilde_inverse}) is traditionally used in the radio astronomy literature, whereas the convention with no factors of $2\pi$ in the complex exponentials (Equations \ref{eq:cosmo_forward} and \ref{eq:cosmo_inverse}) is used in the cosmology literature. As a general rule of thumb, if a quantity is defined in coordinates most suited to describing observations (such as $\theta$ or $\vu$), it is likely that the radio astronomy convention is being employed; if a quantity is defined in terms of comoving cosmological coordinates, the cosmological convention is almost certainly being used.} The inverse transform is
\begin{equation}
\label{eq:cosmo_inverse}
 T(\vr) \equiv \int_{-\infty}^\infty \frac{d^3 k}{(2\pi)^3} e^{i \mathbf{k} \cdot \mathbf{r}}\widetilde{T} (\mathbf{k}).
\end{equation}
The power spectrum $P(\mathbf{k})$ is then defined via the relation
\begin{equation}
\label{eq:PspecDef}
\langle \widetilde{T} (\mathbf{k}) \widetilde{T} (\mathbf{k}^\prime)^* \rangle = (2\pi)^3 \delta^D (\mathbf{k} - \mathbf{k}^\prime) P(\mathbf{k}),
\end{equation}
with $ \delta^D$ being the Dirac delta function, and $\langle \dots \rangle$ signifying an ensemble average, a hypothetical process where one imagines drawing different realizations of our Universe from the underlying statistical distributions that govern it.

While Equation \eqref{eq:PspecDef} is a formal, rigorous definition of the power spectrum, it is not particularly useful for understanding how one computes a power spectrum in practice. This is because it assumes that one is able to take a Fourier transform of $T(\mathbf{r})$ that is infinite in extent. If we imagine a finite (but otherwise still ideal and noiseless) survey, the Fourier modes that we measure are not the true $\widetilde{T} (\mathbf{k})$ modes, but instead are given by
\begin{eqnarray}
\label{eq:Ttildeobs}
\widetilde{T}^\textrm{obs} (\mathbf{k}) &=& \int_{-\infty}^\infty d^3 r e^{-i \mathbf{k} \cdot \mathbf{r}} T(\vr) \phi(\vr) \nonumber \\
&=& \int_{-\infty}^\infty \frac{d^3 k}{(2\pi)^3} \widetilde{T} (\mathbf{k}^\prime) \widetilde{\phi} (\mathbf{k}-\mathbf{k}^\prime ),
\end{eqnarray}
where in the last equality we made use of the convolution theorem, and $\phi(\vr)$ is a function that equals one within the survey region, and zero outside. If we now compute the absolute magnitude squared and take the ensemble average of the result, we obtain
\begin{eqnarray}
\langle |\widetilde{T}^\textrm{obs} (\mathbf{k})|^2 \rangle &=& \int_{-\infty}^\infty \frac{d^3 k_1}{(2\pi)^3} \frac{d^3 k_2}{(2\pi)^3} \langle \widetilde{T} (\mathbf{k}_1) \widetilde{T} (\mathbf{k}_2)^* \rangle \nonumber \\
&& \qquad \times \widetilde{\phi} (\mathbf{k}-\mathbf{k}_1 ) \widetilde{\phi}^* (\mathbf{k}-\mathbf{k}_2 ) \nonumber \\
&=& \int_{-\infty}^\infty \frac{d^3 k_1}{(2\pi)^3} P(\mathbf{k}_1) | \widetilde{\phi} (\mathbf{k} - \mathbf{k}_1 ) |^2
\end{eqnarray}
If the survey volume is large, then $\widetilde{\phi}$ is a relatively narrowly peaked function in comparison to a reasonably smoothly varying $P(\mathbf{k})$. The latter can then be factored out of the integral, and invoking Parseval's theorem then gives
\begin{equation}
\label{eq:PkContinuous}
P(\mathbf{k}) \approx \frac{\langle |\widetilde{T}^\textrm{obs} (\mathbf{k})|^2 \rangle }{V},
\end{equation}
where
\begin{equation}
\label{eq:SurveyVolume}
V = \int d^3 r \phi^2(\vr) = \int d^3 r \phi (\vr)
\end{equation}
is the volume of the survey. Now, with real data one cannot perform an ensemble average. However, if one is willing to make the assumption of isotropy, then the statistical properties of our Fourier modes (including the mean of their squared magnitudes, i.e., their variances) should only depend on the wavenumber $k \equiv | \mathbf{k}|$, and not on the orientation of $\mathbf{k}$.\footnote{This is not true when one includes the effect of \emph{redshift space distortions}, where peculiar velocities are confused for the Hubble flow and lead to an incorrect mapping of observed redshifts to radial distances. Since this sort of mis-mapping only happens radially, it causes a measured power spectrum to depend on the direction of $\vk$, even if the true underlying power spectrum is isotropic and only depends on $k$. See Section \ref{sec:RSD} for an example of an expression for $P(\vk)$ that includes redshift space distortions.} This allows one to replace the ensemble average with an average over direction. If we imagine dividing Fourier space into discrete voxels at discrete spacings of $\mathbf{k}$, then we can write this average as
\begin{equation}
\label{eq:BinningPk}
P(k) \approx \frac{\sum_{\mathbf{k} \in k} |\widetilde{T}^\textrm{obs} (\mathbf{k})|^2}{N_k V},
\end{equation}
where $N_k$ is the number of $\mathbf{k}$ voxels that fall have magnitude between $k-\Delta k / 2$ and $k + \Delta k/2$, where $\Delta k$ is an analyst-defined bin width that is narrow enough to capture any wiggles in the power spectrum. This formula provides a more intuitive view of what a power spectrum measures: it is a measurement of the variance of fluctuations in a field (in our case, $T$) as a function of lengthscale, or more precisely, $k$. Another way to think of the power spectrum is to show (using Equation \ref{eq:PspecDef}) that it is the Fourier transform of the correlation function\footnote{Confusingly, in the nomenclature of statistics, the correlation function as we have defined it here would be more properly termed a covariance function. However, because such terminology is standard in the cosmology literature, we will continue to use it.} $\xi (\vx) \equiv \langle T(\vr) T(\vr-\vx) \rangle$, i.e.,
\begin{equation}
\label{eq:CorrFct}
\xi (\vx) = \int \frac{d^3 k}{(2\pi)^3} P(\vk) e^{-i \vk \cdot \vx},
\end{equation}
thus illustrating that the power spectrum is a measure of position space correlations in our data, but expressed in Fourier space.

\begin{figure}[t]
\centering
\includegraphics[width=0.45\textwidth]{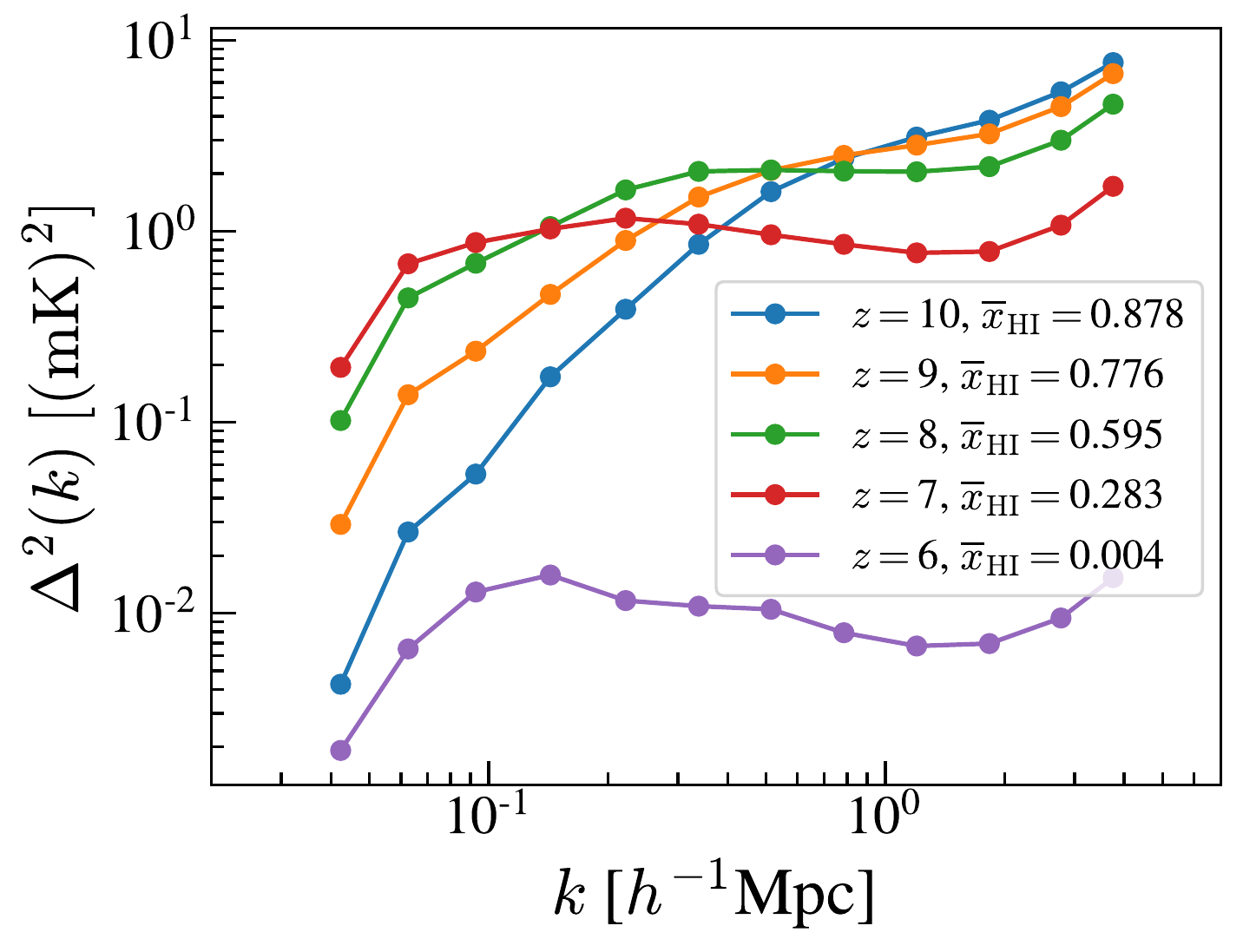}
\caption{Example \tcm power spectra from the reionization epoch, when the global neutral fraction $\overline{x}_\textrm{HI}$ decreases from $\overline{x}_\textrm{HI} \sim 1$ to $\overline{x}_\textrm{HI} \sim 0$. These were generated using \texttt{21cmFAST} \citep{Mesinger2011}, and for the particular set of theoretical parameters chosen, most of reionization happens over the redshift range $6 < z < 10$. Although power spectra do not contain the full information content of an image when non-Gaussianities are present, they do capture many important features of the underlying physics. We note that these curves are intended to be for reference only, as there are considerable modelling uncertainties both within models (due to uncertain values of free parameters) and between models.}
\label{fig:sample_ps}
\end{figure}

In Figure \ref{fig:sample_ps}, we show some example \tcm power spectra from $6 < z < 10$. Instead of plotting $P(k)$, we have opted to instead plot
\begin{equation}
\label{eq:DeltaSqDefinition}
\Delta^2 (k ) \equiv \frac{k^3}{2 \pi^2} P(k).
\end{equation}
This quantity is frequently seen in the literature, and can be motivated by the following argument. Consider the variance of a zero-mean random temperature field at some location. We can write this as
\begin{eqnarray}
\langle T^2 (\vr) \rangle &=&\bigg{\langle}  \left(\int_{-\infty}^\infty \frac{d^3 k}{(2\pi)^3} e^{i \mathbf{k} \cdot \mathbf{r}}\widetilde{T} (\mathbf{k})\right) \nonumber \\
&& \quad \times \left(\int_{-\infty}^\infty \frac{d^3 q}{(2\pi)^3} e^{i \mathbf{q} \cdot \mathbf{r}}\widetilde{T} (\mathbf{q})\right)^* \bigg{\rangle} \nonumber \\
&=& \int_{-\infty}^\infty \frac{d^3 k}{(2\pi)^3} \frac{d^3 q}{(2\pi)^3} e^{i (\mathbf{k}-\mathbf{q}) \cdot \mathbf{r}} \langle \widetilde{T} (\mathbf{k}) \widetilde{T} (\mathbf{q})^* \rangle \nonumber \\
&=& \int_{-\infty}^\infty \frac{d^3 k}{(2\pi)^3} P(k) = \int_0^\infty d \! \ln k \, \Delta^2 (k ),
\end{eqnarray}
where in the penultimate equality we used Equation \eqref{eq:PspecDef} and assumed isotropy. From this, we see that the quantity $\Delta^2 (k )$ may be thought of as the contribution to the position space variance per logarithmic bin in $k$. It is often termed the \emph{dimensionless power spectrum}, but in the case of \tcm cosmology this may be somewhat of a misnomer, since it has dimensions of temperature squared.

While power spectra may not contain the richness of an image, they do capture much of the relevant physics. For example, one sees from Figure \ref{fig:sample_ps} that as reionization proceeds, power moves from small scales (high $k$) to large scales (small $k$) as the characteristic size of bubbles increases,\footnote{This may be a reasonable story for explaining the trends seen in the evolution of the power spectrum, but it is not one that is rigorous. See \citet{Furlanetto2016percolation} for a considerably more well-defined and subtle view of how ionized bubbles grow during reionization.} until the overall global neutral fraction has decreased sufficiently for the power to drop to zero on all scales. Power spectrum measurements can be used to constrain the underlying parameters of a model, and in Section \ref{sec:MCMC} we outline how (in general) one can go about performing this data analysis step. Power spectra can also be used to select between models, and we discuss this in Section \ref{sec:modelselection}.

In addition to the basic power spectrum that we have defined above, there exist variations that can be useful in specific data analysis contexts. If one is dealing not with a three-dimensional volume of data but instead with observations on the surface of a sphere, it is more appropriate to construct an angular power spectrum. Here, one decomposes maps into spherical harmonic modes rather than Fourier modes, so that\footnote{For simplicity, here we assume that one has surveyed the entire sky. A full discussion of the more general case for incomplete sky coverage is a little more involved than it was for rectilinear power spectrum, and so we omit it here. Interested readers may wish to consult resources such as \citet{Alonso2019PseudoCl}.}
\begin{equation}
\label{eq:SphericalHarmonicExpansion}
T(\hat{\vr}) = \sum_{\ell=0}^\infty \sum_{m=-\ell}^{+\ell} Y_{\ell m} (\hat{\vr}) a_{\ell m},
\end{equation}
where $Y_{\ell m}$ denotes a standard spherical harmonic basis function and $a_{\ell m}$ its expansion coefficient. In analogy to Equation \eqref{eq:PspecDef}, we may define the angular power spectrum $C_\ell$ as
\begin{equation}
\label{eq:AngPspec}
\langle a_{\ell m} a^*_{\ell^\prime m^\prime} \rangle = \delta_{\ell \ell^\prime} \delta_{m m^\prime} C_\ell,
\end{equation}
and the analog to Equation \eqref{eq:BinningPk} is
\begin{equation}
\label{eq:CellGuess}
C_\ell = \frac{1}{2\ell + 1} \sum_{m=-\ell}^\ell |a_{\ell m}|^2.
\end{equation}
The angular power spectrum has been the workhorse statistic for decades in the CMB community, since the CMB resides on a two-dimensional surface on the sky. It has been less commonly used in $21\,\textrm{cm}$ cosmology because of the three-dimensional nature of $21\,\textrm{cm}$ surveys. One option would be to create a separate angular power spectrum for each frequency. However, since each frequency is treated in a way that is independent of all others, this method is unable to capture valuable information on correlations in the $21\,\textrm{cm}$ data along the line of sight. An alternative is to compute all possible angular cross-power spectra between frequency channels, i.e.,
\begin{equation}
\label{eq:Cellnunuprime}
C_\ell (\nu, \nu^\prime) \equiv \frac{1}{2\ell + 1} \sum_{m = -\ell}^{+\ell} a_{\ell m} (\nu) a_{\ell m}^* (\nu^\prime),
\end{equation}
where $a_{\ell m} (\nu)$ is the spherical harmonic expansion coefficient for mode $Y_{\ell m}$ of the map at frequency $\nu$ \citep{Santos2005,Datta2007,Bharadwaj2019}. This is essentially a hybrid statistic, being a harmonic space quantity in the angular directions but a correlation function in the frequency/line-of-sight direction. An alternative to $C_\ell (\nu, \nu^\prime) $ that retains its spherical geometry in the angular direction but is expressed in harmonic space in all three directions is the spherical Fourier-Bessel power spectrum \citep{Liu:2016}. To compute it, one decomposes the sky into spherical Bessel functions in the radial direction and spherical harmonics in the angular direction to obtain
\begin{equation}
\label{eq:Tellemdef}
\overline { T } _ { \ell m } ( k ) \equiv \sqrt { \frac { 2 } { \pi } } \int _ { 0 } ^ { \infty } d^3 r \,r ^ { 2 } j _ { \ell } ( k r ) Y_{\ell m}^* (\hat{\vr}) T(\vr) ,
\end{equation}
where $j_\ell$ is the $\ell$th order spherical Bessel function of the first kind. The spherical Fourier-Bessel power spectrum $S_\ell (k)$ is then computed as
\begin{equation}
\label{eq:Sellk}
S_\ell (k) \propto \frac{1}{2\ell + 1} \sum_{m=-\ell}^\ell | \overline{T}_{\ell m} (k) |^2,
\end{equation}
where for clarity we have omitted constants of proportionality. These constants depend on one's survey volume, just like with the analogous expression for the power spectrum in Equation \eqref{eq:BinningPk}. If the constants are included, then one can show that averaging $S_\ell (k)$ over all $\ell$ yields $P(k)$.


Although $S_\ell (k)$ ultimately just gives $P(k)$ when averaged over $\ell$, it is extremely useful as an intermediate data product because it explicitly indexes data by $\ell$. In other words, $S_\ell (k)$ explicitly picks out angular fluctuations, allowing one to focus on only those fluctuations that occur on a particular angular scale. This is extremely useful because $21\,\textrm{cm}$ experiments survey the sky in starkly different ways in the angular and radial directions---fluctuations in the angular direction are probed by different pointings of a single-dish telescope or by an interferometer's sampling of the $uv$ plane, whereas fluctuations in the radial direction are probed by analyzing data at different frequencies (i.e., by looking at the redshift of the \tcm line). Systematics in one's data will therefore often have telltale signatures in a harmonic space that separates out angular from radial fluctuations. This is partially accomplished by $S_\ell (k)$, which separates out the angular fluctuations with $\ell$. But $S_\ell (k)$ retains a dependence on $k$, which describes fluctuations by wavenumber, regardless of orientation. A cleaner separation can be obtained in the flat-sky limit, where the analogous quantity to $S_\ell (k)$ is the cylindrical power spectrum $P(k_\perp, k_\parallel)$. To define $P(k_\perp, k_\parallel)$, we note in the narrow-field, flat-sky limit, the radial direction essentially points along a particular Cartesian direction. This allows us to designate coordinates along this direction as $r_\parallel$, and those along the two directions perpendicular to the line of sight as $\mathbf{r}_\perp$. The corresponding Fourier dual coordinates $\vk_\perp$ and $k_\parallel$ are then defined implicitly via
\begin{equation}
\label{eq:cylind_fourier_def}
\widetilde{T} (\mathbf{k}_\perp, k_\parallel) \equiv \int_{-\infty}^\infty d^2 r_\perp dr_\parallel e^{-i (\mathbf{k}_\perp \cdot \mathbf{r}_\perp + k_\parallel r_\parallel)} T(\vr_\perp, r_\parallel),
\end{equation}
and the cylindrical power spectrum is then
\begin{equation}
\label{eq:BinningPkperpkpara}
P(k_\perp, k_\parallel) \approx \frac{1}{N_{k_\perp, k_\parallel} V} \sum_{\mathbf{k}_\perp \in k_\perp} \sum_{\mathbf{k}_\parallel \in k_\parallel} |\overline{T}^\textrm{obs} (\mathbf{k}_\perp, k_\parallel)|^2.
\end{equation}
Just as Equation \eqref{eq:BinningPk} represents an averaging of power spectrum estimates over spherical shells in Fourier space to yield $P(k)$, Equation \eqref{eq:BinningPkperpkpara} tells us to form $P(k_\perp, k_\parallel)$ by averaging over pairs of rings of radius $k_\perp$ located at $\pm k_\parallel$, with each ring containing $N_{k_\perp, k_\parallel}$ independent $\vk$ modes.

\subsection{Designing an interferometer: Achieving sensitivity to the right modes}
\label{sec:InterferometerSensitivityOptimization}
The cylindrical power spectrum provides a way to separate fluctuations in the maps into fluctuations along the line of sight and fluctuations perpendicular to the line of sight. This can be particularly illuminating for considering the optimal design of an interferometer. Our strategy for this is to compare our interferometer's measurement equation, Equation \eqref{eq:uv_convolution}, to the Fourier modes that we wish to measure. If we take the Fourier transform of Equation \eqref{eq:uv_convolution} along the line of sight, we end up with
\begin{equation}
\label{eq:overline_V}
\widetilde{V} (\vb, \eta) = \int d^2 u  \widetilde{T} (\vu, \eta) \widetilde{A}_p(\mathbf{b} \nu_0 / c- \mathbf{u}, \nu_0) \approx \widetilde{T} ( \mathbf{b} \nu_0 / c, \eta),
\end{equation}
where $\eta$ is the Fourier dual to $\nu$ (although see Section \ref{sec:delayspec} for a discussion of the subtleties regarding this definition). Here, we have made the same approximation as in Section \ref{sec:dish_or_inter}, where we assumed that $A_p$ was reasonably broad, and have defined
\begin{equation}
\widetilde{T} (\vu, \eta) \equiv \int \! d^2 \theta d\nu \, e^{-i 2 \pi (\eta \nu + \vu \cdot \boldsymbol \theta)}  T (\boldsymbol \theta, \nu).
\end{equation}
In going from Equation \eqref{eq:uv_convolution} to Equation \eqref{eq:overline_V} we have neglected the frequency dependence of $\widetilde{A}_p$, evaluating the function at some central frequency $\nu_0$. The intrinsic frequency dependence of the primary beam (i.e., the second argument of $\widetilde{A}_p$) can be minimized in hardware, and this is often a chief consideration in designing one's antenna elements. The dependence arising from the $\vb \nu / c$ term in the first argument (due to the fact that a baseline of fixed physical length samples different angular scales at different frequencies) may be a reasonable approximation for very short baselines, and we will take advantage of this in Section \ref{sec:delayspec}. However, the relaxation of this assumption has profound consequences for avoiding the foreground contaminants that we will describe in Section \ref{sec:fgs}. We will explore this in detail in Section \ref{sec:wedge}.

\begin{figure}[t]
\centering
\includegraphics[width=0.45\textwidth]{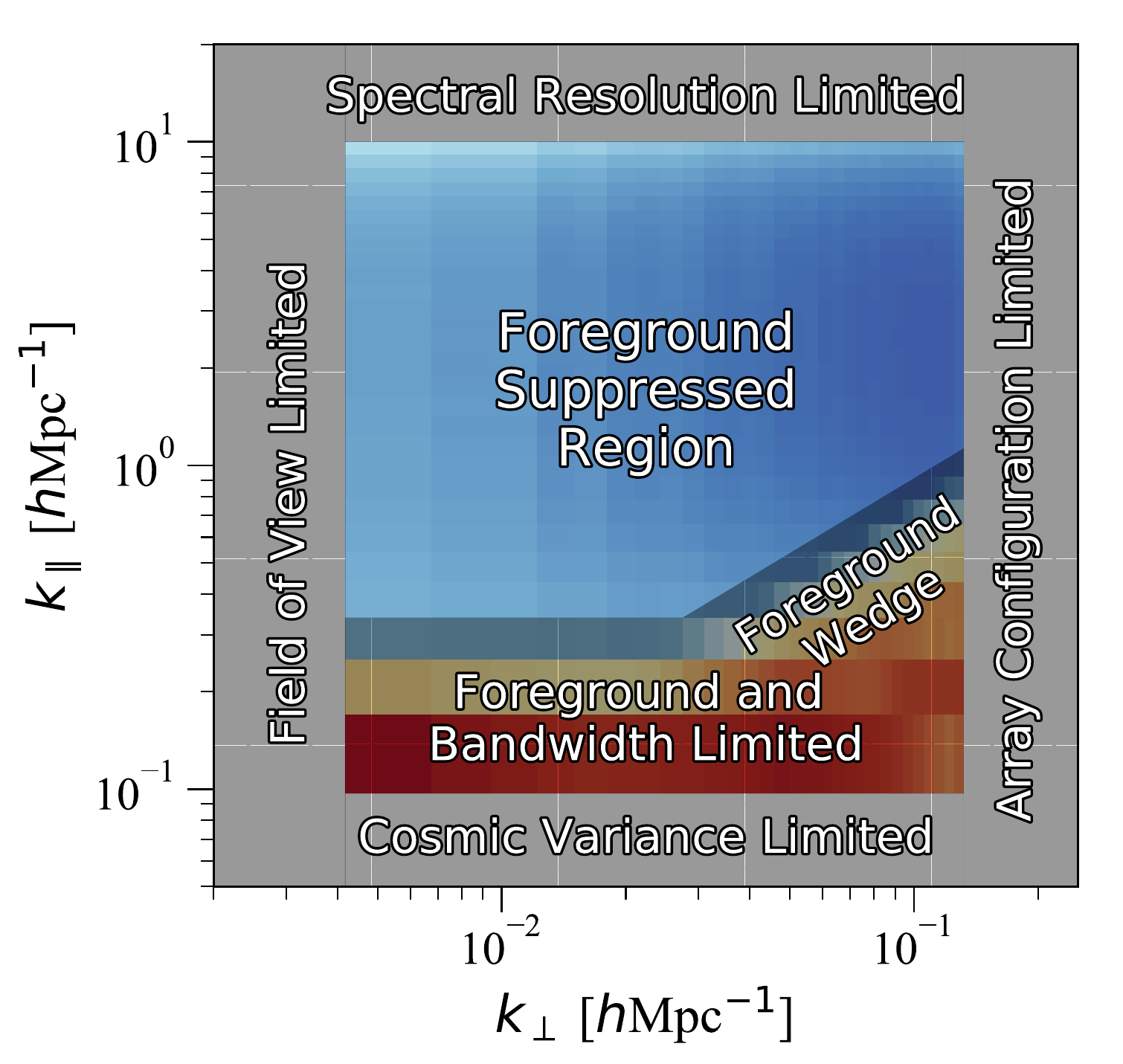}
\caption{A qualitative picture of the region of Fourier space that are accessible to radio interferometers. The largest scale (lowest $k_\perp$ wavenumber) angular Fourier modes are limited by the extent of the survey region. The smallest scales are limited by one's array configuration, since higher $k_\perp$ are probed by longer baselines, and the number of baselines eventually thins out as one goes to longer baselines, even for arrays with a large number of elements. In the line of sight direction, high $k_\parallel$ modes are limited by one's spectral resolution. The low $k_\parallel$ (large scale) modes are dominated by cosmic variance, the radial extent of one's survey (i.e., one's instrumental bandwidth), and foregrounds (see Sections \ref{sec:fgs} and \ref{sec:FgMitigation}). In addition, the inherent chromaticity of interferometers results in a further leakage of foregrounds to higher $k_\parallel$, leading to the \emph{foreground wedge} (see Section \ref{sec:wedge}). The remaining region in Fourier space is sometimes termed the \emph{Epoch of Reionization} window, and is a promising region in which to pursue a first detection of the \tcm power spectrum.}
\label{fig:wedgecartoon}
\end{figure}

For now, however, we can combine our last two equations to give
\begin{equation}
\label{eq:V_overline}
\widetilde{V}( \vb, \eta) \propto \int \! d^2 \theta d\nu \, e^{-i 2 \pi (\eta \nu + \vb \cdot \boldsymbol \theta \nu_0 / c)}  T (\boldsymbol \theta, \nu).
\end{equation}
This is very similar to Equation \eqref{eq:cylind_fourier_def} except that there the relation is written in terms of cosmological coordinates such as $\vr_\perp$ and $r_\parallel$, whereas here our expression is in terms of observational coordinates $\boldsymbol \theta$ and $\nu$. To relate these two sets of coordinates, we begin with the line of sight (radial) comoving distance $D_c$ to redshift $z$ \citep{Hogg1999}
\begin{equation}
\label{eq:LOScomovingdistance}
D _ { c } \equiv \frac { c } { H _ { 0 } } \int _ { 0 } ^ { z } \frac { d z ^ { \prime } } { E \left( z ^ { \prime } \right) } ; \quad E ( z ) \equiv \sqrt { \Omega _ { m } ( 1 + z ) ^ { 3 } + \Omega _ { \Lambda } }
\end{equation}
where $c$ is the speed of light, $H_0$ is the Hubble parameter, $\Omega_m$ is the normalized matter density, $\Omega_\Lambda$ is the normalized dark energy density, and $z$ is the redshift of observation, which is related to the observational frequency $\nu$ via Equation \eqref{eq:RedshiftDef}. Under the small-angle approximation, $D_c$ relates $\boldsymbol \theta$ to $\mathbf{r}_\perp$ via
\begin{equation}
\label{eq:perp_mapping}
\mathbf{r}_\perp = D_c \boldsymbol \theta.
\end{equation}
For relating $\nu$ to $r_\parallel$, one might be tempted to simply use the expression for $D_c$, since $D_c$ gives a radial distance as a function of frequency. However, when computing power spectra from data, one typically splits up the full bandwidth of one's instrument into relatively small bandwidth chunks. This is done in order to avoid the effects of cosmological evolution. Thus, the relevant relation is not the one between frequency and \emph{total} radial distance $D_c$, but the \emph{local} relation (defined relative to the middle of a sub-band) between changes in frequency $\Delta \nu$ and changes in distance $\Delta r_\parallel$. This is given by
\begin{equation}
\label{eq:DistanceDiff}
\Delta r _ { \| } = \frac{\partial D_c}{\partial \nu} \Delta \nu = -\frac { c } { H _ { 0 } \nu _ { 21 } } \frac { ( 1 + z ) ^ { 2 } } { E ( z ) } \Delta \nu.
\end{equation}
If we now assume that both the frequency and the radial distance are measured relative to values appropriate to the approximate centre of one's observational band, we may replace $\Delta r_\parallel$ and $\Delta \nu$ with $r_\parallel$ and $ \nu$, respectively, in a bit of notational convenience. Additionally, we are free to flip the direction of one of our Cartesian axes to rid ourselves of the minus sign in Equation \eqref{eq:DistanceDiff}. Inserting Equation \eqref{eq:perp_mapping} and Equation \eqref{eq:DistanceDiff}, sans minus sign, into Equation \eqref{eq:V_overline} then gives
\begin{eqnarray}
\widetilde{V}( \vb, \eta) \propto &&\int \! d^3r \, T (\boldsymbol \theta, \nu) \nonumber \\
&&\times \exp\left[ -i 2\pi \left( \frac {  H _ { 0 } \nu _ { 21 }E ( z ) } {c ( 1 + z ) ^ { 2 } } r_\parallel + \frac{\vb \cdot \vr_\perp \nu_0 }{ c D_c} \right) \right]  ,\qquad
\end{eqnarray}
and comparing this to Equation \eqref{eq:cylind_fourier_def} reveals that
\begin{equation}
\label{eq:kperpkparamappings}
\mathbf { k } _ { \perp } = \frac { 2 \pi \nu_0 \mathbf { b } } { c D _ { c } } ; \quad  k _ { \parallel } = \frac { 2 \pi \nu _ { 21 } H _ { 0 } E ( z ) } { c ( 1 + z ) ^ { 2 } } \eta.
\end{equation}
We therefore see that an interferometer's baseline distribution determines which angular wavenumbers $\vk_\perp$ it is able to access. In particular, the finest angular scales (largest $k_\perp \equiv | \vk_\perp|$) are limited by the longest baselines. The largest angular scales (smallest $k_\perp$ values) are limited by either the shortest baselines, or if the interferometer is extremely compact (e.g., with nearly touching elements), the survey area. In the line of sight direction, the finest modes (which reside at high $k_\parallel$) are limited by the spectral resolution of one's instrument. The coarsest modes (at low $k_\parallel$) are limited by the bandwidth over which one is able to collect data. These limits are illustrated in Figure \ref{fig:wedgecartoon}, along with a ``wedge" signature from foreground contaminants that is discussed in detail in Section \ref{sec:wedge}. Of course, \emph{within} the accessible region (often denoted the \emph{Epoch of Reionization window}\footnote{The origin of this name is historic, because the geometry of Figure \ref{fig:wedgecartoon} was elucidated in the reionization literature. However, the qualitative structure of the plot applies at all redshifts, even if the precise locations of the various boundaries are different in detail.}) the observations still contain noise that can only be reduced by averaging over independent measurements, for instance by averaging over an extended period of time. We discuss noise in more detail in Section \ref{sec:interferometrypractice}.

To recap, the $k_\parallel$ modes are accessed via an interferometer's spectral information while the pattern of $\vk_\perp$ coverage is determined by the baseline distribution. This leads to some instrument designs for $21\,\textrm{cm}$ cosmology that may seem counterintuitive at first. Consider two qualitatively different scenarios. One where the interferometer is designed to be as compact as possible, with many short baselines, and another where the interferometer has a more spread out configuration. If one were conducting a two-dimensional mapping of the sky at a particular frequency, the latter configuration would have its sensitivity spread out over a greater variety of $\vk_\perp$ modes (or equivalently, $uv$ modes). This also enables one to reach modes with higher $k$, i.e., finer spatial scales, since a more spread out distribution of $\vk_\perp$ modes will inevitably push sensitivity to higher $k_\perp$ values. However, since we are primarily interested in a three-dimensional mapping, there is another way to access small-scale information. Even with a single short baseline that probes low values of $k_\perp$ (large angular scales), one can access small-scale information along the line-of-sight direction. Said differently, if one assumes statistical isotropy, then the power spectrum depends only on $k \equiv (k_\perp^2 + k_\parallel^2)^{1/2}$, and one can access a wide variety of length scales despite only sampling a very narrow range of $k_\perp$.

\begin{figure}[t]
\centering
\includegraphics[width=0.45\textwidth,clip]{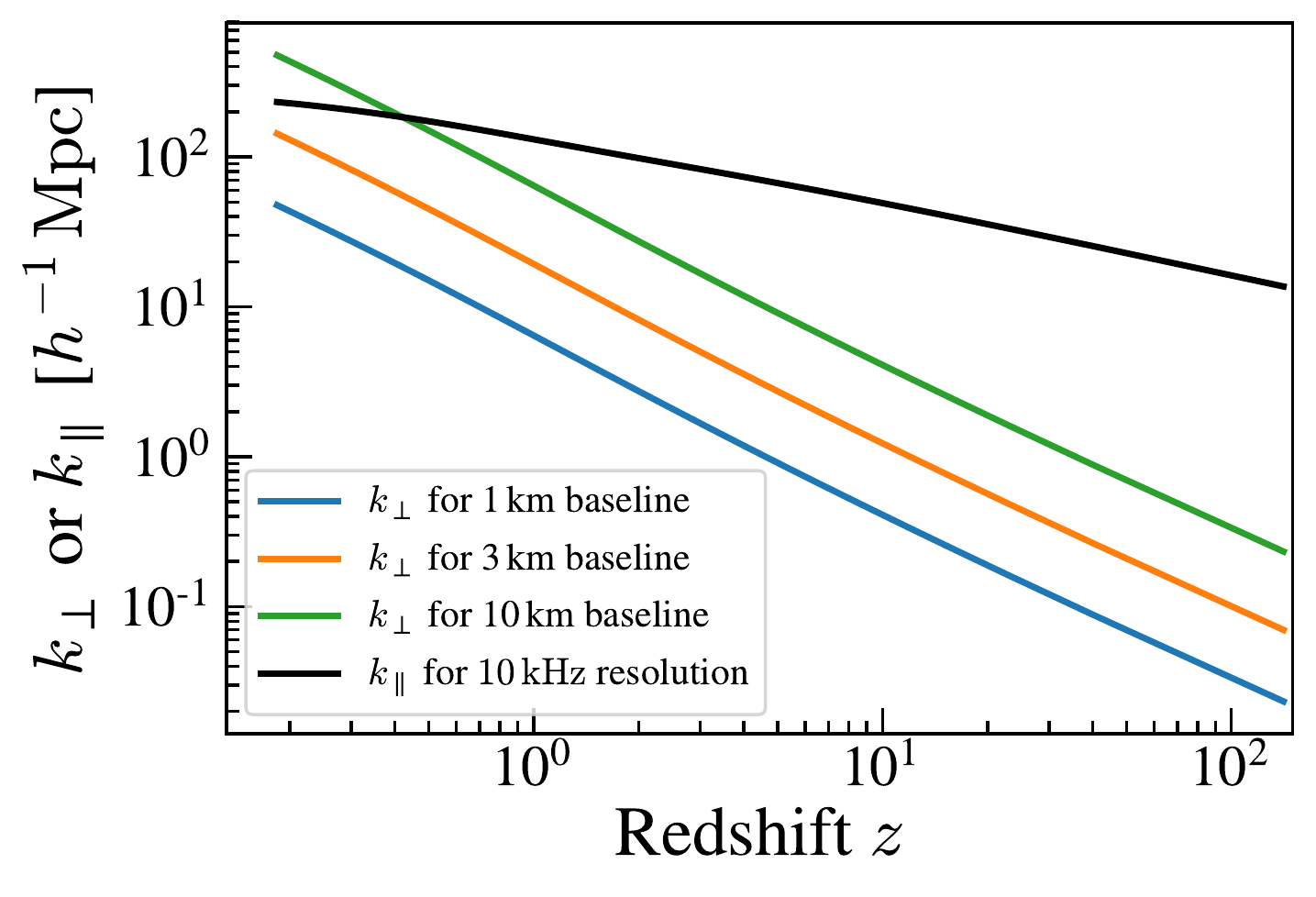}
\caption{Maximum $k_\perp$ (as a function of redshift) accessible to interferometers with baselines of various lengths, compared to the maximum $k_\parallel$ accessible to an interferometer with $10\,\textrm{kHz}$ spectral resolution. At very high redshifts, the only practical way to measure Fourier modes with high wavenumber $k$ is to measure small scale modes along the line of sight (i.e., those with high $k_\parallel$).}
\label{fig:kperpkpara}
\end{figure}

With two qualitatively different ways to access the same Fourier modes, it is natural to ask if one strategy is preferable over the other. In some scenarios, one does not have a choice. For instance, suppose one were interested in measuring fine spatial scales (high $k$) at very high redshifts (e.g., the Dark Ages) using a hypothetical futuristic interferometer. Such a measurement will necessarily require reaching high $k$ values by accessing high $k_\parallel$ modes. To see this, consider Figure \ref{fig:kperpkpara}, where we plot the relations given in Equation \eqref{eq:kperpkparamappings}. One sees that the cosmological scalings are such that to measure Fourier modes of any appreciable wavenumber in the perpendicular direction requires impractically long baselines. This is especially true when one accounts for the fact that measuring ultra high redshift signals will require extremely high sensitivity, so one needs a very large number of long baselines, and not just the small handful that are often used in very long baseline interferometry.

At lower redshifts where the baseline lengths are not prohibitive, one's strategy should be informed by the signal-to-noise ratio. In the low signal-to-noise regime one is limited by sensitivity, and the optimal strategy is to observe with a compact interferometer with antennas that are as closely packed as possible. The reason for this is that compact interferometers will typically contain many identical copies of the same baselines (which therefore sample precisely the same $k_\perp$ modes), especially if the antennas are placed on a regular grid (see Section \ref{sec:designcostssystematics}). Moreover, rotation synthesis causes baselines to rotate into one another on the $uv$ plane, further increasing this redundancy. This effect is particularly pronounced for short baselines, as a shorter ``radius" in $uv$ space causes baselines to spend more integration time in each $uv$ cell. Thus, compact interferometers are ones that concentrate their integration time into a small number of Fourier modes. They measure a few modes to high precision, rather than a large number of modes at low signal to noise.

Concentrating on a select handful of modes results in very high sensitivity to the power spectrum. This is because $N_\textrm{bl}$ baselines integrating for time $t$ on the same Fourier mode results in a \emph{coherent} averaging of $\widetilde{V}$, with the error averaging down as $1/\sqrt{N_\textrm{bl} t}$. The power spectrum error then goes as $1/N_\textrm{bl} t$, since the power spectrum goes as the square of $\widetilde{V}$. This stands in contrast to an interferometer with a spread out configuration that has little redundancy. In such an interferometer, different baselines will tend to sample different $uv$ (and thus different $\mathbf{k}_\perp$) modes. Each of these modes contains different information about the sky, and this information cannot be coherently added. In other words, each baseline measures $\widetilde{V}(\mathbf{b}, \eta)$ at different $\mathbf{b}$, and so one cannot directly average their measurements. The best one can do is to take advantage of \emph{statistical} isotropy to average together statistics like the power spectrum, as suggested in Equations \eqref{eq:BinningPk} and \eqref{eq:BinningPkperpkpara}. Thus, in this regime the power spectrum error (instead of the $\widetilde{V}$ error) is what scales as $1/\sqrt{N_\textrm{bl} t}$, which is a slower scaling than what a redundant array offers.

In the high signal-to-noise regime, however, one's errors are dominated by \emph{cosmic variance}. This is an error contribution that arises because the cosmological field that we seek to measure is random, and thus there is no guarantee that the amplitude of a particular Fourier mode is representative of the true underlying variance (i.e., the power spectrum) from which is it is drawn. Though it is by definition rare, it is entirely possible for a particular realization of a mode to lie in the extreme tails of a probability distribution. In the low signal-to-noise regime this is not a concern, as the contribution to the error budget from measurement uncertainties is larger than that from cosmic variance. Once a particular mode has been measured to a signal-to-noise ratio of order unity, however, it is no longer beneficial to continue integrating on this mode \citep{Tegmark:1997designer}. Instead, one gains by measuring other modes that can be averaged together incoherently to reduce cosmic variance. This favours arrays that are less compact and have fewer repeated redundant baselines.

Similar arguments can be made regarding the observational strategy of a telescope. Just as rare fluctuations may cause a particular Fourier mode to be unrepresentative of the underlying statistics of the sky---and therefore result in a misleading estimate of the power spectrum---the same can be true if one samples only a small (possibly unrepresentative) patch of the sky. Given the finite field of view of a telescope's primary beam, it is therefore necessary to consider the trade-off of whether one should concentrate observations on a small part of the sky or if one should spread out one's integration time over a larger portion of the sky \citep{Trott2014DriftVsPoint}. This optimization is precisely analogous to the optimization problem for interferometer layout: one should integrate on a particular patch of the sky until the patch has been measured to unity signal to noise, at which point one should switch to another patch \citep{Tegmark:1997designer}. And just as with our discussion of interferometer layout, the optimization of observational strategy has consequences for instrument design. Given the rotation of the Earth, the continuous observation of a small patch of sky requires a tracking telescope. On the other hand, if one desires broad sky coverage, a drift-scan telescope suffices.

\subsection{Costs and systematics in instrument design}
\label{sec:designcostssystematics}
Our discussion so far has focused mostly on minimizing uncertainties in our observations, be they uncertainties from the measurements themselves or from cosmic variance. This motivated compact, regular interferometers in the low signal-to-noise regime, and extended interferometers with many unique baselines in the high signal-to-noise regime. The former is also more appropriate for statistical measurements like measurements of the power spectrum, while the latter is more appropriate for imaging.

However, in practice one must contend with issues of cost and instrument stability: an ideal instrument from a sensitivity standpoint may be prohibitively expensive, or may have various practical problems that result in serious systematic errors. As an example, many current-generation \tcm telescopes (see Section \ref{sec:ObsStatus}) are drift-scan telescopes even though they are in a pre-detection low signal-to-noise regime. This design choice obviates the need for a telescope with moving parts, and is therefore commonly made to increase instrument stability in an effort to minimize the possibility of systematics.

Another example of what may (at first glance) be considered a strange choice is the proposal of large, regular interferometric arrays for next-generation arrays that will presumably be deep into the high signal-to-noise regime. Under pure signal-to-noise considerations, one would expect an irregular array that is more suited to imaging. However, regular arrays drive down the hardware cost of producing visibilities from an interferometer. As we shall see in Section \ref{sec:interferometrytheory}, a baseline's measured visibility is computed by cross-multiplying the measured voltages from the baseline's constituent antennas. With $N_\textrm{ant}$ antennas one can form $N_\textrm{ant} (N_\textrm{ant} - 1) / 2$ pairs of antennas, and this is how many baselines worth of visibilities one must compute. Since this goes as $N_\textrm{ant}^2$, the computational cost goes up rapidly with the size of one's array.

Futuristic array designs can circumvent this computational bottleneck in a number of ways. With a regular array, one can in principle reduce the computational cost to $\mathcal{O}(N_\textrm{ant} \log N_\textrm{ant})$. To understand this, suppose the voltages measured by each antenna were laid out on a regular grid, where the grid points correspond to the locations of the antennas. Cross-multiplying every voltage with every other voltage is then equivalent to convolving the gridded function of voltages with itself. This operation is equivalent to computing a series of Fourier transforms, thanks to the convolution theorem, and is therefore realizable in $\mathcal{O}(N_\textrm{ant} \log N_\textrm{ant})$ time using the Fast Fourier Transform (FFT) algorithm \citep{Tegmark2009,TegmarkZaldarriaga2010}. In fact, if the voltages are gridded onto a grid finer than one's antenna spacing, even irregular antenna layouts can take advantage of FFT speedups \citep{Morales2011MOFF,Thyagarajan2017EPIC}. However, this comes at the cost of carrying around extra zeros at grid points where there are no extra antennas, preventing an exact $\mathcal{O}(N_\textrm{ant} \log N_\textrm{ant})$ scaling. Still, the flexibility of arbitrary antenna placement is enticing, and early prototypes of next-generation hardware correlators (whether based on regular antenna grids or not) have shown promising initial results \citep{Beardsley2017EPIC,Kent2019EPIC}.

In closing, we note also that redundant, regular arrays have the advantage of enabling non-traditional calibration strategies. Essentially, regular arrays have multiple copies of the same baseline, and so while there may be $N_\textrm{ant} (N_\textrm{ant} - 1) / 2$ baselines in total, the number of \emph{unique} baselines is considerably smaller. Since identical baselines should yield identical measurements up to noise fluctuations if one's array is correctly calibrated, one can demand this property of mathematical consistency in one's data analysis and work backwards to solve for an array's calibration parameters. We explore the problem of calibration and provide context for why it is necessary in Section \ref{sec:Calib}.

\subsection{A recipe for estimating sensitivity}
\label{sec:Recipe}
As a summary of our discussion on interferometer design, we provide a basic recipe for computing the errors on a power spectrum measurement. For a \emph{tracking} telescope, the recipe is as follows:
\begin{enumerate}
\item Given the antenna locations of an interferometer, compute the $uv$ coordinates of all baselines.
\item Segment a $uv$ plane into discrete cells, each of which has roughly the same width as $\widetilde{A}_p$. Roughly speaking, this width is the inverse of the instantaneous field of view of an antenna. Define a three-dimensional $uv\eta$ space, extending the $uv$ plane into a third dimension of Fourier space. Each cell in this dimension should have an extent of $1/B$, where $B$ is the frequency bandwidth of the observation.
\item Simulate the movement of baselines through the $uv$ cells over a sidereal day of observations (or over however much time the telescope is being operated per sidereal day). Record how much time $t_{\vu}$ each baseline spends in each $uv$ cell.
\item Assuming a roughly constant noise contribution across frequency channels, the uncertainty (standard deviation) from measurement noise for a given $uv\eta$ cell is given by
\begin{equation}
\label{eq:uveta_noise}
\frac{D_c^2 c (1+z)^2}{\nu_\textrm{21} H_0 E(z)} \frac{\Omega_{p}^2}{\Omega_{pp}} \frac{T_\textrm{sys}^2}{2 t_{\vu}},
\end{equation}
where
\begin{equation}
\label{eq:BeamAreaDefs}
\Omega_p \equiv \int \!d\Omega \,A_p (\hat{\vr}); \quad \Omega_{pp} \equiv \int \!d\Omega \,A_p^2 (\hat{\vr})
\end{equation}
and $T_\textrm{sys}$ is the \emph{system temperature}, which is the sum of the sky brightness temperature and the receiver temperature (see Section \ref{sec:interferometrypractice}). This is the uncertainty \emph{for one baseline}. If multiple baselines spend time in a given $uv$ cell, the integration times of each baseline must be added together before being inserted in Equation \eqref{eq:uveta_noise}. The above expressions are derived in \citet{Morales2005,McQuinn2006,Parsons2012b} (albeit with different notations), but see Appendix B of \citet{Parsons2014Limits} for a discussion of the subtleties in defining $\Omega_p$ and $\Omega_{pp}$, which were missed by earlier works.
\item Divide the uncertainty computed above by the total number of sidereal days $t_\textrm{days}$ of observation. Note that the standard deviation scales as $t_\textrm{days}^{-1}$ rather than $ t_\textrm{days}^{-1/2}$ because visibility measurements repeat every day and thus can be coherently averaged prior to the squaring step of forming a power spectrum, as discussed in Section \ref{sec:InterferometerSensitivityOptimization}.
\item Associate each $uv\eta$ cell with its location in $\vk$ space using Equations \eqref{eq:udef} and \eqref{eq:kperpkparamappings}.
\item For high signal-to-noise measurements it is necessary to account for cosmic variance which contributes a standard deviation of $P(\vk)$ (i.e., the cosmological signal itself) to every $uv\eta$ cell. This is added to the instrumental uncertainty.
\item Assuming that different cells in the 3D $\mathbf{k}$ space are averaged together with an inverse variance weighting to obtain $P(k_\perp, k_\parallel)$ or $P(k)$, the corresponding errors can be combined in inverse quadrature. In other words, if $\varepsilon (\vk)$ is the standard deviation for a $\vk$-space cell, then the final averaged error bars are given by $(\sum \varepsilon^{-2})^{-1/2}$, where the sum is over all the $\vk$-space cells that contribute to a particular bin in $(k_\perp, k_\parallel)$ or $k$.
\end{enumerate}
There are several caveats to this (reasonably simple) recipe. The first is that it assumes a tracking telescope. For a drift-scan telescope, one should simulate the movement of baselines (in Step 3 of our recipe) for only the amount of time it takes for Earth rotation to move through one primary beam width. The primary beam width defines the size of a single patch of the sky that one might consider to be an independent observation. The observation of $N_\textrm{patch}$ patches of the sky can then be accounted for by dividing the final errors by $1/\sqrt{N_\textrm{patch}}$. Of course, scaling the errors in this way is an approximation. For a rigorous treatment, one should discard the notion of observing discrete patches and treat the full, curved sky in a single calculation, perhaps using a more suitable basis such as a spherical Fourier-Bessel basis \citep{Liu:2016} or the $m$-mode techniques of Appendix \ref{sec:DriftScans}.

The other big assumption that our recipe makes is that the coherence scale of noise on the $uv$ plane is given by the width of $\widetilde{A}_p$, and that the coherence scale in $\eta$ is given by $1/B$. Measurements smaller than these scales are assumed to be perfectly correlated, while measurements spaced farther apart than these scales are assumed to be entirely uncorrelated. In practice, noise correlations fall off smoothly with the distance between two samples in $uv\eta$ space. A proper treatment therefore requires intentionally over-resolving the space while taking into account error \emph{covariances} (rather than just error variances) between tiny $uv\eta$ cells \citep{Liu:2014a}. Despite these limitations, the recipe outlined above is reasonable for approximate forecasting work, and is implemented (with several convenient additions) by publicly available packages such as \texttt{21cmSense} \citep{Pober2013BAOBAB,Pober2013}.

\section{Current Status of Observations}
\label{sec:ObsStatus}

The last decade has seen a large increase in experimental $21\,\textrm{cm}$ activity. A variety of instruments have been used to place increasingly stringent limits on the $21\,\textrm{cm}$ power spectrum (in the case of experiments targeting Cosmic Dawn and reionization) as well as to detect spatial fluctuations in cross-correlations with traditional galaxy surveys (in the case of post-reionization experiments). Additionally, there has been a recent tentative detection of the sky-averaged (``global") $21\,\textrm{cm}$ signal at $z \sim 17$. We will return to a discussion of this result in Section \ref{sec:GlobalSig}.

For now, we review the instruments that have been used in attempts to measure the spatially fluctuating $21\,\textrm{cm}$ signal. Each instrument is optimized in subtly different ways, which has implications not only for observational capabilities and strategies, but also for analysis methods:
\begin{itemize}
\item The Giant Metrewave Radio Telescope (GMRT). GMRT is a general-purpose low-frequency interferometer located in India approximately $10\,\textrm{km}$ east of the town of Narayangaon. It consists of thirty steerable dishes, each $45\,\textrm{m}$ in diameter. Observations can be made around six frequency bands, with the lowest one centred around $50\,\textrm{MHz}$ and the highest around $1420\,\textrm{MHz}$. \citep{Kapahi1995GMRT}. Being a general-purpose instrument, its layout is not optimized for sensitivity to the $21\,\textrm{cm}$ power spectrum. Nonetheless, the GMRT Epoch of Reionization (GMRT-EoR) project was the first to provide upper limits at redshifts relevant to reionization \citep{Pen2009GMRT,Paciga:2011,Paciga:2013}. The analysis pipeline used to derive these limits employed the pulsar gating calibration technique described in Section \ref{sec:skycal}, and the principal component-based methods for radio frequency interference and astrophysical foreground mitigation described in Sections \ref{sec:RFI} and \ref{sec:fgmodeprojection}, respectively.
\item The Murchison Widefield Array (MWA). The MWA is a custom-built interferometer located in the Shire of Murchison in the Western Australian desert. It is designed for a wide variety of science cases (including Milky Way science, extragalactic studies, solar physics, and radio transients), but with Cosmic Dawn and Epoch of Reionization studies as key motivators \citep{Bowman2013MWA}. Each element of the interferometer consists of a square $4 \times 4$ grid of crossed dipole antennas (forming a \emph{tile}) whose signals are combined electronically. By changing the relative phases with which the signals from each dipole are combined in a process known as \emph{beamforming}, each tile can be electronically ``pointed" at different parts of the sky. Each tile then serves as a single element in a 256-element interferometer operating between $70$ and $300\,\textrm{MHz}$ \citep{Tingay2013MWA}. At any given instant, $30\,\textrm{MHz}$ of bandwidth and 128 elements can be correlated. True to its broad science case, the tile layout of the MWA is a hybrid between a spread out non-regular configuration and two closely packed, hexagonal grids of 36 tiles each \citep{Wayth2018MWAPhaseII}. The MWA has published a number of upper limits over a broad range of redshifts in recent years ranging from Cosmic Dawn redshifts to reionization redshifts \citep{Dillon2014MWALimits,Dillon2015EmpCov,EwallWice2016MWALimits,Beardsley2016MWALimit,Trott2016CHIPS,Jacobs2016comparison,Trott2019b,Trott2019a,Barry2019}. MWA analyses tend to involve multiple pipelines for validation purposes \citep{Jacobs2016comparison}, but most involve some combination of sky-based calibration off of point source catalogs (Section \ref{sec:skycal}), a hybrid approach for foreground mitigation involving the subtraction of bright point sources and foreground avoidance (Section \ref{sec:wedge}), and map-making- based power spectrum estimation methods (whether these maps are traditional images or in Fourier space; Section \ref{sec:pspecmapmaking}). Planned instrument upgrades will target improvements to the electronic hardware, enabling simultaneous correlation of all 256 tiles over a larger instantaneous bandwidth \citep{Beardsley2019}.
\item The Donald C. Backer Precision Array for Probing the Epoch of Reionization (PAPER). PAPER is an interferometer that is optimized for EoR power spectrum measurements. As such, it embraced the idea of maximizing power spectrum sensitivity by arranging 112 dual-polarization dipole antennas in a regular rectangular grid \citep{Parsons2012b}, complemented by 16 antennas that served as outriggers to provide longer baselines. Operating between $100$ and $200\,\textrm{MHz}$, PAPER has published a number of upper limits on the EoR power spectrum \citep{Parsons2010PAPER,Pober:2013,Parsons2014Limits,Ali2015,Jacobs2015Multiz}, but has now been decommissioned, with its site in the South African Karoo desert now being used for the Hydrogen Epoch of Reionization Array (HERA; see below). PAPER's analysis approach uses the redundant baseline calibration algorithm described in Section \ref{sec:RedCal} and bypasses map-making in favour of direct power spectrum estimation using visibilities via the delay spectrum approach of Section \ref{sec:delayspec}, couched in the language of the quadratic estimator framework (Section \ref{sec:pspecestimation}).
\item The LOw Frequency Array (LOFAR). LOFAR is a low-frequency radio interferometer with a dense core of elements centred in the Netherlands complemented by a set of remote international elements \citep{LOFAR}. Each interferometric element is known as a station, and is in fact made of two different types of antennas: a set of high-band antennas (HBAs) cover the $110$ to $250\,\textrm{MHz}$ frequency range and a set of low-band antennas (LBAs) cover the $10$ to $90\,\textrm{MHz}$ range. The number of antennas in each station varies from location to location, but the stations in the Netherlands (which give rise to the shortest baselines and therefore are the ones with greatest sensitivity to cosmology) each have 96 LBAs and 48 HBAs. Whereas each LBA is an individual dipole antenna, each HBA is in fact a tile of 16 antennas (in a similar fashion to the MWA) tied together by an analog beamformer. At each station, the outputs from the LBAs and the HBAs are digitized and then digitally beamformed before the data from all the stations are sent to a central correlator located at the University of Groningen. LOFAR is a multi-purpose observatory that accommodates a wide variety of key science projects, including deep extragalactic surveys, transient phenomena, cosmic rays, solar science, cosmic magnetism, and cosmology. It has recently placed upper limits on both Cosmic Dawn \citep{Gehlot2018Limit} and the EoR \citep{Patil2017}. The analysis pipelines used to derive these limits employed sophisticated direction-dependent calibration schemes (Section \ref{sec:CalibFreqTime}), and use either Generalized Morphological Component Analysis (Section \ref{sec:fgmodeprojection}) or Gaussian Process Regression (Section \ref{sec:NonParam}) for foreground mitigation.
\item The Green Bank Telescope (GBT). GBT is a large $100\,\textrm{m}$ single-dish telescope located in Green Bank, West Virginia \citep{GBT}. Operating between $100\,\textrm{MHz}$ and $115\,\textrm{GHz}$, it is a general purpose instrument. It holds the distinction of having made a detection of $21\,\textrm{cm}$ fluctuations at $z \sim 0.53$ to $1.12$ via cross-correlations with the DEEP2 optical galaxy survey \citep{Chang:2010} and at $z \sim 0.8$ via cross-correlations with WiggleZ \citep{Masui:2013}. Additionally, it has placed upper-limits on the $21\,\textrm{cm}$-only auto spectrum and used the combination of cross- and auto-correlation measurements to place constraints on $\Omega_\textrm{HI} b_\textrm{HI}$ \citep{Switzer2013}. The data used for these results were calibrated using a combination of known bright sources and data obtained from injecting signals from a noise diode into the antenna. Map-making was performed using the maximum likelihood estimator described in Section \ref{sec:LinearMapMakers}, and power spectrum estimation was performed from the resulting maps. Foregrounds were suppressed by projecting out principal component spectra (Section \ref{sec:fgmodeprojection}), with special care taken to compute transfer functions to correct for the possible subtraction of cosmological signal (Section \ref{sec:sigloss}).
\item The Parkes Radio Telescope. Parkes is a general-purpose 64-m single dish telescope \citep{Parkes} operating from $1230$ to $1530\,\textrm{MHz}$. It is located in New South Wales, Australia. In a similar way to GBT, it has been used to measure $21\,\textrm{cm}$ fluctuations in cross-correlation. Combining Parkes data with the 2dF galaxy survey has enabled positive detections in the range $0.057 < z < 0.098$ \citep{Anderson:2018}. Conceptually, the analysis pipelines used for the analyses were very similar to those used for the GBT results.
\item The Owens Valley Long Wavelength Array (OVRO-LWA). OVRO-LWA consists of $288$ dipoles, with 251 of these dipoles located within a $200\,\textrm{m}$-diameter compact core and the remaining $32$ dipoles located farther away to provide baselines of up to $1.5\,\textrm{km}$ in length. The elements in the core are arranged in a pseudo-random fashion, enabling a better point-spread function for imaging. This makes OVRO-LWA a powerful instrument for transient science, for instance in its searches for radio emission from gamma ray bursts \citep{Anderson2018GRBs}, gravitational wave events \citep{OVROLWA2019GW}, or from exoplanets \citep{Anderson2017Exoplanets}. With observations spanning an instantaneous bandwidth from $27\,\textrm{MHz}$ to $85\,\textrm{MHz}$, OVRO-LWA has also produced new results for \tcm cosmology, including a detailed set of low-frequency foreground maps \citep{Eastwood2018} and a upper limit on the \tcm power spectrum at $z \approx 18.4$ \citep{Eastwood2019Pspec}. The pipelines for these results lean heavily on the $m$-mode formalism for map-making (Section \ref{sec:mmodes}), the Karhunen-Lo\`{e}ve mode projection technique for foreground suppression (Section \ref{sec:fgmodeprojection}), and quadratic estimator-based power spectrum estimation (Section \ref{sec:pspecestimation}).
\end{itemize}

\begin{figure*}[t]
\centering
\includegraphics[width=1.00\textwidth,trim={0cm 0cm 0cm 0cm},clip]{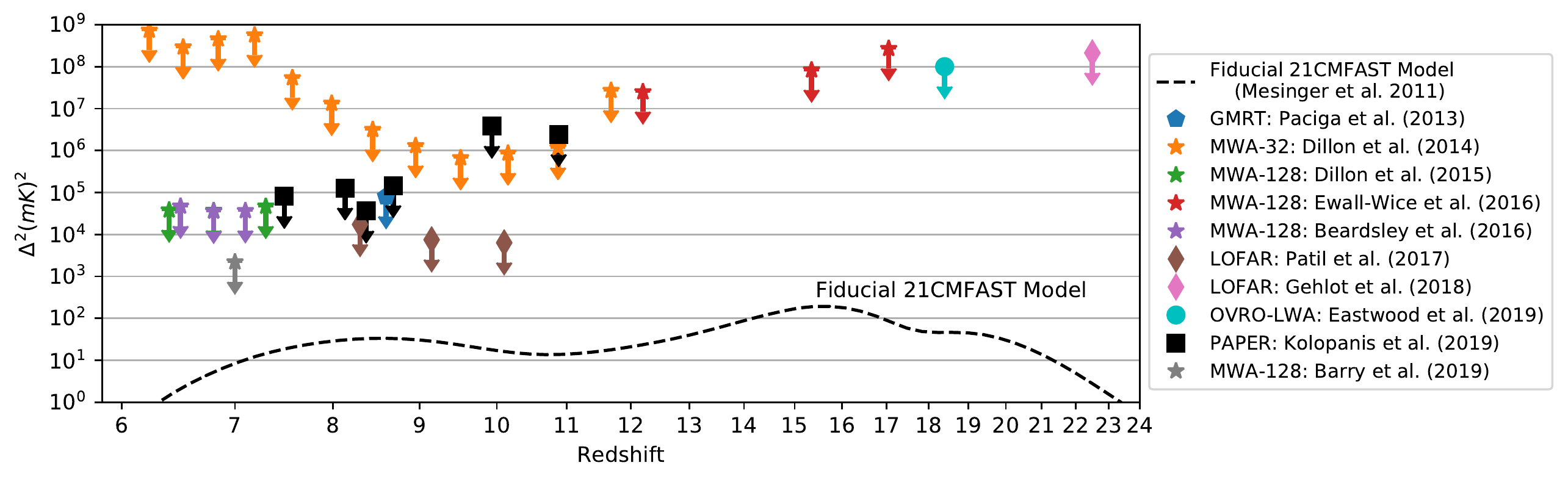}
\caption{A summary of current upper limits on EoR power spectrum measurements. The limits are expressed as $\Delta^2 (k) \equiv k^3 P(k) / 2 \pi^2$ values evaluated at the $k$ bin that gives the most competitive (i.e., lowest) upper limit for each experiment. Thus, different points on the plot come from different $k$ values. A direct comparison between them is thus unfair in principle, and in practice is somewhat reasonable only because the theoretical predictions for $\Delta^2 (k)$ are often fairly flat in the range $0.1\,h\textrm{Mpc}^{-1} < k < 1\,h\textrm{Mpc}^{-1}$ (see Figure \ref{fig:sample_ps}). (Only the LOFAR points fall outside this range). Also shown is a fiducial theoretical model from the \texttt{21cmFAST} semi-analytic code \citep{Mesinger2011}}
\label{fig:limits}
\end{figure*}

Thus far, there has yet to be a detection of the $21\,\textrm{cm}$ auto-power spectrum (i.e., not in cross-correlation with other probes), although upper limits have become increasingly stringent over the years. The experiments listed above were chosen because they have all published such limits, which are illustrated in Figure \ref{fig:limits}. To be sure, there have been setbacks, including revisions and retractions of incorrect upper limits, and we discuss this topic in Section \ref{sec:sigloss}. On the more optimistic side, however, current instruments (e.g., the MWA) are being upgraded and new instruments are being built. These new instruments not only have the sensitivity (at least in principle) to make high-significance detections, but also to diagnose possible systematics with new levels of precision:
\begin{itemize}
\item The Hydrogen Epoch of Reionization Array (HERA). HERA occupies the same South African Karoo desert site that was formerly occupied by PAPER. When complete, it will consist of 350 dishes, each of which has a $14\,\textrm{m}$ diameter \citep{DeBoer:2017}. The dishes are arranged in a compact configuration that is essentially an ``exploded" hexagon: the configuration would be a close-packed hexagon of touching dishes except that the hexagon is split into three parallelograms that are displaced from each other by sub-dish spacings \citep{Dillon:2016}. This produces sub-aperture sampling in the baseline distribution, reducing grating lobes. In addition to 320 dishes in the exploded hexagon configuration, there are 30 outrigger dishes to provide long baselines for better angular resolution. All the elements (including the outriggers) are placed in a such a way that the entire array can be calibrated using the \emph{redundant calibration} method described in Section \ref{sec:RedCal}. HERA is a telescope with no moving parts. It surveys the sky as a drift-scan telescope that is always pointed at zenith. With a custom-designed broadband feed, HERA is in principle capable of observing between $50\,\textrm{MHz}$ and $250\,\textrm{MHz}$. It is designed to have sufficient sensitivity to make a high-significance detection of the \tcm power spectrum both during reionization \citep{Pober2013,Liu2016} and cosmic dawn \citep{EwallWice2016Forecast,Kern2017}.
\item The Canadian Hydrogen Intensity Mapping Experiment (CHIME). CHIME is located at the Dominion Radio Astrophysical Observatory in British Columbia, Canada. It is comprised of four large cylindrical reflectors, each measuring $20\,\textrm{m} \times 100\,\textrm{m}$ (with the parabolic curvature being along the shorter dimension, hence the visual impression of a cylinder). The long axes of the cylinders are oriented north-south, and since it is these axes that are flat and open, the instantaneous field of view is large ($\sim 100^\circ$) in the declination direction \citep{Bandura2014}. In the east-west direction CHIME has a smaller instantaneous field of view, but because the telescope operates as a drift-scan telescope, over time the entire range of right ascension values is covered. In each cylinder are 256 dual-polarization antennas that serve as feeds \citep{Newburgh2014,HincksShaw2015}. Operating between $400$ and $800\,\textrm{MHz}$, CHIME is designed to make precision measurements of the Baryon Acoustic Oscillation scale from $z = 0.8$ to $2.5$ \citep{Shaw2015}. However, in recent years alternate processing backends have been installed on the telescope to enable pulsar monitoring and searches for Fast Radio Bursts (FRBs; \citealt{CHIMEFRBOverview}). The FRB searches, in particular, have revolutionized FRB studies with CHIME discovering an enormous number of new FRBs, including some down to $400\,\textrm{MHz}$ \citep{CHIMEFRBII,CHIMEFRBI}.
\item The Hydrogen Intensity and Real-time Analysis eXperiment (HIRAX). HIRAX is similar to CHIME in its dual science goals (\tcm cosmology from $z \sim 0.8$ to $2.5$ plus transient surveys), but is located in South Africa. With access to the southern sky, HIRAX is particularly well-positioned to take advantage of cross-correlation opportunities, given the large number of galaxy surveys, CMB experiments, and future non-HI intensity mapping surveys with footprints in the south. (See Sections \ref{sec:CrossCorrFGs}, \ref{sec:CrossCorrPossibilities}, and \ref{sec:HigherOrderStatistics} for the power and limitations of cross correlations). When construction is complete, HIRAX will consist of a square $32\times 32$ grid of 1024 dishes operating from $400$ the $800\,\textrm{MHz}$. With 6-m dishes, its instantaneous field of view ($\sim 6^\circ$) is smaller than that of CHIME. However, the HIRAX dishes can be repointed (by hand), enabling large sky coverage. Each pointing maps out a stripe in declination, and with a repointing every $150\,\textrm{days}$ over four years, HIRAX can cover a quarter of the sky with sufficient depth for high-precision cosmological measurements \citep{Newburgh2016CHIME}.
\item The Tianlai experiment. Tianlai is again a post-reionization \tcm intensity mapping experiment, but with a slightly larger frequency range ($400$ to $1420\,\textrm{MHz}$ or $0< z < 2.5$) than CHIME or HIRAX. It is located in the Xinjiang Autonomous Region, China. Its final design is yet to be decided, with the Tianlai team building two separate pathfinder experiments: one is a set of three cylindrical reflectors (each measuring $15\,\textrm{m} \times 40\,\textrm{m}$ and outfitted with $96$ feeds) and the other is a set of $16$ dishes (each $6\,\textrm{m}$ wide). The dishes are arranged in two concentric circles at radii of $8.8\,\textrm{m}$ and $17.6\,\textrm{m}$ around a central dish. Like CHIME and HIRAX, Tianlai is forecasted to make precision cosmological measurements \citep{Xu2015Tianlai}.
\item The BAO from Integrated Neutral Gas Observations (BINGO) experiment. Located in South America, BINGO aims to make measurements in the $0.12 < z < 0.48$ range ($960$ to $1260\,\textrm{MHz}$). As such, it is quite complementary to CHIME, Tianlai, and HIRAX, providing large scale structure measurements at a different set of redshifts. From a technology development standpoint it is also complementary, as it is not an interferometer. Instead it is a single-dish experiment with a two-mirror crossed-Dragone design that is outfitted with a 50-element feed horn array at the focal plane \citep{BINGOI,BINGOII}. The instrument has an instantaneous field of view of $\sim 15^\circ$ and operates in a drift-scan mode to map out a stripe in the sky.
\item The Square Kilometre Array (SKA). The SKA is the largest effort of all upcoming telescopes. It will consist of two separate telescopes, SKA-low and SKA-mid. SKA-low ($50$ to $350\,\textrm{MHz}$) will be located in the Western Australian desert, while SKA-mid ($350\,\textrm{MHz}$ to $15.3\,\textrm{GHz}$) will be located in the South African desert. The former will be comprised of over $\sim 100,000$ antennas grouped into stations, while the latter will be comprised of $197$ dishes. To achieve a small, compact synthesized beam without strong grating lobes, the elements will be arranged in a pseudo-random configuration with a maximum baseline of $65\,\textrm{km}$ for SKA-low and $150\,\textrm{km}$ for SKA-mid. While many details regarding the SKA are still to be determined, it is generally expected to have an expansive science case, including the EoR \citep{SKAI}, post-reionization cosmology \citep{SKAII}, pulsars \citep{SKAIII}, transients \citep{SKAIV}, galaxy formation \citep{SKAV}, magnetism \citep{SKAVI}, planet formation \citep{SKAVII}, and individually detected HI galaxies \citep{SKAVIII}.
\end{itemize}
While we have endeavoured to be reasonably complete in our list of experimental efforts, there will inevitably have been some omissions. For the list of upcoming experiments, we focused on those that are funded (or are reasonably likely to secure funding) and also list \tcm cosmology as one of their main science objectives. A more complete list (at least for post-reionization experiments), including telescopes that could in principle be used for \tcm cosmology (even if they were not designed for it), can be found in \citet{Bull2015}.

While carefully designed hardware is crucial for achieving the high levels of precision necessary for \tcm cosmology, equally important are the analysis choices that one makes. In a sense, \tcm experiments are \emph{software telescopes}. Having summarized the general state of the field, the rest of this paper focuses on $21\,\textrm{cm}$ data analysis in detail, starting with a deeper discussion of interferometry.


\section{Fundamentals of Interferometry}
\label{sec:InterferometryBasics}

In \secref{sec:BasicObs} we introduced the basics of interferometry, defining the key observable, the \emph{visibility}, in the commonly used \emph{flat-sky, unpolarized} limit. While largely adequate for current instruments, the next generation of survey interferometers have large instantaneous fields of view, and need to excise both polarized and unpolarized foregrounds with high fidelity. To appreciate when these approximations are appropriate and when they are not, we need to go back to the start and look at interferometry with no approximations.

In this section we go through the fundamentals of interferometry, starting from the electromagnetism and statistics to explain: what are we measuring? Why do we measure it? And how do we do it in practice? However, later sections can be mostly understood using just the results of \secref{sec:BasicObs}, and so the hurried reader may wish to jump forward to Section \ref{sec:fgs}.

\subsection{Antenna Response}

In radio astronomy we make \emph{coherent} measurements, meaning that an individual antenna directly measures a weighted linear sum of the electric field in the surrounding volume. The antenna generates a voltage proportional to this signal that we can read out. If we consider only emission in the far field, this is equivalent to measuring a direction-weighted summation of incoming electromagnetic plane waves. More precisely, we will model the electric field at a reference point as made up of small vector contributions $\veps(\vrhat, \nu)$  (with the direction of the vector indicating polarization direction) coming from each direction $\vrhat$ and frequency $\nu$
\begin{equation}
    \label{eq:dvEdef}
    d\vE(t) = \veps(\vrhat, \nu) \: e^{2\pi i \nu t} \: d\Omega \, d\nu \; ,
\end{equation}
such that the total electric field that could be measured at time $t$ is the integral over all angles and frequencies.\footnote{It is worth noting that although the sign choice within the exponential \cref{eq:dvEdef} is arbitrary, this choice fixes the sign of all other Fourier exponentials within interferometry. We use the standard convention in radio astronomy and electrical engineering that the transform from frequency to time has a \emph{positive} sign in the exponential.} The electric field at a general point $\vx$ simply adds a direction dependent time delay relative to the reference point
\begin{align}
    d\vE(\vx, t) & = d\vE(t - \vx\cdot\vrhat / c) \\
    &  = \veps(\vrhat, \nu) \: e^{2\pi i \nu (t - \vx\cdot\vrhat / c)} \: d\Omega \, d\nu \; ,
\end{align}
which encodes the fact that as we move towards (or away from) the source of the emission, we receive the emission sooner (or later).

In radio astronomy when we talk about an antenna we can mean many things; here we define it in terms of its operational behaviour. That is that an antenna is the part of the system which maps the emission on the sky into the measured signal. Physically this may consist of many components: a reflector that focuses the radiation onto a dipole, the dipole itself, amplifiers, cables amongst others. These components have many complex interactions within and between signal chains. We abstract over this complexity and simply define the operation of the antenna labelled $i$ (located at position $\vx_i$ on the ground) as the response of the measured output $s_i$ to emission from the sky. That is
\begin{equation}
    \label{eq:primary_beam_def}
    ds_i = \alpha \, \vA_i(\vrhat, \nu) \cdot d\vE(\vx_i, t) \; ,
\end{equation}
where the antenna response $\vA_i$ is more commonly called its \emph{primary beam} or often just its \emph{beam}\footnote{Do not confuse this with the primary beam $A_p$ defined in Section \ref{sec:BasicObs}, which was a \emph{power} primary beam, and a scalar quantity, unlike the vector electric field/voltage primary beam that we are denoting as $\vA$ here. As we will see, the former roughly goes as the square of the latter. Confusingly, both quantities are often referred to simply as the ``primary beam", and it is necessary to understand the context in which the term is being applied.}, and $\alpha$ is an arbitrary scaling we will set later to simplify the notation. Note that as the electric field is a vector quantity, the primary beam must be a vector response and the two multiplied with a dot product to give a scalar voltage output. Provided the properties of the antenna do not depend on time themselves, i.e. the antenna is a stable part of the instrument, the beam depends only on the frequency (a consequence of the Fourier convolution theorem) and the direction of the incoming radiation, which we have used above.

The beam is a function which, for every position on the sky, and at every frequency, describes the response of the instrument. That response is given as a \emph{complex vector}, with the vector being defined as transverse to the direction vector $\vrhat$ (i.e. it is defined in the tangent space to the sphere in that direction). For a physical interpretation of the beam we can think of the amplitude of the vector as telling us how sensitive we are in each direction on the sky; the orientation of the vector tells us what orientation of the incoming electric field we are sensitive to (i.e the polarization); finally, the complex phase tells us about any relative time delay with which we receive the emission.

In integral form we can write the output signal $s_i$ in terms of the beam and emission from the sky as
\begin{equation}
    \label{eq:sky_signal_time}
    s_i(t) = \alpha \iint \! \vA_i(\vrhat, \nu) \cdot \veps(\vrhat, \nu) \: e^{2 \pi i \nu (t - \vx_i \cdot \vrhat / c)} \: d\Omega \, d\nu \; .
\end{equation}

Though the antenna beam $\vA_i(\vrhat, \nu)$ is easy to define, actually measuring it from data, or calculating it using simulations is very difficult and is one of the major challenges facing \tcm cosmology. This is discussed more in \secref{sec:SystematicsPrimaryBeams}.

\subsection{Sky Emission}
\label{sec:SkyEmissionStokes}
To understand how polarization effects our measurements we must first define a reference frame on the sky. As electromagnetic radiation has no component in the direction of propagation, the electric field lies in the two-dimensional plane perpendicular to the direction vector $\vrhat$. Defining an orthonormal basis in this plane $\vehat^a$, the electric field vector can be decomposed into orthogonal polarization states $\eps_a = \vehat^a \cdot \veps$, with the beam decomposing similarly. As we are considering wide-field instruments we need to use a basis which works on the full sky and so we follow the cosmological convention and use the spherical polar unit vectors $\vehat_a = \{\hat{\vec{\theta}}, \hat{\vec{\phi}}\}$, with celestial north aligned along the polar axis. In this convention the vector $\hat{\vec{\theta}}$ points south, while the vector $\hat{\vec{\phi}}$ points to the east.

First, we should remember that the electric field from the sky (that is $\veps$) is a random signal and so is described statistically. As almost all astrophysical radiation is produced by a large ensemble of independent emitters (electrons, atoms, molecules...) by invoking the central limit theorem we can see that the distribution of the observed electric field is Gaussian\footnote{Even searches for non-Gaussian distributed electric fields from highly coherent emitters such as masers \citep{1972PhRvA...6.1643E,2009MNRAS.396.2319D} and pulsars \citep{2003A&A...405..795S} have proved negative.} (with zero mean), and so its statistical properties are encoded entirely within its two-point statistics $\langle \eps_a(\vrhat, \nu) \eps_b^*(\vrhat', \nu') \rangle$ \citep{LFRA3}. If we know that function we know everything of interest about the emission from the sky; however, it is generically a complicated function of multiple positions and frequencies.

Fortunately, on this front we receive some help from our Universe: while our detector is coherent, in radio astronomy the emission from the sky (with rare exception) is spectrally and spatially \emph{incoherent}. That means that there is no correlation between emission from different parts of the sky or at different frequencies. Physically this occurs when the emission is sourced from the sum of many independent systems that have no time/phase correlations between themselves, for example electrons spiralling in a magnetic field emitting synchrotron radiation that have random initial velocities and positions. However, radio emission is often \emph{polarized}. That is, there are non-zero correlations between the orthogonal polarization states. Expressing both of these facts mathematically we have
\begin{multline}
\label{eq:Efield_corr}
\langle \eps_a(\vrhat, \nu) \eps_b^*(\vrhat', \nu') \rangle =\\ \frac{4 k_B Z_0}{\lambda^2} \: T_{ab}(\vrhat, \nu) \: \delta(\vrhat - \vrhat') \delta(\nu - \nu') \; ,
\end{multline}
where $Z_0$ is the impedance of free space and $T_{ab}$ is the coherency matrix describing the polarization of the emission which, as we have defined here, has units of temperature. The polarization of radiation is usually described in terms of the four Stokes parameters each describing how one component of the polarised emission varies over the sky and with frequency. The four components are: $T(\vrhat, \nu)$ which describes the total intensity of the emission; $Q(\vrhat, \nu)$ and $U(\vrhat, \nu)$ which give the linear polarization; and $V(\vrhat, \nu)$ which gives the intensity of circular polarization. In terms of these the coherency matrix is
\begin{equation}
    T_{ab}(\vrhat, \nu) = \frac{1}{2} \begin{pmatrix} T + Q & U + i V \\ U - i V & T - Q \end{pmatrix} \; .
\end{equation}
There are very few astrophysical sources of Stokes V polarization, and those that exist are typically from fast transients like pulsars, and so generally we can assume that Stokes V is zero. We should also remember, that though the underlying electric field is random with no spatial or spectral correlations, the same is not true for its statistics, i.e. the elements of the coherency matrix $T_{ab}$ can be correlated across frequencies and angles (as we will see in \secref{sec:fgs}).

\subsection{Interferometry in theory\ldots}
\label{sec:interferometrytheory}

As the underlying electric field is a Gaussian process, the coherency matrix contains all the information about the emission from the sky. Our goal with interferometry is to try and capture as much information about this as we can.

First, because the emission is spectrally incoherent, it will be more useful to describe our observations in frequencies rather than times. This means that we usually think of our measurements from the antennas as a function of frequency, and so will use their Fourier transforms $s_i(\nu)$, which written explicitly are
\begin{align}
    \label{eq:signalFT}
    s_i(\nu) & = \int s_i(t) \: e^{-2 \pi i \nu t} dt \notag \\
    & = \alpha \int A_i^a(\vrhat, \nu) \, \eps_a(\vrhat, \nu) \: e^{-2 \pi i \nu \vx_i \cdot \vrhat / c} \: d\Omega  \; .
\end{align}
where we have implicitly summed over $a$ using an Einstein summation convention for indices $a$ (and shortly $b$), which we will use for the rest of this section.

As the emission from the sky is Gaussian distributed, and the measurement by the antenna is a linear process, this means that the signal $s_i(\nu)$ is also Gaussian distributed. As all the information about a Gaussian process is captured in the two point statistics, this means that for a fixed set of antennas all the information is contained in the covariance matrix of the signals from the set of antennas. These elements of this covariance matrix are the \emph{visibilities}, the key quantities in radio interferometry. The elements of the matrix are the cross correlations between two feeds, and for a pair $i$ and $j$ are
\begin{align}
 \label{eq:vis_full}
    S_{ij}(\nu) & = \langle s_i(\nu) s_j^*(\nu) \rangle \\
    & = \int A_i^a(\vrhat, \nu) A_j^{b *}(\vrhat, \nu) \: T_{ab}(\vrhat, \nu) \: e^{-2 \pi i \vu_{ij} \cdot \vrhat} \: d\Omega \notag\;
\end{align}
where we define the \emph{baseline} vector $\vb_{ij} = \vx_i - \vx_j$ and $\vu_{ij} \equiv \vb_{ij} / \lambda$, which is the separation between the antennas in wavelengths. We have also set the value of $\alpha = (\lambda^2 / 4 k_B Z_0)^{1/2}$ to eliminate the prefactor found in \eqref{eq:Efield_corr}. Here $S_{ij}$ specifically refers to the part of signal coming from the sky; in the next section we will look at the contribution of noise.

\Cref{eq:vis_full} is the fundamental equation of interferometry, describing how emission on sky generates a signal in our data, and includes exactly all curved sky and polarization effects. Its resemblance to \cref{eq:BasicVis} is apparent, and we will make the connection more closely in \secref{sec:polarization_leakage}.

\subsection{\ldots and in practice}
\label{sec:interferometrypractice}

Above we have presented the idealised picture of an interferometer, but to appreciate fully the challenges of building one and analysing its output we need to delve into how they work in practice. An interferometer can be divided into three distinct parts, the analog chain and two digital systems, the F-engine and the X-engine\footnote{In this work we discuss only FX correlators which are dominant in low-frequency radio astronomy. However, we should note the existence of XF correlators (also known as lag correlators) which perform the frequency transform last. This may have some advantages in terms of spectral line sensitivity and correcting quantisation effects, but requires more computation for interferometers with a large number of antennas. See \cite{TMS2017} for a detailed comparison}. We discuss these below.

One of the key things to    appreciate in how an interferometer works is the separation of the various timescales involved. There are three distinct timescales we need to consider:
\begin{itemize}
    \item $\sim$ \SIrange{1}{10}{\nano\second} --- for \tcm cosmology we are interested in frequencies of around \SIrange{100}{1000}{\mega\hertz}, so we are probing fluctuations in the electric field on roughly nanosecond timescales.

    \item $\sim$ \SI{10}{\micro\second} --- the smallest spectral structures we are interested in are around \SI{100}{\kilo\hertz} which means the longest fluctuations in the electric field we are interested in are $\sim \SI{10}{\micro\second}$.

    \item $\sim \SI{1}{\second}$ --- on time scales longer than a second the change in orientation of the telescope due to rotation of the Earth is significant compared to its resolution.
\end{itemize}
Put together, these give us a practical way to build an interferometer. Between the first two timescales we can transform our data from the time domain into the frequency domain to give ourselves distinct frequency channels. This performs the operation described in \cref{eq:signalFT} and is performed by the F-engine. Between the second and third timescales the samples from each frequency channel should be statistically equivalent, but independent, which allows us to evaluate the ensemble average of \cref{eq:vis_full} by integrating in the domain. This is performed by the X-engine.

\subsubsection{Analog Chain}
\label{sec:analogchain}

The purpose of the analog chain is to transform the electric field coming into the telescope into a signal which can be digitized. At radio frequencies many analog components introduce noise into the system, typically by attenuating the signal we care about and introducing a small amount of thermal noise in its place. The primary goal of an analog chain is to minimize this effect.

There are many stages to the analog chain, the first that we have already discussed is the \emph{antenna} itself (which may also be paired with a reflector), that receives the electric field and turns it into a voltage. At this point the signal is extremely weak and must be quickly amplified so as to not introduce noise. This is performed by a \emph{Low Noise Amplifier} (LNA) which amplifies the signal by many orders of magnitude, at the expense of introducing some noise of its own. The LNA is designed to amplify the signal to well above the thermal noise level in the rest of the system such that any further attenuation does not introduce any extra noise.

At this point there are two more roles for the analog chain:
\begin{itemize}
    \item Removing any unwanted frequencies, i.e. frequencies above the maximum frequency we are interested in and below the minimum, which is important for digitisation (see the following section).
    \item Transporting the signal to the location of the digital system. Although serving a trivial purpose, this particular stage can be the source of several systematics (see \secref{sec:InstrumentSystematics}).
\end{itemize}

At the end of the analog chain we have amplified, and filtered the signal all the while introducing noise into the system. We write the final quantity  passed onto the digital stage as
\begin{equation}
    x_i(t) = s_i(t) + n_i(t) \; ,
\end{equation}
where we have rescaled such that this output is proportional to the input signal. This moves all the effects of amplification and filtering into the final noise term which we write as $n_i(t)$.

\subsubsection{F-engine}

The job of the F-engine is to take the time stream data and turn it into distinct frequency channels (hence the ``F''), however, before that can happen the data must be digitized. This first step is performed by an Analog-Digital Converter (ADC) which produces a discrete series of quantized time samples from the amplified input. The quantization reduces the precision of the input and is typically treated as an extra source of noise \citep{TMS2017} that can be absorbed into $n_i(t)$, although modern correlators typically operate at high enough bit-depth that this contribution is sub-percent \citep{2018JAI.....750008M}. The rate at which the input is sampled is controlled by the range of frequencies we are interested in. For the simplest scenario of standard Nyquist sampling the input must be sampled at $\nu_\text{sample} > 2 \, \nu_\text{max}$ where frequencies above the maximum were filtered out earlier in the signal chain to prevent aliasing of higher frequency signals. However, usually we do not care about frequencies below some $\nu_\text{min}$, in that case we can reduce the sampling rate by either:
\begin{itemize}
    \item Mixing --- using a local oscillator and a mixer, we can shift the frequency range of interest down towards zero, and we then only need to sample at $\nu_\text{sample} > 2 (\nu_\text{max} - \nu_\text{min})$.
    \item Alias sampling --- if the maximum and minimum frequencies are both multiples of some interval $\Delta\nu$ it is possible to alias sample, where we deliberately sample at less than the Nyquist frequency, aliasing our signal down to lower frequencies without the use of a mixer. In this case we need only sample at $\nu_\text{sample} > 2 \Delta\nu$.
\end{itemize}
This gives the shortest timescale we are concerned with: the length of an individual sample $\sim 1 / \nu_\text{sample}$. We will denote the discrete samples taken at some time $t_a = a\,\Delta{t}$ as $x_i[t_a] = x_i(t_a)$.

The next stage in the digital chain is channelization, where we turn the time samples into frequency channels. The simplest way to perform this is to make use of the the Short-Time Fourier Transform. In this process we group the timestream into chunks of $N$ samples, each starting at samples $t_A$ and of length $T = N \Delta{t}$, apply a window function $w[a]$ to account for the non-periodic nature of the chunk and then use a Fast Fourier Transform (FFT) to turn the data into frequency channels centered at $\nu_b = b / T$
\begin{align}
\tilde{x}_i[\nu_b; t_A] & = \sum_a x_i[t_{A + a}] \, w[a] \, \exp{\lp - \frac{2 \pi i a b}{N}\rp} \\
& = \int \tilde{x}_i(\nu; t_A) \, \tilde{w}(\nu - \nu_b) \,  e^{2 \pi i \nu t_A} d\nu \; ,
\end{align}
where we have inserted the definition of $x_i[t_{A+a}]$ in terms of its continuous Fourier transform to show how the measured frequency channels relate to the underlying continuous frequency signal. As expected, to gain higher spectral resolution we simply need to increase the length of the time chunks being transformed. The frequency channel shape is controlled by
\begin{equation}
    \tilde{w}(\nu) = \sum_a w[a] \exp{\lp -2 \pi a \nu \Delta{t}\rp} \; ,
\end{equation}
the Discrete Fourier Transform of the window function $w[a]$ which is typically of width $\sim 1/T$. To choose the window function we need to trade-off sensitivity against localizing the response (i.e. keeping $\tilde{w}(\nu)$ compact around $\nu = 0$), which is important for mitigating RFI (see \secref{sec:RFI}). In practice most interferometers use Polyphase Filter Banks \citep{Price2016} rather than a Short-Time Fourier Transform which combines multiple time chunks (called \emph{taps}). This gives even better localization of the spectral response, but requires more computational power, and introduces slight correlations between spectra close in time. For modern instrumentation the channelization is typically performed by Field Programmable Gate Arrays (FPGAs), integrated circuits that can be programmed at the hardware level \citep{2016JAI.....541005B,2016JAI.....541001H}.

\subsubsection{X-engine}

The role of the X-engine is to take the frequency channelised data and cross correlate all inputs. However, the output of the F-engine is a time series of spectra for each individual input, but for the cross correlation we need to have access to the data for all inputs at the same frequency. This requires a large transpose of the dataset from frequency-ordered to input-ordered. This is a significant bottleneck for large interferometers, and usually requires a combination of fast mesh networks and large pools of memory to perform \citep{Lutomirski2011,Bandura2016}.

After the transpose, all that remains is to calculate the visibilities. This is done by exploiting the separation of timescales and calculating the sample covariance of the frequency channels. By doing this we are forming an estimate of the underlying, true covariance. As it is an estimator, for the moment we write the visibilities as $\hat{V}_{ij}$ given by
\begin{equation}
    \hat{V}_{ij}[\nu_b] = \frac{1}{M} \sum_A \tilde{x}_i[\nu_b; t_A] \tilde{x}_j^*[\nu_b; t_A] \; ,
\end{equation}
though for notational convenience we will just use $V_{ij}$ in subsequent sections. A typical spectral resolution of an interferometer is around $100\,\textrm{kHz}$, meaning we get a new spectrum every $10\,\mu\textrm{s}$. If we average these up to the timescale on which we expect the signal to change by Earth rotation, this implies that we sum over $M \gtrsim 10^5$ spectra. Computationally this step is extremely challenging for large interferometers, requiring $\sim \Delta{\nu} N^2$ operations per second, but is generally easy to parallelize. This is usually performed on FPGAs or GPUs \citep{2014JAI.....350002K,2015arXiv150306202D}.

The quantity $\hat{V}_{ij}[\nu_b]$ can be thought of as an estimator for the underlying ``true'' visibilities\footnote{Technically the estimator has a complex Wishart distribution, that is $(M\, V_{ij}[\nu_b]) \sim \mathcal{W}_C(\bar{V}_{ij}[b], M)$, but for typical interferometers $M$ is large enough that we can treat its statistics as Gaussian.} with expectation
\begin{equation}
   \bar{V}_{ij}[\nu_b] = \int \lv \tilde{w}(\nu - \nu_b) \rv^2 \ls S_{ij}(\nu) + N_{ij}(\nu) \rs d\nu \; ,
\end{equation}
where the noise covariance $N_{ij}(\nu) = \la n_{i}(\nu) n_j^*(\nu) \ra$. The covariance of the estimator $\hat{V}_{ij}$ is
\begin{align}
    \label{eq:sample_noise}
  \Cov{(\hat{V}_{ij}, \hat{V}_{kl})} & = \Cov{(\hat{V}_{ij}, \hat{V}_{lk}^*)} \notag \\
  & = \frac{1}{M} \bar{V}_{ik} \bar{V}_{lj} \; .
\end{align}
\Cref{eq:sample_noise} gives an exact and compact description of the noise of an interferometer. However, we are often in a regime (particularly for low redshift \tcm observations) where $S_{ij} \ll N_{ij}$, and as we can often assume the noise is uncorrelated between feeds $N_{ij} \approx T_\text{recv}^i \delta_{ij}$, we find the standard result that the noise on an interferometric visibility is
\begin{equation}
\label{eq:noise_approx}
  \Var{(\hat{V}_{ij})} = \frac{T_\text{recv}^i T_\text{recv}^j}{M} \; ,
\end{equation}
with distinct visibilities being uncorrelated.

\section{Instrumental Systematics}
\label{sec:InstrumentSystematics}

Modern radio telescopes are complex instruments. Using our data to map the HI distribution in the Universe requires that we first understand how our data is generated by emission from the sky. This requirement is compounded by the presence of extremely bright astrophysical emission (relative to the \tcm signal), that mean we must control and understand our instrument well enough to prevent proliferation of these foreground modes.

Below we talk about the various types of systematics effects generated by radio interferometers. For each of these instrumental properties there is some fraction that we can understand precisely, and some part that we cannot determine. The parts we understand can in principle be added to our analysis, and their effects removed in an optimal manner, though in practice this may be computationally challenging. Even if we can optimally treat the systematic, we may suffer a loss of sensitivity, one which can only be rectified by changing the instrument itself. The parts we do not understand are far more dangerous, as they lead to uncorrected mode proliferation, and thus bias in our results. The key challenge of any \tcm observation program is to devise analysis methods which minimise exposure to undetermined instrumental effects, and, constrain these instrumental unknowns such that they do not appreciably bias our results.

As a rule of thumb for thinking about instrumental systematics, when a systematic needs to be understood for further analysis it must be understood better than the \tcm signal to astrophysical foreground ratio (i.e. the S/F ratio), to avoid leaking foregrounds into our data. This varies depending on where on the sky we are looking and at what frequency band, but is often around $\lesssim 10^{-4}$.

\subsection{Gain Variations}
\label{sec:sys/gains}

Converting the signal arriving at the antenna into one which can be digitized is the job of the analog chain (\secref{sec:analogchain}) and requires a variety of components which amplify, filter and transport the signal. The effect of the analog chain can be described by a \emph{complex} frequency-dependent gain $g(\nu)$ that multiplies each of the received signals.

Generally we can think of the amplitude as describing how much the signal has been amplified by the chain, and the phase describes how it has been been delayed by components along the chain. These are free functions of frequency, though amplitudes are often smooth, and the phase behaviour may be dominated by a single time delay for all frequencies. We should note that absolute phases are typically meaningless, so the phase typically refers to how much one input or frequency has been delayed relative to the others. Learning what these gains are, and thus how to relate our received signal to the true signal from the sky is the process of \emph{calibration} that we discuss in \secref{sec:Calib}.

Gains are typically multiplicative scalars. That is for some true signal $s(\nu)$ the observed voltage $x(\nu) = g(\nu) s(\nu) + \text{noise}$ where the overall gain $g(\nu)$ is the product of all the individual component gains, e.g. $g(\nu) = g_\text{filter}(\nu) \, g_\text{cable}(\nu) \,  g_\text{amplifier}(\nu)$.

Unfortunately, the gain for each signal chain is time dependent. There are many possible sources for this, but the primary source is the thermal environment of the analog chain, two examples of this are:
\begin{itemize}
    \item The level of amplification of an amplifier depends on its temperature. This is typically at the $\sim0.5\%$ per K level, and so over the course of a day we may expect $\sim 5\%$ variations in the amplification along any signal chain \citep{Davis_2012}.
    \item Cables cause both variable attenuation of the signal as the material properties change with temperature, and phase variations due to thermal expansion of the cable increasing the propagation time along it.
\end{itemize}
However this is made even more problematic by the fact that different components experience slightly different thermal environments (are they exposed to the sun? The wind? Are they heated internally?), and also have varying susceptibilities to the temperature they experience. This means the variations in the gain of each signal chain are time variable, (partially) independent, and difficult to predict.

The effect of these gain variations is to modify the measurements we make. Known gains are trivial to correct, as we can divide each signal by its known gains. However, unknown gains limit in the degree to which we can combine different baselines and frequencies to make maps and remove foregrounds with out biasing our results.

It is often useful to divide the gains up into three distinct terms for each signal chain
\begin{equation}
    \label{eq:gain_split}
    g_i(\nu, t) = g_\text{band}(\nu) \: g_i(\nu) \: \lp 1 + \delta{g}_i(\nu, t) \rp
\end{equation}
where
\begin{itemize}
    \item $g_\text{band}(\nu)$ is the \emph{bandpass} of the telescope. A real, time-independent function describing the sensitivity of the whole telescope to each frequency.
    \item $g_i(\nu)$ is a time-independent function, that encodes the input dependent gain applied to the instrument.
    \item $\delta{g}(\nu, t)$ describes time variation of the gains of each channel.
\end{itemize}

One further complexity is that gains are not always scalar quantities. We may have linear mixing between signal paths, generally referred to as \emph{cross talk}, that must be represented by a complex matrix rather than a scalar. This can happen anywhere that the signal paths are electrically ``close'', for instance between polarizations on an antenna \citep{Hamaker1996}, or between inputs on the same analog-to-digital-converter (ADC) chip. This does not fundamentally alter the picture above, but does make correcting for it harder.

\subsection{Primary Beams}

\label{sec:SystematicsPrimaryBeams}

To fully understand the measurements we are making we must understand the primary beams of our antennas $A_i(\vrhat, \nu)$, that describe the spatial and frequency sensitivity of the antennas in the telescope. These directly appear in the measurement equation \cref{eq:vis_full}.

Beams are typically frequency dependent, with complicated spatial structure and polarization response \citep{2017JAI.....650003C,2016ApJ...826..199N,2016ApJ...831..196E,Patra2018HERAdish}, and while some of these effects can be minimised through specific design decisions, they cannot be removed entirely. Additionally, the beam function can vary significantly from feed to feed due to manufacturing/installation errors (e.g. deformations in reflectors, non-identical antennas) and positional effects (e.g., whether the antenna is at the edge of the array or the centre).

As we will discuss later (\secref{sec:FgMitigation}), different analysis techniques place different restrictions on the knowledge and properties of the primary beams. Some technique will internally make linear combinations of different visibilities in order to deconvolve instrument effects, and these are typically very sensitive to our knowledge of the primary beams, requiring a sensitivity similar to the S/F level (see \secref{sec:fgmodeprojection}). There are other techniques (see \secref{sec:wedge}) which avoid combining baselines, and though this has lower requirements on the knowledge of the primary beams, it does place stronger requirements on their properties, in particular their frequency smoothness and polarization purity (see the next section).

An additional systematic effect introduced by primary beams are feed to feed variations, which can introduce non-redundancy on otherwise redundant arrays. This can cause an indirect systematic effect whereby the use of redundant calibration techniques (see \secref{sec:Calib}) can mix the beam non-redundancy into incorrect gain solutions \citep{Orosz2019}.

\subsection{Polarization leakage}
\label{sec:polarization_leakage}

With rare exceptions such as the Ooty Wide Field Array \citep{2017JApA...38...10S}, almost all radio interferometers are constructed from pairs of co-located antennas with orthogonal feed polarizations, generally dipoles oriented at right angles. This gives us sensitivity to the polarization of emission on the sky. However, this sensitivity is imperfect, and can lead to a fundamental entangling of our data where we can't completely separate out the various polarized components. Although within intensity mapping we are generally interested only in the total intensity of the \tcm emission (i.e. Stokes $I$), to remove foregrounds we need to rely on their spectral smoothness (as we will see in \secref{sec:fgs} and \secref{sec:FgMitigation}) and although this is a very good approximation in total intensity they may have significant spectral structure within their polarized emission (see \secref{sec:fgs}). If we cannot disentangle the polarized components we may misestimate the true total intensity, introducing spurious spectral structure and hindering our ability to remove foregrounds.

To see how this entanglement can occur, let us look at constructing the instrumental Stokes I visibility for some baseline. For our telescope, at every location we have two orthogonal dipoles (labelled $X$ and $Y$), which have nearly orthogonal response on the sky around the centre of the beam (this is typical of most telescopes). To construct the Stokes I visibility $V_T$ we form
\begin{align}
    \label{eq:polvis1}
    V_T & = \frac{1}{2}\ls V_{XX} + V_{YY} \rs \\
    & = \frac{1}{2} \int \ls A_X^a(\vrhat) A_X^{b *}(\vrhat) + A_Y^a(\vrhat) A_Y^{b *}(\vrhat) \rs \nonumber \\ &\qquad \times T_{ab}(\vrhat) \: e^{-2 \pi i \vrhat \cdot \vu} \: d\Omega \; .
\end{align}
Let us write the primary beam of each of our feeds as a complex scalar weight multiplied by a unit vector giving the direction of response. That is $A_X^a(\vrhat) = A_X(\vrhat) \hat{x}^a$ and similarly for the $Y$ polarization. Note that $\hat{x}^a$ may have complex components. Using this we can rewrite \cref{eq:polvis1} as
\begin{multline}
    \label{eq:polvis2}
    V_T = \int \bar{A}^2(\vrhat) T(\vrhat) e^{-2 \pi i \vrhat \cdot \vu} \: d\Omega \\ +\frac{1}{2} \int (A_X(\vrhat)^2 \hat{x}^a \hat{x}^{b *} + A_Y(\vrhat)^2 \hat{y}^a \hat{y}^{b *}) P_{ab}(\vrhat) e^{-2 \pi i \vrhat \cdot \vu} \: d\Omega \; .
\end{multline}
where we have defined the polarization averaged beam $\bar{A}^2(\vrhat) = (A_X^2(\vrhat) + A_Y^2(\vrhat)) / 2$, and $P_{ab}(\vrhat)$ is the polarized part of the coherency matrix, i.e.
\begin{equation}
    P_{ab}(\vrhat, \nu) = \begin{pmatrix} Q & U + i V \\ U - i V & - Q \end{pmatrix} \; .
\end{equation}
The first term of \cref{eq:polvis2} is equivalent to the usual unpolarized formalism, whereas the second term describes the leakage of the polarized components into this.

To gain more insight into the source of this leakage, let us look at the effect of small deviations from a perfect system with identical beam amplitudes ($A_X = A_Y$) and orthogonal real polarization orientations across the beam. In this case we can treat the difference in amplitudes $\Delta{A}^2(\vrhat) = A_X^2(\vrhat) - A_Y^2(\vrhat)$ and $\mu(\vrhat) = \vxhat \cdot \vyhat^* = \mu_r(\vrhat) + i \mu_i(\vrhat)$ as small parameters, and expand \cref{eq:polvis2} to linear order. Using this the leakage of polarized intensity into instrumental total intensity $V_{P\rightarrow T}$ is
\begin{multline}
    V_{P\rightarrow T} =  \frac{1}{2} \int e^{-2 \pi i \vrhat \cdot \vu} \, \bigl[ \Delta{A}^2(\vrhat) (\hat{x}^a \hat{x}^b - \hat{y}^a \hat{y}^b) \\ + \bar{A}(\vrhat)^2(\mu(\vrhat)^* \, \hat{x}^a \hat{y}^{b^*} + \mu(\vrhat)\, \hat{y}^a \hat{x}^{b *})\bigr] P_{ab}(\vrhat)  \: d\Omega \; .
\end{multline}
If we the axes of our order polarization basis are are aligned with respect to the zeroth order antenna polarization directions $\vxhat_0$ and $\vyhat_0$ this can be written as
\begin{multline}
    V_{P\rightarrow T} = \int \bigl[ \Delta{A}^2(\vrhat) Q(\vrhat) + \mu_r(\vrhat) \, \bar{A}(\vrhat)^2 U(\vrhat) \\ + \mu_i(\vrhat)\,\bar{A}(\vrhat)^2 V(\vrhat)\bigr] e^{-2 \pi i \vrhat \cdot \vu} \: d\Omega \; .
\end{multline}
showing that in this basis, leakage of feed oriented $Q$ into $I$ is caused by a mismatch in the beam amplitudes whereas leakage of $U$ is caused by real non-orthogonality in the beam responses. Stokes $V$ can leak into instrumental $I$ whenever there is a delay (i.e non zero phase) between the overlap of the two polarisations, though as the sky has very little Stokes V polarization, this term is small. The reverse is also true, in that leakage from $I$ into $V$ is expected to be small, although in practice some leakage is often seen due to calibration errors \citep{Kohn2018}. Assuming no such errors, however, Stokes V measurements are therefore often used as a proxy for noise levels in an observation, provided one can further make the (fairly reasonable) assumption that there are relatively few intrinsically circularly polarized sources on the sky \citep{Patil2017,Gehlot2018}


More generally, we can see that because the polarization leakage terms $\Delta{A}(\vrhat)$ and $\mu(\vrhat)$ are functions of angular position on the sky, we can never construct a linear combination of the polarizations for a visibility that measures only the unpolarized sky. Polarization leakage can only be corrected by combining visibilities from \emph{different} baselines together in a way which allows us to discriminate between different angular positions.

\subsection{Noise Systematics}

Although instrumental noise is typically thought of as a random error, and in a sense ``easy'' to deal with, in a real life radio interferometer, there are still effects which must be understood.

One complication arises from the use of closed packed arrays to maximise sensitivity (\secref{sec:InterferometerSensitivityOptimization}). This leads to many antennas which are close to each other, and may be able to interact directly with each other. In this scenario instrumental noise which is generated within one signal chain can be broadcast out of its antenna and received by a neighbour. This gives a source of noise which is correlated between antennas, leading to the visibilities having a bias due to the correlated noise. This effect tends to diminish as the feed separation increases, but can be problematic for the shortest baselines that are sensitive to the largest scales.

One approach that can work to remove this contribution is to take advantage of the fact that the sky changes over the day and remove an appropriate average of each visibility in order to remove the noise bias. However, this is only as good as the noise properties are stable: like the gains discussed in \secref{sec:sys/gains} the noise level is sensitive to the changing environment of each analog chain and my have substantial variation. Additionally, subtracting the average will also excise the average signal, and this is important to account for (see \secref{sec:sigloss}).


\subsection{A summary of interferometry}

Summarizing the last few sections, we see that individual antenna elements of an interferometer directly sample incident electric fields from the sky, and when pairs of measurements from different antennas are correlated with one another (forming a \emph{baseline} of two antennas), one obtains a \emph{visibility}. The visibility is the fundamental quantity probed by an individual baseline of an interferometer. Ignoring the complications of polarization leakage, the Stokes I visibility is given by
\begin{equation}
\label{eq:SimpleVis}
V_T = \int \bar{A}^2(\vrhat) T(\vrhat) e^{2 \pi i \vrhat \cdot \vu} \: d\Omega,
\end{equation}
where we have replaced the Stokes $I$ intensity with the temperature units used in Section \ref{sec:BasicObs}, assuming that any relevant multiplicative constants have been applied. Under the flat sky approximation, we have
\begin{eqnarray}
\vrhat &=& \left( \theta_x, \theta_y, \sqrt{1 - \theta_x^2 - \theta_y^2} \right)\approx  \left( \theta_x, \theta_y, 1 \right)
\end{eqnarray}
to linear order, where $\theta_x$ and $\theta_y$ are angular coordinates on the sky defined relative to some arbitrary axes centred on an origin known as the phase centre. This then gives (up to some arbitrary constant phase terms)
\begin{equation}
\label{eq:FTvis}
V_T = \int \bar{A}^2(\vtheta) T(\vtheta) e^{2 \pi i \vtheta \cdot \vu_\perp} \: d^2 \theta,
\end{equation}
where $\mathbf{u}_\perp$ is the projection of $\vu$ onto a plane perpendicular to the vector pointing to the phase centre. If we define the phase centre to be the zenith, then this is precisely Equation \eqref{eq:BasicVis}, up to the definition $A_p \equiv \bar{A}^2$ to relate the power beam to the electric field beams. We have thus returned to our starting point in our discussion of interferometry, and the most important piece of intuition continues to be that a visibility is roughly a measure of a Fourier mode of the sky. However, we are now equipped with a better appreciation for the approximations that are required for this to be true, as well as some of the systematic challenges with \tcm cosmology instrumentation. These will all inform the analysis methods that we discuss later in this paper.

\section{Foregrounds}
\label{sec:fgs}
One of the most formidable obstacles to a successful measurement of the \tcm signal is the issue of contamination by astrophysical foregrounds and how they interact with one's instruments. These foregrounds consist of all sources of radio emission within our observing band except for the cosmological \tcm signal of interest. Of course, these signals may be of significant scientific interest in their own right---for instance, data from many \tcm instruments can be used to map the Galaxy (e.g., \citealt{Wayth2015GLEAM,LOFAR2016Survey,Hurley-Walker2017GLEAM,2017MNRAS.472..828S,2017MNRAS.465.3163S}), which is currently not well-surveyed at low frequencies---but, for our present purposes, any emission that is not the \tcm signal can be considered a foreground contaminant, and must be suppressed or removed.

\begin{figure*}[t]
\centering
\includegraphics[width=0.90\textwidth,trim={0cm 0cm 0cm 0cm},clip]{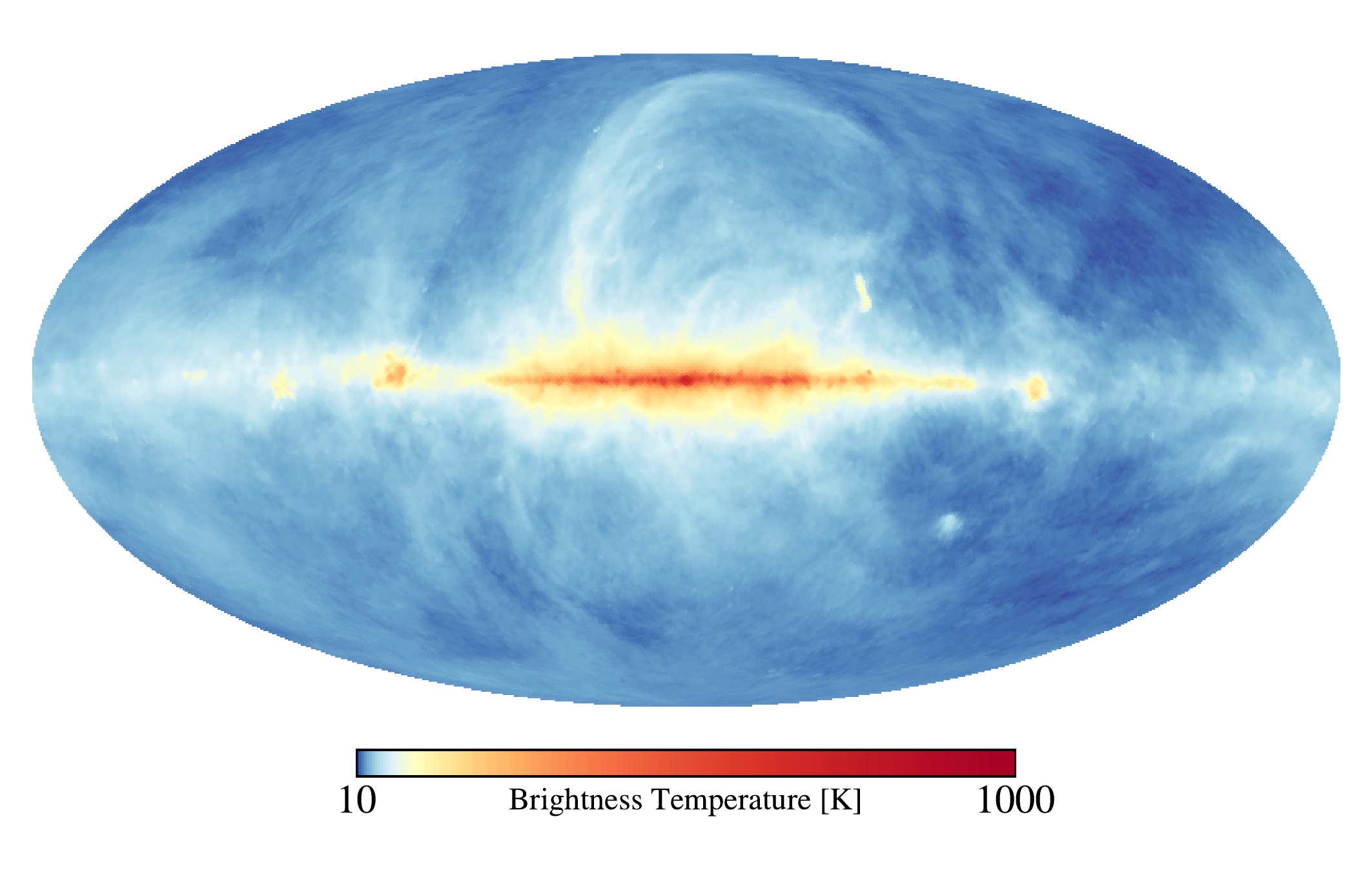}
\caption{The $408\,\textrm{MHz}$ Haslam map \citep{HaslamBackground,HaslamMap,HaslamUpdate} of diffuse synchrotron radio emission in our galaxy. Synchrotron emission is a dominant source of foreground contamination for cosmological \tcm measurements. The emission is clearly brightest in the galactic plane, but the brightness of the foregrounds ($\sim 10$s to $100$s of Kelvin at $408\,\textrm{MHz}$ and stronger at lower frequencies) is much greater than the cosmological \tcm signal everywhere on the sky.}
\label{fig:haslam}
\end{figure*}

The mitigation of foregrounds is not an optional exercise. This is clear from a simple consideration of how much bright the foregrounds are. In Figure \ref{fig:haslam}, we show a map of Galactic radio emission at $408\,\textrm{MHz}$. At such frequencies, the emission is dominated by Galactic synchrotron radiation, and a glance at the typical brightness temperature reveals that foreground contaminants enter at $\sim100$s of K. Towards lower frequencies, synchrotron emission gets brighter, and it is a reasonable approximation to model the spectral dependence as
\begin{equation}
\label{eq:PhenoSynchSpec}
T(\nu) \propto \left( \frac{\nu}{\nu_0} \right)^{-\alpha_\textrm{syn} - \Delta \alpha_\textrm{syn} \ln ( \nu / \nu_0 )},
\end{equation}
where $\nu$ is the frequency, $\alpha_\textrm{syn} \approx 2.8$ is the spectral index, $\Delta \alpha_\textrm{syn} \approx 0.1$ is the running of this spectral index, and $\nu_0$ is an (arbitrary) pivot frequency that is degenerate with the overall amplitude of emission in a particular pixel \citep{Wang2006}. In addition to synchrotron radiation, there are several other important sources of foregrounds. These include free-free emission, bright radio point sources, and unresolved point sources, with dust emission generally negligible at the low frequencies of interest for \tcm cosmology \citep{Tegmark2000foregrounds,Planck2016X,Planck2018IV}. While many of these emission mechanisms may be subdominant to Galactic synchrotron radiation at the relevant frequencies, they are nonetheless important contaminants to consider, as they can still be much brighter than the cosmological signal. For reference, experiments targeting the EoR are aiming to measure cosmological signals on the order of $\sim 10$s of mK, while those targeting lower redshifts seek to measure a $\sim 0.1$ mK signal. At higher frequencies, the latter has dimmer foregrounds, but in both cases, the foreground-to-signal ratios are a formidable $\sim 10^5$ in temperature.

In some respects, \tcm observations are similar to observations of the CMB in that the removal of foregrounds is a prerequisite for precision measurements. However, the foreground challenge in \tcm is different from that of the CMB in two important ways. First, the aforementioned foreground-to-signal ratio is much worse for \tcm cosmology. With the CMB, the cosmological signal is in fact the dominant source of emission far away from the Galactic plane (e.g., at the Galactic poles). Thus, exquisite foreground removal is needed only for precision measurements and the desire to map large portions of the sky, and not for a simple detection. With \tcm cosmology, foregrounds dominate in all parts of the sky, and robust removal strategies are required even for a detection-level measurement.

The second crucial difference between foreground removal in the CMB and \tcm cosmology comes from the fact that the former is mostly a measurement of angular anisotropies, while the latter aims to make three-dimensional tomographic maps. Although CMB observations are beginning to target spectral distortions from a pure blackbody spectrum (e.g., with thermal Sunyaev-Zel'dovich measurements and future experimental proposals for measuring $\mu$- or $y$-type distortions; e.g., \citealt{Chluba2018}), most CMB constraints to date have relied on measurements of angular anisotropies. When measuring angular anisotropies, the different frequency channels of a CMB observation simply provide redundant information: the cosmological signal follows a blackbody spectrum while the foregrounds do not. Thus, extracting the cosmological signal amounts to discarding non-blackbody contributions to the measurement, and the redundancy allows (in principle) a perfect separation of foregrounds from the cosmological signal provided. This redundancy is lost in \tcm cosmology because every frequency channel of an observation carries unique cosmological information, since they each correspond to a different redshift. Thus, unlike with the CMB, there is no mathematical method to solve for the cosmological signal self-consistently from the data. With $N$ frequency channels, $N$ pieces of cosmological information to solve for, and (up to) $N$ foreground degrees of freedom to simultaneously constrain, one has too many variables for too few measurements.

To successfully eliminate foreground contaminants, then, requires reducing the number of variables that one needs to solve for. One solution, in principle, is to simply eliminate the need to solve for any of the foreground degrees of freedom by having exquisitely precise models of the foregrounds that can be directly subtracted from the raw data. Essentially, one would then be able to devote all $N$ frequency channels to solving for $N$ cosmological degrees of freedom. In practice, it is impossible to write down an \emph{a priori} foreground model that is precise to the $10^5$ dynamic range between foregrounds and cosmological signal. It thus follows that one cannot hope to constrain all $N$ modes of the cosmological signal; instead, there will in general be some modes where foregrounds will dominate, with the cosmological signal irretrievably buried within the bright contaminants. Foreground mitigation is therefore tantamount to identifying some way in which the foregrounds can be robustly sequestered to a limited set of modes.

In Section \ref{sec:FgMitigation}, we provide a detailed discussion of strategies that have been proposed for foreground mitigation. Here, we provide a quick qualitative discussion of how one might go about sequestering foregrounds. One popular proposal is to consider the spectral properties of foregrounds. As one sees from Equation \eqref{eq:PhenoSynchSpec}, the spectrum of Galactic synchrotron emission is a relatively smooth function of frequency. Other sources of foregrounds are expected to behave in a reasonably similar way \citep{Tegmark2000foregrounds,Planck2016X,Planck2018IV}. Decomposing spectral data into modes that vary rapidly with frequency and modes that vary slowly with frequency might then allow foregrounds (which dominate the latter) to be separated from the cosmological signal (which dominate the former). The foreground modes are then simply discarded, while the cosmological modes are retained for further analysis.

The use of spectral smoothness to enact foreground separation is a promising approach, and indeed, it is the central assumption of most foreground mitigation methods described in Section \ref{sec:FgMitigation}. However, there are two key caveats that are important to bear in mind. First, note that formulae such as Equation \eqref{eq:PhenoSynchSpec} are merely phenomenological fits, and at the high precision necessary for \tcm cosmology, there may be non-smooth components to foreground emission. This is certainly a possibility that one should consider, particularly since future datasets may provide evidence suggesting this. But for now, empirical models suggest that foregrounds are relatively simple both spectrally and spatially \citep{GSM,Liu:2012,Zheng:2017}. In addition, studies of physical models for the underlying emission processes come to similar conclusions \citep{GMOSS2017a,GMOSS2017b}. For example, \citet{Petrovic2011} and \citet{Bernardi2015} find that with reasonable assumptions regarding the energy distribution of the electrons powering synchrotron radiation, it is hard to produce spectra with large curvature in frequency. This conclusion is partly driven by the fact that the total observed synchrotron spectrum is the convolution of the electron energy distribution and the spectrum of each electron in the distribution. This convolution serves to further smear out spectral features. Similar conclusions can be made for the sea of unresolved extragalactic point sources \citep{Liu:2012}. To be fair, there do exist some sources of foregrounds that are not spectrally smooth. Radio recombination lines are a prime example of this. However, both \citet{OhMack2003} and \citet{Petrovic2011} find that they are unlikely to be bright enough to be of concern.

We note that foregrounds, particularly the diffuse synchrotron emission from our own galaxy, can be highly polarized \citep{Kogut2007} with the direction of the polarization related to the local magnetic field. As this emission propagates through the magnetized ISM towards us, it is subject to Faraday rotation which rotates the polarization vector $\propto \textrm{RM}\,\lambda^2$, where $\textrm{RM}$ is the rotation measure along the line of sight. \cite{Oppermann2012} find that typical values of $\textrm{RM}$ can be $\sim$ \SIrange{1}{100}{\radian \metre^{-2}} which corresponds to a full rotation of the polarization vector within $\sim \SI{10}{\mega\hertz}$ for frequencies around \SI{300}{\mega\hertz}. This means that there is tremendous spectral variation in the polarized intensity signal which has no counterpart in the total intensity \citep{2019AJ....158...44W}. In combination with polarization leakage within in the instrument (see \secref{sec:polarization_leakage}) this may irreversibly contaminate the observed total intensity with spectral structure from the polarized emission, limiting our ability to remove foregrounds.

Another complication with using spectral smoothness to sequester foregrounds is that observationally, one sees these foregrounds through one's instrument. Thus, even if the intrinsic properties of the foregrounds are such that they can be limited to a few smooth spectral modes, instrumental effects can cause \emph{mode proliferation}, where foregrounds leak into modes where they might not be expected \citep{Switzer:2014}. This is of serious concern. In Section \ref{sec:InstrumentSystematics}, we discussed the systematics that may give rise to such leakage, and we will provide a few suggestions for how to deal with some of them in later sections. It would be fair to say, however, that the interplay of foreground contamination and instrumental systematics is currently the chief obstacle in \tcm cosmology, with the strict requirements on calibration, for example, driven by the high dynamic range between foregrounds and the cosmological signal. This is a problem that has yet to be solved definitively, and the near-term progress of the field will be intimately tied to progress on this front.

\section{Interlude: from our Universe to visibilities and back}
\label{sec:Interlude}
In the preceding sections, we have described how radiation from our Universe (encoding information about fundamental physics as well as astrophysics) is transformed by an interferometer into a set of visibilities. We have also discussed how these visibilities are additionally corrupted by foreground contaminants and instrumental systematics. In the next few sections, our goal is to undo the aforementioned processing, in order to extract the scientific insights that are encoded by our Universe in the \tcm line. Section \ref{sec:Calib} describes the problem of calibrating one's interferometer. Once calibration is complete, one may choose to make a map of the sky. This process is described in Section \ref{sec:mapmaking}, although power spectrum estimation can occur with or without a preceding map-making step, as we discuss in Section \ref{sec:pspecestimation}. Section \ref{sec:FgMitigation} examines the mitigation of foreground contamination. While many foreground suppression algorithms operate on the data \emph{prior} to power spectrum estimation, we choose to describe power spectrum estimation first because it often informs how one goes about removing foregrounds. The process of going from power spectra to constraints on theoretical models is described in Section \ref{sec:MCMC}, and a discussion of statistics beyond the power spectrum follows in Section \ref{sec:BeyondPspec}.

\section{Calibration}
\label{sec:Calib}

As discussed in Section \ref{sec:InstrumentSystematics}, in a real instrument one does not necessarily measure ideal visibilities as described by (for example) Eq. \eqref{eq:BasicVis}. Instead, the visibility $V^\textrm{meas}_{ij}$ measured by a baseline consisting of the $i$th and $j$th antennas might be given by
\begin{equation}
\label{eq:CalibEqn}
V^\textrm{meas}_{ij} = g_i g_j^* V^\textrm{true}_{ij} + n_{ij},
\end{equation}
where $g_i$ and $g_j$ are complex-valued gain factors that are unknown \emph{a priori}, $n_{ij}$ is the noise on this baseline, and $V^\textrm{true}_{ij}$ is the true visibility that we would measure with a perfectly calibrated noiseless interferometer. (For notational cleanliness, for most of this section we avoid writing quantities as functions of frequency and time, even though the dependence is always implicitly there. We explicitly address calibration as a function of frequency and time in Section \ref{sec:CalibFreqTime}.) It is $V^\textrm{true}_{ij}$ that we ultimately want, but to get to it, we need to somehow solve for the gain/calibration factors $g_i$ and $g_j$. Doing so is what we mean by calibration.

In its most general form, we cannot solve Eq. \eqref{eq:CalibEqn} for what we want. With $N_\textrm{ant}$ antennas, there are $N_\textrm{ant} (N_\textrm{ant}-1)/2$ baselines from which we measure visibilities (excluding baselines of length zero, i.e., those formed by forming the autocorrelation of an antennas with itself, as discussed in Section \ref{sec:dish_or_inter}). But there are also $N_\textrm{ant} (N_\textrm{ant}-1)/2$ true visibilities that we would like to solve for along with $N_\textrm{ant}$ gain factors. We thus have more variables to solve for than we have measurements with which to constrain them, which is of course mathematically impossible.

To proceed, we must back away from the most general form of Eq. \eqref{eq:CalibEqn} and make assumptions about the sky, our instrument, or both. In the next few sections, we explore various options for making such assumptions.

\subsection{Calibration by fitting to a sky model}
\label{sec:skycal}
Consider first the option of making assumptions about the sky. Suppose, for example, that the sky consisted of just a single bright point source with brightness $I_0$. If we work under the flat-sky approximation to simplify this toy example, then the true visibility $V^\textrm{true} (\mathbf{b})$ measured by a baseline $\mathbf{b}$ is
\begin{equation}
V^\textrm{true} (\mathbf{b}) \propto I_0 \exp\left( -i 2\pi \frac{ \mathbf{b} \cdot \boldsymbol \theta_0 }{\lambda} \right),
\end{equation}
which we obtained by inserting a Dirac delta function centred on source position $\boldsymbol \theta_0$ into the $I(\boldsymbol \theta)$ term of Eq. \eqref{eq:BasicVis}. Using the fact that the baseline vector $\mathbf{b}$ connecting the $i$th and $j$th antennas (located at $\vx_i$ and $\vx_j$, respectively) can be written as $\vb = \vx_i - \vx_j$ we have
\begin{equation}
V^\textrm{true}_{ij} \propto I_0 \exp\left[ -i 2\pi \frac{ (\vx_i - \vx_j) \cdot \boldsymbol \theta_0 }{\lambda} \right],
\end{equation}
and inserting this into Eq. \eqref{eq:CalibEqn}, we obtain
\begin{equation}
V^\textrm{meas}_{ij} = g_i g_j^* I_0 + n_{ij},
\end{equation}
where we have absorbed the phase factors into $g_i$ and $g_j$. With this restricted form of the equation, we have $N_\textrm{ant} (N_\textrm{ant}-1)$ measurements (treating the real and imaginary parts of the visibilities as separate measurements) but only $2N_\textrm{ant}+1$ numbers to solve for (real and imaginary parts of our $N_\textrm{ant}$ complex gains plus the amplitude of our single bright point source). Our calibration problem has now become possible to solve.

In reality, the sky does not consist of a single bright point source and nothing else. For this to even be a good approximation, one must observe a bright source using an instrument with a field of view that is narrow enough to exclude other bright sources. Moreover, the angular resolution of the instrument must be sufficiently low for the source to be a point source, and not an extended source whose spatial structure is resolved. While the angular resolution requirement might in some cases be satisfied given the low angular resolution of compact interferometers that are used for $21\,\textrm{cm}$ cosmology, many of these instruments have fields of view that are too wide to enable an easy isolation of a single point source.

Despite its shortcomings, our toy single-point-source model illustrates the general principle that if the number of degrees of freedom describing the sky are limited, then it is possible to solve for calibration parameters. Generalizing our model slightly, one can imagine that there are sky models that are more complicated than just a single point source, but are still simple enough for calibration. For example, one might model the sky as a sum of many point sources of varying brightness. Calibration then amounts to minimizing the quantity
\begin{equation}
\label{eq:CalibChiSq}
\chi^2 =\sum_{i=1}^{N_\textrm{ant}} \sum_{j=i+1}^{N_\textrm{ant}} \frac{ |V^\textrm{meas}_{ij} - g_i g_j^* V^\textrm{model}_{ij}|^2}{\sigma_{ij}^2},
\end{equation}
where $\sigma_{ij}^2$ is the noise variance on the baseline formed from the $i$th and $j$th antennas, and $V^\textrm{model}_{ij}$ is a simulated visibility for this baseline based on our sky model. A common way to minimize the $\chi^2$ shown here is to use an iterative approach. One begins by assuming various values for the free parameters of one's sky model. This then allows the simulation of ideal visibilities $V^\textrm{model}_{ij}$. With these visibilities in hand, Eq. \eqref{eq:CalibEqn} can be used to solve for the complex gains (ignoring the noise term, which we cannot do anything about, except perhaps to average it down in time; see Section \ref{sec:CalibFreqTime}). The complex gains can then be used to solve for an improved set of ideal visibilities, again using Eq. \eqref{eq:CalibEqn}. The entire process can then be iterated, going back and forth between solving for calibration parameters and sky model visibilities until convergence. Note that because this is a non-linear minimization problem, there is the potential for noise to bias the results, in the sense that time-averaged solutions need not necessarily converge to the correct answer. Having said that, if the starting point of the iterations are relatively close to the truth, there is a reasonable expectation that the bias will be small. These issues are discussed in \citet{Liu2010}, albeit in the context of a different calibration strategy (the redundant calibration algorithm described in Section \ref{sec:RedCal}), but the ideas are similar.

Naturally, the more realistic a sky model, the better one's calibration solution \citep{Patil2016}. In pipelines like those described in \citet{sullivanFHD}, \citet{Jacobs2016comparison}, \citet{Line2017PUMA}, and \citet{Noordam2004LOFARCalibration}, high-quality point source catalogs (e.g., \citealt{Carroll2016}) are routinely used to model the sky. A key challenge in this area is that catalogs will never be complete, for there will always exist many faint sources that are unaccounted for. This has been shown to result in calibration errors with artificial spectral features that make foreground mitigation difficult \citep{Barry2016Calibration}. (See Sections \ref{sec:fgs} and \ref{sec:FgMitigation} for extended discussions of the foreground problem). Another challenge is the inclusion of model components beyond point sources, such as diffuse Galactic emission. One potential solution to this problem is to use \emph{pulsar gating}, as has been demonstrated in GMRT  measurements \citep{Pen2009GMRT,Paciga:2011}. The key idea is to difference two observations that are slightly separated in time, one when the pulsar is ``off" and one when the pulsar is ``on". Assuming that all other sources of sky emission (and possibly RFI) are constant during the short separation in time, one can then make the approximation that the sky flux is dominated by a single source (the pulsar) in the differenced data. In principle, this obviates the need for complicated sky modelling. In practice, pulsar gating may not always be a viable technique, given that it not only requires the availability of a suitable pulsar, but also imposes particular requirements on one's instruments, such as a reasonably narrow field of view (so that other transient phenomena are not picked up) and good time resolution (to resolve the on/off behaviour of the pulsar).

In general, then, a good sky model beyond point sources is necessary. At the low radio frequencies of $21\,\textrm{cm}$ cosmology, diffuse sky emission is often quite bright and non-negligible for the required levels of calibration precision. Precisely how one should parametrize and include diffuse models in calibration is still very much of an open question. For now, many analyses attempt to circumvent this by performing calibration using only long baselines. The concept here is that because diffuse sky emission is---by definition---on spatially extended scales, its influence should only dominate the visibilities of baselines that are short. In contrast, localized bright point sources contribute to the visibilities of all baselines. This means that the approximation of a point source-dominated sky is better for long baselines than for short baselines. Thus, one should in principle be able to obtain better calibration solutions by performing a point source model calibration using only long baselines, and then transferring the calibration solutions to all baselines by using the fact that a particular antenna will in general be part of both short and long baselines. However, consistency checks following such procedures have often revealed suspicious discontinuities in data quality at the (arbitrarily defined) boundaries between ``short" and ``long" baselines, suggesting that further work is necessary in this area \citep{Gehlot2018Limit}.

\subsection{Calibration using self consistency}
\label{sec:RedCal}
An alternative to sky-based calibration is to calibrate an interferometer by requiring that a properly calibrated dataset must be internally self consistent. Recall from Section \ref{sec:InterferometerSensitivityOptimization} that in order to boost sensitivity to the power spectrum, many interferometers have their antennas arranged in a regular grid. This results in multiple copies of the same baseline, whose visibilities---once calibrated---should be identical up to noise fluctuations. Turning this consistency into a requirement allows one to work backwards and to calibrate one's interferometer using \emph{redundant calibration algorithms}\footnote{The phrase ``redundant calibration" is a rather unfortunate one, for it seems to imply that our calibration is unnecessary! While a name like ``identical baseline calibration" might be more appropriate, the term ``redundant calibration" is fairly widespread in the literature, and so we employ it in this paper.} \citep{Wieringa1992, Liu2010}.

For an interferometer with redundant baselines, $V_{ij}^\textrm{model}$ depends only on the difference between $\vr_i$ and $\vr_j$. This is another way to satisfy the requirement from Section \ref{sec:skycal} of reducing the number of degrees of freedom that one must solve for. For example, with a rectangular $n_x \times n_y$ array of regularly spaced antennas, there are $n_x n_y (n_x n_y - 1)/2$ total baselines (and therefore measured visibilities $V^\textrm{meas}$), but only $2 n_x n_y - n_x -n_y$ unique visibilities. Adding to this are $n_x n_y$ gain parameters, yielding $3 n_x n_y - n_x -n_y$ that are unknown and must be solved for. This number, though, is in general smaller than the number of measured visibilities, and redundant calibration solves for both the unique visibilities and the gain parameters at the same time by minimizing Equation \eqref{eq:CalibChiSq} (with our model for the visibilities simply being the set of unique visibilities). This minimization can either be performed numerically in a non-linear way using gradient descent algorithms \citep{Marthi2014,Grobler2018} or by linearizing the expressions analytically in some way and solving the resulting linear equations \citep{Liu2010}.

The advantage of redundant calibration is that although it makes assumptions regarding the instrument (see Section \ref{sec:HybridCal} for some examples), it is relatively free of assumptions regarding sky emission. By solving directly for the unique visibilities, one need not decide ahead of time whether the sky is dominated by point sources or diffuse emission or (most likely) is some mixture of both. However, it is crucial to note that redundant calibration is not \emph{entirely} free of modelling assumptions. This is because demanding self consistency in a dataset only allows one to solve for a \emph{relative} calibration between antennas, not an \emph{absolute} one. For example, redundant calibration is unable to fix an absolute scale for the calibration parameters and unique visibilities that it solves for. This is because one can always double all gain parameters while dividing all unique visibilities by four with no change to the $\chi^2$ of the fit, since two gains are multiplied onto the true visibility in Equation \eqref{eq:CalibEqn}. Similarly, the absolute phase cannot be obtained from redundant calibration, because the transformation $g_i \rightarrow g_i \exp(i \varphi)$ (where $\varphi$ is some real number) leaves the equations unchanged. Finally, the transformations $g_i \rightarrow g_i \exp(i r_{i,x} \psi_x)$ and $g_i \rightarrow g_i \exp(i r_{i,y} \psi_y)$ (where $\psi_x$ and $\psi_y$ are constants, and $r_{i,x}$ and $r_{i,y}$ are the $x$ and $y$ positions of the $i$th antenna, respectively) also leave the goodness of fit unchanged. This linear phase gradient can be intuitively thought of (in the narrow-field limit) as a rotation of the sky. Since redundant calibration relies only on the self-consistency of one's data, the system of equations that govern the calibration procedure cannot be sensitive to details about the sky \citep{Liu2010}.

We thus have four degrees of freedom (global amplitude, global phase, and two linear phase gradients) that are unconstrained by redundant calibration. In fact, there are more such degenerate degrees of freedom if we consider more than just the calibration of a single polarization, which is implicitly what we have assumed in our discussion of redundant calibration so far. With multiple polarizations, there are four degeneracies per polarization direction if data from each polarization are treated independently. If assumptions about the sky signals are made in order to relate the different polarization directions (e.g., if one assumes that the sky does not contain any circularly polarized sources), then the number of degenerate parameters can be reduced \citep{Dillon2018}. In any case, to fix the degenerate parameters, one must appeal to a minimal amount of sky modelling \citep{Li2018}. Thus, redundant calibration is never truly sky-independent, and many of the problems discussed in Section \ref{sec:skycal} with sky-based calibration (such as sky model completeness) end up resurfacing in the context of redundant calibration \citep{Byrne2019}.

\subsection{Calibration using hybrid techniques}
\label{sec:HybridCal}
Given the importance of calibration to the precision operation of an interferometer, it is worthwhile to combine some of the calibration techniques discussed above. At the simplest level, sky-based calibration is needed to fix the degenerate parameters of redundant calibration. Going a little further, a rough sky-based calibration (even with an incomplete sky model) can be used to provide a starting point in parameter space for gradient descent algorithms that attempt to minimize the redundant calibration $\chi^2$ \citep{ZhengMITEoR2014,Li2018}.

More generally, it is possible to write down calibration algorithms that naturally combine sky model information and the self-consistency requirement of redundant calibration. Because redundant calibration does not depend on the sky (up to the fixing of degenerate parameters), it is essentially marginalizing over all possible skies, including those that are unphysical or unlikely to be correct, given prior knowledge. In effect, one is attempting to solve a harder problem than is necessary, because the fact that our knowledge of the sky is imperfect should not mean that all sky information is discarded! Similarly, the redundancy of two baselines should not be regarded as an all-or-nothing proposition. Two baselines of an array may not be perfectly redundant in reality, perhaps due to primary beam differences or positional uncertainties in antenna placement (among many other possibilities), but their data will likely still be highly correlated. Treating the visibilities measured by two such partially redundant baselines as being completely independent will likely result in greater uncertainty in calibration solutions. On the other hand, simply ignoring problems such as antenna position offsets leads to biases in calibration solutions \citep{Joseph2018}.

One approach to including partial redundancy is to do so perturbatively within the framework of redundant calibration \citep{Liu2010}. An example of a type of non-redundancy where this can be done is positional errors in antenna placement. The effect of small positional errors in antenna placement is to shift nominally redundant baselines slightly on the $uv$ plane. As a result, these baselines no longer lie perfectly on top of each other on the $uv$ plane, but instead have some small spread about fiducial points on the plane. The visibilities measured by each baseline in the cluster can then be written as a Taylor expansion, with the zeroth order term being the visibility that would be measured at the fiducial point, and the first order corrections being determined by gradients in the two directions of the $uv$ plane and the ``distance" between a visibility and the fiducial point on the plane. Solving for the sky is then tantamount to solving for the fiducial visibility and the two gradients for each cluster. With an array layout that is sufficiently regular, there are still enough measurements and few enough degrees of freedom to enable well-defined solutions.

The weakness of the aforementioned approach is that it can only be used to account for certain types of perturbations, and moreover, is still agnostic to the possible form of the sky. A better solution is the \texttt{CorrCal}\footnote{\url{https://github.com/sievers/corrcal}} algorithm \citep{Sievers2017}. In \texttt{CorrCal}, the visibility data from the different baselines are modelled as random variables drawn from a correlated higher-dimensional Gaussian distribution.\footnote{In reality, the statement that the data is Gaussian is certainly not rigorously true. However, \citet{Sievers2017} argues that the forms of non-Gaussianity typically appearing in interferometric data are unlikely to cause \emph{bias} in one's calibration solutions, and should only result in an increased \emph{variance}.} Let $\vv^\textrm{true}$ be a vector storing the true visibilities that would be measured if all gain parameters were unity [i.e., it is a vectorized version of $V_{ij}^\textrm{true}$], and similarly for $\vv^\textrm{meas}$. The two vectors are related by a diagonal gain matrix $\mG$, such that
\begin{equation}
\vv^\textrm{meas} = \mG \vv^\textrm{true} + \vn,
\end{equation}
where $\vn$ is a similarly vectorized version of the noise. Since we are modelling the data as being Gaussian distributed, their probability distribution is given by $p(\vv^\textrm{meas}) \propto \exp \left( - \chi^2 / 2 \right)$, where
\begin{equation}
\chi^2 = \vv^\textrm{meas} (\mN + \mG \mC \mG^\dagger)^{-1} \vv^\textrm{meas},
\end{equation}
with $\mC$ being the covariance of the true visibilities and $\mathbf{N} \equiv \langle \mathbf{n} \mathbf{n}^\dagger \rangle$ being the noise covariance matrix. It is $\mC$ that allows us to encode redundancy and sky information. For example, suppose we only have data from two baselines. If they are perfectly redundant, then we might model $\mC$ as
\begin{equation}
\label{eq:PerfectCorrCovar}
\mC =
\alpha^2
 \begin{pmatrix}
1 & 1 \\
1 & 1
 \end{pmatrix} ,
\end{equation}
where $\alpha$ is a parameter controlling the amplitude of the sky signal. The ones in this matrix ensure that this sky signal is perfectly correlated (i.e., identical) between the redundant baselines. Having written down a $\chi^2$ for our measurement, our calibration procedure then amounts to performing gradient descent in the parameter space spanned by elements of $\mG$. \citet{Sievers2017} provides a discussion of how this can be done in a computationally tractable way for large arrays.

The strategy of encoding all the details of our measurement in the covariance is one that reproduces redundant calibration as a special case. This can be proven by re-interpreting $\alpha$ as an adjustable parameter. Taking $\alpha \rightarrow \infty$ is equivalent to insisting that the redundancy of the array be perfectly respected, and with such a limit the $\chi^2$ becomes equivalent to one for a redundant calibration algorithm \citep{Sievers2017}. But crucially, \texttt{CorrCal} allows redundancy to be relaxed. For example, if two baselines are expected to be similar but not perfectly redundant, one can include this information by slightly depressing the off-diagonal elements in Eq. \eqref{eq:PerfectCorrCovar}. In general, the precise level of correlation between two baselines can be modelled, and this information can be inserted into the covariance. An example of this would be the imperfect (but significant) correlation between two baselines that are nominally redundant but fail to be because of primary beam differences. Assuming that the sky corresponds to a Gaussian field (as \texttt{CorrCal} does), the correlation between two such baselines can be readily computed. In this way, partial correlation can be accounted for in one's calibration solution.

\texttt{CorrCal} also allows one to encode priors on sky information, because this information manifests itself as a particular correlation structure between different baselines. For instance, as we discussed in Sec. \ref{sec:skycal}, a sky dominated by a single point source gives rise to visibilities with constant complex magnitude across all baselines, while diffuse emission gives rise to visibilities that decrease in amplitude with increasing baseline length. By including this correlation structure between baselines in the covariance matrix, one changes the $\chi^2$ function to allow it to favour fits to the data that are consistent with one's prior on the sky. Precisely what prior information one wishes to include can be done fairly flexibly. For example, when including information about known bright point sources, one may choose to include strong priors on their positions but not their fluxes. This may be prudent because positions are generally better known than fluxes, particularly since existing radio surveys may not cover the same frequency bands as one's current survey. \texttt{CorrCal} thus does not waste prior information just because it is incomplete, and in our example of bright source positions, this can substantially improve the quality of one's phase calibration.

\subsection{Calibration solutions as a function of frequency and time}
\label{sec:CalibFreqTime}
In our discussion of calibration so far, we have made no mention of the frequency or time dependence of our calibration solutions or our sky models; every algorithm that we have outlined can be applied on a frequency-by-frequency, time-by-time basis. From a purely mathematical standpoint, this might even be considered necessary, as an instrument's response could in principle fluctuate greatly as a function of time and frequency. In practice, instruments are generally designed to be as stable as possible, making it less important that a calibration solution be able to respond to sudden changes.

While it is true that a smooth function of frequency and time is simply a special case of a flexible function that is free to vary arbitrarily, there can be a considerable downside to the latter. A calibration solution that is derived using only data at a single timestamp and frequency channel may be quite noisy. Applying such a calibration solution to the data would then result in an incorrectly calibrated dataset that is spectrally unsmooth, even if both the instrument response and the true sky were in fact spectrally smooth. In other words, by overfitting the noise (or biases due to sky-model incompleteness) during the calibration process, false spectral and temporal signatures can be imprinted in the data \citep{Barry2016Calibration}. This can negatively impact one's ability to measure the cosmological $21\,\textrm{cm}$ signal, for as we will discuss in Section \ref{sec:FgMitigation}, most foreground mitigation algorithms identify foreground contaminants by their spectral smoothness, which is destroyed if one's calibration algorithm imprints unsmooth spectral structure on the data.

To avoid unwarranted structure in one's calibration solutions, there are a number of strategies that can be employed. One strategy is to limit the exposure of the calibration solution to pieces of data that are likely to, say, be chromatic. For example, \citet{EwallWice2017} and \citet{Orosz2019} have found that false chromaticity can be mitigated by reducing the weight of data from long baselines when solving for a calibration solution. This is because long baselines are intrinsically more chromatic, since $\vu = \vb / \lambda$, which means that long baselines sample a larger variety of $\vu$ modes as a function of frequency. The result is that when data from a variety of baselines are brought together to solve for a calibration solution, spectrally unsmooth artifacts can ``leak" from long baselines to short baselines. Excluding long baselines excludes the source of unsmooth artifacts, improving the overall solution. Note however, that while this may work well for mitigating spectral structure, one must be cognizant of the fact that it is somewhat in conflict with the suggestion of Section \ref{sec:skycal}, where the exclusion of \emph{short} baselines was highlighted as a way to alleviate problems arising from sky model incompleteness.

Another strategy that can be employed to avoid excessive frequency structure is to extend \texttt{CorrCal}'s system of equations to include all frequency channels at once. In other words, rather than writing (and solving) a separate set of equations for each frequency, one can concatenate the $\vv^\textrm{meas}$ vectors from different channels into a longer vector and to employ \texttt{CorrCal} as a single calibration step that calibrates all frequencies at once. This allows one to encode covariance information between not just the measurements of different baselines, but also measurements between different frequencies. Thus, just as one can impose priors on the sky's spatial structures during calibration (as discussed in the previous section with the example of point sources), one can impose priors on its frequency structure. This enables the imposition of smoothness requirements, allowing one to solve for a smooth bandpass calibration.

Of course, extending a system of calibration equations to perform a simultaneous calibration across all frequencies is a computational demanding venture, since the relevant matrices are now much bigger. In some cases, calibration routines have resorted to large-scale distributed computing. A notable example of this is the \texttt{SAGECal-CO} algorithm. The original incarnation of this calibration, \texttt{SAGECal}\footnote{\url{http://sagecal.sourceforge.net/}} \citep{Yatawatta2008,Kazemi2011}, is essentially an iterative sky model-based scheme for performing direction-dependent calibration, using the space alternating generalized expectation (SAGE) maximization for which it is named. The \texttt{-CO} variant brings in the technique of consensus optimization \citep{Yatawatta2015,Yatawatta2016,Yatawatta2017,Yatawatta2018}. In the specific case of multi-frequency radio interferometric calibration, one demands that the calibration solutions are well-fit by smooth functions of frequencies (e.g., polynomials). The consensus optimization algorithm allows this to be achieved by a multi-stage iterative process. First, each distributed computational system calibrates data from one specific frequency. The results from all the different frequencies are then brought together and fit to the chosen smooth parametrization. The best-fit smoothed calibration solutions are then sent back to the distributed systems for further iteration. Upon convergence, one has effectively solved a constrained optimization that can be practically scaled to large distributed computational systems.

As an alternative to computationally intensive calibration schemes that explicitly optimize over frequency, one may pursue the simple strategy of simply smoothing one's calibration solutions after an initial calibration. In fact, this may be the most computationally viable strategy when considering the fact that ultimately, calibration solutions should have some levels of smoothness not just in frequency, but also in time. Smoothing a calibration solution in frequency and/or time can be done in an \emph{ad hoc} way, or by averaging/filtering the solution in a way that is informed by some prior expectations (or observed performance) of an instrument. One way to achieve the latter is to study one's instrument in detail. For example, \citet{Barry2018PhD} performed a thorough comparative study of calibration solutions as a function of different types of cables used within the MWA system, as well as how these calibration solutions vary with measured temperatures on the telescope site. Alternatively, one can try to model the behaviour of the calibration solutions statistically. \citet{ZhengMITEoR2014}, for example, study the MITEoR instrument by computing the power spectrum of their calibration solutions in frequency and in time. The resulting power spectra (of their calibration solutions) have the form of a white noise floor (i.e., a flat power spectrum) plus an increase in power towards short timescales and over smooth frequency modes, indicative of a power spectrum of true calibration variations. The identification of a two-component calibration ``signal" plus noise model then enables the construction of a Wiener filter (essentially a ``signal over signal plus noise" weighting in the Fourier dual space to time and frequency). Applying the resulting Wiener filter to the calibration solutions smooths out the solutions, in principle providing the closest possible estimate to the true calibration variations as possible amidst noise.

Having described various techniques for ensuring smooth calibration parameters, we note that this pursuit of smoothness is predicated on the assumption that the inherent instrumental response is spectrally smooth and stable in time. If, for example, one's instrumental response is a complicated function of frequency, the calibration should reflect this and inherit the same complication. One might think, then, that with current hardware designs that often still show non-trivial fluctuations, one should avoid imposing a smoothness prior. However, with experiments currently still in the regime of setting upper limits, imposing smoothness on one's calibration solutions is generally a conservative choice. By doing so, one is essentially leaving residual spectral structure in the data, which generally increases the level of the measured power spectrum. This gives rise to a conservative result, as long as the increased power level is interpreted as an upper limit and not a detection of the \tcm signal. For a detection, it is likely that extensive end-to-end simulation suites will be required to show that the measured power spectra have not been biased by calibration errors. Ideally, these simulations should also capture the covariance between calibration errors and the power spectrum errors, such that one can marginalize over the former in a way that self-consistently inflates the latter.

%

\section{Map-making}
\label{sec:mapmaking}

Once calibrated, we may regard the output data produced by a correlator to be a visibility sample for every pair of inputs, at every frequency channel as a function of time. As each visibility is approximately a Fourier mode of some part of the sky, they are not by themselves particularly easy to interpret. Turning the raw data into something more useful is the operation of map-making.

However, we should emphasize that while map-making is an essential tool, it is not a prerequisite for further analysis of \tcm datasets, particularly for areas like power spectrum estimation. In fact, analysis from maps can in some cases be more problematic than proceeding directly from the underlying visibilities. We will elaborate on this point later in the section and in Section \ref{sec:pspecestimation}.

The predominant method of map-making in radio astronomy are variants on the \emph{CLEAN} algorithm \citep{Hogbom}. However, certain properties of CLEAN make it problematic for our purposes, so we will start our discussion by looking at \emph{linear} map-makers which are common in CMB cosmology before returning to CLEAN in \secref{sec:CLEAN}.

Generally, map-making does not couple frequencies, so for the rest of this section we will drop the frequency index and have it be implicit that the operations are performed on a frequency by frequency basis.

To construct a linear map-maker we first need to define the forward model. In our case we need to first define a discretization of our visibility data, which is typically the data for every antenna-pair integrated into samples of fixed length time. We will write this as a vector $\vv$. We also need to discretize the sky, usually into individual pixels in some scheme (e.g. HEALPix; \citealt{HEALPIX}), which we denote as the vector $\vs$. We describe the mapping between the two by
\begin{equation}
    \label{eq:discrete_vis}
    \vv = \mB \, \vs + \vn
\end{equation}
which is the discretized version of Equation \eqref{eq:vis_full}, where
\begin{align}
    \vv_{(ijt)} & = V_{ij}(t) \\
    \vs_{(pab)} & = T_{ab}(\vrhat_p) \\
    \lp\mB\rp_{(ijt)(pab)} & = A_i^a(\vrhat_p; t) A_j^{b *}(\vrhat_p; t) e^{-2 \pi i \vu_{ij}(t) \cdot \vrhat_p} \\
    \vn_{(ijt)} & = n_{ij}(t) + \text{discretization errors}
\end{align}
where the compound indices $(ijt)$ and $(pab)$ index over all the feed pairs $ij$ and time samples $t$, in the first group, and all pixels $p$ and polarization $ab$. As Equation \eqref{eq:discrete_vis} is discrete, for the fixed data to be equivalent to the true continuous case, we must absorb a discretization error into the definition of the discrete noise $\vn$. For a well chosen discretization scheme this contribution should not bias the output be and significantly smaller than the instrumental noise; in practice, achieving this this means choosing a pixelization with much higher resolution than the smallest scales resolvable by the telescope. Provided the scheme does not depend on the input data we would not expect it to bias the output maps.

\subsection{Linear Map-makers}
\label{sec:LinearMapMakers}

Having discretized our problem we can proceed to build a map-maker. Conceptually this is straightforward---we simply need to do a \emph{statistical inversion} of Equation \eqref{eq:discrete_vis} to recover the sky (represented by $\vs$) from our measurements ($\vv$). This inversion is statistical rather than exact, because we must account for the fact our data is noisy, and that in general the forward projection matrix $\mB$ is not invertible.

As this is a statistical inference, we need to account for the properties of the noise, which we assert to be Gaussian with zero mean and a known covariance $\mN = \la \vn \vn^\hconj \ra$. This allows us to write down a likelihood function for the sky
\begin{equation}
\label{eq:mapmaking_likelihood}
    \mathcal{L}(\vs) = \textrm{Pr}(\vv | \vs) = \calG_C(\vv - \mB \vs, \mN)
\end{equation}
where $\textrm{Pr}(\vv | \vs)$ is the probability distribution for $\vv$ given $\vs$, and $\calG_C(\vx, \mC)$ is a complex circularly symmetric Gaussian in $\vx$ zero mean and covariance $\mC$.\footnote{We ignore any \emph{phase dependent} couplings coming from the non-zero \emph{relation matrix} on short baselines \citep{Myers2003}.}

We will start with the easiest statistically meaningful map that we can make. Starting with Equation \eqref{eq:discrete_vis}, we apply the matrix $\mNh$ to \emph{pre-whiten} the noise,\footnote{By the negative square root we mean any factorization of the inverse noise matrix such that $\mN^{-1} = (\mNh)(\mNh)^\hconj$. This can be found either by Cholesky decomposition or, by eigendecomposition and replacing the eigenvalues with their square root.} giving
\begin{equation}
    (\mNh \vv) = (\mNh \mB) \vs + (\mNh \vn) \; .
\end{equation}
By pre-whitening we have transformed all the measured modes in $\mNh \vv$ to have the same noise level. As all modes are now equal in the eyes of statistics, to get back to something which looks like a map, we can simply project back with the \emph{transpose} of the pre-whitened forward matrix, to give
\begin{align}
    \label{eq:map_dirty}
    \hat{\vs}_\text{dirty} & = (\mNh \mB)^\hconj (\mNh \vv) \\
    & = \mB^\hconj \mN^{-1} \vv \; .
\end{align}

This quantity is called the \emph{dirty map}, and in radio interferometry this operation (when performed in the $uv$ plane) is also called \emph{gridding}.
We can interpret this operation as taking the amount of signal observed in the data, and placing it back onto the sky with the (complex conjugate) of the transfer function it was measured with. Though this is the simplest way of transforming the data back into a map like quantity, in general this does not yield a map which accurately reflects the sky. For an interferometer we typically think of the sky as having been convolved with the \emph{dirty beam}, which is essentially the Fourier transform of the $uv$-plane sensitivity.

Despite this, the dirty map is the basis for almost all map-making algorithms because it is a \emph{sufficient statistic} for the true sky $\vs$ (meeting the Fisher-Neyman criterion). The dirty map is a \emph{biased estimator}: by substituting Equation \eqref{eq:discrete_vis} into Equation \eqref{eq:map_dirty} and taking the expectation we find that
\begin{equation}
    \label{eq:dirty_exp}
    \la \hat{\vs}_\text{dirty} \ra = \mB^\hconj \mN^{-1} \mB \vs \; ,
\end{equation}
where not only can we expect the morphology to look different, but the map is not even dimensionally correct (having dimensions of inverse temperature).

However, the form of Equation \eqref{eq:dirty_exp} goes give us a hint of how to make an unbiased estimator for the sky signal. We could simply apply the inverse of the prefactor in Equation \eqref{eq:dirty_exp} to the $\hat{\vs}_\text{dirty}$. In the case where there are no unmeasured modes on the sky, this prefactor is in fact invertible, giving a new estimate
\begin{equation}
\label{eq:ml_cmb}
    \hat{\vs}_\mathrm{ML} = \lp \mB \mN^{-1} \mB^\hconj \rp^{-1} \mB^\hconj \mN^{-1} \vv \; .
\end{equation}
In actual fact, this is the maximum likelihood estimate for $\vs$, which we could have found more directly by maximisation of Equation \eqref{eq:mapmaking_likelihood}. The inverse term $(\mB \mN^{-1} \mB^\hconj)^{-1}$ attempts to \emph{deconvolve} the dirty map into an accurate map of the sky. Equation \eqref{eq:ml_cmb} is typically used for CMB map-making, although it has also been used in \tcm cosmology \citep{Masui:2013}. 

In the case that there are completely unmeasured modes of the sky, the matrix $\mB \mN^{-1} \mB^\hconj$ is singular and cannot be inverted.  To avoid this we can write the maximum likelihood solution in terms of the Moore-Penrose pseudo-inverse (denoted with a $^+$)
\begin{equation}
    \hat{\vs}_\mathrm{ML} = \lp \mN^{-1/2} \mB \rp^+ \mN^{-1/2} \vv \; .
\end{equation}
The distinction between this and Equation \eqref{eq:ml_cmb} is that singular modes are regularised by the Moore-Penrose pseudo-inverse to have zero power.

Though the maximum likelihood map is an unbiased and optimal estimate of the sky, in many ways it is not a practical map-maker. As modes are essentially divided by their sensitivity, in its pure form it will severely upweight noise-dominated modes in real data. This has dramatic effects when making a map and can add large amounts of noise below the resolution limit of the telescope, which may dominate any real signal on larger scales. To resolve this we must find a way to \emph{regularize} the inversion of these low sensitivity modes.

If we have some statistical knowledge about the sky in many cases we can produce a much better map-maker by performing a Bayesian inference of the sky. In the simplest case let us assume we do not know exactly what the sky looks like, but that we do have some knowledge of the two point statistics, which we can encode in a pixel-pixel covariance matrix $\mS$. This allows us to place a Gaussian prior $\textrm{Pr}(\vs)$ on the sky maps. Bayes' theorem tells us that
\begin{equation}
    \textrm{Pr}(\vs | \vv) \propto    \textrm{Pr}(\vv | \vs) \textrm{Pr}(\vs),
\end{equation}
where $\textrm{Pr}(\vs | \vv)$ is the posterior distribution that we seek. Since both the prior and the likelihood (Equation \ref{eq:mapmaking_likelihood}) are Gaussian, the posterior is also Gaussian, which after some extensive manipulation we can write as
\begin{equation}
    \textrm{Pr}(\vs | \vv) = \calG(\vs - \mC_\mathrm{W} \mB^\hconj \mN^{-1} \vv, \mC_\mathrm{W})
\end{equation}
where the covariance $\mC_\mathrm{W}$ is defined by
\begin{equation}
    \label{eq:wiener_covariance}
    \mC_\mathrm{W}^{-1} = \mS^{-1} + \mB^\hconj \mN^{-1} \mB \; .
\end{equation}
Taking the expectation of this distribution (or equivalently the maximum a posteriori point) gives us a new map-maker
\begin{equation}
    \hat{\vs}_\mathrm{W} = \ls \mS^{-1} + \mB^\hconj \mN^{-1} \mB \rs^{-1} \mB^\hconj \mN^{-1} \vv \; .
\end{equation}
This map-maker is a \emph{Wiener filter} and has regularized the inversion by only deconvolving modes where we expect the signal amplitude to be larger than the noise (for these modes it is equivalent to the maximum-likelihood estimator). Modes where this is not true are suppressed by their signal to noise ratio. We note that the Tikhonov regularisation \citep{Eastwood2018}, where $\mS^{-1}$ is set to $\varepsilon \mathbf{I}$, can be viewed as a special case where we impose a white noise prior on the sky signal. Alternatively, it can also be viewed as a tradeoff between two extremes: if $\varepsilon$ is zero, $\hat{\vs}_\mathrm{W}$ reduces to Equation \eqref{eq:ml_cmb}, and one is fully deconvolving the instrument's response at the expense of incorporating noise modes; if $\varepsilon$ is large, the noisy modes are suppressed, but the resulting point spread functions relating the true sky and the obtained map are wider \citep{ZhengBruteForce2017}.

One of the major advantages of linear map-makers is that their statistics are straightforward to calculate. For example for the Wiener filter map-maker, the total covariance $\mC_\mathrm{W}$ is defined in Equation \eqref{eq:wiener_covariance} (accurate when our model of the signal covariance is correct), and the noise covariance in map space is
\begin{equation}
    \mN_\mathrm{W} = \mC_\mathrm{W} \mB^\hconj \mN^{-1} \mB \mC_\mathrm{W} \; .
\end{equation}

\subsection{CLEAN}
\label{sec:CLEAN}

The most commonly used map-making techniques within radio astronomy are variants of the CLEAN algorithm which use a non-linear process to perform the deconvolution of a dirty image.

CLEAN is an iterative algorithm with the (simplified) loop at iteration $i$ being
\begin{enumerate}
    \item Form a new dirty image $\vd^i$ from the set of visibilities at the previous iteration $\vv^{i-1}$ (if this is the first iteration use the raw visibilities $\vv$). That is
    \begin{equation}
        \vd^i = \mB^\hconj \mN^{-1} \vv^{i-1} \; .
    \end{equation}
    \item Select the brightest pixel in the dirty image $\mu = \argmax{\vd^i}$ and add a fraction $f$ of its flux into the same pixel in the clean image $\vs^i$. Mathematically
    \begin{equation}
        s^i_\alpha = s^{i - 1}_\alpha + \delta_{\alpha,\mu} \, f \, d^i_{\mu} \; .
    \end{equation}
    \item Update the residual visibilities by subtracting the projection of the cleaned image to define
    \begin{equation}
        \vv^i = \vv - \mB \vs^i \; .
    \end{equation}
\end{enumerate}
This loop is processed until a sufficient number of \emph{components} of have been added to the clean image.

Schematically, CLEAN works by assuming that the sky is a collection of point sources and constructing a model of them by iteratively removing them from the dirty image. At radio frequencies there are a significant number of point sources, and given that these have been a major area of research, a point-source prior (combined with its algorithmic and computational simplicity) has been very successful. However, this assumption means that it is much more effective at producing maps of point sources than diffuse components like galactic synchrotron and \tcm emission, though extensions to the algorithm can do better \citep{1984AJ.....89.1076S,MultiCLEAN}. This is in contrast to the Wiener map-maker which struggles to deconvolve point sources \citep{Berger2017}.

The non-linear and iterative nature of CLEAN (which comes from finding the maximum pixel in Step 2 above), means it is particularly hard to analyze its statistical properties. Constructing a covariance for the map output is not an easy process and may be best tackled by Monte-Carlo simulation.

\subsection{Noise Covariance}
\label{sec:MapmakingNoiseCovar}
For maps to provide more than a simple visualization of our data, we need to understand the uncertainties within them. This means that in addition to the map we need to be able to produce a covariance matrix. For linear map-makers, such as the Wiener filter, the noise covariance is easy to calculate and does not depend on the map itself. In contrast the non-linear process used by CLEAN makes it challenging to assess the noise covariance of the maps without large suites of Monte-Carlo simulations.

Even where it is possible to calculate the noise covariance in the map basis it may be desirable more to produce power spectra and other statistics from visibilities instead of from maps. As the noise covariance is typically diagonal in visibility space between frequencies, times and baselines (or at the very least has short ranged correlations between baselines) it is compactly represented. After we have transformed our data into maps this is not the case and the noise will be correlated amongst pixels; typically, this coupling is between nearby pixels, though for some telescope configurations such as redundant arrays with large spacings this correlation can be non-local.

\subsection{Practicalities of map-making}

\label{sec:MapmakingPracticalities}

In the above discussion we have given short shrift to the actual implementation of a realistic map-maker, focussing on the theoretical aspects of the various map-makers. In reality these form only a small number of the factors that must be considered.

One major aspect is computational complexity. For mapping large areas or working at high resolution the number of degrees of freedom we are solving for may be large as the number of pixels $N_\text{pix}$ is potentially in the billions. Superficially any single projection such as $\mB \vs$ is an $O(N_\text{pix}^2)$ and so this can be problematic both from a computational and storage perspective. This can be mitigated by taking advantage of the flat sky approximation to perform the matrix-vector products using Fast Fourier Transforms (see Equation \ref{eq:FTvis}). For the linear map-makers there are still large matrix operations we must perform (i.e. multiply by the matrix inverse for the deconvolution); if the full computation required for this is prohibitive, one option is to use conjugate gradient to determine the inverse using only the forward operations that can be applied via FFT \citep{Myers2003}. In contrast CLEAN-like algorithms only require forward projections and so are likely to be computationally more straightforward.

A second aspect is the observational strategy of the telescope. In particular, does it track a fixed location on the sky, tile multiple regions, or does it point at a fixed location in the local frame and let the sky rotate overhead (in other words, is the telescope \emph{drift scanning})? In the latter two cases we are forced to stitch together observations from different (but overlapping) pointings into a single map, either by mosaicking \citep{1979ASSL...76...61E}, or more specialized techniques such as the $m$-mode formalism, which we consider in the next section.

Finally, above we noted that CLEAN is particularly suited for imaging radio point sources whereas the Wiener map-maker is more suited to diffuse imaging. In reality the sky is a mix of both components, and so map-making for \tcm cosmology may have to involve a mixture of CLEAN-like point source peeling and linear diffuse imaging.

\subsection{Map-making with m-modes}
\label{sec:mmodes}

In this section we describe the $m$-mode formalism, a recent technique that allows wide-field polarimetric imaging to be performed computationally efficiently. However, these gains are at the expense of generality: the $m$-mode formalism works only for \emph{drift scan} telescopes, although these are predominant in upcoming \tcm experiments (see \secref{sec:ObsStatus}). A full treatment of the formalism can be found in \citet{Shaw2015}; we give an abbreviated version here, starting from where we left off in \secref{sec:InterferometryBasics}.

\subsubsection{Introduction}

In \secref{sec:InterferometryBasics} we described the sky emission in terms of the coherency matrix $T_{ab}(\vrhat)$; for the m-mode formalism it will be useful to separate this explicitly into the individual fields for each Stokes parameter. We can do that by writing
\begin{equation}
    \label{eq:coherency_expansion}
    T_{ab}(\vrhat) = \sum_X \calP_{ab}^X \, X(\vrhat)
\end{equation}
where $X \in \ls T, Q, U, V \rs$ represents the different Stokes parameters in the sky. The polarization tensors $\calP_{ab}$ are simply the Pauli matrices
\begin{align}
\calP^T_{ab} & = \frac{1}{2}\begin{pmatrix} 1 & 0 \\ 0 & 1\end{pmatrix},
&
\calP^Q_{ab} &= \frac{1}{2}\begin{pmatrix} 1 & 0 \\ 0 & -1\end{pmatrix},
\notag \\
\calP^U_{ab} &= \frac{1}{2}\begin{pmatrix} 0 & 1 \\ 1 & 0\end{pmatrix},
&
\calP^V_{ab} &= \frac{1}{2}\begin{pmatrix} 0 & i \\ -i & 0\end{pmatrix}.
\end{align}

Using Equation \eqref{eq:coherency_expansion} we can rewrite the visibility measurement equation Equation \eqref{eq:vis_full} as
\begin{equation}
    \label{eq:vis_time}
    V_{ij}(t) = \sum_X \int B_{ij}^X(\vrhat; t)\, X(\vrhat) \, d\Omega + n_{ij}(t)
\end{equation}
where we define the beam transfer function $B_{ij}^X(\vrhat; t)$ for each Stokes field as
\begin{equation}
    B_{ij}^X(\vrhat; t) = A_i^a(\vrhat; t) A_j^{b *}(\vrhat; t) \calP_{ab}^X e^{-2\pi i \vrhat\cdot\vu_{ij}(t)} \; .
\end{equation}

For drift scan telescopes, excepting noise for the moment, the time stream is periodic on the sidereal day and we can simply map time $t$ into hour angle $\phi$. The presence of this instrumental noise breaks the exact periodicity, but if we average our time streams over sidereal days, the periodicity is approximately recovered.

To map this into the $m$-mode formalism we are going to make a series of transformations described in detail in \cite{Shaw:2014,Shaw:2015}, which we can summarise as
\begin{itemize}

    \item Fourier transform the time stream with respect to hour angle $\phi$. This transforms $V_{ij}(\phi) \rightarrow V_{ij;m}$, where $m$ indexes the different Fourier modes.

    \item Take the spherical harmonic transform of the beam transfer functions and sky temperature and polarization. This maps $X(\vrhat) \rightarrow a_{lm}^X$, and $B_{ij}(\vrhat; \phi)^* \rightarrow B_{ij;lm}(\phi)^*$ and transforms the sky integration to summation of $l$ and $m$.

    \item Apply the restriction that the pointing only changes by Earth rotation. In terms of the Beam transfer matrices that is $B_{ij;lm}(\phi) = B_{ij;lm}(\phi=0) e^{i m \phi}$. Integrating over $\phi$ now generates a Kroenecker delta that we can sum over $m$.

    \item Map the $Q$ and $U$ polarization into $E$ and $B$ mode polarization.

\end{itemize}
After applying these transformations we can rewrite \eqref{eq:vis_time} as
\begin{equation}
    \label{eq:vis_m_nearly}
    V_{ij; m} = \sum_{l,Y} B_{ij; lm}^Y a_{lm}^Y + n_{ij; m}
\end{equation}
where the polarization index $Y \in \ls T, E, B, V \rs$. This is an exact (for transit telescopes) mapping from the sky, described by multipoles $a_{lm}^Y$, to our measurements, given by the Fourier coefficients of each visibilities time stream. The key thing to note here is that the Fourier coefficient and the spherical harmonic order are the same, i.e. there is no summation over $m$.

Our visibilities make complex valued measurements of the sky but the sky itself is real valued. This means that for any $m > 0$, while the coefficients $V_{ij;m}$ and $V_{ij;-m}$ are independent, the multipoles $a_{l,m}^Y$ and $a_{l,-m}^Y$ are not. To make this clear, we will group together the positive and negative $m$ measurements together by defining
\begin{align}
V_{ij; m}^+ & = V_{ij; m} & V_{ij; m}^- & = V_{ij; -m}^* \\
B_{ij; lm}^{Y, +} & = B_{ij; lm}^Y & B_{ij; lm}^{Y, -} & = (-1)^m B^{Y *}_{ij; l, -m} \\
n^{+}_{ij; m} & = n_{ij; m} & n^{-}_{ij; m} & = n^{*}_{ij;-m} \; ,
\end{align}
and now the $m$ index only ranges for $m \ge 0$.

Using this Equation \eqref{eq:vis_m_nearly} becomes
\begin{equation}
    \label{eq:vis_m}
    V_{\alpha; m}^\pm = \sum_{l,Y} B_{\alpha; lm}^{Y, \pm} a_{lm}^Y + n_{\alpha; m}
\end{equation}
for $m \ge 0$. This is the $m$-mode measurement equation, describing how the polarised sky observed through our telescope produces our measured visibilities.

\subsubsection{Application to map-making}
\label{sec:mmodeapptomapmaking}
To use the $m$-mode formalism to construct a map-maker in the style of \secref{sec:LinearMapMakers} we need to appropriately discretize and vectorize Equation \eqref{eq:vis_m}.

First, in the m-mode formalism the degrees of freedom are naturally discrete as they are spherical harmonics of the sky. Second, as the telescope has fixed size, the sensitivity to large values of $l$ and $m$ decreases exponentially beyond a certain scale, and so we only need consider a finite range of each. We take $l \lesssim 2\pi D_\text{max} / \lambda$ and $m \lesssim 2 \pi D_\text{EW} / \lambda$ where $D_\text{max}$ is the largest dimension of the array, and $D_\text{EW}$ is the total size in the East-West direction. See \cite{Shaw:2015} for a more detailed discussion. This means that for any given telescope we can easily represent all the quantities in Equation \eqref{eq:vis_m} as finite-length vectors and matrices.

We can define our vectorisation as
\begin{align}
    \label{eq:mmode_vec}
    \ls \vv \rs_{(m\,ij\,\pm)}  & = V_{ij; m}^\pm \; , & \ls \mB \rs_{(m\,ij\,\pm)(m' Y l)} & = B^{Y,\pm}_{ij; lm} \delta_{m m'} \; , \notag \\
    \ls \vs \rs_{(m Y l)} & = a^Y_{l m}     \; ,     & \ls \vn \rs_{(m \alpha)}         & = n_{\alpha; m} \; .
\end{align}
where the sky-like degrees of freedom have a compound index $(m Y l)$ comprised of $m$-mode, polarization, and spherical harmonic $l$, and the visibility-like degrees of freedom are indexed as $(m\,ij\,\pm)$, which is comprised of $m$-mode, feed pair, and positive or negative degree of freedom. Using this we can write Equation \eqref{eq:vis_full} as
\begin{equation}
    \vv = \mB \, \vs + \vn
\end{equation}
from which we can apply all the results of \secref{sec:LinearMapMakers}. For instance we can use the Wiener map-maker, which we give again here
\begin{equation}
    \label{eq:wiener2}
    \hat{\vs}_\mathrm{W} = \ls \mS^{-1} + \mB^\hconj \mN^{-1} \mB \rs^{-1} \mB^\hconj \mN^{-1} \vv \; .
\end{equation}

Despite its complexity, there are significant advantages to the $m$-mode formalism. First, for drift scan telescopes it is essentially exact, making it naturally wide-field and fully polarized. While that was also true for the various linear map-makers in \secref{sec:LinearMapMakers}, as we noted in \secref{sec:MapmakingPracticalities} the extremely large matrices involved may be completely impractical to invert or decompose. In contrast, for the $m$-mode formalism these operations are often completely tractable.

To see why, let us note that the matrix $\mB$ defined in Equation \eqref{eq:mmode_vec} is block diagonal, i.e. the interferometer itself does not couple $m$ values. That means that if in Equation \eqref{eq:wiener2} the noise and sky covariances were also diagonal in $m$, the whole equation could be evaluated on an $m$-by-$m$ basis. Let us look at these two contributions in turn:
\begin{itemize}
    \item Noise --- if the noise is \emph{stationary} i.e. the correlations in the noise depend only on the time separation and not on the actual time itself
    \begin{equation}
        \la n_{ij}(t) n_{kl}^*(t') \ra = N_{(ij)(kl)}(t - t') \;,
    \end{equation}
    then the noise in $m$-mode space is also diagonal in $m$.

    \item Sky --- if the statistics of the sky do not depend explicitly on $m$, for instance if we model the sky as a Gaussian random field
    \begin{equation}
        \la a_{lm}^Y a_{l'm'}^{Y'\, *}\ra = C_l^{Y Y'} \delta_{l l'} \delta_{m m'}
    \end{equation}
    then the sky statistics are explicitly diagonal in $m$. In practice this is a good approximation for the \tcm field, but may not be for the galactic foregrounds.
\end{itemize}
If we can make both these approximations, the operations in Equation \eqref{eq:wiener2} can be applied $m$-by-$m$. This gives huge computational savings. For any matrix-vector multiplication we save a factor of $O(m_\text{max})$ in computation, and for any matrix-matrix operation such as eigendecomposition or inversion we save a factor of $O(m_\text{max}^2)$. For many instruments $m_\text{max} \gtrsim 10^3$ and thus we can save $\gtrsim 10^6$ in computation.

In practice, the approximation of noise stationarity is poor at low frequencies where the amplifier noise is small compared to the contribution from the sky (making \cref{eq:noise_approx} invalid). In this limit the noise is determined primarily by the brightness of the sky which depends strongly on where the telescope is pointing. However, for map-making, neglecting stationarity does not add any bias to the maps, it simply makes the estimation non-optimal. In many cases the significant computational savings may enable a more thorough analysis and outweigh the increased noise (over the theoretical optimum) of the estimator.

\section{Power spectrum estimation}
\label{sec:pspecestimation}

\begin{figure}[t]
\centering
\includegraphics[width=0.45\textwidth,clip]{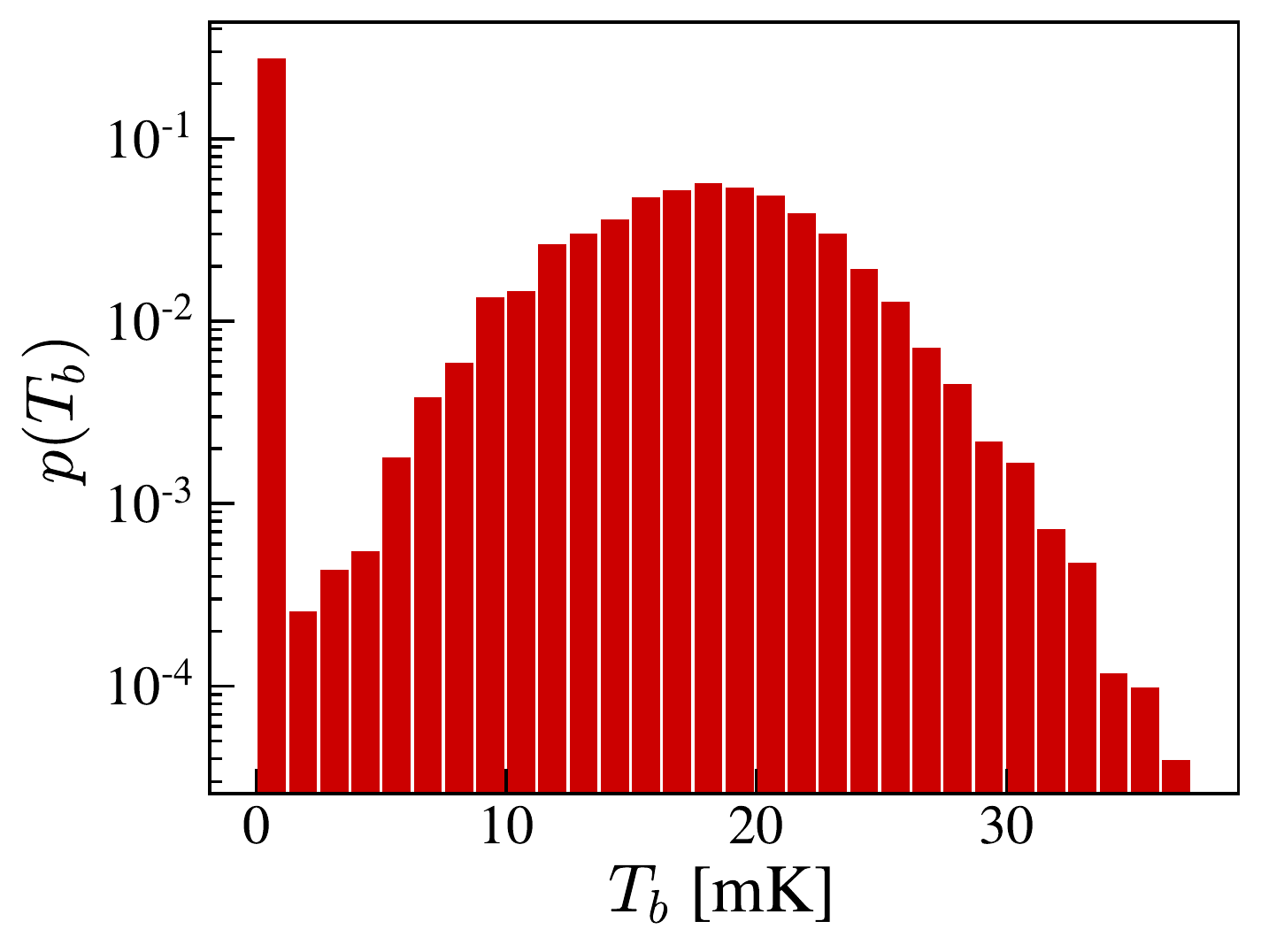}
\caption{A probability distribution function of the \tcm brightness temperature image shown in Figure \ref{fig:21cmFAST}. The distribution is clearly non-Gaussian: one sees not only a spike at $T_b = 0$ (due to patches of our Universe that have been reionized) but also a skewness in the rest of the histogram. These non-Gaussian features are not captured by measuring a power spectrum.}
\label{fig:reionization_histogram}
\end{figure}

The ultimate richness of \tcm cosmology is in its ability to produce full, three-dimensional tomographic maps over a wide range of redshifts. This is in principle achievable using the methods that we just discussed in Section \ref{sec:mapmaking}. However, due the sensitivity and systematic concerns, current experiments have tended to focus on statistical characterizations of the cosmological signal instead. A prime example of this is the power spectrum, defined by Eq. \eqref{eq:PspecDef}. The power spectrum quantifies the variance (``power") of spatial fluctuations in the \tcm brightness temperature field as a function of various length scales (``spectrum"). Measuring the power spectrum is a goal of all currently operating \tcm instruments that are focused on spatial fluctuations. This emphasis is partly for historical reasons, given the success of other cosmological probes such as the CMB and galaxy surveys in extracting information from their respective power spectra. However, it is important to stress that the power spectrum is an exhaustive summary of a data set's information content only if the fluctuations follow Gaussian statistics. In the case of \tcm fluctuations, strong non-Gaussianities are present, particularly during Cosmic Dawn and the EoR. This is illustrated in the clearly non-Gaussian probability distribution function of \tcm brightness temperatures shown in Figure \ref{fig:reionization_histogram}. Thus, while measuring the power spectrum represents an excellent start, there is in principle some richly non-Gaussian information that can be extracted from data sets beyond the power spectrum. We explore such possibilities in Section \ref{sec:HigherOrderStatistics}.

In this section, however, we focus on measurements of the power spectrum. Heuristically, Eq. \eqref{eq:PspecDef} suggests that the power spectrum can be estimated by casting the measured \tcm field in Fourier space and then squaring the result. While this is qualitatively correct, here we set up the problem of power spectrum estimation in a more systematic fashion, which will enable a rigorous quantification of error properties. For concreteness, we illustrate how this works using the three-dimensional power spectrum $P(\vk)$ as an example (deferring a discussion of other types of power spectra to the examples in Appendix \ref{sec:OQEexamples}). We first discretize the continuous quantity $P(\vk)$ into a series of piecewise constant \emph{bandpowers}, which are then grouped into a vector $\vp$, with each component signifying the power at a different value of $\vk$. We then write a generic \emph{quadratic estimator} $\vphat$ of the power spectrum $\vp$:
\begin{equation}
\label{eq:generic_phat}
\phat_\alpha \equiv \vx^\dagger \mE^\alpha \vx,
\end{equation}
where $\vx$ is the input data (discretized in some way and stored as a vector), $\mE^\alpha$ is a matrix whose form is chosen by the data analyst, and $\alpha$ indexes different components of $\vp$, i.e., $P(\vk)$ in different bins/bands of predefined thickness in $\vk$-space.

The estimator that we have written is generic for any choice of basis for our input data. For example, one could imagine first preprocessing one's data into a three-dimensional image cube of the \tcm field. The vector $\vx$ would then be a list of voxel values in some predefined order, with each element of the vector corresponding to one location in the cube. The matrix $\mE^\alpha$ might then take the (rough) form
\begin{equation}
\mE^\alpha_{ij} \sim e^{i \vk_\alpha \cdot \vr_j} e^{-i \vk_\alpha \cdot \vr_i},
\end{equation}
so that when inserted into Equation \eqref{eq:generic_phat}, the result is that the two copies of $\vx$ are each Fourier transformed before they are multiplied together to accomplish a squaring of the Fourier coefficients---precisely what our heuristic recipe for power spectrum estimation suggested. As another example, imagine that the input data has already been Fourier transformed, such that different elements of $\vx$ contain the Fourier coefficients for different $\vk$ vectors. In such a basis, our matrix might take the form
\begin{equation}
\mE^\alpha_{ij} \sim \delta_{i\alpha} \delta_{j\alpha},
\end{equation}
where $\delta_{ij}$ is the Kronecker delta function, equal to $1$ if $i=j$ and zero otherwise. With this $\mE^\alpha$ matrix, one is essentially picking out the $\alpha$th Fourier coefficient to square.

In general, our data will not be written in bases that are as easily relatable to the power spectrum as a three-dimensional image cube or a set of Fourier coefficients. For a general quadratic estimator of the power spectrum, then, we require a formal link between the vector space of our input data and the Fourier space of our output. To do so, we define the covariance matrix $\mC$ of the data $\vx$ to be
\begin{equation}
\label{eq:CovDef}
\mC \equiv \langle \vx \vx^\dagger \rangle,
\end{equation}
where (as with Equation \ref{eq:PspecDef}) the angle brackets $\langle \dots \rangle$ signify an ensemble average.\footnote{Of course, an ensemble average is not an operation that can be performed on real data. If the statistical properties of the data are well-understood, the covariance can be computed either analytically or via a large suite of simulations. Empirical estimations from the data are possible, for instance by assuming stationary statistical properties in time and replacing the ensemble average with a time average. However, such approaches are not without risk, as we describe in Section \ref{sec:sigloss}. In this section, though, it suffices to consider the covariance an abstract theoretical quantity that is central to the framework of power spectrum estimation.} The covariance matrix (in whatever basis it is written) can be related to the power spectrum via the relation
\begin{equation}
\label{eq:CQPlink}
\mC = \mC^{(0)} + \sum_\alpha p_\alpha \mQ^\alpha,
\end{equation}
which essentially makes the claim that the covariance matrix is a linear function of the power spectrum. The matrix $\mC^{(0)}$ is a constant term that does not depend on the power spectrum, containing portions of the covariance that do not depend on the sky (e.g., an instrumental noise covariance).\footnote{In the way we have defined it, $\mathbf{C}^{(0)}$ does not contain astrophysical foregrounds. This means the estimated power spectrum will not just be the power spectrum of the cosmological signal. It becomes an effective power spectrum of total sky emission. Alternatively, one can choose to define $\mathbf{C}^{(0)}$ such that it contains astrophysical foregrounds. The end result is the same, however, because then Equation \eqref{eq:GeneralP-hatExpectation} below acquires an additive foreground bias in addition to a noise bias. This foreground bias adds to the cosmological power spectrum, again giving an effective power spectrum of total sky emission.} The matrix $\mQ^\alpha$ is defined as
\begin{equation}
\label{eq:QmatrixdCdp}
\mQ^\alpha \equiv \frac{\partial \mC}{\partial p_\alpha}
\end{equation}
and is the response of the covariance matrix to the $\alpha$th power spectrum bandpower. That there is a linear relation between the data covariance matrix $\mC$ and the power spectrum is unsurprising, as the two quantities are just ensemble-averaged two-point statistics in different bases: the former in the native basis of the measurement and the latter in a discretized Fourier basis. In fact, the quadratic estimator formalism that we are developing here is not limited to the estimation of the rectilinear power spectrum $P(\vk)$ as defined by Equation \eqref{eq:PspecDef}. Any two-point statistic can be estimated, including the angular power spectrum $C_\ell$ (Equation \ref{eq:AngPspec}), the cross-angular power spectrum $C_\ell (\nu, \nu^\prime)$ (Equation \ref{eq:Cellnunuprime}), and the spherical Fourier-Bessel power spectrum $S_\ell (k)$ (Equation \ref{eq:Sellk}). The abstract generality of the quadratic estimator formalism is what gives it flexibility. However, this abstractness can make the rest of this section difficult to follow. Thus, we encourage readers who prefer concrete examples to frequently consult Appendix \ref{sec:OQEexamples}, where we explicitly work out some toy models of power spectrum estimation.

We can proceed here, however, with our quest for a general power spectrum estimator now that we have a precise relation between the data covariance and the power spectrum bandpowers. If we take the expectation value of Equation \ref{eq:generic_phat}, we obtain
\begin{equation}
\label{eq:GeneralP-hatExpectation}
\langle \hat{p}_\alpha \rangle = \sum_{ij} \mE^\alpha_{ij} \langle  x_j x_i^* \rangle = \sum_\beta  \textrm{tr}\left[ \mE^\alpha \mQ^\beta \right] p_\beta + \textrm{tr} \left[ \mE^\alpha \mC^{(0)} \right].
\end{equation}
The second term is an additive bias term that must be eliminated. If $\mC^{0}$ represents an instrumental noise covariance, for example, this bias term will not represent true sky power, but will instead be an instrumental noise bias. To understand its origin, recall that a power spectrum is a quantity that is quadratic in the data. Thus, even if the noise is equally likely to have positive or negative contributions in the data $\vx$, the squaring operation inherent in estimating a power spectrum makes the noise appear as a positive bias in the power spectrum estimate. To eliminate this bias, one must either model and subtract it off, or avoid incurring it in the first place. The former approach requires having an exquisitely accurate model of $\mC^{0}$, and is therefore generally not favoured. Instead, one often opts for the latter approach, which can be accomplished by modifying Equation \eqref{eq:generic_phat}. Instead of having two copies of $\vx$, one can instead compute an estimator of the form $\vx_1 \mE^\alpha \vx_2$, where $\vx_1$ and $\vx_2$ are two datasets with identical sky contributions but different realizations of the noise. For instance, if a telescope observes the same parts of the sky day after day, one might form $\vx_1$ from data taken on odd-numbered days, while $\vx_2$ is formed from data taken on even-numbered days. Assuming that the noise is uncorrelated across different days of observations, the cross-covariance $\langle \vx_2 \vx_1^\dagger \rangle$ between will not contain any noise terms, and the Equation \eqref{eq:GeneralP-hatExpectation} will similarly not be additively biased.

\begin{figure*}[t]
\centering
\includegraphics[width=1.0\textwidth]{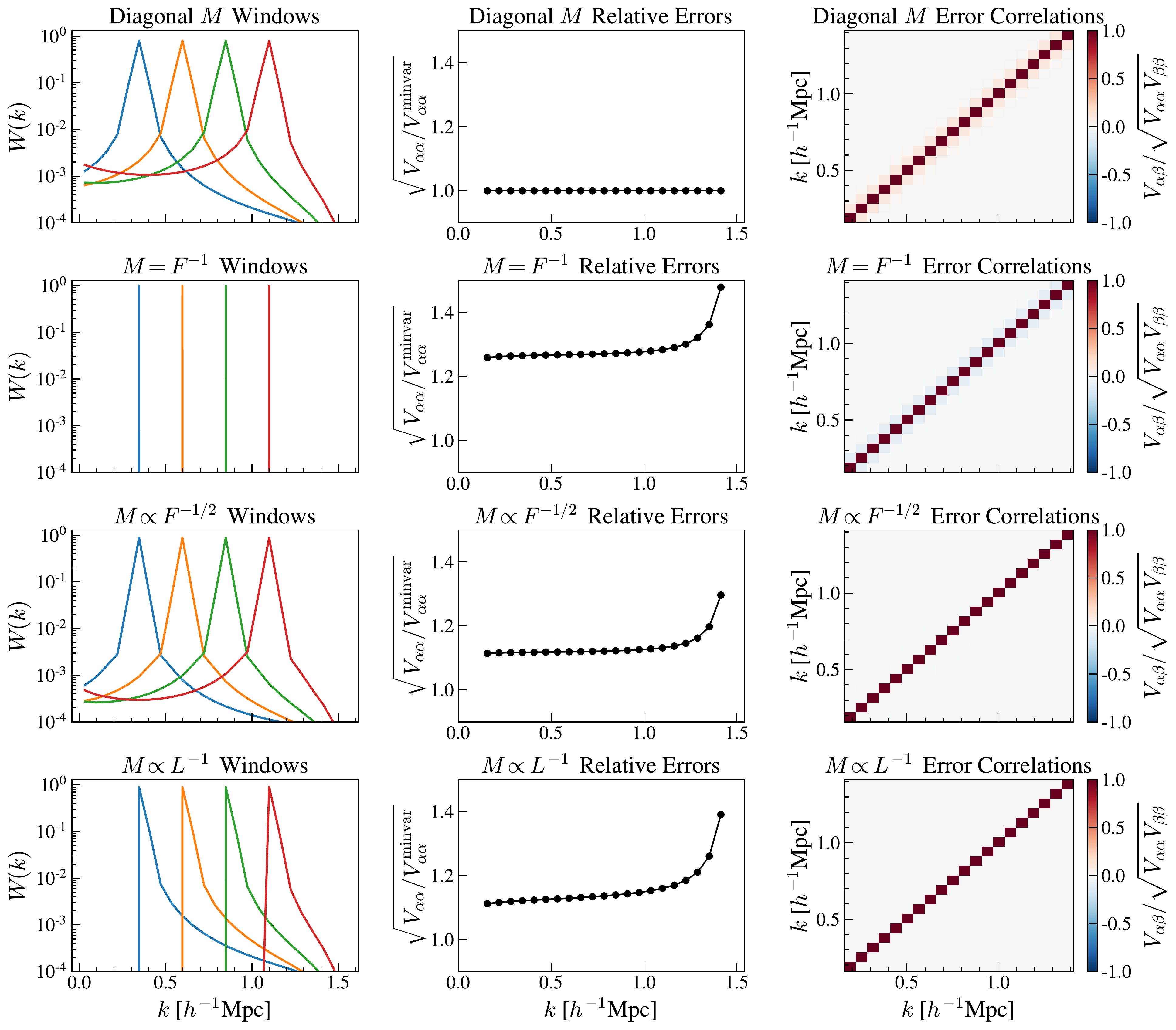}
\caption{Window functions, error bars, and error correlations for a minimum-variance estimator of the power spectrum (top row; Equation \ref{eq:minvarpalpha}), the $\mM = \mF^{-1}$ estimator that has the slimmest possible window functions (second row), the estimator with uncorrelated errors on the bandpower where each row of the normalization matrix $\mM$ is proportional to $\mF^{-1/2}$ (third row), and the Cholesky-based estimator discussed in Section \ref{sec:decorrWedge} that is designed to minimize foreground contamination leakage from low $k$ to high $k$ (fourth row). The errors (middle column) are all plotted relative to the errors obtained from the minimum-variance estimator. In general, one sees that estimators with overly narrow window functions (second row) have the highest errors and neighbouring bins that have negative error correlations (right column) between neighbouring bins. The minimum variance error estimator achieves the smallest possible errors at the price of positive error correlations. The last two estimators strike a balance between the widths of their window functions and the sizes of their error bars, and are thus able to deliver power spectrum estimates with uncorrelated errors.}
\label{fig:windows_and_error_covariances}
\end{figure*}

Assuming that the bias term has been eliminated, the expectation value of our estimator can be written as
\begin{equation}
\label{eq:Wind1}
\langle \hat{p}_\alpha \rangle = \sum_{\beta} W_{\alpha \beta} p_\beta,
\end{equation}
where
\begin{equation}
\label{eq:Wab}
W_{\alpha \beta} \equiv \textrm{tr}\left( \mE^\alpha \mQ^\beta \right).
\end{equation}
For notational convenience, we may write all of this as
\begin{equation}
\label{eq:phatWp}
\langle \mathbf{\hat{p}} \rangle= \mW \vp
\end{equation}
where we have grouped the different bandpowers of the true power spectrum and the estimated power spectrum into the vectors $\vp$ and $\mathbf{\hat{p}}$, respectively, and have grouped $W_{\alpha \beta}$ values into a \emph{window function matrix} $\mW$. Each row of the window function matrix represents the window function for an estimate of a particular bandpower (i.e., bin in $\mathbf{k}$), which quantifies the linear combination of different $\mathbf{k}$ bins of the \emph{true} power spectrum that are being probed. A sensible power spectrum estimator should be constructed to give bandpowers with window functions that are relatively sharply peaked around their target $\mathbf{k}$ bins. In the left column of Figure \ref{fig:windows_and_error_covariances}, we show the window functions for a (completely artificial) survey of a one-dimensional universe that is $200\,h^{-1}\textrm{Mpc}$ in extent. (This is essentially a one-dimensional version of the example that is worked out in Appendix \ref{sec:Toy3DPk}). The different rows of the figure show the window functions for different power spectrum estimators (in other words, a different choices of $\mE^\alpha$) that we will introduce below. For clarity, we show only four window functions, i.e., four rows of $\mW$, but in reality there is of course a separate window function for every $k$ bin that is probed.

Continuing with our development of the formalism, we stress that the vector space inhabited by $\vp$ , $\mathbf{\hat{p}}$, and $\mathbf{W}$ is \emph{not} the same as the vector space inhabited by $\vx$ and $\mC$. As illustrated in the examples above, the latter vector space is described by whatever basis one chooses to describe the input data. The former vector space, on the other hand, is the output Fourier space of bandpowers. Linking the two spaces is $\mQ^\alpha$, via Equation \eqref{eq:CQPlink}. As an example of this link\footnote{Again, for more concrete examples, see Appendix \ref{sec:OQEexamples}.}, we note that if the data is written in position space and is free of practical observational considerations such as instrumental noise, then Equation \eqref{eq:CQPlink} is simply a discretized version of the standard cosmological result that the correlation function is the Fourier transform of the power spectrum. We therefore expect that in a sensible power spectrum estimator, $\mE^\alpha$ should involve $\mQ^\alpha$.

In order to arrive at a specific form for $\mE^\alpha$, one must impose several constraints on the problem. The first is that our power spectrum estimate be correctly normalized. This can be accomplished by examining Equation \eqref{eq:phatWp}, which specifies the relation between our power spectrum estimate $\mathbf{\hat{p}}$ and the true power spectrum $\vp$. As written, it shows that each bandpower estimate is a weighted sum of the true bandpowers; for our estimate to be correctly normalized, the weights must be subject to some constraint. One reasonable choice would be to impose the condition that the bandpowers are peak normalized, so that $W_{\alpha \alpha} = 1$. Another sensible choice would be to insist that our weighted sum of true bandpowers be a weighted \emph{average}. Mathematically, this would translate to the condition $\sum_\beta W_{\alpha \beta} = 1$. While the latter is more common in analyses of real data, we pick the former for simplicity in the derivation that follows.

Aside from an overall normalization constraint, we would ideally also want a power spectrum estimator that has the smallest possible error bars (``optimal"). To obtain error bars, we compute the covariance of $\hat{p}_\alpha$, which is given by
\begin{eqnarray}
\label{eq:Vgeneral}
V_{\alpha \beta} &\equiv& \langle \hat{p}_\alpha \hat{p}_\beta \rangle - \langle \hat{p}_\alpha\rangle \langle \hat{p}_\beta \rangle \nonumber \\
&=& \sum_{ijkm} \mE^\alpha_{ij} \mE^\beta_{km} \left( \langle x_j x_i^* x_m x_k^* \rangle - \langle x_j x_i^* \rangle \langle x_m x_k^* \rangle \right). \quad
\end{eqnarray}
To further simplify this expression, one must make assumptions about the underlying probability distribution function of the data $\vx$. First, we assume (again for simplicity) that the data consist of real-valued numbers. Essentially identical results can be obtained with the more general assumption of complex-valued data. Second, we assume that $\vx$ is drawn from a multivariate Gaussian field.
Making this assumption allows one to invoke Isserlis' theorem/Wick's theorem \citep{Isserlis1918,Wick1950} to express the four-point function of $\vx$ (i.e., the expectation value containing four copies of $\vx$ in Equation \ref{eq:Vgeneral}) as a sum of products of two-point functions. Phrased differently, all higher moments of a Gaussian distribution can be expressed in terms of its variance. For instance, one can show that
\begin{eqnarray}
\label{eq:Wick}
\langle x_i x_j x_k x_m \rangle = \langle x_i x_j \rangle \langle x_k x_m \rangle + \langle x_i x_k \rangle \langle x_j x_m \rangle \nonumber \\
+ \langle x_i x_m \rangle \langle x_j x_k \rangle.
\end{eqnarray}
This allows the covariance between the $\alpha$th and $\beta$th bandpower to be written as
\begin{equation}
\label{eq:VCECE}
V_{\alpha \beta} = 2 \textrm{tr} \left( \mC \mE^\alpha \mC \mE^\beta \right).
\end{equation}
The covariances can be collected into a matrix $\mV$ (which resides in the output Fourier space of bandpowers) whose elements are given by $V_{\alpha \beta}$. The error bars on our estimated power spectrum bandpowers correspond to the square root of the diagonal of this matrix, while the off-diagonal elements measure the extent to which errors on different bandpowers are correlated. In the middle and right columns of Figure \ref{fig:windows_and_error_covariances} we illustrate the diagonal and off-diagonal structures, respectively, for our toy one-dimensional example. The diagonal errors are normalized to those of the minimum variance estimator that we are about to derive.

Continuing towards our goal of deriving an optimal estimator for the power spectrum, we use a Lagrange multiplier $\lambda$ to minimize $V_{\alpha \alpha}$ subject to the constraint that $W_{\alpha \alpha} = 1$. That is, we minimize
\begin{equation}
\mathcal{L} = 2 \textrm{tr} \left( \mC \mE^\alpha \mC \mE^\alpha \right) - \lambda \textrm{tr}\left( \mE^\alpha \mQ^\alpha \right).
\end{equation}
Differentiating this with respect to $\mE^\alpha$ gives
\begin{equation}
\frac{\partial \mathcal{L}}{\partial \mE^\alpha} = 4 \mC \mE^\alpha \mC - \lambda \mQ^\alpha,
\end{equation}
where we assumed that $\mC$, $\mE^\alpha$, and $\mQ^\alpha$ are all symmetric matrices. (Note that more generally, if $\vx$ were complex the relevant assumption here would be that the matrices are Hermitian). Setting the derivative equal to zero gives
\begin{equation}
\label{eq:OQEFormOfE}
 \mE^\alpha \propto \mC^{-1} \mQ^\alpha \mC^{-1},
\end{equation}
which in turns means that
\begin{equation}
\label{eq:palphaunnormed}
\hat{p}_\alpha \propto  \vx^\dagger \mC^{-1} \mQ^\alpha \mC^{-1} \vx.
\end{equation}
Fixing the constant of proportionality can be done by imposing our normalization constraint. This gives
\begin{equation}
\label{eq:minvarpalpha}
\hat{p}_\alpha = \frac{1}{2 F_{\alpha \alpha}} \vx^\dagger \mC^{-1} \mQ^\alpha \mC^{-1} \vx,
\end{equation}
where $F_{\alpha \alpha}$ is the $\alpha$th diagonal element of a more general object known as the \emph{Fisher matrix}, which has elements
\begin{equation}
\label{eq:Fisher}
F_{\alpha \beta} = \frac{1}{2} \textrm{tr} \left( \mC^{-1} \mQ^\alpha \mC^{-1} \mQ^\beta \right).
\end{equation}
Equations \eqref{eq:minvarpalpha} and \eqref{eq:Fisher} together define a power spectrum estimator that is (by construction) a minimum variance estimator.

Our minimum variance estimator is in fact just one example of a broader class of optimal quadratic estimators. In deriving it, we assumed that the normalizing constant was a single scalar number. More general, suppose we defined
\begin{equation}
\label{eq:qalpha}
\hat{q}_\alpha \equiv \frac{1}{2} \vx^\dagger \mC^{-1} \mQ^\alpha \mC^{-1} \vx
\end{equation}
as unnormalized bandpowers. Grouping these into a vector living in the output Fourier space (just as we did with $\vp$ , $\mathbf{\hat{p}}$, and $\mathbf{W}$), we may normalize our bandpowers in a more general way by defining a normalization \emph{matrix} $\mM$, such that
\begin{equation}
\label{eq:pmq}
\hat{\vp} = \mM \hat{\vq}.
\end{equation}
Written in this way, our previous minimum variance estimator is one where $\mM$ was chosen to be diagonal. Another possible choice for $\mM$ would be to set $\mM = \mF^{-1}$. Such an estimator has the attractive property that $\langle \mathbf{\hat{p}} \rangle= \vp$ (i.e., $\mW = \mI$). This can be seen by observing that
\begin{equation}
\langle \hat{q}_\alpha \rangle \equiv  \frac{1}{2} \sum_\beta \textrm{tr} \left( \mC^{-1} \mQ^\alpha \mC^{-1} \mQ^\beta \right) p_\beta = \sum_\beta F_{\alpha \beta} p_\beta,
\end{equation}
that is, $\langle \hat{\vq} \rangle = \mF \vp$. As a result, one obtains $\langle \hat{\vp} \rangle = \mM \mF \vp$, and comparing this to Equation \eqref{eq:phatWp} gives
\begin{equation}
\label{eq:WMF}
\mW = \mM \mF.
\end{equation}
One then sees that, indeed, picking $\mM = \mF^{-1}$ gives $\mW = \mI$. This is illustrated graphically in the second row of Figure \ref{fig:windows_and_error_covariances}, where the window functions are seen to be spikes at one $k$ bin when $\mM = \mF^{-1}$.

That the $\mM = \mF^{-1}$ normalization gives an estimator with $\langle \mathbf{\hat{p}} \rangle= \vp$ is of theoretical significance. The result that $\langle \mathbf{\hat{p}} \rangle= \vp$ means that the estimator is one that has no multiplicative bias whatsoever. This is true regardless of whether one is referring to a scalar multiplicative bias (where the overall scale of the power spectrum is biased) or a matrix multiplicative bias (where the overall scale may be correct but different bandpowers of the true power spectrum may be mixed together in the estimator for a particular bandpower). Under the constraint of an unbiased estimator, the Cramer-Rao inequality guarantees that the optimal estimator of the power spectrum has an error covariance $\mV$ that is given by the inverse of the Fisher matrix, $\mF^{-1}$. Such an estimator is optimal in the sense that no other estimator can have smaller error bars. Frustratingly, the Cramer-Rao inequality does not provide a recipe for how to construct this estimator. Fortunately, we have already inadvertently written it down: our quadratic estimator with $\mM = \mF^{-1}$ is the optimal estimator of the power spectrum. To see this, note that  in general, our quadratic estimator is defined by
\begin{equation}
\mE^\alpha \equiv \frac{1}{2} \sum_{\beta} M_{\alpha \beta} \mC^{-1} \mQ^\alpha \mC^{-1},
\end{equation}
and combining this with Equation \eqref{eq:VCECE} yields
\begin{equation}
\label{eq:VMFM}
\mV = \mM \mF \mM^t,
\end{equation}
from which one can immediately see that $\mV = \mF^{-1}$ if the normalization matrix $\mM$ is chosen to be $\mF^{-1}$.

While the $\mM = \mF^{-1}$ estimator has theoretically attractive properties, in practice it has several shortcomings. For instance, although the Cramer-Rao inequality guarantees that our estimator has the smallest possible error bars \emph{under the constraint that $\langle \mathbf{\hat{p}} \rangle= \vp$}, it is possible to abandon this constraint in order to obtain an estimator with smaller errors. Indeed, Equation \eqref{eq:minvarpalpha} (i.e., a quadratic estimator with a diagonal $\mM$) is precisely such an estimator, since it was derived by explicitly minimizing bandpower variances. The cost of this is that one tends to obtain broader window functions, with $\mW$ no longer equal to $\mI$, as seen in Figure \ref{fig:windows_and_error_covariances}. Intuitively, one can make an analogy to the binning of data. Narrow window functions (the extreme example being $\mW = \mI$) essentially amount to narrow bins in $k$. Each bin contains very little information, and thus the errors are large. This is illustrated in the middle column of Figure \ref{fig:windows_and_error_covariances}, where we show the expected errors relative to the minimum-variance case. As expected, the narrower the window functions, the larger the errors. Broad window functions are equivalent to wide bins in $k$ that average over many independent pieces of information, yielding small error bars on the binned data. There is thus a trade-off between an estimator's resolution in $k$ and the error bars on each bandpower; in fact, this observation can be formalized into a relation analogous to uncertainty principles found in quantum mechanics \citep{Tegmark1995Uncertainty}.

Another consideration in selecting an $\mM$ matrix is the resulting correlation structure of errors. With its wide window functions, the minimum-variance diagonal $\mM$ estimator tends to give positive off-diagonal elements for $\mV$, and thus neighbouring bandpowers are positive correlated. This can be seen in the right column of Figure \ref{fig:windows_and_error_covariances}, where we plot the correlation matrix of errors, where the $(\alpha, \beta)$th element is given by $V_{\alpha \beta} / \sqrt{V_{\alpha \alpha} V_{\beta \beta}}$. One can see that light pink off-diagonal bands, signifying a positive correlation between bins. On the other hand, $\mM = \mF^{-1}$ generally gives negatively correlated bandpowers (notice the light blue off-diagonal bands). A middle ground is to pick an $\mM$ matrix that diagonalizes $\mV$. From Equation \eqref{eq:VMFM}, it is clear that one way to achieve this is to pick an $\mM$ matrix where each row is proportional to $\mF^{-1/2}$ (where the proportionality constants\footnote{Note the plural, because each \emph{row} may have a different proportionality constant. As such, it is incorrect to write $\mM \propto \mF^{-1/2}$ (which would require every row to have the same proportionality constant. Unfortunately, this particular abuse of notation is fairly common in the literature, and for compactness we perpetuate this in Figure \ref{fig:windows_and_error_covariances}.} are chosen to satisfy the proper normalization of the window functions). With such an $\mM$ matrix, the $\mV$ matrix of Equation \eqref{eq:VMFM} becomes diagonal, and the estimated bandpowers have uncorrelated errors. As one might expect, the window functions are not as narrow as they were with $\mM = \mF^{-1}$, nor are they as broad as they were with a diagonal $\mM$. Similarly, the error bars on the bandpowers are intermediate in size.

%
%
%
%

In summary, we have derived a powerful framework for the general problem of power spectrum estimation. With knowledge of the statistical properties of the data, via its covariance matrix $\mathbf{C}$, the optimal quadratic estimator that we derived above is guaranteed to deliver the smallest possible error bars. The key is the application of the inverse covariance $\mathbf{C}^{-1}$ to the data (e.g., in Equation \ref{eq:minvarpalpha}). An additional choice must be made regarding the normalization matrix $\mM$. There is thus a whole family of optimal quadratic estimators, all of which weight the data by $\mathbf{C}^{-1}$, but (owing to different choices for $\mM$), have slightly different trade-offs between error bars, error covariances, and window functions. Note, however, that these quadratic estimators are all optimal so long as $\mM$ is chosen to be an invertible matrix (which is the case in all the examples discussed above), since the multiplication of an invertible matrix does not change the information content of a vector.

However, there are several important caveats to the approach outlined here. First, certain portions of our derivation required the assumption that $\vx$ is drawn from a Gaussian distribution. In particular, while Equation \eqref{eq:Vgeneral} for the bandpower covariance $\mV$ does not assume Gaussianity, it does not simplify to Equation \eqref{eq:VCECE} unless Gaussianity is assumed. In many $21\,\textrm{cm}$ applications, this assumption is violated. For example, the $21\,\textrm{cm}$ brightness temperature field is highly non-Gaussian during the EoR, thanks to the complicated network of ionized bubbles. The presence of non-Gaussian fluctuations means that Equation \eqref{eq:VCECE} may not give strictly correct error information \citep{Mondal2015nonGauss,Mondal2016,Mondal2017,Shaw2019nonGauss}. Additionally, our so-called optimal power spectrum estimator may not be strictly optimal with non-Gaussianities. However, an inverse covariance weighting of the data is often still a good idea in practice. Moreover, it should be emphasized that even if our power spectrum estimator ceases to be an optimal one (and thus is not guaranteed to have the smallest possible error bars), it remains a perfectly valid estimator. Our estimator will continue to be correctly normalized, because a correctly normalized power spectrum is simply one where the window functions are correctly normalized. Being able to correctly normalize our estimator is then tantamount to being able to accurately calculate our window functions in the face of non-Gaussianity. Fortunately, Equation \eqref{eq:Wab} satisfies this requirement, since its derivation depended only on quadratic (i.e., variance) statistics, and not higher-order statistics (like the four-point function of Equation \ref{eq:Wick}) that require extra assumptions about Gaussianity in order for our algebraic simplifications to apply.

A second caveat to this is that our optimal estimator assumes that the covariance matrix $\mC$ is known. This is frequently not a good assumption, since $\mC$ includes both instrumental noise and the contributions from the sky. The former can be characterized to some extent using laboratory and \emph{in situ} measurements, but the latter is particularly uncertain, given that the low-frequency radio sky has historically been poorly explored. (Although of course this situation is changing with the advent of large scale \tcm experiments!) One possible approach is to model the covariance empirically from the data itself \citep{Dillon2015EmpCov}. However, such empirical modelling must be performed with extreme care, for it has the potential to cause signal loss \citep{Cheng2018}, i.e., a power spectrum estimate that is biased low. We revisit this issue in much more depth in Section \ref{sec:sigloss}.

A third challenge is that of computational cost. A brute force application of Equation \eqref{eq:palphaunnormed} results in rather large matrices. To be concrete, consider the length of the data vector $\vx$ for a modern \tcm interferometer. Such an interferometer might consist of several hundred antennas. For concreteness, consider a $300$-element interferometer, which would then have visibility data from up to $300 \times 299 / 2 = 44850$ baselines. With each baseline taking data over $\mathcal{O}(1000)$ frequency channels, $\vx$ ends up being a vector of length $\sim 50,000,000$ \emph{per time integration}! A typical time integration might be $\sim 1\,\textrm{s}$ long, and to achieve the required sensitivities, \tcm experiments require a year to several years of total integration. One thus sees that naively, a brute force approach would result in $\vx$ vectors that are rather unwieldy to store or manipulate. The problem becomes even worse when one realizes that $\mC$ has size equal to the length of $\vx$ squared, and needs to be inverted in Equation \eqref{eq:palphaunnormed}!

To be fair, there are several generic ways in which the length of $\vx$ can be reduced. First, even though an interferometer with $N$ antennas has $N(N-1)/2$ baselines, these baselines may not be unique. For example, instruments like CHIME, HIRAX, and HERA have their antennas laid out on regular grids, which means that there are multiple baselines that in principle measure the same signals (up to some noise variations). Assuming that the data from these baselines have been correctly calibrated, the visibilities from identical copies of a given baseline type can be averaged together, dramatically reducing the data volume. Another way in which the data volume can be reduced is to recognize the fact that typically, one does not analyze the full frequency bandwidth of one's dataset at once. This is because the full frequency range of a typical instrument maps to a large change in redshift, violating assumptions about stationary statistics that are inherent in power spectrum estimation\footnote{In the definition of the power spectrum (Equation \ref{eq:PspecDef}), notice the presence of the Dirac delta function, which says that different Fourier modes are independent. This is the result of assuming that the statistics of our Universe are stationary, i.e., the statistical properties are the same in any location. To see why this is, consider how one would construct a feature---say, an anomalously high-density spike---at some special location in our Universe. To create such an outlier, a large number of the underlying Fourier modes would have to all peak at the chosen location. For such a conspiracy to occur, the Fourier modes must be correlated. We thus see that the assumption of independent Fourier modes (which is central to the definition of the power spectrum) is intimately tied to the assumption of stationary statistics.}.  Finally, as the Earth rotates over a full period of a day, the cosmological measurements repeat themselves. Thus, some averaging can be done by folding multiple sidereal days into one.

Despite these reductions, power spectrum estimation continues to be a computational challenge. Differing approaches to this challenge have led to multiple styles of power spectrum estimation that can be viewed as special cases of our general quadratic estimator from Equation \eqref{eq:generic_phat}, if not our \emph{optimal} estimator of Equation \eqref{eq:palphaunnormed}. We now highlight the advantages and disadvantages of some common approaches.

\subsection{Power spectrum estimation via map-making}
\label{sec:pspecmapmaking}
One way to deal with large data volumes is to compress the raw time series into maps of the sky. This can be accomplished using the methods outlined in Section \ref{sec:mapmaking}. If done in this manner, reducing a time series into a map is a lossless form of information compression \citep{Tegmark:1997mapmaking}, in the sense that any error bars on parameter constraints derived from the maps are provably just as good as those achievable from the raw time series. Note that the resulting maps need not be images in configuration space (i.e., with the axes of the maps being comoving distances in three dimensions). Maps can easily well be in harmonic space (e.g., spherical harmonic coefficients in the angular directions and spherical Bessel function coefficients in the radial direction; \citealt{Liu:2016}) or even some hybrid space where some directions are expressed in Fourier space while others are in configuration space \citep{Trott2016CHIPS}. This last possibility is particularly popular for interferometric map-making over small patches of the sky \citep{Dillon2015Mapmaking}, since in the flat-sky approximation an interferometer is essentially sampling in Fourier space in the angular directions but configuration space in the frequency/radial direction. In all of the above, map-making is lossless.

Following map-making, data volumes are typically small enough that if one simply desires some non-optimal estimate of the power spectrum, the procedure is conceptually straightforward. For example, if one had a three-dimensional map in configuration space, one could simply implement Equation \eqref{eq:PkContinuous}: Fourier transform all three directions, square the result, and normalize appropriately (with some binning during the process if desired). However, as we alluded to above, this is a suboptimal process that does not produce the smallest possible errors. And importantly, the rigorous estimation of one's error bars on the power spectrum is itself a challenge, perhaps even more so than the estimation of the power spectrum itself.

Said differently, implicitly accompanying the data $\vx$ (now in the form of a map) is its covariance $\mC \equiv \langle \vx \vx^\dagger \rangle$, which contains statistical error information that must be propagated through to the final power spectrum estimate. In addition to covariances that arise from the measurement process, the map-making process itself may introduce extra correlations. Typically, even with the compressed size of a map compared to raw interferometric data, $\mC$ is too large to be stored or manipulated. Thus, assumptions must be made regarding the structure of the $\mC$:
\begin{itemize}
\item Suppose the total covariance matrix $\mC$ is decomposed into the sum of a sky signal covariance matrix $\mS$ and an instrumental noise covariance matrix $\mN$. This is a reasonable decomposition as the instrumental noise is generally uncorrelated with the sky signal. Had this not been the case, there would be cross-covariance terms between the signal and the noise.
\item The noise covariance is typically assumed to be diagonal in frequency. This is reasonable because the visibility data coming out of an interferometer is uncorrelated between different frequency channels, and the map-making process usually proceeds independently for each frequency (unless the output of one's map-making pipeline is in harmonic space radially). It is important to emphasize, however, that while the noise is independent, it is generally not white. Mathematically, this is equivalent to saying that the noise covariance is diagonal but not proportional to the identity. This means that once the data are Fourier transformed along frequency, the noise will cease to be uncorrelated, as the identity matrix is the only matrix that is diagonal in all bases.
\item In the angular directions, different Fourier modes are frequently assumed to be uncorrelated. This approximation is made particularly often when performing theoretical forecasts of an interferometer's sensitivity to the power spectrum. The motivation is that an interferometer makes maps from visibilities, which are probes of a localized region in $uv$ space. Rewriting Equation \eqref{eq:FTvis} using the Convolution Theorem, we have
\begin{equation}
V_I = \int \widetilde{A}_p(\vu - \vu^\prime) T(\vu^\prime)  \: d^2 u^\prime,
\end{equation}
which suggests that if one's bin size on the $uv$ plane is coarser than $\widetilde{A}_p$, then each visibility provides an independent probe of a single $uv$ pixel. We stress, however, that this is a crude approximation, and that in practice one ought to use as fine a $uv$ pixelization as is feasible, and then to quantify the (strong) covariances between nearby pixels.
\item Some pipelines take advantage of the fact that the effective point spread function of an interferometer (the ``synthesized beam" in interferometric parlance) is reasonably narrow once all baselines of an interferometer have been included in the analysis. Thus, there is limited correlation between widely separated parts of the sky, and one can split the sky into a small number of facets that are assumed to be independent \citep{Cornwell1992,TegmarkZaldarriaga2010,Dillon2015Mapmaking}. In this manner, one is essentially making the approximation that one's covariance matrices are block diagonal. Alternatively, if a set of redundant overlapping facets are used, this is equivalent to assuming band diagonality of the covariance matrices. It is important to note, however, that the assumption of independent facets can be dangerous for interferometers with very regular antennas layouts. This is because aliasing induced by strong grating lobes of the synthesized beam may cause significant inter-facet correlations.
\end{itemize}

\subsection{Power spectrum estimation via delay spectra}
\label{sec:delayspec}
While a map-making approach uses visibility data from all baselines of an array (minus any that are deemed to be irrecoverably corrupted by systematics) to form a map before proceeding to a power spectrum, a delay spectrum approach does the opposite. Introduced by \citet{Parsons2012} for \tcm power spectrum estimation, the delay spectrum approach is one where data from a single baseline (or possibly a group of identical redundant baselines) is pushed through to a single power spectrum estimate before combining data from multiple non-identical baselines.

The delay spectrum is named after the delay transform, which is simply the Fourier transform of the frequency spectrum of a single baseline. The delay transformed visibility $\widetilde{V}_b (\tau)$ is thus given by
\begin{equation}
\label{eq:DelayTransformDef}
\widetilde{V}_b (\tau) \equiv \int d\nu V(\nu) \phi(\nu) e^{-i 2 \pi \nu \tau},
\end{equation}
where $\tau$ is the delay, and has dimensions of time. The function $\phi(\nu)$ is an optional tapering function. If no taper is applied, this is simply a tophat function that is zero outside the observed frequency band. The delay-transformed visibility is a convenient quantity to work with for two reason. First, the delay-transformed visibility can be interpreted geometrically. If we consider a limiting case where the sky and the primary beam are frequency independent, one sees that
\begin{eqnarray}
\widetilde{V}_b (\tau) &=& \int  d\Omega A_p(\vrhat) T(\vrhat) \left[ \int d\nu e^{-i 2 \pi \nu (\tau + \vb \cdot \vrhat /c) } \phi(\nu) \right] \qquad  \label{eq:VtildeIntermediate} \nonumber \\
&= &\int  d\Omega A_p(\vrhat) T(\vrhat) \widetilde{\phi} (\tau + \vb \cdot \vrhat /c).
\end{eqnarray}
If the observational band and the tapering function are reasonably broad, $\widetilde{\phi}$ becomes narrow. Under such an approximation, $\widetilde{\phi}$ is zero unless $\tau \approx - \vb \cdot \vrhat /c$, so each value of delay receives its contribution only from parts of the sky that satisfy $\tau \approx - \vb \cdot \vrhat /c$. This also motivates the term \emph{delay}, because for a given baseline $\vb$, the quantity $ \vb \cdot \vrhat /c$ is the time delay between the arrival of incoming radiation from direction $\vrhat$ at the first and the second antennas of the baseline. A particular delay mode thus receives radiation from a ring of constant delay in the sky; taking a delay transform therefore enables visibility data to be interpreted geometrically, essentially allowing data from a single baseline to be ``imaged" in 1D. Of course, this result is rigorously true only if the primary beam and the sky emission are both frequency independent. Neither assumption holds in strict detail. However, it is frequently a design goal of \tcm experiments to produce a frequency independent (or at least a spectrally smooth) primary beam as much as possible. The sky emission is also a reasonably slowly varying function of frequency, since it is in practice dominated not by the redshifted \tcm signal that we seek, but instead, by foreground contaminants such as Galactic synchrotron radiation (which we know from Section \ref{sec:fgs} are spectrally smooth). The delay spectrum is therefore a useful tool for diagnosing and mitigating foregrounds. We discuss this further in Section \ref{sec:FgMitigation}.

In addition to being helpful for foreground mitigation, delay-transformed visibilities are natural stepping stones towards power spectrum estimates from single baselines. Conceptually, the idea is again centred on the fact (discussed in Section \ref{sec:InterferometerSensitivityOptimization}) that a single baseline's visibility is a probe of a single $\vk_\perp$ mode on the sky, with the mapping between the baseline vector and the $\vk_\perp$ vector given by Equation \eqref{eq:kperpkparamappings}. Since the different frequencies of the visibility map to different radial distances, a visibility is therefore a quantity that resides in a hybrid space where the angular directions are expressed in Fourier space while the radial direction is in configuration space. Thus, a single Fourier transform of visibility data in frequency---a delay transform---should be sufficient for recasting the data in the three-dimensional Fourier space where it can then be squared and binned to produce a power spectrum.

Importantly, however, one must recognize that this approach is only an approximate one. The reason for this is that Equation \eqref{eq:kperpkparamappings} is only approximate. A more exact expression for the mapping between the baseline vector $\vb$ and $\vk_\perp$ is given by
\begin{equation}
\label{eq:moreexactkperp}
\vk_\perp =  \frac{2 \pi \nu \vb}{c D_c},
\end{equation}
which is identical to Equation \eqref{eq:kperpkparamappings} except the expression involves $\nu$, the frequency of a visibility, rather than $\nu_0$, the central frequency over some larger bandwidth. Our new expression reveals that a single baseline in fact drifts through multiple $\vk_\perp$ modes as a function of frequency. This means that to properly perform a Fourier transform along the line of sight, one requires multiple baselines that have the same $\nu \vb$ value across the radial extent of the transform.\footnote{Of course, this problem only exists because we are using $P(\vk)$ as an example in this section, which requires a Fourier transform along the line of sight to estimate. It is not a problem for statistics such as $C_\ell (\nu)$, for which only single frequency channels are needed. Indeed, visibility-based estimators for $C_\ell (\nu)$ can be written down for this in a reasonably direct manner \citep{Choudhuri2014}.}

The delay-spectrum approach to power spectrum estimation avoids the complications of combining data from multiple baselines. It assumes that for short baselines and relatively short frequency ranges, $\nu \vb$ does not vary substantially. In effect, one is assuming that Equation \eqref{eq:kperpkparamappings} is a good approximation. If this is the case, we can make the connection between the delay spectrum and the power spectrum precise by writing Equation \eqref{eq:VtildeIntermediate} in terms of the three-dimensional Fourier transform $\widetilde{T}(\vu, \eta)$ of the sky emission, so that (in the flat-sky approximation)
\begin{eqnarray}
\label{eq:freqindepvis}
\widetilde{V}_b (\tau) &=& \int d^2\theta d\nu A_p (\vtheta) e^{-i 2 \pi \nu (\tau + \vb \cdot \vtheta /c) } \phi(\nu) \nonumber \\
&& \times \int d^2 u d\eta e^{i 2\pi (\eta \nu + \vu \cdot \vtheta)} \widetilde{T} (\vu, \eta) \nonumber \\
&=& \int d^2 u d\eta  \widetilde{T} (\vu, \eta) \int d\nu e^{-i 2\pi \nu (\tau-\eta )} \phi(\nu) \widetilde{A}_p(\nu \vb / c  - \vu) \nonumber \\
& \approx & \int d^2 u d\eta  \widetilde{T} (\vu, \eta)  \widetilde{\phi}(\tau-\eta ) \widetilde{A}_p(\nu_0 \vb / c  - \vu),
\end{eqnarray}
where in the last line we made the approximation discussed above, where $\nu \vb$ is taken to be roughly constant and is evaluated at some representative frequency $\nu_0$. One sees from this expression that the delay-transformed visibility for a short baseline is roughly the Fourier amplitude of a $3$D Fourier mode, because $\widetilde{\phi}$ and $\widetilde{A}_p$ are relatively compact. The integral is thus a sampling of $\widetilde{T} $ at $\eta \approx \tau$ and $\vu \approx \nu_0 \vb / c$.\footnote{Conventionally, $\eta$ is used to denote the true Fourier dual to frequency (constructed using data from multiple baselines), whereas $\tau$ is the Fourier dual to the frequency spectrum on a single baseline. The statement that $\eta \approx \tau$ is therefore equivalent to saying that the delay approximation (i.e., that $\nu \vb \approx \nu_0 \vb$) is a good one in the context of power spectrum estimation.}  Given this result, one expects that the absolute square of the delay transform should be proportional to the power spectrum (at least in expectation). This is indeed the case:
\begin{eqnarray}
\langle |\widetilde{V}_b (\tau) |^2 \rangle &= &\int \! d^2 u_1 d^2 u_2 d\eta_1 d\eta_2 \widetilde{\phi}(\tau-\eta_1) \widetilde{\phi}^*(\tau-\eta_2) \nonumber \\
&&\times \widetilde{A}_p\left(\nu_0 \vb / c  - \vu_1\right) \widetilde{A}^*_p\left(\nu_0 \vb / c  - \vu_2\right)  \nonumber \\
&&\times  \langle \widetilde{T}(\vu_1, \eta_1) \widetilde{T}(\vu_2, \eta_2)  \rangle \nonumber \\
&=& \int d^2 u d\eta \big{|} \widetilde{A}_p(\nu_0 \vb / c  - \vu)\big{|}^2 \big{|}  \widetilde{\phi}(\tau-\eta ) \big{|}^2 P(\vu, \eta) \nonumber \\
& \approx & P\left(\vu = \nu_0 \vb / c, \eta = \tau \right) \nonumber \\
&& \times \int d^2 u d\eta \big{|} \widetilde{A}_p(\nu_0 \vb / c  - \vu)\big{|}^2 \big{|}  \widetilde{\phi}(\tau-\eta ) \big{|}^2 \nonumber \\
&=& P\left(\vu = \frac{\nu_0 \vb }{ c}, \eta = \tau \right) \int d^2 \theta d\nu \big{|} A_p(\theta) \big{|}^2 \big{|} \phi(\nu) \big{|}^2, \label{eq:delayscalar} \qquad\,
\end{eqnarray}
where in the last equality we invoked Parseval's theorem. In the preceding line, we made the approximation that the power spectrum varies slowly over the scales where $|\widetilde{A}_p|^2$ and $|\widetilde{\phi}|^2$ are non-zero. With this assumption, we were able to evaluate $P(\vu, \eta)$ at $(\vu, \eta) = (\nu_0 \vb / c, \tau)$ and to factor the term out of the integral. This is mathematically reminiscent of the Feldman-Kaiser-Peacock (FKP) approximation that is commonly employed in galaxy surveys \citep{Tegmark1998FutureSurveys}.

With this derivation, one sees that an estimate of the power spectrum can be obtained by squaring delay-transformed visibilities from a single baseline and dividing by the integral on the right hand side of Equation \eqref{eq:delayscalar}. While our derivation required an ensemble average, its omission in practice simply results in a power spectrum estimate that does not necessarily deliver precisely the true theoretical power spectrum, but instead has some error bar associated with it (as it must).

The delay spectrum approach provides a power spectrum estimation method that stays close to the primary data output of an interferometer: the visibilities from individual baselines. This has several advantages. First, it means that cuts to the data (such as those due to malfunctioning baselines or radio frequency interference; see Section \ref{sec:RFI}) can happen near the end of one's analysis process. In addition, because power spectra are estimated from individual baselines before they are averaged together, calibration requirements are in principle less stringent. For instance, suppose some calibration error caused each baseline's visibilities to be multiplied by a different complex phase. Such an error would have no effect on a delay-based power spectrum, since each baseline's delay spectrum is individually squared, causing complex phases to cancel out. On the other hand, this type of calibration error (in essence, a \emph{relative} calibration error between different baselines) would affect a power spectrum estimation method that is based on map-making, since map-making typically requires the combination of data from multiple baselines into one map. \citet{Morales2019diverseanalyses} provide a discussion of the different calibration errors that arise in power spectrum pipelines that do not employ the delay approximation versus those that do.

The disadvantage of a delay spectrum approach is that it is often difficult to extract the full sensitivity of an array. Since baselines are treated individually, any information about the power spectrum that resides in the cross multiplication of two different baselines is lost. Baselines that are significantly different will have visibilities that are sensitive to significantly different Fourier modes of the sky; thus, their cross multiplication will not provide much extra sensitivity to the power spectrum. On the other hand, baselines that are similar in length and orientation (including those that become similar due to the rotation synthesis effect discussed in Section \ref{sec:dish_or_inter}) will be sensitive to the power spectrum. It is certainly possible to generalize the delay spectrum estimates between two non-identical baselines---one essentially repeats the derivation that led to Equation \eqref{eq:delayscalar}, but with the cross multiplication $\langle \widetilde{V}_{b_1} \widetilde{V}_{b_2}^*  \rangle$ between $b_1$ and $b_2$ in place of $\langle | \widetilde{V}_{b} |^2 \rangle$. The result is simply a slightly different normalization integral on the right hand side of Equation \eqref{eq:delayscalar} that must be divided out to obtain the power spectrum \citep{Zhang2018nearredundant}. However, by mixing information from different baselines, the aforementioned calibration advantages are lost.

As presented here, the delay spectrum approach also does not take full advantage of a long time integration to average down noise. Implicit in the derivations here are that $V(\nu)$ refers to the visibility at a single instant in time. Typically, a power spectrum is estimated for each time instant, and the resulting collection of power spectra are then averaged together. The delay spectrum approach is therefore a ``square then average" method, whereas a map-making-based approach would be an ``average then square" method. The former represents an incoherent averaging of the signal which does reduce noise, but the latter is a coherent averaging that reduces noise more quickly. More concretely, suppose that there are $N_t$ number of time samples in one's data set. If $N_t$ independent power spectra are formed, averaging them together should reduce error bars by $\sim 1/ \sqrt{N_t}$. On the other hand, if these $N_t$ data samples are first combined coherently into a map of the \tcm brightness temperature, it is the map's error that goes down by $\sim 1 / \sqrt{N_t}$. Since power spectra are quadratic functions of the brightness temperature field, the resulting averaged power spectrum has errors that scale roughly as $\sim 1 / N_t$, which is a considerably faster scaling. To recover the lost sensitivity from a ``square then average" algorithm, several methods have been proposed. For telescopes that track a particular field over time, \citet{2016SourabhTrackingDelay} compute cross delay spectra of all possible pairs of time integrations over a long observation. For drift-scan telescopes, one can employ techniques such as fringe-rate filtering, which performs a weighted sliding average over a visibility time series. If the weights of this average are carefully optimized, fringe-rate filtering can recover the full sensitivity of an ``average then square" power spectrum estimation \citep{ParsonsFringeRate2016}. However, the result is a series of visibilities that have strong noise correlations between them, and care must be taken to ensure that these correlations are properly propagated into one's error budget.

In closing, we address some of the approximations made in deriving the delay transform power spectrum estimator. In deriving the normalizing scalar in Equation \eqref{eq:delayscalar}, we made use of the flat-sky approximation. While many instruments designed for \tcm cosmology have a large field of view (and therefore violate the strict regime of validity of the approximation), in practice one finds that the curved sky corrections to the derivation are negligible \citep{Liu:2016}.

Another approximation is the FKP-like approximation of assuming that the power spectrum varies relatively slowly as a function of $\vu$ and $\eta$. This assumption is typically violated if foreground contaminants are present in one's data, because foregrounds are orders of magnitude brighter than the cosmological signal (see Section \ref{sec:fgs}) and preferentially occupy certain Fourier modes (see Section \ref{sec:FgMitigation}). As a result, when one moves away from the foreground-dominated regions of Fourier space, the power spectrum varies rapidly as $\vu$ and/or $\eta$ changes. Despite this issue, one can show that the normalizing scalar derived in Equation \eqref{eq:delayscalar} still correctly normalizes the power spectrum. In other words, a power spectrum estimated using the delay spectrum approach of this section remains a perfectly valid estimator, provided its error properties are properly accounted for. In particular, the FKP-like approximation in Equation \eqref{eq:delayscalar} is tantamount to assuming that our power spectrum's window functions are delta functions, since the square of the delay spectrum---once normalized---is equal to the power spectrum at a specific point in Fourier space, rather than being an integral over a finite region. Backing off from this approximation to the previous line, we see that the square of the delay spectrum is an integral over the power spectrum, and thus the window functions (with the FKP-like approximation) can be read off. In fact, because our estimator is a quadratic function of the data, the machinery of the quadratic estimator can be employed to provide a rigorous quantification of the statistics associated with the delay spectrum \citep{Liu:2014b}. The quadratic estimator formalism can also be used to quantify the effects of the delay approximation. This is accomplished by constructing an estimator in the delay \emph{basis}, but avoiding the invocation of the delay \emph{approximation} by computing window functions that show how a delay spectrum estimate samples linear combinations of the true rectilinear Fourier modes \citep{Parsons2012,Liu:2014a}.

\subsection{Power spectrum estimation via $m$-mode analysis}

A third way to achieve a computationally feasible power spectrum estimation pipeline is to use the $m$-mode formalism that was presented in Section \ref{sec:mmodes}. Recall that the $m$-mode formalism provides substantial computational savings by recognizing that for a drift scan telescope, if one Fourier transforms a set of visibilities in time, the $m$th Fourier coefficient $V_m$ of the result depends only on spherical harmonic mode coefficients $a_{\ell m}$ with the same $m$ indices. Suppressing polarization indices to focus our discussion on power spectrum estimation, we have
\begin{equation}
V^i_m (\nu) = \sum_\ell B_{\ell m}^i (\nu) a_{\ell m} (\nu) + n^i_m (\nu),
\end{equation}
where $i$ is an index specifying the baseline, $n^i_m$ is an instrumental noise contribution, and $B_{\ell m}^i $ is the beam transfer matrix from Section \ref{sec:mmodes}. One sees that whereas there is a sum over $\ell$, there is no sum over $m$. Different $m$ modes can therefore be solved for independently of one another. This is also a conclusion that can be seen in the flat-sky derivation of Appendix \ref{sec:DriftScans}, where we bridge the intuition between the curved-sky case and the $uv$ plane. There, we see that only in the north-south direction is there a convolution of different Fourier modes together by the primary beam, whereas in the east-west direction each Fourier mode remains independent. This is the flat-sky analog of different $m$ modes being independent.

The advantages of analyzing drift scan telescope data in the $m$-mode formalism in fact extend beyond map-making. This is important because as we remarked in Section \ref{sec:MapmakingNoiseCovar}, if one's end goal is not a map (but instead is, say, a power spectrum), then there are advantages to dealing directly with visibilities in both foreground subtraction and power spectrum estimation. That the different $m$ modes do not couple to one another makes many foreground subtraction algorithms presented in Section \ref{sec:FgMitigation} easier to implement.

The situation becomes more complicated in the context of power spectrum estimation, because further assumptions must be made in order for different $m$ modes to be independent. In particular, if we look back at Equation \eqref{eq:palphaunnormed}, we see that we need both $\mC$ and $\mQ^\alpha$ to possess the same block-diagonal, $m$-decoupled structure. In Section \ref{sec:mmodeapptomapmaking}, we in fact already argued that $\mC$ has the right structure, but we will do so again here in a more explicit way, in order to be able to compute $\mQ^\alpha \equiv \partial \mC / \partial p^\alpha$ (Equation \ref{eq:QmatrixdCdp}).

Consider first the covariance matrix of the data in the $m$-mode basis. This is given by
\begin{eqnarray}
\mC_{AB} &\equiv& \langle \langle V^i_m (\nu) V^{i^\prime}_{m^\prime} (\nu^\prime)^* \rangle \nonumber \\
&=& \mN_{AB} + \sum_{\ell \ell^\prime} B_{\ell m}^i B_{\ell^\prime m^\prime}^{i^{\prime *}} \langle a_{\ell m} (\nu) a_{\ell^\prime m^\prime} (\nu^\prime)^* \rangle, \qquad
\end{eqnarray}
where we have collapsed triplets of baseline, $m$ mode, and frequency indices into upper case Latin indices, and $\mN_{AB} \equiv \langle n^i_m (\nu) n^{i^\prime}_{m^\prime} (\nu^\prime)^* \rangle$. One sees that elements of $\mC$ with $A$ and $B$ corresponding to $m \neq m^\prime$ are not necessarily zero, depending on the statistical properties of the spherical harmonic coefficients on the right hand side. The covariance matrix therefore mixes different $m$ modes. This means that when enacting the $\mC^{-1}$ weighting of the $\vx$ (as demanded by, e.g., Eq. \ref{eq:palphaunnormed}), one loses the computational savings of working in an $m$ mode basis, since all $m$ modes must be simultaneously considered.

To regain our computational savings, it is necessary to assume that the sky is statistically isotropic. With this assumption, we can use the fact that
\begin{equation}
\langle a_{\ell m} (\nu) a_{\ell^\prime m^\prime} (\nu^\prime)^* \rangle = C_\ell (\nu, \nu^\prime) \delta_{\ell \ell^\prime} \delta_{m m^\prime},
\end{equation}
where $C_\ell (\nu, \nu^\prime)$ is the cross power spectrum between frequencies $\nu$ and $\nu^\prime$. With this simplification, the covariance becomes
\begin{equation}
\mC_{AB} = \mN_{AB} + \delta_{m m^\prime} \sum_{\ell} B_{\ell m}^i B_{\ell m}^{i^{\prime *}} C_\ell (\nu, \nu^\prime).
\end{equation}
The independence of different $m$ modes is now manifest in the second term. The first term will exhibit the same dependence provided the noise contributions are statistically stationary in time. Provided these assumptions are satisfied, we may proceed by relating $C_\ell (\nu, \nu^\prime)$ to the power spectrum. In the flat sky approximation, for instance, one has
\begin{eqnarray}
C_\ell (\nu, \nu^\prime) &=& \frac{1}{\pi D_c D_c^\prime} \int_0^\infty dk_\parallel \cos(k_\parallel \Delta D_c) P(\mathbf{k}, \overline{z}) \nonumber \\
&\approx& \sum_\alpha \left[ \frac{\Delta D_c \Delta k_\parallel}{\pi D_c D_c^\prime} \cos(k_\parallel \Delta D_c) \right] p_\alpha
\end{eqnarray}
where $D_c$ and $D_c^\prime$ are the comoving line of sight distances corresponding to $\nu$ and $\nu^\prime$, respectively, $\Delta D_c$ is their difference, and $\overline{z}$ is an effective (weighted mean) redshift between $\nu$ and $\nu^\prime$ \citep{Datta2007,Shaw:2014}. A similar expression can be written down for a full curved-sky treatment \citep{Datta2007}, and \citet{Eastwood2019Pspec} provides insight into when the flat-sky approximation is appropriate. For now, though, continuing with our flat sky expression and inserting it into $\mC_{AB}$ yields an equation of the same form as Eq. \eqref{eq:CQPlink}, with
\begin{equation}
\mQ^\alpha_{AB} = \delta_{m m^\prime} \sum_{\ell} B_{\ell m}^i B_{\ell m}^{i^{\prime *}} \left[ \frac{\Delta D_c \Delta k_\parallel}{\pi D_c D_c^\prime} \cos(k_\parallel \Delta D_c) \right].
\end{equation}
We thus see that the matrices needed for power spectrum estimation (i.e., $\mQ^\alpha$ and $\mC$) all satisfy the property that they do not couple different $m$ modes. One can therefore use each $m$ mode to form its own estimate of the power spectrum, and then to average the different estimates together. In conclusion, the computational savings of an $m$-mode analysis are preserved not only for the map-making exercises of Section \ref{sec:mapmaking}, but also for power spectrum estimation.

To arrive at our conclusion, it was necessary to assume a statistically isotropic sky. This assumption is violated by foreground emission, and implies a coupling of different $m$ modes. Foreground emission should be included in the modelling of $\mC$, for even if foreground subtraction has been performed prior to power spectrum estimation, the subtraction will not have been perfect, and a residual covariance will be present. To see why foreground covariances generally couple different $m$ modes, consider the fact that foreground emission in the Galactic plane is substantially stronger than that near the Galactic poles. To produce such a pattern of emission, the spherical harmonic modes that comprise the sky emission must constructively interfere at the Galactic plane, which cannot be accomplished if the different $m$ modes are uncorrelated as the different modes must collectively ``know" to peak near the Galactic plane. Some advanced $m$-mode treatments allow limited relaxations of statistical isotropy (e.g., by including the modelling of bright point sources; \citealt{Berger2017}), but have yet to be shown to be free of artifacts that could compromise the spectral smoothness of the data. In any case, it should be noted that while the application of the above expressions for $\mQ^\alpha$ and $\mC$ to a statistically anisotropic sky is non-optimal (in the sense of delivering the smallest possible error bars), it does not result in biased power spectrum estimates. This can be shown by repeating the derivations following Equation \ref{eq:OQEFormOfE}, but with generic weighting matrices instead of $\mathbf{C}^{-1}$. This essentially mimics an incorrectly modelled covariance, and one can show that even in such a scenario, the resulting estimators satisfy $\langle \mathbf{\hat{p}} \rangle= \vp$, i.e., they are unbiased.\footnote{Our claim is true provided the covariance modelling is not done empirically (i.e., based on the data itself). See Section \ref{sec:sigloss} for discussions of this subtlety.}
\subsection{Bayesian power spectrum estimation}
\label{sec:BayesPspec}
As a final example of a power spectrum estimation method, we discuss a Bayesian approach. Up until now, our discussions have centred on \emph{frequentist} approaches to power spectrum estimation, where one follows a deterministic recipe for computing power spectra from the data. A Bayesian approach, on the other hand, begins with a procedure for forward modelling some data given a possible underlying power spectrum. This forward-modelling process is then performed many times for a wide variety of different possible power spectra [i.e., a wide variety of possible forms for $P(\vk)$], and the resulting data are checked against the actual observed data. By exploring the space of possible power spectra and how well each one fits the observed data, a Bayesian approach in principle provides not just a best-fit answer to what the power spectrum is, but also a full probability distribution of possible answers, thereby capturing a richer statistical description of one's results than what a simple set of error bars provide.

Mathematically, the key to pursuing the aforementioned approach is Bayes' theorem. Suppose we group a set of hypothetical power spectrum values into a vector $\boldsymbol \theta$ (with each element storing the value of the power spectrum in a different $k$ bin, i.e., a bandpower).  Computing the power spectrum is then tantamount to computing the probability distribution of $\boldsymbol \theta$ \emph{given} some data $\mathbf{d}$, which we denote $\textrm{Pr}(\boldsymbol \theta | \mathbf{d})$. To compute this distribution, one invokes Bayes' theorem, which states that
\begin{equation}
\label{eq:Bayes}
\textrm{Pr}(\boldsymbol \theta | \mathbf{d}, \mathcal{M}) =\frac{ \mathcal{L} (\mathbf{d} | \boldsymbol \theta, \mathcal{M}) \textrm{Pr}(\boldsymbol \theta, \mathcal{M})}{\textrm{Pr}(\mathbf{d}| \mathcal{M})},
\end{equation}
where $\mathcal{M}$ signifies the theoretical model, $\textrm{Pr}(\mathbf{d}| \mathcal{M})$ is known as the \emph{evidence} (which we return to in Section \ref{sec:modelselection}), $\textrm{Pr}(\boldsymbol \theta)$ is the \emph{prior} on the parameters, and $\mathcal{L} (\mathbf{d} | \boldsymbol \theta)$ is the likelihood function. The prior represents our previous belief (before folding in the current data) in the distribution of the bandpowers. This is updated by the likelihood function, which is where the forward modelling aspect of Bayesian inference enters. It is how details of the observation are incorporated into the analysis. As a toy example, suppose we were ``observing" a set of simulated Fourier coefficients of a Gaussian random sky drawn from a power spectrum $P(\vk)$. These Fourier coefficients are then grouped into our data vector $\vd = [\widetilde{T} (\vk_1), \widetilde{T} (\vk_2), \dots]$, and the parameter vector that we wish to infer is a vector of bandpowers, i.e., $\boldsymbol \theta = [P(\vk_1), P (\vk_2), \dots]$. With no noise or instrumental effects, the likelihood takes the form
\begin{equation}
\label{eq:PspecLikelihood}
\mathcal{L} (\mathbf{d} | \boldsymbol \theta) = \frac{1}{\sqrt{(2\pi)^{N_d } \det [ \mP(\boldsymbol \theta) ]}} \exp \left( -\frac{1}{2} \mathbf{d}^t [\mathbf{P} (\boldsymbol \theta)]^{-1} \mathbf{d}  \right),
\end{equation}
where $N_d$ is the length of $\vd$, and $\mP$ is a diagonal matrix with $\boldsymbol \theta$ along the diagonal. This likelihood is simply a reflection of the fact that in this example our fluctuations are Gaussian, plus the fact that by definition the power spectrum is the variance of Fourier coefficients, as one can see from Equation \eqref{eq:PspecDef}.

If one thinks of the likelihood as a function of $\mathbf{d}$ with $\boldsymbol \theta$ fixed, then it is a properly normalized probability distribution for the observed data. However, if one interprets the likelihood as a function of $\boldsymbol \theta$ with $\mathbf{d}$ fixed at the observed values, then one does not in general have a properly normalized probability distribution, even if the constants of proportionality are included in Equation \eqref{eq:PspecLikelihood}. Fortunately, Bayes' theorem is precisely what we need to obtain a probability distribution for $\boldsymbol \theta$: one sees from Equation \eqref{eq:Bayes} that one just needs to multiply by the prior, and then to divide by the evidence $\textrm{Pr}(\mathbf{d} | \mathcal{M})$. But since the evidence does not depend on the parameters $\boldsymbol \theta$, it is simpler to think of this division as just a normalization of the posterior distribution.

Conceptually, then, obtaining probability distributions for the $\boldsymbol \theta$ is simple. The posterior is simply evaluated over a wide variety of different $\boldsymbol \theta$ values using Bayes' theorem. Computationally, however, such a brute-force method quickly becomes infeasible, since the amount of parameter space volume that must be covered grows exponentially with the number of parameters (i.e., with the length of the vector $\boldsymbol \theta$). To deal with this, one generally employs a Markov Chain Monte Carlo (MCMC) analysis. With an MCMC, one jumps around parameter space with a predetermined set of rules, evaluating the likelihood after each jump. The rules are chosen such that the resulting \emph{chain} of parameter space locations are distributed in a special way. In particular, forming a histogram of these parameter space locations gives the posterior distribution $\textrm{Pr}(\boldsymbol \theta | \mathbf{d})$. Sampling the distribution in this way mitigates the ``curse of dimensionality", where bad computational scalings with the number of dimensions make an explicit computation of the posterior prohibitive.

An MCMC analysis also makes it straightforward to marginalize over parameters that are not of interest. Algebraically, if one is not interested in, say, parameter $\theta_0$, one simply averages over all plausible values of it (``marginalizing over it") by integrating the posterior from $-\infty$ to $\infty$ in $\theta_0$. In other words, the quantity
\begin{equation}
\textrm{Pr}(\theta_1, \theta_2,  \dots | \mathbf{d} ) = \int_{-\infty}^\infty \textrm{Pr}(\theta_0, \theta_1, \theta_2,  \dots | \mathbf{d} ) d\theta_0
\end{equation}
is the correct marginalized probability distribution for the remaining parameters. Taking this to the extreme, if one is interested in the distribution of a single parameter (for instance, in order to quote a single error bar on the measurement), one marginalizes over all parameters but the parameter of interest. An MCMC analysis makes this trivial, as marginalization is equivalent to simply ignoring various parameters when constructing histograms.

Because MCMC analyses are ubiquitous in cosmology (and indeed, other fields as well), we will not review the details here, instead referring readers to resources such as \citet{MackayBook, Hogg2018}. We now return to the specific problem of \tcm power spectrum estimation and provide a very brief and simplified introduction to the Bayesian methods pioneered by \citet{Sims2016,Sims2019a, Sims2019b}.

While the likelihood function that we wrote down in Equation \eqref{eq:PspecLikelihood} is a good start, it assumes a set of perfectly measured Fourier coefficients $\{ \widetilde{T} (\vk_i) \}$. If we had such a set of uncorrupted coefficients, power spectrum estimation would be trivial. The challenge is to relate the power spectrum directly to data coming off an interferometer. In other words, for our Bayesian inference problem, we would like to define $\vd$ to be something closer to the raw data. For example, we could let $\vd$ be a set of visibilities that have been accumulated over time on the $uv$ plane. \citet{Sims2019a} then take the approach of inferring $\va \equiv [\widetilde{T} (\vk_1), \widetilde{T} (\vk_2), \dots]$ and $\vp \equiv [P(\vk_1), P (\vk_2), \dots]$ jointly.\footnote{This is in fact a simplified version of the \citet{Sims2019a} approach, which includes a few extra parameters to account for spatial fluctuations that are on length scales beyond the extent of the survey volume being analyzed.} Using Bayes' theorem gives
\begin{eqnarray}
\textrm{Pr}(\va, \vp | \vd, \mathcal{M}) &\propto & \mathcal{L} (\vd | \va, \vp, \mathcal{M}) \textrm{Pr}(\va, \vp | \mathcal{M}) \nonumber \\
&\propto & \mathcal{L} (\vd | \va, \mathcal{M}) \textrm{Pr}(\va | \vp, \mathcal{M}) \textrm{Pr}( \vp | \mathcal{M}), \quad
\end{eqnarray}
where in the second line we used the fact that $\textrm{Pr}(\va, \vp) = \textrm{Pr}(\va | \vp) \textrm{Pr}( \vp)$. We also dropped the $\vp$ argument from the likelihood because the data does not explicitly depend on $\vp$, since knowing $\va$ (the actual Fourier coefficients of our sky) completely determines the data $\vd$, up to instrumental noise. Written in this way, we have all the pieces that we need to evaluate the posterior: $\textrm{Pr}( \vp | \mathcal{M})$ is simply the prior on our power spectrum, $\textrm{Pr}(\va | \vp, \mathcal{M})$ is the probability distribution of the Fourier coefficients given the power spectrum (which is precisely Equation \ref{eq:PspecLikelihood}, despite the different notation), and $\mathcal{L} (\vd | \va, \mathcal{M})$ can be written as
\begin{equation}
\mathcal{L} (\vd | \va, \mathcal{M}) \propto \exp \left[ -\frac{1}{2} (\mathbf{d}-\mT \va)^\dagger \mN^{-1} (\mathbf{d}-\mT \va)\right],
\end{equation}
assuming that the noise in the visibilities is Gaussian with instrumental noise covariance matrix $\mN$. Here, $\mT$ is a matrix that transforms Fourier coefficients $\va$ into visibilities. This process can be written as the multiplication of a matrix because going from $\{ \widetilde{T} (\vk_i) \}$ to a configuration space map $T(\vr)$ is a linear operation, as is Equation \eqref{eq:BasicVis} for going from the map to visibilities.

With a full expression for $\textrm{Pr}(\va, \vp | \vd, \mathcal{M}) $, one can sample from this distribution using MCMC techniques. Obtaining what we are ultimately interested in---the final posterior $\textrm{Pr}(\vp| \vd)$ for the power spectrum $\vp$ given the data $\vd$---then just requires marginalizing over $\va$. \citet{Sims2019a} explore two different ways to do this. One uses a Hamiltonian Monte Carlo method to make the process computationally tractable. The other performs an analytic marginalization over $\va$. Either way, the end result is a full set of probability distributions for power spectrum bandpowers. In fact, \citet{Sims2019b} goes further and incorporates a foreground model into the likelihood, unifying power spectrum estimation with foreground mitigation. In doing so, one advantage of this Bayesian approach (over many of the approaches discussed in Section \ref{sec:FgMitigation}) is that it permits a statistically disciplined treatment of uncertainties in our understanding of foregrounds.

%

\section{Foreground mitigation}
\label{sec:FgMitigation}
Having established the general principles behind foreground contamination and power spectrum estimation, we now give a sampling of various specific mitigation methods that have been proposed in the literature. While our implicit context will typically revolve around foreground mitigation for a power spectrum measurement, many of the methods we outline below are also applicable to other data products, such as images. We begin with an examination of \emph{astrophysical} foreground contamination, as those represent contaminants that are unavoidable for any experiment
\subsection{Astrophysical contaminants}
\subsubsection{Parametrized fits}
\label{sec:paramfits}
Given that foreground sources are generally expected to be spectrally smooth, one may hope that foregrounds can be mitigated simply by fitting---and subtracting---smooth\footnote{We note that when we say ``smooth", we are \emph{not} referring to the mathematician's definition of ``smooth", i.e., infinitely differentiable. Rather, we mean \emph{slowly varying}.} functions from the spectra. Some proposals involve first imaging the data (e.g., by using some of the map-making techniques outlined in Section \ref{sec:mapmaking}), and then performing a pixel-by-pixel fit of smooth functions to the spectra \citep{DiMatteo2002,Santos2005,Wang2006}. However, when taking the resulting (hopefully foreground-mitigated) data through to a power spectrum, one finds that this method performs rather poorly on fine angular scales \citep{Liu2009,Bowman2009}. The reason is that fine angular scales are probed by the long baselines of an interferometer, and with many interferometric array configurations, this is precisely the regime where baseline coverage becomes sparse. The consequences of this can be understood by thinking about the $uv$ plane. Recall from Equation \eqref{eq:udef} that $|\mathbf{u}| \sim b \nu / c$. Now consider a single $uv$ pixel on a part of the plane that is so sparse that it is measured by just a single baseline, with no other baselines in the vicinity. While a baseline may probe this $uv$ pixel for this frequency, at a different frequency it probes a slightly different $uv$ pixel. This means that for a fixed $uv$ pixel, the baseline coverage is intermittent as a function of frequency, and the alternation between zero and non-zero information content imprints unsmooth spectral features that are difficult to fit out using smooth functions. To some extent, this can be mitigated by performing the smooth spectral fits in a pixel-by-pixel manner not in image space, but in $uv$ space \citep{Liu2009b}. This allows certain $uv$ pixels to be given zero weight in the spectral fits at frequencies where there are no measurements. This partially alleviates the problem, but fundamentally, it is a symptom of the intrinsic chromaticity of interferometry, which we alluded to in our discussion of delay spectra in Section \ref{sec:delayspec}. In the context of foregrounds, it makes spectrally smooth foregrounds appear spectrally unsmooth when viewed through the instrument. We discuss this in more detail in Section \ref{sec:wedge}.

In fitting smooth functions to spectra, one must decide on what smooth functions to fit. A popular choice is to decompose the spectra into a linear combination of orthogonal polynomials. The most slowly varying polynomials are then discarded. These fits can be performed on intensity data that is expressed as a function of frequency or in log-log space. The latter has the advantage that log-log polynomials are power laws (with corrections that are small if the higher order terms are subdominant), which are a good fit to both empirical observations and simple physical models of foreground emission (see Section \ref{sec:fgs}). However, in practice this can be difficult, because as we discussed in Section \ref{sec:dish_or_inter}, interferometers discard the zero mode (i.e., the angular mean) of the data. Thus, the spectra are not guaranteed to be positive, which motivates avoiding log-log space.

\subsubsection{Non-parametric fits}
\label{sec:NonParam}
The aforementioned parametric fits are open to the criticism that they require an arbitrary parameterization, i.e., an arbitrary choice of basis functions. On one hand, one requires only that the basis functions span the space of possible spectra. In other words, they simply need to form a proper basis. On the other hand, an unwise choice of basis could result in inefficient foreground removal, where a large number of modes need to be removed before the data are relatively foreground-free. This then runs the risk of removing the signal too, since in each mode there (in principle) resides both foregrounds and cosmological signal. In expanding one's spectra in terms of a set of pre-defined (but arbitrary) basis functions that are not connected to physical models of the foregrounds, one is simply hoping that two assumptions hold. First, that the foregrounds only occupy a small handful of the modes. Second, that they completely dominate the cosmological signal in these modes. Collectively, these assumptions---if true---minimize the probability of accidentally projecting out the cosmological signal, which is a subject that we will return to in Section \ref{sec:sigloss}.

Although it does not eradicate the aforementioned concerns, an alternative approach to the problem is to use non-parametric fitting techniques. Such techniques do not impose \emph{a priori} functional forms to describe the foregrounds. Instead, they essentially provide a quantitative criterion by which to judge the ``smoothness" of the spectra, thus allowing the smooth portions of the data to be fit out. For example, \citet{Harker:2009} suggest a scheme known as Wp smoothing. In the Wp scheme, a curve is considered unsmooth (and therefore unlikely to contain foregrounds) if it has large \emph{changes} in curvature over the frequency range of interest. The reason for selecting this criterion rather than the actual curvature is that a function with persistent moderate levels of curvature integrated over a large frequency range would be considered an unsmooth function, even though real foregrounds could easily exhibit such behaviour. The Wp method identifies foregrounds by performing least-squares fits to the data, but with a penalty for large changes in curvature. The fitting process is thus a balance between two competing demands: the desire for a good fit to the data (which would motivate the fit to pass through all data points even if this means a final spectrum with many wiggles) and the desire to minimize changes in curvature (which drives the fit to have fewer wiggles). The relative weights given to these demands are adjusted to enable the best-possible identification of foregrounds in simulations. Once the smooth (hopefully foreground-dominated) fit has been obtained, it is subtracted from the original data to give (hopefully cosmological signal-dominated) residuals.

As another example of a non-parameteric approach, we consider the proposal of \citet{Cho:2012}, which takes direct advantage of the differing spectral coherence lengths of the foregrounds versus the cosmological signal. Suppose one formed a stack of $uv$ plane ``images" at the different frequency channels of observation. If these images are then averaged across frequency, the slowly varying foregrounds will accumulate signal to noise more quickly than the rapidly varying (and random) cosmological signal. Thus, the averaged map has a suppressed cosmological signal, and can be used as an angular template for the foregrounds (either as an actual map or as a statistical model in the form of an angular power spectrum). The template can then be multiplied by an overall scaling factor at every frequency and subtracted off the data, with the scaling factors determined by minimizing the residuals of the fit.

A final example of a non-parametric approach is to use Gaussian Process regression (GPR; see \citealt{Rasmussen2006} for an introduction). Recently proposed in \citet{Mertens:2018} for the \tcm cosmology, the idea behind GPR foreground removal is to model the observed $N_\textrm{freq}$ data points of a spectrum as being drawn from an $N_\textrm{freq}$-dimensional Gaussian distribution. The discrete $N_\textrm{freq} \times N_\textrm{freq}$ covariance matrix of the distribution is then assumed to be a discrete sampling of a continuous covariance $K(\nu, \nu^\prime)$, which takes on a particular functional form that is chosen by the data analyst. With knowledge of this continuous covariance, one has constraints on the correlations between the measured data points and a (hypothetical) new data point anywhere on frequency axis. This enables the prediction of a series of new data points as a function of frequency, thus producing a fit to the data.

For the application to \tcm, the overall covariance function is often modelled as the sum of covariance functions that describe each of the components. Thus, one might have $K(\nu, \nu^\prime) \equiv K_{21} + K_\textrm{fg}$, where $K_{21}$ and $K_\textrm{fg}$ are covariance functions for the \tcm signal and the foregrounds, respectively. A popular model for these functions are the Matern kernels. The Gaussian is a special case of a Matern kernel, and qualitatively, the Matern kernels peak at $\nu = \nu^\prime$ and decay away from this equality. The characteristic width of this decay is an adjustable parameter, and is typically adjusted so that $K_\textrm{fg}$ has a longer coherence length than $K_{21}$, reflecting the fact that the foreground spectra are smoother than the cosmological signal. With both the kernels defined, one can use the foreground kernel to write down the posterior distribution for new data points \emph{given} the measured data, which are assumed to be distributed based on the sum of the two kernels. The result is a prediction---with quantified uncertainties---of the foreground spectrum, which is then subtracted off the data to yield a hopefully cleaned dataset.

While GPR is a non-parametric approach to fitting in the sense that it does not assume fixed functional forms for its basis functions, one does have to make a choice about the form of the covariance functions. This is necessary to make the problem a well-conditioned one---it is simply impossible to fit data without making assumptions! However, to avoid making assumptions that are too stringent (which could come dangerously close to declaring that we know \emph{a priori} precisely what the foreground spectra are), the kernels used in most GPRs possess tuneable parameters that are fit for as part of the data fitting exercise. The uncertainties in these parameters can then be marginalized over in the final predictions of the foreground spectra. While this does not change the fact that a form for the covariance had to be chosen, it does allow the data to drive the precise parameters of the covariance. For example, in our application we can allow the widths of the foreground and cosmological kernels to be learned from the data. In principle, one can (and should) even go one step further and use statistical model selection techniques (such as the Bayesian evidence; see Section \ref{sec:modelselection}) to confirm that the chosen covariance models are sensible choices for describing the data.

\subsubsection{Mode projection}
\label{sec:fgmodeprojection}
Mode projection is yet another way in which foregrounds can be removed. By this, we mean expressing the data in some basis, and then zeroing out the contribution from selected basis components that are believed to be dominated by foregrounds. The polynomial subtractions of Sec. \ref{sec:paramfits} are in fact an example of this, since fitting out low-order polynomials is equivalent to expressing the spectra in terms of orthogonal polynomials, and then projecting out the lowest order terms. Here, we examine proposed mode projection schemes that use information beyond the simple observation that foregrounds tend to be spectrally smooth.

One example of going beyond spectral smoothness is the removal of bright point sources. Implicitly, the motivation for this is the (extremely reasonable) prior that anomalously bright pixels are almost certainly foreground sources. In CMB observations, bright point sources can be removed simply by masking out the pixels where they reside (e.g., \citealt{Planck2018IV}). In the case of \tcm interferometers, the synthesized beam of an instrument has considerable structure, which makes point source removal much more complicated \citep{Pindor2011}. Thus, the projecting out of point sources generally requires a forward-modelling effort, where point source catalogs are propagated through a simulation of instrument visibilities, which are then subtracted from the data \citep{Bernardi2011ptsrc,sullivanFHD}.

Another strategy is to use the fact that however the foregrounds behave, they are likely to be large in amplitude compared to the cosmological signal. Thus, the dominant modes of a data are likely to be dominated by foregrounds, making it in principle possible to extract information about the foregrounds from the data itself. For instance, one could estimate the frequency-frequency covariance matrix from the data, and then perform an eigenvalue decomposition to identify the strongest modes in frequency. These modes can then be projected out. This is essentially a Principal Component Analysis (PCA; \citealt{Liu:2012}), which is often implemented in practice using a Singular Value Decomposition \citep{Paciga:2013}. Compared to a simple projecting out of, say, low-order polynomials, there are both pros and cons to this approach. One advantage is that the projected modes are simply whatever modes are dominant in the data, whether they are spectrally smooth or not. This enables the method to capture the fact that even if the intrinsic foregrounds are smooth, they may no longer be smooth once they are viewed through the instrument. However, it is important to emphasize that such an approach is not assumption free. Consider, for example, how one might go about estimating a frequency-frequency covariance matrix. From a theoretical standpoint, the covariance is given by
\begin{equation}
C(\nu, \nu^\prime) \equiv \langle x(\nu) x^*(\nu^\prime) \rangle,
\end{equation}
where $x(\nu)$ is a visibility spectrum. Implicit in this definition is the ability to take an ensemble average, signified by $\langle \dots \rangle$. In practice, of course, one is not able to take true ensemble averages, and one must employ some proxy. One could, for example, replace the ensemble average with a time average as one's telescope surveys different parts of the sky. However, in doing so one is essentially averaging over the sky, and thus the resulting covariance captures only the average frequency-frequency correlations. Any differences in statistical behaviour between, say, the Galactic plane and the Galactic poles is not modelled by this procedure.

The scheme that we have just described is a blind, data-driven scheme. This has the advantage that it can capture unknown characteristics in the instrument and/or the sky. However, this is also a disadvantage because without an underlying model for the data, it is difficult to guard against over subtracting and destroying part of the signal \citep{switzer:2015}. Some of this can be slightly mitigated by more advanced techniques like the Karhunen-Lo\`{e}ve (KL) transform. With the KL technique, rather than ordering modes by their contribution to the total (cosmological signal plus foreground plus noise) variance in the data, one models the signal and the contaminant (foreground plus noise) covariance matrices, which are then used to form a set of modes that are ordered by their signal-to-contaminant ratio \citep{Shaw:2014}. In this way, one has greater confidence that the first few modes (i.e., the ones that are projected out of the data) have negligible cosmological signal in them, minimizing the possibility of signal loss. However, there is still little guidance as to precisely how many modes ought to be projected out, and as one increases the number of modes, one inevitably begins to incur signal loss. Another difficulty is that in principle, the formation of the KL modes requires good models for both the signal and contaminants. This is easier in some contexts than in others. For instance, since the post-reionization \tcm signal traces the dark matter distribution, it can generally be modelled reasonably well (perhaps with some uncertainties revolving around the bias). This is more difficult for reionization experiments, since the astrophysics of the era is largely unconstrained by observations. In practice, though, the KL projection algorithm seems to be relatively robust even if the covariance matrices are slightly mismodelled, so long as the most general properties (such as the spectral smoothness of foregrounds) are accounted for \citep{Shaw:2014}.

Further variants are possible for identifying dominant (and therefore hopefully foreground) modes in the data. In general, the problem can be phrased as one where a series of sky maps at different frequencies are arranged into an $N_\textrm{freq}$ by $N_\textrm{pix}$ array $\mathbf{D}$, where $N_\textrm{freq}$ is the number of frequency channels and $N_\textrm{pix}$ is the number of sky pixels. The goal is then to decompose this data array into the product of two arrays $\mathbf{C}$ and $\mathbf{M}$:
\begin{equation}
\label{eq:DCM}
\mathbf{D} \approx \mathbf{C} \mathbf{M},
\end{equation}
where $\mathbf{C}$ is an $N_\textrm{freq}$ by $N_\textrm{comp}$ array and $N_\textrm{comp}$ by $N_\textrm{pix}$, where $N_\textrm{comp}$ is the number of independent components used to model the data. We thus have spectral modes encoded in $\mathbf{C}$, and spatial modes encoded in $\mathbf{M}$. In general, there is no unique solution to this problem, since one can define a new set of matrices $\mC^\prime$ and $\mM^\prime$ that are able to produce an identical fit to the data (i.e., $\mC^\prime \mM^\prime = \mC \mM$) provided $\mC^\prime \equiv \mC \boldsymbol \Psi$ and $\mM^\prime \equiv \boldsymbol \Psi^{-1} \mM$, where $\boldsymbol \Psi$ is an arbitrary invertible matrix \citep{Zheng:2017}. Moreover, if the data require a large number of components $N_\textrm{comp}$ to be properly modelled, then the number of parameters in the model ($N_\textrm{freq} N_\textrm{comp} + N_\textrm{comp} N_\textrm{pix}$) may exceed the number of measurements ($N_\textrm{freq} N_\textrm{pix}$). Assumptions or priors must therefore be made to regularize the problem. The PCA treatment described earlier in this section is equivalent to populating columns of $\mathbf{C}$ with a small handful of eigenvectors of $\mathbf{D}^t \mathbf{D}$. One then solves (or rather, fits) Eq. \eqref{eq:DCM} for $\mM$. \citet{Zheng:2017} takes this one step further in an iterative approach: once $\mM$ has been solved for, it can be held constant in another fit of Eq. \eqref{eq:DCM} to solve for a refined $\mathbf{C}$. The process can then be repeated until convergence. Other possibilities for regularizing this general problem include Independent Component Analysis (ICA; \citealt{chapman:2012,Wolz:2014,Wolz:2017}) and Generalized Morphological Component Analysis (GMCA; \citealt{chapman:2013}). In ICA, one maximizes the non-Gaussianity in the derived modes. The reasoning is that if an identified mode is implicitly a sum of multiple independent components, the central limit theorem ensures that this sum will be closer to Gaussian-distributed than the independent components. Maximizing non-Gaussianity is therefore a way to constrain the underlying independent modes. In GMCA, the spatial structure of foregrounds are considered in a wavelet basis, and one performs a fit to Eq. \eqref{eq:DCM} while imposing a sparsity prior.

\subsubsection{Mode weighting}
\label{sec:ModeWeightingForegrounds}
An alternative to the schemes outlined above is the idea of \emph{downweighting} modes rather than projecting them out completely. The motivation here is that even if foregrounds are dominant in certain modes, the cosmological signal is typically not precisely zero. There is thus useful information to be extracted from all modes, which is something that cannot be done if any modes have been projected out completely.

One way to implement a downweighting algorithm is to simply employ the optimal quadratic estimator formalism. Because $\mC$ contains a covariance contribution from foregrounds, the $\mathbf{C}^{-1}$ step of Eq. \eqref{eq:qalpha} has the effect of downweighting foregrounds \citep{Liu:2011}. Essentially, one is treating the foregrounds as a very correlated form of noise, and mitigating this noise by inverse covariance weighting. As a toy example, imagine that the non-foreground portion of the covariance is given by the identity matrix $\mathbf{I}$, and that the foreground covariance has rank 1:
\begin{equation}
\mC = \mI + \lambda \vu \vu^\dagger,
\end{equation}
where $\lambda$ is some parameter controlling the strength of the foregrounds, and $\vu$ is a unit vector of the same length as the data vector (in whatever basis one has chosen) and might be some template for the foregrounds. The Woodbury identity then shows that applying $\mathbf{C}^{-1}$ is the same as
\begin{equation}
\mC^{-1} \approx \mI - \frac{\lambda \vu \vu^\dagger}{1+\lambda}.
\end{equation}
One therefore sees that $\mC^{-1}$ has the effect of subtracting off a portion of $\vu$, but that this subtraction does not become a complete projection unless $\lambda \rightarrow \infty$. To gain more intuition for how this procedure is able to suppress foregrounds, note the following. If $\vu$ took precisely the form of a cosmological Fourier mode, then applying $\mC^{-1}$ would suppress this foreground contribution in the quantity $\mC^{-1} \vx$, but pushing through to the end of the quadratic estimator, one would find that the normalization matrix $\mM$ of Eq. \eqref{eq:pmq} undoes this suppression. This is simply the statement that if the foregrounds look precisely like the cosmological signal, foreground mitigation becomes impossible. On the other hand, suppose the foregrounds have different statistical properties than the cosmological signal. For instance, over small bandwidths, the cosmological signal possesses stationary statistics (i.e., its statistical properties are translation-invariant), whereas the foregrounds behave differently at different frequencies, being typically brighter at low frequencies. The $\mC^{-1}$ method takes advantage of this by downweighting the data from low frequencies when estimating a particular cosmological Fourier mode. This is possible because any particular Fourier mode receives contributions from every frequency, and thus its amplitude can be estimated even if certain frequencies are suppressed.

Of course, for modes where foregrounds are strongly dominant over the signal, the cosmological information content gain from recovering those modes will be negligible, and mode projection algorithms are an excellent approximation to the optimal treatment. For modes where the foregrounds and signal are more comparable, mode weighting can in principle allow improved science constraints from power spectrum measurements. However, a potential challenge with this approach is the modelling of foreground covariances. Good foreground covariance models are difficult to construct in an \emph{a priori} fashion. One is therefore driven to semi-empirical constraints. Modelling the low-frequency sky can be difficult, however, given that surveys of the relevant frequencies are substantially incomplete (although the situation is rapidly improving with the advent of \tcm cosmology! See, e.g., \citealt{LOFAR2016Survey,2017MNRAS.465.3163S,Eastwood2018}). An alternative to this is to use the data themselves to model the foregrounds, although as we discuss in Section \ref{sec:sigloss}, this carries with it the substantial risk that part of the cosmological signal can be destroyed in the process. For initial detection-era level experiments, then, it may be advisable to opt for a clean projection of contaminated mode rather than a riskier downweighting of a possibly inaccurate model. In general, the latter approach should only be chosen if it is deemed necessary to recover specific modes for one's scientific constraints.

\subsubsection{Avoiding foregrounds in the foreground wedge}
\label{sec:wedge}

Because of modelling difficulties, an alternative to the explicit removal of foregrounds is to simply \emph{avoid} them. If foregrounds can be shown to be sequestered to certain parts of Fourier space, then one could simply choose to exclude those contaminated regions when fitting one's measured power spectra to theoretical models.

In principle, the foreground avoidance paradigm is a very natural one to pursue, since the foregrounds are in fact naturally sequestered in Fourier space. Consider the $P(k_\perp, k_\parallel)$ power spectrum defined in Equation \eqref{eq:BinningPkperpkpara}, where $k_\perp$ is the Fourier wavenumber for modes perpendicular to the line of sight, and $k_\parallel$ for modes parallel to the line of sight. Since the line of sight direction corresponds to the frequency axis, spectrally smooth foregrounds in principle map to the lowest $k_\parallel$ bins of a power spectrum measurement. Thus, the foregrounds are naturally compact in the $(k_\perp, k_\parallel)$ Fourier space, and can be avoided simply by looking away from the lowest $k_\parallel$ modes.

In practice, the subtleties of interferometry complicate the simple picture we have just painted, and the foregrounds proliferate beyond the lowest $k_\parallel$ bins. The problem is particularly acute for interferometers with a sparse set of baselines.  As an extreme example of a sparse array, consider the simplest possible case of a sparse interferometer---a single baseline with a baseline vector $\textbf{b}$. From Section \ref{sec:InterferometerSensitivityOptimization}, we have that in the flat-sky limit this baseline measures
\begin{equation}
\label{eq:Vbnu}
V_b (\nu) =  \int \!\!d^2 r_\perp T(\vr_\perp, \nu) A_p (\vr_\perp, \nu) e^{-i 2 \pi \nu \vb \cdot \vr_\perp / c D_c },
\end{equation}
where for convenience in this section, we have implicitly applied Equation \eqref{eq:perp_mapping} (and suppressed the resulting factors of $D_c$) to write this integral in terms of $\vr_\perp$ rather than $\boldsymbol \theta$. As argued before, this baseline probes a particular angular scale on the sky, and by Fourier transforming over frequency, one accesses Fourier modes along the line of sight of the observation. However, our previous treatment neglected that a baseline of a given fixed physical length and orientation $\vb$ probes multiple scales, since the fringe pattern of an interferometer is frequency dependent. Mathematically, the scale probed by a baseline $\vb$ is $\vk_\perp \sim 2 \pi \vb \nu / c D_c$, which is proportional to $\nu$. This means that if we were to, say, estimate the power spectrum using the delay spectrum approach, then the required Fourier transform over frequency would not be independent of the Fourier transform given by Equation \eqref{eq:Vbnu}. We neglected this in Equation \eqref{eq:freqindepvis} of Section \ref{sec:delayspec} by hiding the frequency dependence of the exponent in Equation \eqref{eq:Vbnu} under the definition of a (presumed frequency-independent) variable $\vu \equiv \nu \vb / c$. Treating this frequency dependence properly in the Fourier transform over frequency, we compute
\begin{eqnarray}
\widetilde{V}_b (\tau) &\equiv &\int \! d\nu e^{-i 2\pi \nu \tau} V_b(\nu) \nonumber \\
& = & \int \! d\nu d^2  r_\perp T(\vr_\perp, \nu) A_p(\vr_\perp, \nu) e^{-i 2 \pi \nu \left(\frac{ \vb \cdot \vr_\perp }{ c D_c} + \tau\right) }.\quad
\qquad
\end{eqnarray}
To proceed, we make the approximation that the primary beam $A_p$ is frequency independent. In general, this is a better approximation than neglecting the frequency dependence of the fringe pattern because the primary beam often changes slowly with frequency, unlike the fringe pattern, whose phase often goes through multiple periods of $2\pi$ within one's observational bandwidth. Making this approximation and expressing the sky intensity in terms of its Fourier wavenumbers then gives
\begin{eqnarray}
\widetilde{V}_b (\tau) &\approx& \int \frac{d^3 k}{(2\pi)^3} e^{i k_\parallel r_*} \widetilde{T} (\vk) \int d\nu d^2 \vr_\perp A_p (\vr_\perp) \nonumber \\
&& \times e^{i \vk_\perp \cdot \vr_\perp - i \alpha k_\parallel \nu -i 2 \pi \nu \left(\frac{ \vb \cdot \vr_\perp }{ c D_c} + \tau\right)} \nonumber \\
&=& \int \frac{d^3 k}{(2\pi)^3} e^{i k_\parallel r_*} \widetilde{T} (\vk) \int d^2 \vr_\perp e^{i \vk_\perp \cdot \vr_\perp} \nonumber \\
&& \times A_p (\vr_\perp) \delta^D \left( \frac{\vb \cdot \vr_\perp}{c D_c} + \tau +\frac{ \alpha k_\parallel}{2\pi} \right),
\end{eqnarray}
where we have made the approximation that the mapping between frequency and radial distance (Equation \ref{eq:LOScomovingdistance}) is approximately linear over the observational band. This enables us to write $D_c \approx r_* - \alpha \nu$, where
\begin{equation}
\label{eq:alphadistanceconversion}
\alpha \equiv \frac{1}{\nu_{21}} \frac{c}{H_0} \frac{(1+z)^2}{E(z)},
\end{equation}
with $r_*$ being a constant with dimensions of length, and the cosmological factors having the same meanings as they did in Section \ref{sec:InterferometerSensitivityOptimization}. This quantity $\alpha$ is approximated as frequency independent, as is the the factor of $D_c$ in the integral over $\nu$. With a single baseline, we may say without loss of generality that the baseline vector is aligned with the $x$ direction, where $\vr_\perp \equiv (x,y)$. Evaluating the $\vr_\perp$ integral then gives
\begin{eqnarray}
\widetilde{V}_b (\tau) &\approx& \int \frac{d^3 k}{(2\pi)^3} e^{i k_\parallel r_*} e^{-i \frac{k_x c D_c}{ 2\pi b}(\alpha k_\parallel + 2\pi \tau)}\widetilde{T} (\vk)  \nonumber \\
&& \times \int dy e^{ik_y y} A_p \left[ \frac{c D_c}{ 2\pi b}(\alpha k_\parallel + 2\pi \tau), y \right],
\end{eqnarray}
where in this step we implicitly assumed $180^\circ$ rotational symmetry in the beam, such that $A_p (\mathbf{r}_\perp) = A_p (-\mathbf{r}_\perp) $, simply to avoid notational clutter of extra minus signs. (If one desires, the same final conclusions can be reached without assuming any symmetries). Proceeding to a power spectrum by squaring  $\widetilde{V}_b (\tau)$ and taking the ensemble average, we obtain
\begin{eqnarray}
\label{eq:modeprolif}
\widehat{P}(\mathbf{k}) &&\propto \langle |\widetilde{V}_b (\tau)|^2 \rangle \nonumber \\
&&= \int \!\! \frac{d^3 k}{(2\pi)^3} P(\mathbf{k}) \nonumber \\
&& \qquad \times \Bigg{|} \int \! dy e^{i k_y y} A_p \left[ \frac{cD_c}{2\pi b} (\alpha k_\parallel + 2 \pi \tau), y \right]\Bigg{|}^2.\qquad
\end{eqnarray}
In the parlance of Section \ref{sec:pspecestimation}, the squared term inside the integral correspond to the (unnormalized) window functions of the delay power spectrum estimator: they inform us of the specific linear combination (i.e., the integral, in this continuous case) of the true power spectrum $P(\mathbf{k})$ that we probe when forming delay-based power spectrum estimators. The window functions therefore quantify leakage on the Fourier plane, and are key to investigating how foregrounds proliferate in Fourier space in particular.

\begin{figure*}[t]
\centering
\includegraphics[width=1.0\textwidth]{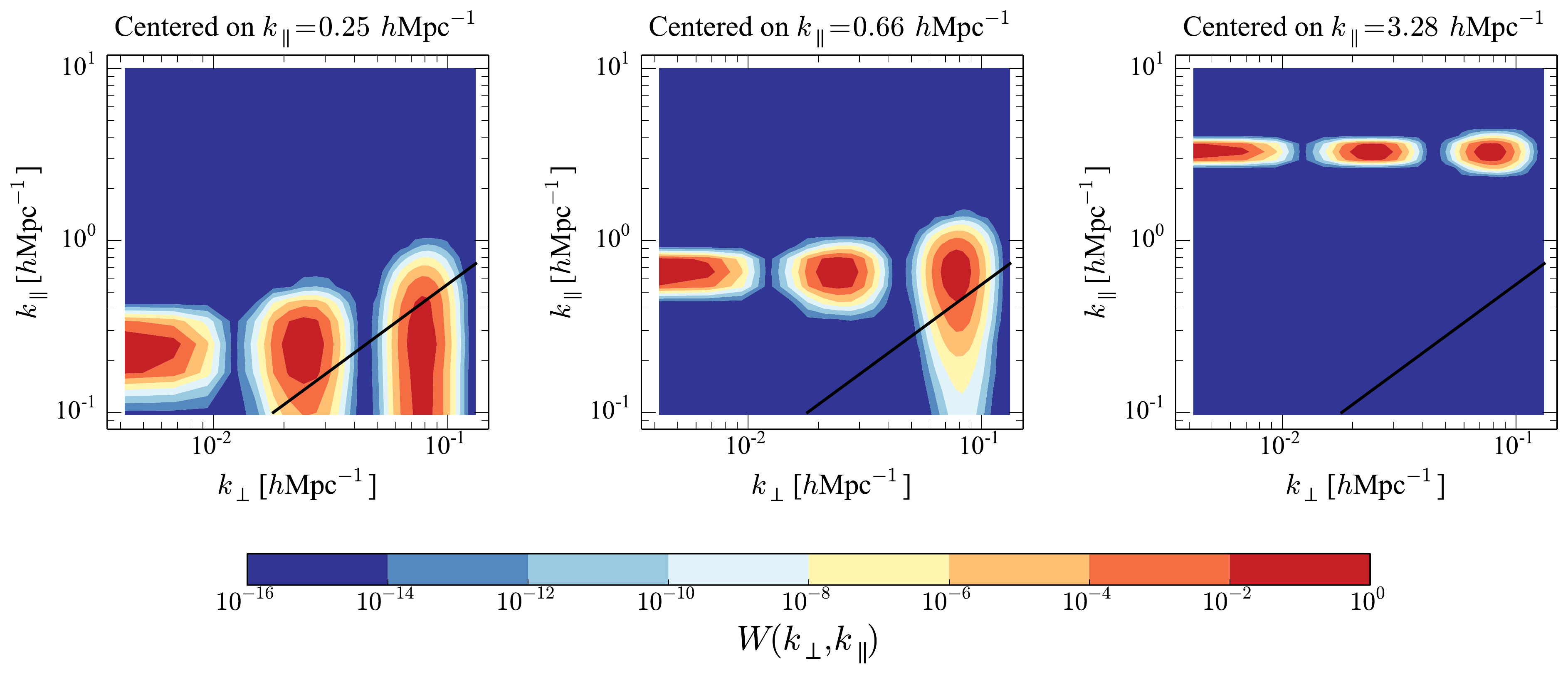}
\caption{An example set of window functions for a power spectrum estimator for a typical radio interferometer. Each window function is centred on a different point on the $(k_\perp, k_\parallel)$ plane. When one constructs a power spectrum estimator for a particular point, the measurement is in fact a linear combination of Fourier modes near that point. This linear combination is given by the relevant window function. One sees that the window functions are generally quite compact, but towards high $k_\perp$, they develop long tails towards low $k_\parallel$. The result is a leakage of foregrounds towards higher $k_\perp$, which gives rise to the foreground wedge feature seen in Figure \ref{fig:wedgecartoon}.}
\label{fig:windowfcts}
\end{figure*}

As a simple toy model for foreground emission, suppose that $P(k) \propto \delta^D (k_\parallel)$.\footnote{For a similar calculation that uses a more sophisticated foreground model, see \citet{Murray2017BetterModelForFG}.} Such a power spectrum might represent a population of dim unresolved point sources that are unclustered (hence the $k_\perp$-independent form that would make a picture of the sky look like white noise), but with flat (i.e., frequency-independent) spectra. Inserting this into our expression and simplifying the result yields
\begin{equation}
\label{eq:wedgePkderiv}
\widehat{P}(\mathbf{k}) \propto \int dy A^2_p \left[ \frac{cD_c \tau}{b}, y \right] \equiv \overline{A^2_p} \left[ \frac{k_\parallel}{k_\perp} \frac{c}{H_0} \frac{(1+z)}{E(z)} \right],
\end{equation}
where $\overline{A^2_p}$ is the squared primary beam profile integrated over the direction perpendicular (``$y$", in our notation) to the baseline vector, and we have inserted the fact that a baseline of length $b$ mostly probes $k_\perp \approx 2 \pi \nu b/D_c c$, and the delay mode $\tau$ mostly probes $k_\parallel \approx 2 \pi \tau / \alpha$ (refer back to Equations \ref{eq:moreexactkperp} and \ref{eq:alphadistanceconversion}, plus the delay approximation $\tau \approx \eta$ of Section \ref{sec:delayspec}). What this shows is that even with an input foreground sky where all emission resides in $k_\parallel = 0$ modes, the resulting power spectrum $\hat{P}(\mathbf{k})$ estimate contains power in a whole variety of $k_\parallel \neq 0$ modes. This has been termed \emph{mode mixing} in the literature \citep{Morales:2012}. Because of the inherently chromatic response of an interferometer, foregrounds have leaked from $k_\parallel = 0$ to other modes. This can be seen by plotting the window functions of our power spectrum estimator (Figure \ref{fig:windowfcts}). One sees that when measuring power spectra high $k_\perp$, low $k_\parallel$ power (i.e., foreground power) is inadvertently mixed in.

Mode mixing is both a problem and an opportunity. The problem is that the foreground modes have proliferated \citep{Switzer:2014}, contaminating more modes than one would naively expect. However, from Equation \eqref{eq:wedgePkderiv}, we see that this proliferation has a clear signature in Fourier space---the contamination decreases as one moves towards higher $k_\parallel$, with the falloff taking the form of the squared primary beam profile. Lines of constant contamination on the $k_\perp$-$k_\parallel$ plane are given by $k_\perp \propto k_\parallel$. This gives rise to the characteristic shape known as the \emph{foreground wedge} \citep{Datta2010wedge,Morales:2012,Parsons2012,Vedantham2012wedge,Trott2012,Hazelton2013,Pober:2013,Thyagarajan2013,Liu:2014a,Liu:2014b}, which is illustrated in Figure \ref{fig:wedgecartoon}. With detailed simulations going beyond our toy model (including realistic instrument beams and realistic foregrounds), the detailed profile of the wedge becomes more complicated (such as the \emph{pitchfork effect} for some instrument and foreground combinations; see \citealt{Thyagarajan:2015b,Thyagarajan:2015a} for details). However, the overall shape remains a robust prediction.

\begin{figure}
    \begin{center}
        \includegraphics[width=\linewidth]{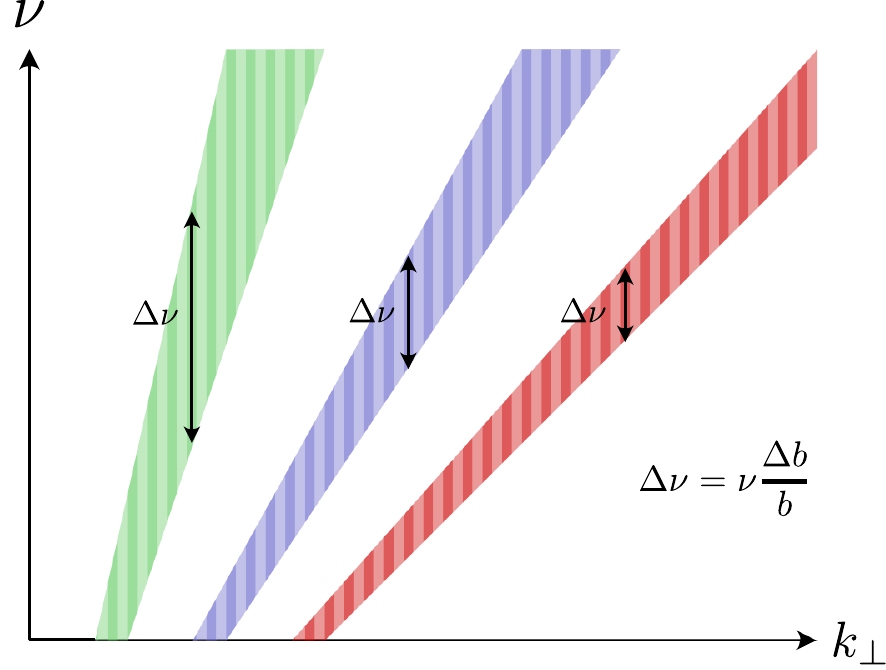}
    \end{center}
    \caption{Each baseline (shaded as different colors in the above plot), integrates over a distinct and linearly increasing set of $k_\perp$ modes at each frequency. The foregrounds are very highly correlated in frequency, but the \emph{apparent} correlation length for each baseline can only be as long as $\Delta\nu$, the distance in frequency that a single foreground Fourier mode is observed for (i.e. the height of each shaded region at fixed $k_\perp$). This distance $\Delta\nu$ is inversely proportional to the baseline length $b$. When performing the delay transform we observe a spread in delay of length $\Delta\tau \sim 1 / \Delta\nu \propto b$ because of the finite correlation in frequency. This spread is what we call the foreground wedge. Noting that $k_\parallel \propto \tau$ and $k_\perp \propto b$ we find the standard prediction that $\Delta k_\parallel \propto k_\perp$, and if we carefully trace through the coefficients in this argument, we reproduce Equation \eqref{eq:horizonline} exactly.}
    \label{fig:wedgeinterp}
\end{figure}

A simple way to interpret the wedge is that since each foreground Fourier mode is observed by a decreasing range of frequencies as we move to higher $k_\perp$, this decreases the observed correlation length, and thus increases the spread in delay (or equivalently $k_\parallel$). We illustrate this in Figure \ref{fig:wedgeinterp}.

That the foreground wedge has such a characteristic shape is an opportunity, because it limits how badly foregrounds can proliferate. To see this, note that the primary beams of most instruments fall off from their pointing centres, reaching zero at an angle $\theta_0$ away (which, for many antenna designs, may not be until the horizon). Thus, the proliferated foreground power goes to zero when the argument of $\overline{A^2_p}$ in Equation \eqref{eq:wedgePkderiv} is equal to $D_c \theta_0$ (recall our unusual convention in this section where we express angles on the sky in terms of the perpendicular comoving distances that they subtend). This means that the foregrounds do not extend above the line
\begin{equation}
\label{eq:horizonline}
k_\parallel = k_\perp \frac{H_0 D_c  E(z) \theta_0}{c (1+z) },
\end{equation}
which is a conclusion that turns out to be almost exact (up to slight ambiguities about how to define $\theta_0$) even in the presence of full curved-sky effects that our flat-sky derivation ignored \citep{Liu:2016}. Equation \eqref{eq:horizonline} enables a strategy of \emph{foreground avoidance}, where one simply makes measurements outside the wedge in what is sometimes termed the \emph{Epoch of Reionization window}.

A strategy of avoidance is conceptually attractive in that one does not need to worry about an exquisite modelling and subtraction of the relevant foregrounds. However, it is not a panacea for the formidable foreground problem of $21\,\textrm{cm}$ cosmology. One limitation is that while the observational evidence suggests that foregrounds do tend to be mostly sequestered to the wedge \citep{Pober:2013}, there is usually some low-level leakage beyond the horizon line of Equation \eqref{eq:horizonline}. Because the foregrounds are orders of magnitude brighter than the cosmological signal, even a little leakage can make cosmological measurements prohibitive. This leakage can occur for a variety of reasons. First, in our derivation we assumed that foregrounds were perfectly flat in frequency. In practice, the foreground spectra (while smooth) will not be perfectly flat. This means that their intrinsic spectra extend beyond the $k_\parallel = 0$ mode, and the resulting wedge moves up to higher $k_\parallel$, reducing the available Fourier space for cosmological measurements. Second, our derivation assumed a systematics-free instrument. In practice, calibration errors can cause leakage beyond the wedge, as can instrumental effects such as cable reflections (\citealt{EwallWice2016MWALimits}; Kern et al., submitted). As a result of these concerns, it may be advisable to pursue hybrid strategies where foreground subtraction is used in conjunction with foreground avoidance \citep{Kerrigan:2018}. Precisely how successful such a hybrid approach might be is still an open question. In forecasting papers in the literature, one often sees the success of foreground subtraction parametrized in terms of $\theta_0$. One where foreground subtraction does not allow one to push down to lower $k_\parallel$ than the line indicated in Equation \eqref{eq:horizonline} with $\theta_0 \sim \mathcal{O}(1)$ (representative of the horizon) is often termed a ``horizon wedge" scenario. A more optimistic scenario of a ``primary beam wedge" is often assumed, corresponding to a situation where $k_\parallel$ modes can be cleaned down to the line where $\theta_0$ is about the primary beam width. Even if foreground subtraction is not this successful (i.e., it does not allow a recovery of modes inside the horizon wedge), it may still be useful. This is because it may still reduce the foreground amplitude in contaminated modes, such that low-level leakages of these reduced modes outside the horizon end up being below the amplitude of the cosmological signal.

Another downside to a strategy of avoidance is that one gives up on measuring certain Fourier modes. This results in lost sensitivity \citep{Chapman2016}, particularly since one is preferentially losing modes at low $k$, which is where signal to noise is expected to be highest. This is because the instrumental noise tends to be fairly flat as a function of $k_\parallel$, which is often approximately $k$, since $k_\parallel$ is much larger than $k_\perp$ for many instruments, especially those probing reionization redshifts. In contrast, the signal drops by orders of magnitude in $P(k)$ as one moves to higher $k$. It is for this reason that instruments such as HERA (which was designed with explicit consideration of the foreground wedge; \citealt{Dillon:2016,DeBoer:2017}) have so many short baselines. Such short baselines primarily probe the small $k_\perp$ parts of Fourier space, where the wedge does not extend to such high $k_\parallel$ (and therefore $k$), preserving the opportunity to measure high sensitivity low-$k$ modes. Even so, the mere act of cutting out the wedge modes results in a loss of sensitivity. Additionally, it is important to note that even without the excision of the wedge, sensitivity can be reduced by error correlations between different Fourier modes \citep{Liu:2014a}. There is thus a reduction in the number of independent measurements, and this occurs throughout Fourier space because the physics of mode mixing applies to all modes [e.g., Equation \eqref{eq:modeprolif} shows that modes proliferate for all $\mathbf{k}$]. Its effect is just more apparent at low $k$ because of the large dynamic range between the foregrounds and the cosmological signal.

In limiting one's measurements to modes above the wedge, one is taking advantage of the statistical isotropy of our Universe to probe a given $k$ mode by measuring low $k_\perp$ modes where $k = (k_\perp^2 + k_\parallel^2)^{1/2} \approx k_\parallel$ \citep{Morales:2004}. However, recall from Section \ref{sec:ImageVsStat} that even if the intrinsic power spectrum is isotropic, peculiar velocities induce redshift-space distortions that make the measured power spectrum anisotropic. This anisotropy contains cosmological information (see Equation \ref{eq:RSD}), and so measuring it is of considerable interest. The signature of redshift space distortions is a dependence of the power spectrum on the orientation of one's $\mathbf{k}$ mode, which is often parametrized using the parameter $\mu \equiv k_\parallel / k$ so that one is constraining $P(k, \mu)$ rather than $P(k)$. Excluding wedge modes therefore reduces the range of $\mu$ values that are available for constraining the functional form of $P(k, \mu)$ [or more typically, for fitting parametrized forms for $P(k, \mu)$]. This can reduce signal to noise in redshift space distortion investigations \citep{Pober:2015}, as well as introduce biases in the spherically averaged power spectrum $P(k)$ \citep{Jensen:2016}. Additionally, it can make it difficult to employ reconstruction techniques \citep{Seo:2016} for undoing non-linear evolution that have been successfully used to sharpen BAO signatures in galaxy redshift surveys.

\subsubsection{Decorrelating the foreground wedge}
\label{sec:decorrWedge}
The prevalence of interferometric approaches to \tcm cosmology means that the wedge occupies a central role in the foreground mitigation literature. In the past, there has been considerable confusion as to whether the phenomenon of the wedge is fundamental or not. More precisely, one can ask if it is possible to undo the effects of the wedge, putting foregrounds back into the low-$k_\parallel$ Fourier modes that they intrinsically occupy. Whether the foreground wedge can be undone is a question whose answer depends on both instrumental design and analysis strategy. In principle, with the right instrument and the right analysis, the wedge can be undone. However, as a practical matter, it is typically difficult to entirely rid oneself of the effects of the wedge.

Consider first some proposed analysis strategies mitigating the wedge. One possible method would be to take advantage of the fact that picking $\mM = \mF^{-1}$ in Equation \eqref{eq:pmq} gives delta function window functions (see Figure \ref{fig:windows_and_error_covariances} for a one-dimensional toy demonstration), where the power spectrum estimated for a particular Fourier mode is sourced only by power that intrinsically resides there. This can be used in the choice $\mM \equiv \mathbf{F^\prime}^{1/2} \boldsymbol \Pi \mF^{-1}$, where $\mF^{-1}$ first attempts to sequester foreground power to their native low-$k_\parallel$ modes (essentially undoing the integral in Equation \ref{eq:modeprolif}), followed by a projection matrix $\boldsymbol \Pi$ that zeros out such modes before $\mathbf{F^\prime}^{1/2}$ performs what is effectively a smoothing operation on the Fourier plane to avoid the anti-correlation and high-error problems that were highlighted in Section \ref{sec:pspecestimation} with a pure $\mF^{-1}$ treatment. (We write $\mathbf{F^\prime}^{1/2}$ rather than $\mathbf{F}^{1/2}$ to denote the minor adjustment that must be made to account for the fact that in projecting out low-$k_\parallel$ modes, a small amount of cosmological signal will be lost, which biases the final answer if no adjustments are made).

In simulations, wedge decorrelation using the $\mM$ matrix has had mixed success \citep{Liu:2014b}. On one hand, the technique does appear to reduce foreground contamination at high $k$. However, numerical stability is an issue, since the decorrelation relies on small differences in the long tails of different window functions on the $k_\perp$-$k_\parallel$ plane. In addition, the notion that $\mF^{-1}$ should be able to undo window function effects is strictly true only if one assumes that the measured temperature field has stationary statistics, i.e., it has translationally invariant statistical properties and is therefore describable by a power spectrum. This is violated by foregrounds, which typically have a higher amplitude at lower frequencies.

One variant of $\mM$ matrix decorrelation is to use a judicious choice of $\mM$ not to completely undo the effects of the window functions, but to instead limit leakage beyond the wedge. Suppose that one were to order the elements of $\hat{\vp}$ such that the bandpowers in the vector go from modes that are suspected to be heavily contaminated (i.e., those in the wedge) to those that are less contaminated (typically those outside the wedge). Recalling that $\mW = \mM \mF$, if we select an $\mM$ matrix such that $\mW$ becomes upper triangular, then the window functions will possess long tails away the wedge region, but cut off abruptly towards the wedge. In other words, predominantly cosmological power is permitted to leak from outside the wedge to inside the wedge, but predominantly foreground power cannot leak from inside the wedge to outside the wedge. This can be accomplished if we first perform a Cholesky decomposition on $\mF$, such that $\mF \equiv \mathbf{L} \mathbf{L}^\dagger$, where $\mathbf{L}$ is a lower triangular matrix. Then, one sees that if $\mM \propto \mathbf{L}^{-1}$ (with a row-by-row normalization to make sure that rows of $\mW$ sum to unity), then $\mW \propto \mathbf{L}^\dagger$, giving the upper-triangular, wedge-isolating form that we desire. This is demonstrated for a toy one-dimensional survey in the bottom row of Figure \ref{fig:windows_and_error_covariances}. In one dimension, the wedge-contaminated modes are simply the low-$k$ modes, and one sees that the window functions permit leakage of power from high $k$ values to low $k$ values, but not the other way around.

An alternative to picking $\mM$ wisely is to attempt to undo the effects of the wedge earlier in the pipeline, prior to the squaring step of forming power spectra. Conceptually, the wedge arises because interferometers are inherently chromatic instruments, with chromatic point spread functions. In principle, one could attempt to undo these point spread functions in an intermediate map-making step of one's pipeline. In Equation \eqref{eq:ml_cmb} from Section \ref{sec:mapmaking}, for example, the inverse term undoes the point spread function of an instrument. If this inverse is computable, it will therefore have the effect of undoing the wedge. However, the inverse does not always exist. For sparse interferometers with poor $uv$ coverage, the inverse generally does not exist. To undo the effects of the wedge, a necessary condition is to have an instrument with sufficiently dense baseline coverage that the true Fourier transform along the line of sight can be taken, using multiple baselines with equal $\nu \vb$ (see Figure \ref{fig:wedgeinterp} and Section \ref{sec:delayspec}). Thus, removing or suppressing the wedge requires not just an appropriate analysis pipeline, but also a suitable instrument such as a single dish telescope (which can be thought of as an interferometer with a fully sampled $uv$ plane up to the size of the aperture) or an appropriately optimized interferometer \citep{Murray2018}.

\subsubsection{Cross correlations}
\label{sec:CrossCorrFGs}
Another way to combat foregrounds is to avoid measuring the $21\,\textrm{cm}$ auto power spectrum, but instead to measure the cross power spectrum with some other tracer of cosmic structure. The key idea is that while each tracer may have its share of foreground contaminants, the sources of these contaminants is different amongst the different tracers. Thus, the foregrounds will be uncorrelated with one another, and if $\mathbf{x} \equiv \mathbf{x}_s + \mathbf{x}_f$ is the measurement from the first tracer and $\mathbf{y} \equiv \mathbf{y}_s + \mathbf{y}_f$ is the measurement from the second tracer, heuristically we have
\begin{equation}
P_\textrm{cross} \sim \langle \vx \vy \rangle \sim \langle \vx_s \vy_s \rangle + \langle \vx_f \vy_f \rangle \sim \langle \vx_s \vy_s \rangle.
\end{equation}
One sees that only the signal portion remains. While appealing, this result should be interpreted with caution. First, note that it only applies in expectation. In practice one can use the common cosmological practice of replacing the ensemble average with a spatial average. However, the quality of foreground mitigation is now tied to parameters such as one's survey area, since spatial averages are prone to unlucky fluctuations when the spatial area is small. Moreover, while cross correlations can in principle yield results that are free of foreground \emph{bias}, they can still suffer from inflated \emph{variance} compared to a hypothetical foreground-free survey. This is because the variance of the measured power takes the form
\begin{eqnarray}
\textrm{Var} (P_\textrm{cross}) &\sim& \langle \vx^2 \vy^2 \rangle \nonumber \\
& \sim & \langle \vx_s^2 \vy_s^2 \rangle + \langle \vx_s^2 \vy_f^2 \rangle + \langle \vx_f^2 \vy_s^2 \rangle + \langle \vx_f^2 \vy_f^2 \rangle, \quad \qquad
\end{eqnarray}
and thus we see that the foregrounds do contribute to the variance (and therefore the errors). Performing foreground subtraction on each data set prior to cross correlation does reduce this variance, but the point remains that the effect of foregrounds cannot be ignored simply because one is performing cross correlations.

Our argument here is exactly analogous to ones that we made earlier regarding noise. In Section \ref{sec:dish_or_inter} we argued that because interferometric visibilities are formed by cross-multiplying two independent antennas, they do not possess a noise bias. However, they certainly do possess a noise variance (otherwise, interferometric measurements would have no uncertainties associated with them!) Similarly, in Section \ref{sec:pspecestimation}, we argued that a noise bias does not arise when forming a cross-correlation power spectrum between two datasets taken at different times. But again, the power spectrum still ends up with a noise error bar.

Despite these caveats, cross correlations have been used to great effect in the past. For example, to date the only detections of cosmological $21\,\textrm{cm}$ fluctuations have come from cross correlating GBT and Parkes data with galaxy survey data \citep{Anderson:2018,Chang:2010,Masui:2013}. For higher redshifts there are fewer candidates for cross correlating with $21\,\textrm{cm}$ measurements, but a promising avenue might be [CII] intensity mapping surveys \citep{gong_et_al2012,Chang2015SKA,Kovetz2017whitepaper,Beane2018Bispectrum,Dumitru2019,Kovetz2019whitepaper}. Although the aforementioned challenges would have to be dealt with in the interpretation of a cross-correlation measurement, a consistent [CII]-HI measurement would be highly complementary to an HI auto power spectrum, boosting one's confidence in an initial detection.

\subsection{Interloper lines}

In our discussions so far, we have focused on mitigating the effect of broadband foregrounds, i.e., those that emit across a wide range of frequencies. In principle, foreground contamination can also occur because of discrete line emission from other transition lines. More concretely, two different lines (say, with rest frequencies $\nu_1^\textrm{rest}$ and $\nu_2^\textrm{rest}$) will appear at the same observed frequencies in our data if they are emitted at two different redshifts (denoted $z_1$ and $z_2$) such that
\begin{equation}
\frac{\nu_1^\textrm{rest}}{1+z_1} = \frac{\nu_2^\textrm{rest}}{1+z_2}.
\end{equation}
This particular form of the foreground problem is often called the problem of \emph{interloper lines}.

In the case of HI \tcm mapping, interloper lines are generally not a problem---in a fortunate accident, there are essentially no strong lines with the right wavelengths emitting at the right redshifts to contaminate the \tcm signal. However, this is not the case for other lines that are used in intensity mapping. For example, an experiment targeting emission from the $157.7\,\mu\textrm{m}$ [CII] line at $z=7$ might easily confuse the [CII] emission with the ladder of CO rotational lines that emit at frequencies $J \times 115 \,\textrm{GHz} $ as CO transitions from its $(J+1)$th rotational state to its $J$th state. Estimates suggest that these rotational lines are sufficiently strong at $z \sim 0.9$ (for the $4\rightarrow 3$ transition), at $z \sim 1.4$ (for $5 \rightarrow 4$), at $z \sim 1.9$ (for $6 \rightarrow 5$), and at $z \sim 2.4$ (for $7 \rightarrow 6$) \citep{LidzTaylor2016}.

Several solutions to the problem of interloper lines have been proposed. One is to mask strong point sources, which can be substantial contributors to the interloper problem \citep{breysse_et_al2015, yue_et_al2015, silva_et_al2015}. Another possibility is to match putative detections of a line with companion lines that are also expected to be present at the same redshift, given reasonable assumptions about the physics of galaxies \citep{kogut_et_al2015}. Taking a statistical approach, one might also cross correlate with external data sets \citep{visbal_and_loeb2010, gong_et_al2012, gong_et_al2014,Switzer:2017,Switzer2019}. Another attractive option is to use the fact that a misidentified spectral line will be misplaced when converting to a cosmological coordinate system, but only in the radial direction and not in the angular directions. The presence of interloper lines will therefore induce statistical anisotropies with predictable signatures that can be fit for in data analyses \citep{cheng_et_al2016, LidzTaylor2016,Liu:2016}.

\subsection{Ionosphere}
\label{sec:Ionosphere}
Closer to Earth than astrophysical emission sources is the ionosphere. The ionosphere is a turbulent layer of our atmosphere that is ionized by incoming solar radiation \citep{1985SSRv...41...91K}. The existence of the ionosphere impacts low-frequency radio observations in many ways. In fact, many interferometers that were designed for cosmology have turned out to be excellent instruments for ionospheric studies (e.g., \citealt{Loi2015b, Loi2015a, Loi2016, Mevius2016})!

In the context of cosmology, the ionosphere affects observations in several ways. First, there is some level of absorption \citep{Vedantham2014GlobalSignalIonosphere}. More troublesome than the absorption are the scintillation effects. As an ionized plasma, the ionosphere acts as a screen for radio waves, imprinting extra time- and position-dependent phases. This can cause refractive shifts in the apparent locations of radio sources on the sky (known as the \emph{tip-tilt} effect in optical astronomy). It can also cause amplitude scintillations (\emph{twinkling} in the optical parlance) by focussing and defocussing the wavefronts. The apparent structure of astronomical sources can also change (i.e., the effective point spread function of a telescope can broaden and acquire extra structure; \emph{seeing} in the optical parlance). Finally, the ionized nature of the ionosphere causes the polarization direction of radio waves to be Faraday rotated. This can be troublesome for foreground subtraction because it can cause spectrally smooth foreground emission to appear spectrally unsmooth. To see this, recall from Section \ref{sec:polarization_leakage} that in general, the visibility measured by a baseline of a radio interferometer exhibits polarization leakage, where linearly polarized emission can leak into what one interprets as unpolarized Stokes $I$ emission. Now, Faraday rotation is a frequency-dependent phenomenon, where the polarization direction of linearly polarized emission is rotated by an angle proportional to $\lambda^2$. Since polarization angles must be between $0$ and $\pi$, the result of this is that after Faraday rotation, a measurement that is projected along one linear polarization axis can acquire a strongly oscillating frequency dependence. When this leaks into one's Stokes $I$ measurement, the result is a spectrally unsmooth foreground that may be difficult to distinguish from the cosmological signal \citep{Moore2013,Kohn2016}. Together, the various effects of the ionosphere can result in non-negligible biases in measured \tcm power spectra, particularly when the ionosphere is exhibiting strong activity \citep{Trott2018Ionosphere}.

Fortunately, each of the aforementioned ionospheric effects can be mitigated, at least in principle. Ionospheric absorption is expected to relatively spectrally smooth \citep{Vedantham2014GlobalSignalIonosphere}, making it a relatively benign influence on \tcm measurements. (Provided one is observing above the ionosphere's plasma frequency of $\sim 1$ to few MHz, below which the ionosphere is opaque). The situation is a little different for the issue of refractive shifts in position. This is typically viewed as problem for direction-dependent calibration. For example, one can use known positions of catalogued bright point sources to simultaneously solve for antenna gains and ionospheric phase shifts \citep{Mitchell2008, Jordan2017,Gehlot2018}. In some cases, Global Positioning Satellite measurements can be used to provide supplemental information to enhance solutions obtained under such a scheme \citep{Arora2015}. Note that even if one is able to obtain reasonable direction-dependent calibration solutions, residual ionospheric effects will remain, partly due to the imperfections in the calibration process and partly because the ionosphere has variations on timescales shorter than the typical temporal cadence at which one calibrates. These residual effects manifest themselves in the data as an extra ionospherically sourced scintillation noise \citep{Vedantham2015a}. The effect of this scintillation noise depends on the array configuration, because of the interplay between baseline lengths and the spatial and temporal coherence lengthscales of the ionosphere. Widely separated antennas, for example, are receiving incoming radiation that has been scattered by different patches of the ionosphere. \citet{Vedantham2016} find that for minimally redundant array configurations, scintillation noise enters at approximately the level as thermal noise, and is confined to the same wedge-like region in Fourier space (see Section \ref{sec:wedge}) as foregrounds are in a power spectrum analysis. Thus, one might reasonably expect that foreground avoidance techniques that take advantage of the wedge will successfully mitigate scintillation noise. For compact, redundant array configurations with a high filling factor, scintillation noise may be larger. However, the noise tends to be spatially and spectrally coherent \citep{Vedantham2016}, again enabling foreground mitigation techniques to do double duty and suppress scintillation noise along with foregrounds.\footnote{Note that many of these results were derived in the \emph{weak-scattering} regime, where the ionospheric phase fluctuations are assumed to be small over the first Fresnel zone. Observationally, the weak-scattering approximation seems to be a reasonable one except during short periods of high solar activity (H. Vedantham 2019, private communication).}

The issue of polarization leakage can also be dealt with by performing exquisite polarized calibration. Recall again from Section \ref{sec:polarization_leakage} that whereas polarization leakage cannot be properly corrected for using information from a single baseline, it can---at least in principle---be corrected for if one is able to combine visibilities from a sufficiently large collection of different baselines. In other words, in the limit where an array is configured in the right way to \emph{image} the sky, one has access to information as a function of angular position, and so one can correct for polarization leakage, which is also a function of angular position. This is not a luxury that all telescopes have. For example, arrays that are configured to have a sparse set of regular, repeated baselines cannot pursue such a strategy. However, one mitigating factor is that if observations are repeated night after night, the astrophysical signal should remain constant while the specific realization of the random ionospheric fluctuations is different every night. Thus, folding multiple days of data into an averaged dataset can suppress polarized leakage systematics from the Faraday rotation caused by the ionosphere \citep{Moore2017,Martinot2018}. Ironically, then, the randomness of the ionosphere helps to partially solve the problems it has created!

\subsection{Radio Frequency Interference}
\label{sec:RFI}
In addition to contaminating signals arising from astrophysical sources and the ionosphere, low-frequency radio instruments must also contend with terrestrial contaminants. These contaminants are collectively known as radio frequency interference (RFI). There are a variety of ways that RFI can occur. Radio stations, for instance, broadcast strongly in many frequency ranges of interest to \tcm cosmology, a prime example being the FM broadcast band from $88$ to $108\,\textrm{MHz}$ (corresponding to $z \sim 12$ to $15$). Satellite or aircraft communications also contribute to RFI, as do radio transmitters being operated on the ground. In many cases, these RFI sources need not be local. For example, ionized meteor trails are known to reflect RFI from faraway locations. At the other extreme, self-generated RFI can be produced by insufficiently shielded electronics in the telescopes themselves. In general, all these sources of RFI can be orders of magnitude brighter than sources of astrophysical origin (including foregrounds), and therefore must be mitigated.

The most basic strategy for RFI mitigation is to avoid the RFI in the first place by building telescopes in remote locations. Some sites are legally protected as radio quiet zones. Examples include the Dominion Radio Astrophysical Observatory in British Columbia, Canada,  the Green Bank Observatory in West Virginia, USA, and the Square Kilometre Array sites in the South African Karoo desert and the Western Australian desert \citep{Bowman2010RFI,SokolowskiRFI2016}. Of course, these sites are typically only protected against RFI in certain frequency ranges, so a site that is appropriate for post-reionization \tcm measurements may not be appropriate for reionization measurements and vice versa. Other popular locations are not necessarily legally protected, but are simply isolated or have natural geographical features (such as favourable combinations of mountains and valleys). Examples of sites like this include the Owens Valley Radio Observatory (in California, USA; \citealt{Eastwood2018}), Marion island (located halfway between South Africa and Antarctica; \citealt{PRIZM2019}), the Forks (in Maine, USA; \citealt{ZhengMITEoR2014}), the Timbaktu Collective (in southern India; \citealt{SinghSARAS2018}), the Ladakh site (in the Himalayas; \citealt{SinghSARAS_overview2018}), and the Castrillon quarry (a defunct gold mine located in northern Uruguay; \citealt{BINGO_update2016}). Looking ahead, there even exist futuristic proposals to place radio arrays on the far side of the Moon, avoiding both RFI and the ionosphere \citep{LazioMoon2009}.

\begin{figure*}[t]
\centering
\includegraphics[width=1.0\textwidth,trim={0cm 6cm 0cm 6cm},clip]{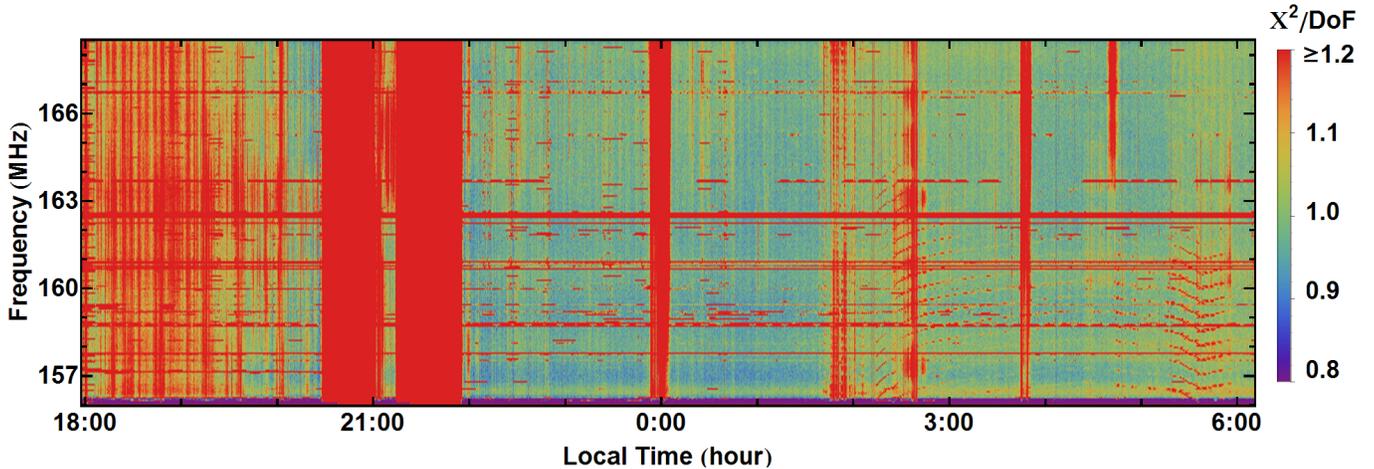}
\caption{A qualitative representation of RFI in typical radio interferometric data. Because RFI (in red) tends to show up as outliers in data, one way to visualize it is to fit a model to the data (and one's calibration solution) assuming no RFI, and to look for parts of the dataset where the $\chi^2$ per degree of freedom is too high. Here we show this quantity as a function of time and frequency using data from the MITEoR telescope. One sees clear indications of broadband RFI localized to specific times, persistent RFI localized to specific frequencies, and intermittent RFI sources. For more details regarding the observations, please see \citet{ZhengMITEoR2014}.}
\label{fig:RFI}
\end{figure*}

Beyond site selection, RFI must also be removed during data analysis. A common strategy is to identify RFI based on its characteristics as a function of frequency, time, amplitude, position, polarization, and any other axes over which observations are taken \citep{EkersRFI2002}. Mobile radio transmitters, for example, tend to be bright but might be used only sporadically. This leads to a signal that is localized in time, enabling an RFI mitigation strategy that involves \emph{flagging} (and subsequently ignoring) time integrations that are extreme outliers in amplitude. Radio stations, on the other hand, tend to be persistent in time, but only transmit at specific narrow ranges in frequencies. One can thus flag specific outlier frequency channels, an example of which can be seen in Figure \ref{fig:RFI}. Generally, this is reasonably effective, but care must be taken to ensure that the brightness of the RFI does not push one's electronics into a non-linear regime, which can cause narrowband RFI to leak into other frequencies \citep{MoralesWyitheReview2010,ZhengMITEoR2014}.

There are also more sophisticated algorithms than simple threshold flagging to combat RFI. As a first extension, we can note that very often RFI is not perfectly isolated to single times/frequencies. Thus, rather than simply threshold flagging on a channel-by-channel or time-by-time basis, one might also consider additional flagging if neighbouring times/frequencies both exceed a lower threshold. This is the basis behind algorithms such as \texttt{SumThreshold} \citep{OffringaSumThreshold2010}. As an extension of these algorithms, there also exist post-processing methods where flagged regions/shapes on the frequency-time plane are expanded to include neighbouring times and frequencies (e.g., as applied in \citealt{Winkel2007}). Such methods are sometimes referred to as \emph{morphological} algorithms, since they expand flagged shapes in time and frequency. Taking advantage of morphology can be preferable to lowering one's thresholds as a strategy for catching additional RFI. This is because overly aggressive (i.e., low) amplitude thresholds can result in a large number of false positives and additionally skew the statistics of the true signal. For example, suppose our data consisted of Gaussian-distributed signal plus some extremely bright RFI spikes. With a high threshold, one removes only the RFI and preserves the signal. As the threshold is lowered, however, one begins to mistakenly flag samples that reside in the extreme tails of the Gaussian-distributed data. By eliminating these data, one introduces an artificial skew in the data distribution. Thus, morphological methods may be particularly well suited for detecting low-level RFI. An advanced example of a morphological algorithm is the one implemented in \texttt{AOFlagger}, which expands RFI shapes using a \emph{scale-invariant rank} (SIR) operator \citep{OffringaSIR2012}. The SIR operator is scale invariant in the sense that any expansions or contractions (whether as a function of frequency or time) in some RFI pattern results in an identical expansion or contraction in the pattern of flags. The key idea here is that RFI often appears with similar shapes at a variety of frequency and time scales (i.e., it is often scale invariant) depending on the details of the emissions sources and the instrument in question. The algorithm for detecting such RFI should therefore respect this symmetry.

Another powerful method for identifying RFI is Principal Component Analysis (PCA). Suppose we considered a set of raw visibilities coming from $N_\textrm{bl}$ baselines, each having recorded data over $N_\textrm{freq}$ frequency channels for $N_t$ time samples. Forming an $N_t \times N_\textrm{bl} N_\textrm{freq}$ array $\mathbf{X}$ with the data, one can estimate a covariance matrix by computing $\mathbf{X} \mathbf{X}^\dagger$. In doing so, one is producing a covariance estimate by averaging over time to yield an $N_\textrm{bl} N_\textrm{freq} \times N_\textrm{bl} N_\textrm{freq}$ matrix. Performing an eigenmode decomposition of this covariance, the most dominant eigenmodes will tend to be sourced by persistent sources of RFI, which survive the time-averaging operation. The eigenvectors paired with the largest few eigenvalues can then serve as templates for how RFI appears in the data. Armed with such templates, one possibility would be to project out these RFI modes. Alternatively, note that a single eigenvector is a vector of length $N_\textrm{bl} N_\textrm{freq}$, and thus can be thought of as a visibility dataset in its own right (albeit without the time axis). This set of visibilities can be imaged, providing a near-field image of the positions of RFI sources. This provides a literal real-world map that can used to physically locate RFI sources, which can then be dismantled before further data is taken. (Provided one has permission to do so!) Such a strategy has been successfully deployed in \citet{Paciga:2011}.\footnote{In that paper, the PCA is implemented using a Singular Value Decomposition, but the principle is the same.}

Recently, machine learning techniques have also been brought to bear on the problem of RFI flagging. Using techniques borrowed from the computer vision community, one can treat a waterfall plot (i.e., data on the time-frequency plane, such as what is shown in Figure \ref{fig:RFI}) as an image with RFI features to be identified. Neural networks can then be trained with example data (either from simulations where one knows precisely where RFI was inserted, or from real data that have been flagged with a trusted algorithm) until they correctly identify RFI. By using waterfall plots of \emph{both} visibility amplitude and phase as inputs to the neural network, \citet{KerriganRFI2019} find that they are able to train neural networks that are generally competitive with state-of-the-art flagging algorithms. An important exception is that their neural networks often miss RFI ``blip" events that are localized to single pixels in time-frequency space. However, machine learning RFI flaggers are still in their infancy, and thus many improvements are likely in the next few years. Moreover, they tend to be orders of magnitude quicker than traditional algorithms in computational speed (once the neural networks are trained), and can therefore yield substantial computational savings for large next-generation telescopes that have enormous data volumes.

In closing, we point out that while RFI is in general a nuisance to be avoided, there are some limited situations where it can be helpful. For example, one would generally consider the ORBCOMM satellite constellation (which broadcasts at $137\,\textrm{MHz}$) to be an irritating source of RFI for \tcm measurements targeting reionization. While this is true, the known broadcast frequency of these satellites enables them to serve as calibrator sources that can be used to provide empirical mappings of the primary beam patterns of an antenna. With 30 such satellites in low-Earth orbits that precess over time, these sources transit through an antenna's field of view quickly and frequently, enabling a reasonably densely sampled mapping of the primary beam \citep{NebenOrbcomm2015,Line2018}. Similar techniques are being developed using drones that are outfitted with radio transmitters \citep{JacobsDrone2017}. For drift-scan telescopes, known astrophysical point sources can also be used. However, with the motion of the sources governed entirely by the Earth's rotation relative to the celestial sphere, one cannot in general solve for a full beam profile without making additional symmetry assumptions \citep{PoberBeamMapping2012}.

\subsection{Signal loss}
\label{sec:sigloss}

With our need to subtract, suppress, or otherwise avoid strong contaminants in our foregrounds, one is essentially in the business (whether directly or indirectly) of subtracting two large numbers to obtain a small number, i.e., the cosmological signal. One of the key challenges to this is that one may not be able to perform foreground mitigation well enough, leaving residuals that are small but still dominate the cosmological signal. However, equally problematic (and perhaps more pernicious) is the problem of oversubtraction, where the cosmological signal itself is suppressed in the foreground subtraction process. Often termed \emph{signal loss} in the literature, this can lead to overly aggressive---and therefore incorrect---upper limits on the strength of the \tcm signal.

Signal loss is particularly harmful when it is undiagnosed. Already, this has led to multiple revisions of upper limits on the \tcm signal that have been published in the literature. As two particularly instructive case studies, consider \citet{Paciga:2011} versus \citet{Paciga:2013} and \citet{Ali2015} versus \citet{Ali2018}. In the former case, \citet{Paciga:2011} attempted to remove foregrounds from GMRT data by first modelling the foregrounds using what amounts to a moving boxcar averaging of the data in the spectral direction. This extracts the spectrally smooth component of the data, which is then subtracted from the data. Upon further analysis, however, \citet{Paciga:2013} found that the subtracted model substantially overlapped with the Fourier modes where limits on the cosmological signal were being placed. In other words, the cosmological signal was being subtracted along with the foregrounds. In \citet{Ali2015}, foreground suppression was performed by the inverse covariance weighting method described in Section \ref{sec:ModeWeightingForegrounds}. However, without \emph{a priori} covariance models, an empirical model for the covariance matrix was constructed by replacing the ensemble average in the definition of the covariance matrix, $\mC \equiv \langle \vx \vx^\dagger \rangle$, with a time average. If an infinite series of time samples had been available, this should have converged to the true covariance matrix. However, with a finite number of time samples, there can be errors in the empirical covariance estimate \citep{dodelson_schneider2013,taylor_joachimi_etal2014}. In the case of \citet{Ali2015}, this issue was exacerbated by the fringe rate filtering described in Section \ref{sec:delayspec}, where visibilities were pre-averaged together in order to increase signal to noise, at the expense of decreasing the number of independent data samples for covariance estimation. To make matters worse, having few independent samples not only results in an inaccurate covariance estimate, but also one where the estimate retains ``memory" of the random fluctuations in the actual data. With the covariance matrices now a strong function of the specific realization of $\vx$ itself (as opposed to just the statistical properties of $\vx$), the so-called quadratic estimate of the power spectrum in Eq. \eqref{eq:palphaunnormed} ceases to be a quadratic form. As a result, the usual normalization of the power spectrum in Eq. \eqref{eq:minvarpalpha} does not apply, as it was derived assuming a quadratic form for the power spectrum estimator. If the usual normalization is used anyway, the resulting power spectrum estimate can be shown to be typically biased low \citep{Cheng2018}, which is equivalent to saying that signal loss has occurred.

Signal loss is in principle not a problem if it is understood and corrected for. Since signal loss is tantamount to employing a biased estimator for the power spectrum, an accurate quantification of the bias enables corrections. Often this quantification takes the form of simulations, where one is essentially computing a transfer function that accounts for the signal loss due to foreground subtraction (or in principle, any other form of loss). One subtlety with such simulations is that in many foreground subtraction schemes, the cleaned data is not a linear function of the input data. (This is equivalent to the observation we made above, where the resulting power spectrum estimator---which involves a squaring of the data---is no longer a quadratic function of the initial data). Thus, it is insufficient to simulate only the cosmological signal portion of the data through the foreground cleaning process, even though it is ultimately the attenuation of this signal that we're interested in. Doing so misses the impact of correlations between the cosmological signal and any foreground models estimated from the data, which include the cosmological signal \citep{Paciga:2013, switzer:2015, Cheng2018, Ali2018}. Another complication with quantifying transfer functions through simulations is that at low radio frequencies, even state-of-the-art foreground models (and for that matter, state-of-the-art instrument models) are unlikely to be particularly accurate, making it hard to know what to simulate. One approach that has been employed to get around this is to perform cosmological signal injection on real data, effectively using the real data as a template for the foregrounds. However, interpreting such simulations can be difficult \citep{Cheng2018}, and thus in general it is preferable to employ algorithms that are less susceptible to signal loss in the first place. Alternatively, it is (at least in principle) possible to sidestep some of these issues via the Bayesian approach for joint foreground and power spectrum estimation \citep{Sims2019b} that we briefly discussed in Section \ref{sec:BayesPspec}. Fundamentally, signal loss can occur during foreground mitigation because at least some part of the cosmological signal can be traded for foregrounds in one's model without sacrificing the goodness of fit to the data. This degeneracy is fully captured in a Bayesian approach, which produces full joint probability distributions of all parameters in the model (whether they are foreground or signal parameters). When marginalizing over foreground parameters as nuisance parameters, the error bars on the power spectrum bandpowers will increase to reflect degeneracies between foregrounds and signal. This fully captures what we have been calling signal loss, although in this probabilistic framework ``signal loss" is a slight misnomer as nothing has really been ``lost"---we have simply produced a more realistic estimate of the uncertainties in our final power spectrum given our (considerable) uncertainties in the behaviour of the foregrounds.
With these larger errors, the power spectrum is in principle unbiased, because the posterior distributions fully capture the range of possible outcomes within the foreground-signal degeneracy, although care must be taken to ensure that any summary statistics are not defined in a way that gives the mistaken impression of bias in the presence of significantly asymmetric posteriors.

In closing, we note that foreground mitigation is not the only culprit in the problem of signal loss. Calibration errors, for example, can very easily result in signal loss. Continuing the discussion from Section \ref{sec:CalibFreqTime}, imagine a calibration scheme that allows the gains of an instrument to be adjusted frequency channel by frequency channel. Allowing such freedom in one's data analysis means that the spectrally fluctuating cosmological signal can be perfectly fit out and eliminated by misestimated calibration parameters, particularly in the case of direction-dependent calibration \citep{MouriSardarabadi2019}. The spectral smoothing of calibration solutions described in Section \ref{sec:CalibFreqTime} can in principle mitigate this problem, but to ensure that this is indeed the case, one should ideally employ end-to-end simulations that capture as much of one's analysis pipeline as possible (see \citealt{sullivanFHD,Lanman2019astropaper,Lanman2019josspaper} for examples of progress towards this). Another valuable approach along this theme is to test multiple analysis pipelines against one another \citep{Jacobs2016comparison}. In general, it is clear that whatever one's approach for calibration, foreground suppression, map-making, and power spectrum estimation, a careful \emph{validation} of the chosen analysis approach will be crucial for precision \tcm science.

\section{Power spectra to parameters/science}
\label{sec:MCMC}
Measuring the power spectrum is not an end in itself but rather a means to an end. The ultimate goal is to use measurements to place constraints on (or better yet, rule out) theoretical models. Power spectra should therefore be thought of as a compressed version of the data that can be used to distinguish between different theoretical scenarios. If the fluctuations that are being measured are Gaussian-distributed, then this compression is lossless. If not, then the compression is lossy, and the power spectrum contains less information than the original data. Regardless of whether this is the case, the exercise of quantitatively comparing theoretical and observational power spectra is generally still a fruitful one that enables the measurement of key parameters that govern physical models.

The process of extracting parameters from power spectrum measurements is generally tackled by Bayesian inference. Again, we may invoke Bayes' theorem (Equation \ref{eq:Bayes}), as we did in Section \ref{sec:BayesPspec} to constrain the power spectrum. However, whereas before the data vector $\vd$ consisted of interferometric visibilities and the parameters of interest $\boldsymbol \theta$ were the power spectrum bandpowers, here the data $\vd$ are the power spectrum bandpowers and the parameters $\boldsymbol \theta$ are those in the physical models. With this setup, if the errors on the power spectrum are assumed to be Gaussian,\footnote{In general, the errors on the power spectrum are not expected to be Gaussian, but Gaussianity may be a reasonable approximation if the errors are small. Alternatively, if the Bayesian methods for power spectrum estimation from Section \ref{sec:BayesPspec} are used, one can incorporate information from the non-Gaussian power spectrum posterior into the likelihood here.} the likelihood takes the form
\begin{equation}
\label{eq:PspecParamLikelihood}
\mathcal{L} (\mathbf{d} | \boldsymbol \theta) \propto \exp \Bigg{\{} -\frac{1}{2} [ \mathbf{d} - \mathbf{d}_M (\boldsymbol \theta)]^t \mathbf{V}^{-1} [ \mathbf{d} - \mathbf{d}_M (\boldsymbol \theta) ] \Bigg{\}},
\end{equation}
where $\mathbf{d}_M (\boldsymbol \theta)$ is the prediction of our theoretical model for the power spectrum bandpowers, and $\mathbf{V}$ is the covariance matrix of the observed bandpowers, as defined in Equation \eqref{eq:VMFM}. Sampling from this distribution using an MCMC algorithm and multiplying by a prior then produces a sampling of the posterior distribution $\textrm{Pr}(\boldsymbol \theta | \vd, \mathcal{M})$ of the parameters given the data and the model (see Section \ref{sec:BayesPspec}). This is a fairly standard way to constrain model parameters, but in the rest of this section, we highlight several subtleties that arise in \tcm data analysis that may not be important in other cosmological analyses.

\subsection{Light cone effects}

One of the key features of \tcm surveys is their ability to probe large cosmological volumes. While this of course results in statistically powerful scientific constraints, it can result in subtleties in analysis that arise because of cosmological evolution efforts. In particular, the survey volumes have the property that the lowest frequency edge of the volume may be at a significantly higher redshift than the highest frequency edge. In other words, cosmological evolution is typically non-negligible in intensity mapping surveys, and one is surveying a past light cone of our Universe rather than a three-dimensional volume at fixed time.

With significant evolutionary effects, our surveyed temperature fields are no longer statistically translation invariant. This implies that the power spectrum no longer captures all the information contained in the fields. To see this, consider Equation \eqref{eq:PspecDef}, which defines the power spectrum. If one imagines a discrete version of the equation where there are a finite and discrete set of $\mathbf{k}$ values, then the left hand side is essentially a covariance matrix between every ``pixel" in Fourier space and every other pixel in Fourier space. The right hand side says that if the field is statistically translation invariant, then this covariance matrix is diagonal, with the diagonal elements given by the power spectrum. Thus, with assumptions of translation invariance (and Gaussianity; see Section \ref{sec:HigherOrderStatistics}), the power spectrum contains all the information about the field. If translation invariance is broken, however, then the covariance matrix is no longer diagonal, and the power spectrum alone does not capture everything.

There are several solutions to this problem. One is to simply capture the relevant off-diagonal elements. This only needs to be done along the frequency direction, since it is redshift evolution that is responsible for destroying translation invariance. Thus one possibility is to compute a quantity such as the cross-angular power spectrum from every pair of frequency channel in one's data, i.e., $C_\ell (\nu, \nu^\prime)$ \citep{Santos2005,Datta2007,Bharadwaj2019}. This quantity records full correlation information in the frequency direction, and in the angular direction assumes statistical isotropy to justify the use of angular power spectra. Such a formalism is particularly fruitful for low-redshift \tcm cosmology, where (at least on linear scales) the evolutionary effects can be relatively straightforwardly modelled by a growth factor that independently multiplies each Fourier mode by some scalar factor. This makes it simple to relate the measured quantity, $C_\ell (\nu, \nu^\prime)$, to theoretical parameters, in a way that is not true for experiments targeting reionization and Cosmic Dawn redshifts.

One potential weakness of computing quantities such as $C_\ell (\nu, \nu^\prime)$ is that it may be too general of a way to capture the statistics of non-translation invariant fields. By capturing the cross-correlation between any possible pair of frequencies, one can encode arbitrary violations of translation invariance. This gives up on a powerful constraint on the problem: the fact that for $\nu$ and $\nu^\prime$ that are close to one another, the field should be approximately translation invariant. One way to incorporate this information is to try to search for an alternative to the Fourier basis, essentially searching for a basis where the two-point covariance is diagonal. This is a difficult problem to solve because doing so in rigorous generality requires modelling the exact nature of the cosmological evolution. However, in the spirit of this approach, wavelet transforms have been shown to be a promising possible way forward \citep{Trott2016}.

Of course, one can simply choose to \emph{ignore} the light cone effect. For small enough frequency ranges, this is likely a fair approximation. In any case, even if light cone effects are significant, the power spectrum is still a reasonable statistic to compute---it is simply not one that captures all the information. Additionally, the light cone effect is arguably a higher order effect that is to be tackled only after high signal to noise measurements of the power spectrum are available. Indeed, this has been the stance taken with recent upper limits that have been placed on the \tcm power spectrum. One simply splits up the entire observational bandwidth into smaller sub bands, measuring an evolution-averaged power spectrum in each of many redshift bins. However, it should be noted that in principle, light cone effects do have the potential to affect the interpretation of even upper limits, since they can potentially depress \tcm power on a some scales during reionization \citep{Datta2012lightcone,LaPlante2014}.

\subsection{Computationally expensive model predictions}
\label{sec:emulators}
A key assumption in the use of MCMCs to constrain model parameters is that one has a way to generate model predictions quickly once the model parameters are provided. In other words, MCMCs allow an efficient exploration of parameter space \emph{provided} the evaluation of $d_M (\boldsymbol \theta)$ at any given parameter space location $\boldsymbol \theta$ is quick. In many cases, this is not the case. At post-reionization redshifts, complicated models are in principle required to model the total HI content and distribution relative to dark matter, although some of these complications can be omitted if the focus is restricted to linear scales or robust signatures like BAOs. At reionization redshifts and above, astrophysical effects on various scales have a complex interplay with cosmological fields, necessitating either radiation-hydrodynamical simulations or semi-analytic treatments for any theory predictions. These can be slow to run, with even semi-analytic codes potentially taking several hours per run to obtain predictions at the wide variety of redshifts needed to compare to predictions.

\begin{figure*}[t]
\centering
\includegraphics[width=1.0\textwidth]{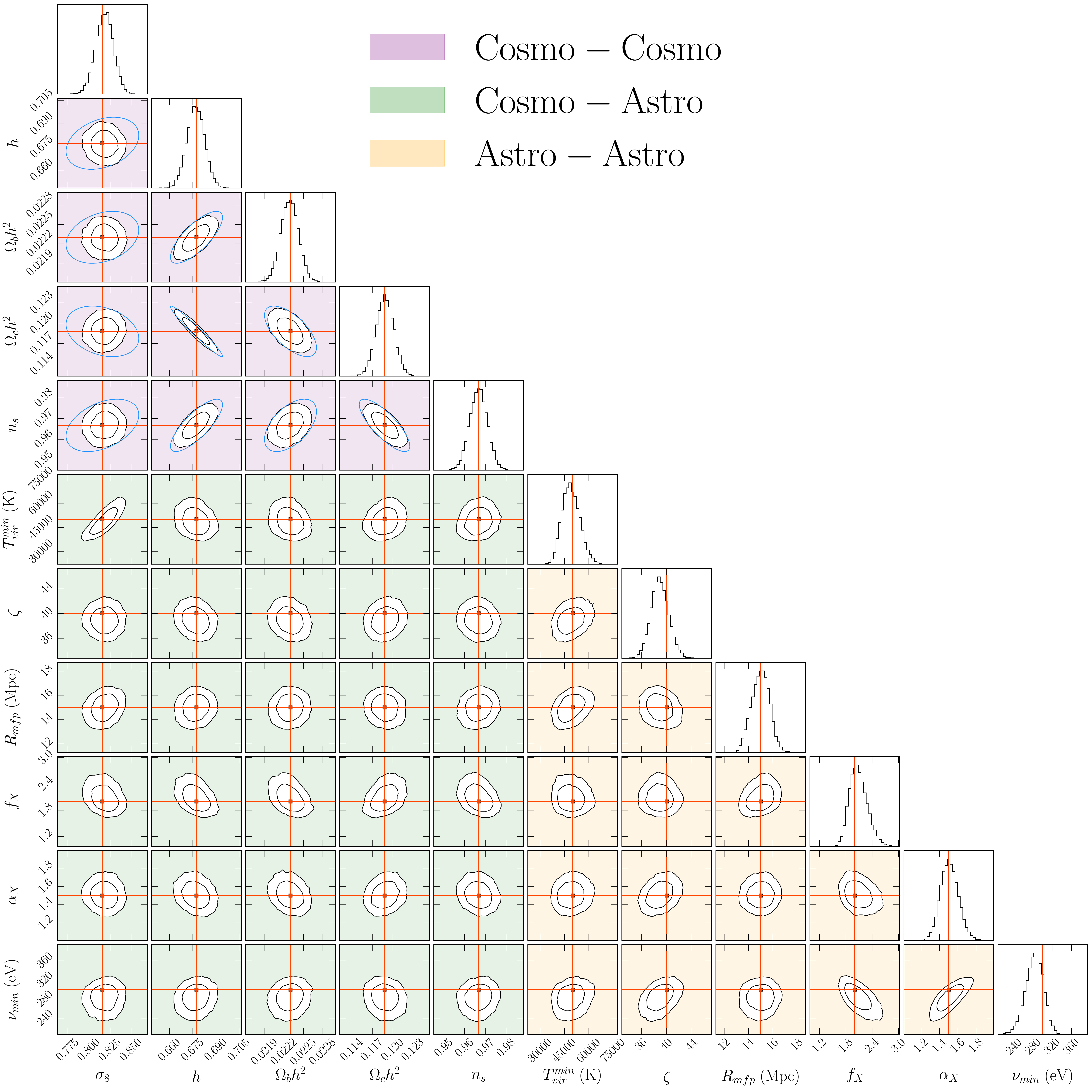}
\caption{Forecasted HERA constraints on theoretical model parameters from a hypothetical $z =5$ to $z=25$ power spectrum measurement. Forecasts (and eventually, fits to the real data) like these that involve both cosmological and astrophysical parameters are made possible by the emulators discussed in Section \ref{sec:emulators}. The cosmological parameters here are the root-mean-square amplitude of density fluctuations on $8\,h^{-1}\textrm{Mpc}$ scales, $\sigma_8$; the dimensionless Hubble parameter, $h$; the normalized baryon density, $\Omega_b$; the normalized cold dark matter density, $\Omega_c$; and the spectral index of the primordial scalar fluctuations, $n_s$. The astrophysical parameters are the minimum virial temperature of ionizing halos, $T_\textrm{vir}^\textrm{min}$; the ionizing efficiency, $\zeta$; the mean free path of ionizing photons in ionized regions, $R_{mfp}$; the luminosity of X-ray sources, $f_X$; the spectral index of X-ray emission, $\alpha_X$; and the minimum photon energy for escape into the intergalactic medium, $\nu_{min}$. Note that there is considerable freedom as to how one chooses to parametrize reionization; this is simply one possible parametrization, and others are possible (e.g., see \citealt{Park2019}). For details on the forecasting methodology and detailed instrumental assumptions, please see \citet{Kern2017}.}
\label{fig:triangle}
\end{figure*}

One solution to this problem is to forgo the exact evaluation $d_M (\boldsymbol \theta)$ in favour of a \emph{surrogate} or \emph{emulated} model, $d_M^\textrm{em} (\boldsymbol \theta)$, that is quick to evaluate. Such an emulator can be constructed by evaluating the exact model at various carefully chosen grid points in parameter space, and fitting to them to create $d_M^\textrm{em} (\boldsymbol \theta)$ as a higher dimensional interpolation function. This approach has been successfully employed in the past in the CMB and galaxy survey communities \citep{Heitmann2006,Fendt2007,Heitmann2009,Heitmann2014,Habib2007,Schneider2011,Aslanyan2015}, and has recently been demonstrated in forecasts of \tcm measurements at reionization and Cosmic Dawn redshifts \citep{Kern2017,Schmit2018}. An example of this from \citep{Kern2017} is shown in Figure \ref{fig:triangle} for a hypothetical HERA power spectrum measurement in the range $5 < z < 25$. With emulators, it is possible to perform MCMC forecasts (and eventually data analyses) over parameter spaces that include not just astrophysical parameters but also cosmological parameters. With the high sensitivities of current and upcoming experiments, this has been shown to be necessary for accurate constraints \citep{Liu2016}.

\subsection{Uncertainties in the underlying theoretical models}
\label{sec:modelselection}
In the discussion above, we have implicitly assumed that there is an underlying theoretical model $\mathcal{M}$ that is known to be correct, and that only their adjustable parameters are unknown and need to be constrained. Of course, this is in general not true. For example, at $z > 6$ the astrophysics of reionization and Cosmic Dawn is extremely uncertain, to the extent that it is not just model parameters that are unconstrained, but also the underlying frameworks. At $z<6$ the situation is slightly better, since the HI gas density is expected to be a biased tracer of matter. However, the HI bias (defined as the square root of the ratio of the HI power spectrum to the matter power spectrum) becomes non-linear at small scales \citep{Wang2019}, and only recently have robust theoretical frameworks been developed to model this behaviour \citep{Villaescusa-Navarro2018}. While these frameworks were strongly motivated by detailed hydrodynamical simulations (and therefore agree with them reasonably well), it would be fair to say that revisions and refinements may be necessary as \tcm measurements mature.

With the underlying theoretical frameworks uncertain, one option is to build models that are general and flexible, and some semi-analytic codes are headed in this direction \citep{Park2019}. An alternative is to perform model \emph{selection} prior to model \emph{fitting}. One way to do this is to simply examine which models are able to fit the data well. However, this tends to reward models that have an excessive number of parameters, some of which may not be physically well-motivated and exist simply to improve the fit. A possible alternative to this would be to compute the Bayesian evidence. Recall from Section \ref{sec:MCMC} that the evidence is given by $\textrm{Pr}(\mathbf{d} | \mathcal{M})$, i.e., the probability of the observed data $\mathbf{d}$ given the theoretical model $\mathcal{M}$. This can be computed by noticing that because Equation \eqref{eq:Bayes} must be a properly normalized posterior probability distribution, marginalizing over all the parameters in the numerator gives us the denominator, and thus
\begin{equation}
\textrm{Pr}(\mathbf{d} | \mathcal{M})= \int  \mathcal{L} (\mathbf{d} | \boldsymbol \theta, \mathcal{M}) \textrm{Pr}(\boldsymbol \theta, \mathcal{M}) d^{n} \theta,
\end{equation}
where $n$ here is the number of parameters. In model selection we are ultimately interested in a closely related quantity: the probability of the theoretical model given the observed data, $\textrm{Pr}(\mathcal{M} | \mathbf{d}  )$. This is a quantity that we can obtain using Bayes' theorem, which gives
\begin{equation}
\textrm{Pr}(\mathcal{M} | \mathbf{d}  ) = \frac{\textrm{Pr}(\mathbf{d} | \mathcal{M}) \textrm{Pr}(\mathcal{M}) }{\textrm{Pr}(\mathbf{d})},
\end{equation}
where $p(\mathcal{M})$ is the prior on the model. If we have two competing theoretical models, $\mathcal{M}_1$ and $\mathcal{M}_2$, we may compare them by computing the ratio of their probabilities. This is known as the \emph{Bayes factor}, and is given by
\begin{equation}
\frac{\textrm{Pr}(\mathbf{d} | \mathcal{M}_1) \textrm{Pr}(\mathcal{M}_1) }{\textrm{Pr}(\mathbf{d} | \mathcal{M}_2) \textrm{Pr}(\mathcal{M}_2)}.
\end{equation}
We thus see that in selecting between different models, one computes the ratio of the evidences, adjusting for one's priors on the plausibility of each model. Which model is ultimately favoured depends on whether the ratio is greater or less than unity; precisely how far the ratio needs to be from unity for one model to be favoured over the other is in some ways a matter of taste, although a rough guide that is frequently used is the one given by \citet{KassRaftery1995}.

The Bayesian evidence has so far been employed in forecasting exercises for potential power spectrum measurements, suggesting that upcoming instruments should have the sensitivity to distinguish between different classes of reionization models \citep{BinniePritchard2019}. It has also been proposed as a way to select between different parameterizations of the foregrounds (\citealt{Harker2015}; Sims et al., in prep.), and has been applied to real global signal data (described in Section \ref{sec:GlobalSig}) to evaluate different models for the cosmological signal, foregrounds, and systematics.

\section{Power spectrum alternatives and variants}
\label{sec:BeyondPspec}
For the bulk of this paper, we have focused on measurements of the \tcm power spectrum. The primary reason for this is that the power spectrum is what most of the current \tcm experiments are attempting to detect and characterize (with the exception of a small number of \emph{global signal} experiments described in Section \ref{sec:GlobalSig}). There is no deep, fundamental reason this. The power spectrum is simply a convenient summary statistic for describing the spatial fluctuations being measured. Moreover, power spectra have a long history in cosmology, and have the advantage of being well-defined quantities that are relatively straightforward to compute from theory and/or simulations.

However, the \tcm power spectrum is ultimately limited as a summary statistic, for both theoretical and practical reasons. Theoretically, the power spectrum may not contain all the information contained in the \tcm field (or the underlying quantities that determine the \tcm brightness temperature). Practically, it is prudent to include other summary statistics when reducing \tcm data, since these alternatives may have complementarity systematics. In this section, we highlight a few possible power spectrum alternatives and variants.

\subsection{Cross correlations}
\label{sec:CrossCorrPossibilities}

As a first variation on the \tcm power spectrum, we consider cross power spectra between the \tcm line and some other tracer of structure in our Universe. The cross power spectrum is defined in much the same way as the \tcm power spectrum, i.e., as a two-point covariance function in Fourier space as defined in Equation \eqref{eq:PspecDef}, except with one copy of the \tcm brightness temperature field and one copy of the other tracer rather than two copies of the \tcm field. There are a large number of tracers that are suitable candidates for such a cross correlation, especially at $z < 6$. For instance, the first detections of the cosmological \tcm line were made in cross correlation with traditional galaxy surveys at $z \sim 0.8$. At slightly higher redshifts, CO rotational lines and $[$CII$]$ hyperfine transitions have been proposed for cross-correlations. It is possible that CO may be insufficiently abundant at $z \gtrsim 6$ due to photo dissociations \citep{Lidz2011}, but $[$CII$]$ intensity mapping observations have been planned for up to $z \sim 9$ \citep{Stacey2018CCATp}. At even higher redshifts, there is the potential of cross-correlating with Lyman-$\alpha$, H$\alpha$, $[$OII$]$, and $[$OIII$]$ intensity maps that will be provided by satellite missions such as the Spectro-Photometer for the History of the Universe, Epoch of Reionization, and Ices Explorer (SPHEREx; \citealt{SPHEREx}), and possibly by futuristic proposed missions such as the Cosmic Dawn Intensity Mapper \citep{CDIM} and the Origins Space Telescope \citep{Battersby2018,Meixner2018}.

For a cross-correlation measurement to have high signal to noise, it is necessary but not sufficient for the two participating surveys to overlap in configuration space (i.e., for the surveys to look at the same part of the sky and the same redshifts); it is also necessary for the surveys to overlap in Fourier space. In other words, the surveys must be sensitive to the same length scales. This can be achieved either by having telescopes that have comparable angular and spectral resolutions, or by ensuring that the finer dataset also be sensitive to large-scale fluctuations that the coarser dataset is exclusively sensitive to. This is not simply a matter of hardware design; care must also be taken to make sure that one's data analysis does not destroy Fourier modes that are important for cross correlation. As an example of this, note that any foreground subtraction algorithm that relies on suppressing smooth modes along the line of sight will inevitably produce maps that destroy power at low $k_\parallel \sim 0$ modes. Because of this, a cross correlation of such maps with any integral field (i.e., one that is projected along the line of sight, such as any CMB map) will have very low signal to noise. One way around this would be to simply skip foreground subtraction, and hope that the foregrounds are uncorrelated between the two surveys and thus will not contribute. However, as we discussed in Section \ref{sec:CrossCorrFGs}, this is only the case \emph{in expectation}, and residual foreground variance may remain. It is therefore likely that foreground mitigation will play a role even for cross-correlation studies, depressing the final signal-to-noise ratio in the final measurement.

\subsection{Higher order statistics}
\label{sec:HigherOrderStatistics}
To increase the signal to noise on a cross correlation between an integral field and a foreground-suppressed \tcm map, one possibility is to go beyond a two-point cross-correlation between two fields and to instead use higher order $n$-point statistics. If the signals in question were Gaussian, higher order statistics would not yield any new information: $n$-point functions would vanish for odd $n$ by symmetry and reduce to powers of the two-point functions for even $n$. Moreover, statistical translation invariance guarantees that different Fourier modes are not correlated with one another, giving rise to a two-point function that is non-zero only if the two ``points" coincide in Fourier space, which we recall from Equation \eqref{eq:PspecDef} is what we denote the power spectrum. If non-Gaussianities are present, different Fourier modes couple to one another and higher order statistics contain new information. For example, the \emph{bispectrum} $B(\mathbf{k}_1, \mathbf{k}_2, \mathbf{k}_3)$ can be defined as
\begin{equation}
\label{eq:bispectrumdef}
B(\mathbf{k}_1, \mathbf{k}_2, \mathbf{k}_3) \equiv (2\pi)^3 \delta^D (\vk_1 + \vk_2 + \vk_3) \langle \widetilde{T}(\vk_1) \widetilde{T}(\vk_2) \widetilde{T}(\vk_3) \rangle,
\end{equation}
where one sees that as long as $\vk_1 + \vk_2 + \vk_3 = 0$ (i.e., the three vectors form a closed triangle), different Fourier modes can be correlated with one another.

One way in which mode correlations can occur is through nonlinear gravitational evolution, and it is through such a mechanism that it may be possible to cross-correlate a \tcm map with an integral field. Non-linear evolution results in a coupling between large-scale tidal fields and small-scale density fields. To access the information contained in this coupling, one can either try to directly measure the higher-point statistics or attempt tidal reconstruction. With the latter approach, small-scale measurements can in principle be used to reconstruct large-scale modes \citep{Zhu2016TidalReconstruction}, recovering cosmological information that was lost in data analysis. Such tidal reconstruction schemes can in principle be applied to a variety of different cross-correlations between \tcm data and other measurements. Examples include the CMB (via Integrated Sachs-Wolfe or kinetic Sunyaev-Zel'dovich contributions) or photometric galaxy surveys (which behave in a similar way to integrated fields since large redshift uncertainties mean that often only angular correlation information is trustworthy) \citep{Zhu2018fgmoderecovery,Li2018kSZtidal}. While tidal reconstruction forecasts vary in their levels of optimism regarding the approach (see, e.g., \citealt{GokselKaracayli2019} for a more pessimistic evaluation), its enormous scientific potential clearly justifies further investigation.

Another mechanism for inducing mode correlations---and therefore another probe of the underlying physics---is gravitational lensing by large scale structure. Statistical characterizations of lensing have been used to great effect in the CMB community, for instance in placing neutrino mass constraints \citep{Planck2018params} or as one possible way to break the various \emph{geometric degeneracies} in earlier CMB experiments, such as that between the normalized curvature parameter $\Omega_K$ and $H_0$ \citep{Efstathiou1999,Howlett2012}. In \tcm cosmology, similar lensing measurements are in principle possible, although there are some crucial differences. For one, in CMB lensing there is just a single source plane at high redshifts, whereas in \tcm cosmology there are a series of source planes. Having multiple source planes can increase the signal to noise of a lensing detection, but care must be taken to ensure that correlations between the source planes are taken into account, since the different source planes are part of a larger correlated three-dimensional field \citep{ZahnZaldarriaga2006}. An additional complication with \tcm lensing is that the source planes themselves probe matter fluctuations that contain mode couplings due to the aforementioned gravitational evolution effects. These intrinsic mode couplings can bias statistical estimators of lensing, which are designed to detect lensing by looking for correlations between modes. It is possible, however, to write down \emph{bias hardened} estimators that are protected against such contamination \citep{Foreman2018}.

In addition to the effects described above (which affect all redshifts), there are also strong non-Gaussian signatures that are intrinsic to the \tcm field during reionization \citep{Iliev2006, Mellema2006}. This arises because (among other factors) the \tcm brightness temperature involves the product of the neutral fraction and the density. Even if we restrict ourselves to scales where the density field is Gaussian, the neutral fraction field will necessarily be non-Gaussian given that it is restricted to take on values between $0$ and $1$. Aside from this line of purely mathematical reasoning, there are also physical reasons to expect non-Gaussianity. In particular, galaxies form at the peaks of the density field (i.e., they are biased tracers of density), and thus the ionization structures they produce around them are caused by the high-end tails of the mass function \citep{Wyithe2007galaxyioncorr}, resulting in ionized structures that do not depend linearly on overdensity \citep{Wyithe2007nonG}. This leads to strong non-Gaussianities \citep{Lidz2007nonlinearpspec}. Similar effects are present at even higher redshifts (prior to reionization) due to inhomogeneities in X-ray heating \citep{Watkinson2015,Watkinson2019}.

To access the non-Gaussian information contained in the high redshift \tcm field, one possibility is to simply measure the bispectrum \citep{Watkinson2017}. Recent studies have pointed out that sensitivity to the bispectrum scales rather favourably with time (scaling as $t^{-3/2}$ rather than $t^{-1}$ for the power spectrum, as discussed in Section \ref{sec:Recipe}), and that foregrounds may not be prohibitive for certain $(\vk_1, \vk_2, \vk_3)$ triangle configurations \citep{TrottBispectrum2019}. Theoretical predictions have suggested that the bispectrum may break certain parameter degeneracies that arise in analyses utilizing only the power spectrum \citep{Shimabukuro2017}. The bispectrum also contains distinctive signatures that mark the progression of reionization in the form of sign changes \citep{Majumdar2018}. Such sign changes occur because the \tcm bispectrum can be written as the sum of all possible bispectrum combinations between density and ionization (e.g., three-point functions of density or three-point functions consisting of two factors of density and one factor of neutral fraction, and so on). Some of these combinations may probe fields that are correlated to one another, while others probe an anti-correlated set of fields. This results in sign changes in the total bispectrum as different combinations dominate the total bispectrum as reionization progresses. Additional interesting signatures can be obtained by forming the bispectrum not from the three-point function of the Fourier modes themselves, but their \emph{phases}, where each copy of $\widetilde{T}$ in Equation \eqref{eq:bispectrumdef} is replaced by $\widetilde{T}/ |\widetilde{T}|$. Such a bispectrum of phases has been shown in theoretical studies to be a probe of the characteristic length scales of ionized regions during reionization \citep{GorcePritchard2019}.

Although theoretically promising, the experimental feasibility of bispectrum measurements is still an open question. While current and planned instruments may have the sensitivity to potentially detect the bispectrum \citep{Yoshiura2015}, most studies assume that foregrounds can be removed to extremely high precision. Moreover, calibration errors are typically not included in theoretical studies. One way to (in principle) sidestep the need for high precision calibration is to measure the phase of the bispectrum (in contrast to the bispectrum of phases described above). In particular consider the phase of the \emph{angular} bispectrum, where the bispectrum is computed using interferometric data from a single frequency channel. In the flat sky approximation, each baseline of an interferometer probes a different angular Fourier mode (see Equation \ref{eq:kperpkparamappings} and Section \ref{sec:ImageVsStat}), which means that the bispectrum is simply the three-point correlation function of three interferometric visibilities whose baseline vectors form a closed triangle. The triangular requirement is imposed by the Dirac delta function of Eq. \eqref{eq:bispectrumdef}, and can be satisfied by using the visibilities from baselines that connect three antennas. Computing the phase of the angular bispectrum is then equivalent to computing the phase of $V^\textrm{meas}_{12} V^\textrm{meas}_{23} V^\textrm{meas}_{31}$ where $V^\textrm{meas}_{ij}$ is the measured visibility from a baseline formed between the $i$th and $j$th antennas. This has the attractive property of being immune to per-antenna calibration effects considered in Section \ref{sec:Calib}, since we can insert Eq. \eqref{eq:CalibEqn} to see that
\begin{eqnarray}
\label{eq:BispectrumPhase}
\phi &=& \textrm{arg} (V^\textrm{meas}_{12} V^\textrm{meas}_{23} V^\textrm{meas}_{31} ) \nonumber \\
&=& \textrm{arg} ( g_1 g_2^* V^\textrm{true}_{12} g_2 g_3^* V^\textrm{true}_{23} g_3 g_1^* V^\textrm{true}_{31} )\nonumber \\
&=& \textrm{arg} ( |g_1|^2 |g_2|^2 |g_3|^2 V^\textrm{true}_{12}  V^\textrm{true}_{23}  V^\textrm{true}_{31} )\nonumber \\
&=& \textrm{arg} (  V^\textrm{true}_{12}  V^\textrm{true}_{23}  V^\textrm{true}_{31} ),
\end{eqnarray}
where $\textrm{arg} (\cdots)$ signifies the phase of a complex number, and in the last equality we used the fact that multiplying a complex number by a real number does not affect its phase. We see that our bispectrum phase $\phi$ (which is also known as the \emph{closure phase} in more traditional radio astronomy contexts; \citealt{Cornwell1999,TMS2017,Carilli2018closure}) is independent of per-antenna gain calibration factors, making it particularly attractive a quantity to measure with real data. Of course, aside from calibration errors one must also contend with foreground contamination. Simulations have suggested though, that subsequently computing the delay spectrum of the $\phi$ may potentially allow foregrounds to be separated from the cosmological signal \citep{Thyagarajan2018closurephase}.

\subsection{Probability distribution functions}

In principle, to capture the full effects of non-Gaussianity, one must compute an infinite series of higher-order correlation functions. For instance, beyond the bispectrum there is the trispectrum (or equivalently, the four-point correlation function). One alternative to this is to simply compute the probability distribution function (PDF) of voxels in a \tcm map \citep{Wyithe2007nonG,Harker2009PDFs,Watkinson2014,Shimabukuro2015}, which then captures the full extent of non-Gaussianity in one statistic. The weakness of this approach is that it does not capture correlation information \emph{between} pixels, since the value of each pixel is simply binned into a histogram (i.e., a PDF) without regard for other pixels. Some correlation information can be recovered by forming two-point PDFs that give the joint probability of measuring various brightness temperature values in two pixels. Closely related to this is the idea of difference PDFs, where one quantifies the probability distribution function of differences between the values of two voxels separated by a given distance \citep{BarkanaLoeb2008differencePDF,GluscevicBarkana2010,PanBarkana2012}.

Many of these PDFs have shown promise in capturing important features of the underlying physics. However, they have yet to be shown to be practical in the face of foreground contamination. The foreground challenge is greater for PDFs than for statistics such as the power spectrum. This is because the foregrounds do not enter in an additive manner in the PDFs; instead, the PDF of foregrounds is \emph{convolved} with the PDF of the cosmological signal, since the PDF of a sum of two random variables is given by the convolution of the constituent PDFs. Typically, the foreground brightness temperature will vary with a larger amplitude from pixel to pixel (up to 100s of K in variation, depending on the frequency) than the cosmological signal does (which might have $\sim$mK-level variations). This means that the foreground contribution makes the measured (i.e., foreground contaminated) PDF very broad, and the foregrounds must be known to exquisite precision in order for their PDF to be deconvolved out to leave the narrow cosmological signal PDF. Having such precision in our knowledge of foregrounds is difficult at the low radio frequencies that are relevant. However, recent work has suggested that with multiple tracers (e.g., galaxy surveys in conjunction with \tcm maps), it may be possible sidestep this requirement in the foreground PDF deconvolution \citep{Breysse2019pdf}.

To increase signal to noise, one might also consider measuring the higher order moments (e.g., the skewness and/or the kurtosis) of PDFs. Condensing non-Gaussianity signatures into a few moments means that there are fewer numbers to measure, thereby increasing sensitivity. Studies have suggested that even though the skewness and the kurtosis do not capture the full richness of a PDF, they are able to capture some interesting non-Gaussian effects of reionization \citep{Wyithe2007nonG,BoomThesis}. These statistics are potentially measurable even in maps that have been preprocessed to filter out the foreground wedge region of Fourier space described in Section \ref{sec:wedge} \citep{BoomThesis}. The filtering, however, does affect the moments (beyond just removing the foregrounds), and so a correct interpretation of the results will likely require simulation-based forward models.

\subsection{Imaging}

Capturing the full richness of spatial fluctuations in the \tcm line---including its cosmological evolution, probability distribution and higher-order correlations---is an image of the \tcm field. However, the information-completeness of images comes at a cost. First, images will generally have lower signal to noise than averaged/binned statistics such as the power spectrum, necessitating larger and more sensitive arrays. Second, an image is not a statistical property of the underlying field; instead, it is a specific \emph{realization} of the underlying statistics.

Having said this, the fact that an image is a realization rather than a statistical property can be an advantage. Consider, for example, the possibility of comparing \tcm images with the locations of high-redshift galaxies during reionization. The \tcm images can in principle provide context for the detected high-redshift galaxies. A galaxy's location within an ionized bubble, for instance, might indicate whether the galaxy resides in a newly ionized region of our Universe, or one where reionization happened long ago. In practice, there is unfortunately a mismatch in angular scales: by design, intensity mapping surveys such as \tcm surveys do not seek to resolve individual galaxies, unlike with high-redshift galaxy observations, which in turn tend to have fairly narrow fields of view. As an example of this severe mismatch of scales, we note that the entire Great Observatories Origins Deep Survey South field (GOODS South field; covered by the Hubble Space Telescope, the Spitzer Space Telescope, and the Chandra X-ray Observatory) is approximately the same size as the resolution of a single pixel in maps produced by HERA! Future widefield surveys covering sky areas comparable to what is envisioned for the Wide Field Infrared Survey Telescope (WFIRST) would be better matched for the scales probed by \tcm instruments. However, the importance of line-of-sight fluctuations in \tcm maps means that one ideally requires spectroscopic galaxy redshifts to avoid the washing out of radial fluctuations that results from photometric redshift estimates. Additionally, the low signal-to-noise of a high redshift \tcm map would likely necessitate a statistical cross correlation rather than an analysis of the maps themselves.

Perhaps more promising for the direct analysis of \tcm maps is to use them to provide \emph{probabilistic} models of the context in which high-redshift galaxies reside. In particular, although resolution effects make it difficult to definitively determine whether a high-redshift galaxy resides in an ionized bubble or not, it is possible to use \tcm maps to predict the probability that a pixel is ionized or not \citep{Beardsley2015}. Importantly, such a scheme works reasonably well even if one assumes that foreground mitigation causes certain Fourier modes to be irretrievably lost within the wedge-shaped region described in Section \ref{sec:wedge}.

Another way to use images is to use them to identify extremes. For reionization studies, this can be helpful both at the beginning and the end of reionization. Towards the end of reionization, recent studies have suggested that reasonably large neutral islands can persist even when our Universe is almost completely ionized \citep{Becker2015,Malloy2015,Kulkarni2019,Keating2019}. Conversely, luminous quasars are able to generate large ionized bubbles even if our Universe is mostly neutral \citep{WyitheLoeb2014}. These large features are potentially detectable by current-generation arrays, particularly using analysis techniques such as matched filtering \citep{MalloyLidz2013}. However, further forecasts need to be made to definitively quantify the effect of foregrounds and instrumental systematics on such measurements.

Matched filtering is an example of an analysis scheme where a specific feature of the \tcm maps is searched for, with the feature selected by the analyst ahead of time because of its potential to constrain the underlying physics. An alternative to this is to use machine learning, where the features are not selected by hand, but instead are the result of an optimization process (``learning") that searches for features that are best able to constrain the ultimate quantities of interest. Suppose that one is interested in extracting a parameter vector $\boldsymbol \theta$ from an image whose voxel values are stored in a data vector $\mathbf{d}$. With a machine learning approach, we seek a function $f$ that provides good estimates $\hat{\boldsymbol \theta}$ of our parameter vector, i.e., we seek a function where
\begin{equation}
\boldsymbol \theta \approx \hat{\boldsymbol \theta} \equiv f (\mathbf{w}, \mathbf{d}),
\end{equation}
with $\mathbf{w}$ being a set of adjustable parameters. Starting with a general form for $f$, the adjustable parameters are determined by requiring that $\boldsymbol \theta \approx \hat{\boldsymbol \theta}$ over a series of \emph{training} examples of $\mathbf{d}$ where the true parameters are known.\footnote{In this paper, we will only discuss \emph{supervised learning} algorithms where the correct answers are known in the training data, as opposed to \emph{unsupervised learning} algorithms, where this is not the case.} Precisely what it means for $\boldsymbol \theta \approx \hat{\boldsymbol \theta}$ is quantified by a cost function. If $\boldsymbol \theta$ were to be a set of parameters governing some theoretical model (like we saw in Section \ref{sec:MCMC}, for example), then one possible choice might be the mean squared error between the recovered parameters and the true parameters. The training process is then tantamount to finding $\mathbf{w}$ such that
\begin{equation}
\sum_i | \boldsymbol \theta - f (\mathbf{w}, \mathbf{d}_i)|^2
\end{equation}
is minimized, where the index $i$ runs over the different training datasets. Once the best $\mathbf{w}$ vector is found, it is fixed and $f$ can be applied to either test datasets (for validation) or to real data to provide parameter estimates.

In machine learning, one typically selects forms for $f$ that are extremely flexible. A particularly popular choice is to have $f$ take on the form of a Convolutional Neural Network (CNN; see, e.g., \citealt{Goodfellow-et-al-2016}). In a CNN, the input data (i.e., the input image) is passed through a series of convolutions, downsamplings, and linear transformations. After each of these steps the data is additionally passed through a non-linear function that is predetermined by the data analyst. The $\mathbf{w}$ vector encapsulates the details of the various stages of the CNN. For instance, some of its elements might capture the shapes of the convolution kernels that the data is put through. With a CNN, one can in principle mimic \emph{any} reasonably well-behaved higher-dimensional function \citep{Cybenko1989}. It is for this reason that they are powerful, although care must be taken to ensure that they are not \emph{too} powerful, in the sense that one must avoid overfitting. When a CNN is overfit to the training data, it is essentially memorizing the training dataset
and its corresponding parameter values. Fortunately, there are well-established techniques such as dropout \citep{Srivastava2014dropout} to prevent these problems.

In several proof of concept studies, CNNs have shown some promise as ways to analyze \tcm data. CNNs have been shown in simulations to be able to take in \tcm images and successfully recover theoretical parameters (such as the ionizing efficiency of the first galaxies during reionization; \citealt{Gillet2019}) or phenomenological parameters (such as the duration of reionization; \citealt{LaPlanteNtampaka2018}) directly from \tcm images. CNNs can also potentially perform model selection, determining whether a \tcm image favours a galaxies-driven reionization scenario or an AGN-driven reionization scenario even when power spectrum measurements are unable to distinguish between the two \citep{Hassan2019}. Despite its immense promise, the application of CNNs to \tcm cosmology is still in its infancy, and further studies (particularly regarding the presence of foregrounds and systematics in one's images) are necessary before CNNs can be robustly applied to real data.

\section{Global signal measurements}
\label{sec:GlobalSig}

For most of this paper, we have focused on the mapping of the spatial \emph{fluctuations} of the cosmological signal. However, there exists a complementary signal---the \emph{global} \tcm signal---which is also of interest. The global signal refers to the mean brightness temperature of the \tcm line as a function of redshift, averaged over all angular directions of the sky. In the language of spherical harmonics, the global signal is the $(\ell, m) = (0,0)$ monopole mode, which contains independent information. An example theory prediction for what the global \tcm signal might look like can be seen in the middle panel of Figure \ref{fig:EOS}.

\subsection{Science applications of the global \tcm signal}
\subsubsection{The low redshift ($z \lesssim 6$) global signal}
\label{sec:RSD}
At different redshifts, the global signal enters analyses in different ways. For low-$z$ intensity mapping, the global signal is often regarded as a nuisance factor. It essentially sets the overall normalization of the power spectrum. Of course, this normalization is not known \emph{a priori}, which can cause degeneracies that prohibit constraints on parameters of interest. For example, suppose one is interested in using redshift-space distortions to constrain the growth of structure. In such studies, one often attempts to constrain the quantity $f \sigma_8$, where $\sigma_8$ is the root mean square amplitude of matter fluctuations smoothed on $8h^{-1}\,\textrm{Mpc}$ scales, $f \equiv d \ln D / d \ln a$ is the dimensionless growth rate, $D$ is the growth function of perturbations in linear perturbation theory, and $a$ is the scale factor. With redshift space distortions, one measures (to linear order in perturbation)
\begin{equation}
\label{eq:RSD}
P (\vk) = \overline{T}_b^2  (b_\textrm{HI} \sigma_8 + f \sigma_8 \mu^2)^2 P_m(k) / \sigma_{8,\textrm{fid}}^2,
\end{equation}
where we have omitted all noise terms for simplicity, $\mu \equiv k_\parallel / k$, $\overline{T}_b^2$ is the global signal, $b_\textrm{HI}$ is the HI bias, $\sigma_{8,\textrm{fid}}$ is a fiducial value for $\sigma_8$. One sees that there is a perfect degeneracy between $f\sigma_8$ and $\overline{T}_b^2$, since they enter as a multiplicative combination and can therefore be perfectly traded off for one another. Fortunately, recent theoretical work has demonstrated that by going to the mildly non-linear regime, the next-order terms go as higher powers of $f \sigma_8$, while $\overline{T}_b^2$ remains an overall normalization. This allows the degeneracy to be broken, and forecasts suggest that competitive constraints for $f\sigma_8$ can be obtained \citep{CastorinaWhite2019}. Another possibility is to use multi-tracer techniques, where information from futuristic galaxy surveys can be combined with \tcm surveys to break the degeneracy \citep{Fialkov2019}.

\subsubsection{The high redshift ($ z \gtrsim 6$) global signal}

\begin{figure}
    \includegraphics[width=0.45\textwidth]{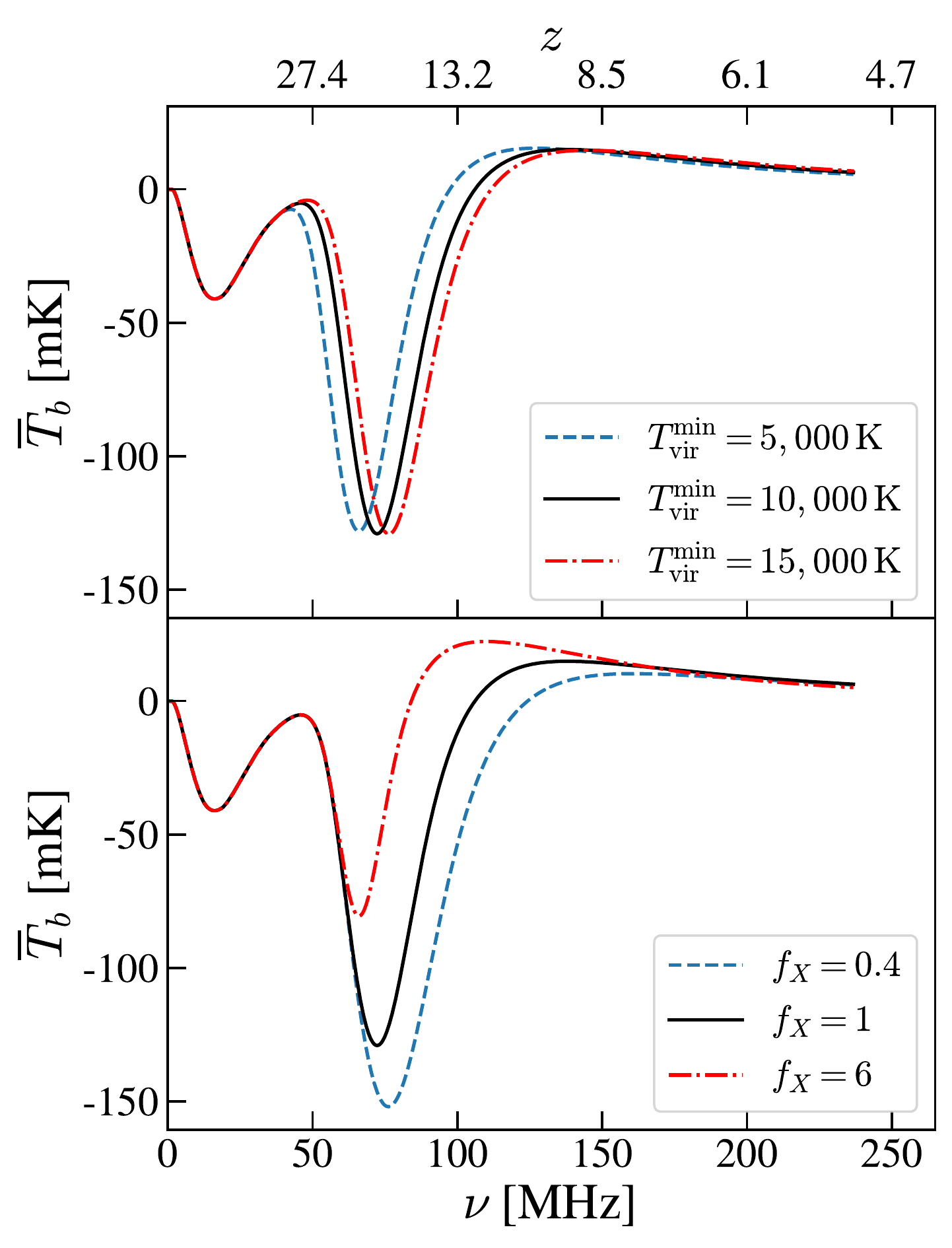}
    \caption{Example global \tcm signal curves. The top panel shows the effect of changing $T_\textrm{min}^\textrm{vir}$, the minimum virial temperature of ionizing halos. The bottom panel shows the effect of changing the proportionality constant $f_X$ between X-ray luminosity and star formation rate. These example curves were generated using the Accelerated Reionization Era Simulations (ARES) code \citep{Mirocha2015,Mirocha2017}.}
    \label{fig:global_sig}
\end{figure}

At high-$z$, the global signal is seen as a particularly interesting signal in its own right. As Figure \ref{fig:global_sig} illustrates, the global signal is an incisive probe of the various phenomena during Cosmic Dawn. During the reionization epoch, the evolution of the global signal is essentially a measure of the ionization history, with the signal going slowly to zero with the neutral fraction of our Universe. At higher redshifts, the global signal is driven by X-ray heating and Ly$\alpha$ coupling, and serves as a probe of energy injection into the IGM.

\subsection{Measuring the global \tcm signal in practice}

Because of differences in their target signals, experiments pursuing the spatial fluctuations of the \tcm line tend to be quite different from those attempting to measure the global signal. The latter, by definition, does not need to resolve spatial features on the sky. Thus, global signal experiments are most commonly conducted using single antennas that have wide beams, although creative approaches using interferometry have been proposed and attempted \citep{Mahesh2014,LOFARMoon2015,Presley2015,Singh2015,Venumadhav2016,McKinley2018}. For the single-element experiments, their wide beams integrate instantaneously over a substantial fraction of the sky, which provides an excellent approximation to the monopole, since the typical correlation lengths of the anisotropies are on much smaller angular scales than the beam size. These experiments do not have the extreme sensitivity requirements of those seeking to measure spatial fluctuations. This is because global signal experiments seek only to capture the relatively gradual cosmic evolution (and therefore frequency evolution) of the signal, and not to resolve the fine spectral features caused by the spatial fluctuations along the line of sight. Global signal experiments can therefore employ relatively coarse frequency channels. As a comparison, a typical experiment targeting spatial fluctuations during Cosmic Dawn might have a spectral resolution of $\Delta \nu \sim 50 \,\textrm{kHz}$, whereas a global signal experiment designed for the same redshifts might have $\Delta \nu \sim 1\,\textrm{MHz}$. With the ability to integrate over larger bandwidths, noise levels quickly beat down in less than a day to levels much lower than the expected amplitude of the cosmological signal. Of course, allowances must be made for non-ideal observing conditions, but even so, far less time is needed to reach acceptably low noise levels for global signal experiments than for power spectrum experiments.

Despite the obvious appeal of their conceptual simplicity, global \tcm experiments are still challenging to perform because the systematics are formidable. Just as with the spatial fluctuation experiments, global signal experiments need to contend with foregrounds \citep{Shaver1999,Bernardi2015}. Most experiments pursue a similar strategy to the fluctuation experiments, in that they base their foreground separation strategy on the spectral smoothness of the foregrounds. This may be more difficult for global experiments, because the signal itself can be quite smooth. For example, consider an experiment targeting the reionization epoch, when the amplitude of the global signal is expected to transition from non-zero to zero, tracing the evolution of the IGM from neutral to ionized. If reionization is a relatively extended process, then the expected global signal is one that smoothly and monotonically decreases as a one goes from low frequencies (high redshifts) to high frequencies (low redshifts). This is essentially the same behaviour as one sees for astrophysical foregrounds, making foreground mitigation extremely difficult. It is for this reason that global \tcm signal experiments observing the reionization epoch may be confined to ruling out very rapid reionization. This is also why many recent efforts have instead targeted the absorption feature during Cosmic Dawn, taking advantage of its non-monotonicity to separate it from the foregrounds.

The crux of the global signal foreground problem is that global signal measurements are, in a sense, too simple. By reducing the entire measurement to a single spectrum, there are far fewer degrees of freedom in the measurement, limiting the number of ways in which the cosmological signal can differ from the foregrounds. Because of this, there have been proposals for using more than just spectral information in the measurement of the monopole. For instance, if one surveys the sky with reasonably fine angular resolution, the parts of the sky with the brightest foregrounds can be selectively avoided, rather than having the entire sky of foregrounds averaged into our global spectrum. This is equivalent to saying that because we seek to measure the monopole of the cosmological signal, anything appearing in the higher multipoles must be a foreground, providing us with another mechanism for foreground rejection beyond the spectral information. Note, however, that the foregrounds will still have a monopole component, and thus spectral methods are still required. By avoiding the brightest parts of the sky, one is simply reducing the amplitude of the foregrounds from being the average over the whole sky to being its minimum value. To go beyond this and to take full advantage of the spatial information to actively \emph{subtract} foregrounds requires a detailed knowledge of its \emph{correlations} and heteroscedasticity between different parts of the sky; if this information is available, foregrounds can in principle be distinguished from the signal even if their spectral signatures are extremely similar \citep{LiuGlobal2013}.

An alternate way to take advantage of spatial foreground information is to use polarization information. Consider a linear polarization-sensitive telescope that is pointed at a celestial pole. Anisotropic foregrounds appear in the two linear polarizations at different strengths due to projection effects. As the Earth rotates, this induces polarization signals that are modulated in time \citep{NhanPol2017}. This provides a way to distinguish anisotropic foregrounds from the isotropic foregrounds and the cosmological signal without requiring large telescopes that resolve the characteristic scales of the anisotropies. Initial progress has been made in implementing such an approach using the Cosmic Twilight Polarimeter \citep{NhanPol2018}.

Yet another proposal for distinguishing foregrounds from the global \tcm signal is to use velocity information \citep{Slosar2017dipole,Deshpande2018dipole}. In particular, the Solar System's motion relative to the cosmic rest frame induces a dipole pattern in what would otherwise be a pure monopole signal. It can be helpful to consider the dipole as being the sum of two terms. The first is a boosting of the signal intensity, analogous to the CMB dipole. The second is a frequency dependent effect. Consider motion towards the CMB, which results in the blueshifting of photons received by one's telescope. If the global signal were independent of frequency, this would have no effect, since for a given frequency channel there would be just as many photons blueshifting \emph{into} the channel from lower frequencies as there are photons blueshifting \emph{out} of the channel into higher frequencies. The size of the effect is therefore proportional to the \emph{slope} of the global signal curve, and because there are some reasonably large gradients in typical predictions for the global signal (see Figure \ref{fig:global_sig}), it is this second frequency-dependent effect that dominates the dipole signature. In absolute terms the dipole \tcm signal is still small compared to the monopole, but it is---crucially---different from the dipole caused by the modulation of the foreground monopole, since our velocity vector relative to the Milky Way is not aligned with our velocity vector relative to the cosmic rest frame. This can potentially provide a clean signature for further separating foregrounds from the cosmological signal, although other techniques would still have to be employed in order to isolate extragalactic foregrounds.

Aside from astrophysical foregrounds, global signal experiments must also contend with RFI and the ionosphere \citep{Vedantham2014GlobalSignalIonosphere,Datta2016ionosphere}. Current experiments deal with this by careful data excision and calibration (often using similar techniques to those described in Sections \ref{sec:Ionosphere} and \ref{sec:RFI}), but there have also been proposed experiments such as the Dark Ages Radio Explorer (DARE; \citealt{Burns2012DARE, Burns2017DARE}) and the Dark Ages Polarimetry PathfindER (DAPPER; \citealt{Tauscher2018DAPPER}) that propose going to the far side of the Moon to avoid these problems entirely.

Finally, global signal experiments require exquisite control of systematics \citep{RogersBowman2012}, such as those due to antenna beam chromaticity \citep{Vedantham2014GlobalSignalIonosphere,Bernardi2015,MozdzenBeam2016,SinghSARAS_overview2018}, cable reflections \citep{Monsalve2017} and environmental conditions \citep{BradleyPlane2018}. In many ways, the calibration requirements for global signal experiments are more extreme than those for interferometric experiments, since single-element experiments possess a noise bias (as discussed in Section \ref{sec:dish_or_inter}) that must be well-characterized and subtracted. To achieve this level of calibration, experimental groups tend to rely on a combination of laboratory measurements \citep{RogersBowman2012,Patra2013,Monsalve2016,Monsalve2017}, \emph{in situ} measurements \citep{Patra2017}, and principal component-based modelling from the actual data \citep{Tauscher2018}. Typically, the instrumental effects are calibrated out of the measurement by introducing nuisance calibration parameters. However, given the limited number of degrees of freedom in a global signal experiment, one must take care to design experiments in a way that avoids the need for a large proliferation of calibration parameters \citep{Switzer:2014}.

\subsection{The current status of global signal experiments}

Most global signal experiments have so far focused on Cosmic Dawn and reionization, given the richer phenomenology accessible in those epochs compared to the post-reionization era.

Early global experiments provided measurements of the spectral index of the diffuse radio emission at low frequencies (i.e., the $\alpha_\textrm{syn}$ parameter of Equation \ref{eq:PhenoSynchSpec}; \citealt{RogersBowman2008,PatraSpectralIndex2015,MozdzenSpectralIndex1,MozdzenSpectralIndex2}). More recently, the timing of reionization has been constrained by the Shaped Antenna measured of the background RAdio Spectrum (SARAS; \citealt{SinghSARAS_overview2018}) experiment and the Experiment to Detect the Global Epoch of reionization Signature (EDGES; \citealt{Bowman2008EDGES}). SARAS has ruled out rapid reionization in the interval $6 < z < 10$ where the rate of change $d\overline{T}_b / dz$ of the global signal $\overline{T}_b$ with redshift is less than $114\,\textrm{mK}$ \citep{SinghSARAS2017,SinghSARAS2018}. Similar results have come out of EDGES, but in terms of a phenomenological model where reionization is parameterized in terms of a midpoint redshift $z_\textrm{mid}$, a characteristic duration $\Delta z$, and the globally averaged neutral fraction $\overline{x}_\textrm{HI}$ given by
\begin{equation}
\overline{x}_\textrm{HI} (z) = \frac{1}{2} \left[ \tanh \left( \frac{z-z_\textrm{mid}}{\Delta z} \right) + 1\right].
\end{equation}
Early EDGES data ruled out the possibility that reionization was more abrupt than $\Delta z \sim 0.06$ \citep{Nature2010}. Additional data sharpened that constraint, ruling out $\Delta z \lesssim 1$ if the midpoint of reionization occurred at $z_\textrm{mid} \sim 8.5$, and $\Delta z \lesssim 0.4$ if it occurred anywhere in the range $6.5 < z_\textrm{mid} < 14.8$. Combining some of this data with EDGES observations centred on Cosmic Dawn redshifts as well as constraints from the CMB and high redshift quasars has favoured a minimum virial temperature $T_\textrm{vir}$ in the range $10^{4.5} M_\odot$ to $10^{5.7} M_\odot$ \citep{MonsalveGalaxies2018}, a CMB optical depth $\tau_e$ of less than $0.063$, and a mean free path $R_{mfp}$ of ionizing photons that is greater than $27.5\,\textrm{Mpc}$ \citep{2019ApJ...875...67M}.

Of considerable recent interest are constraints on Cosmic Dawn. SARAS has ruled out a variety of models with an extremely cold IGM that would result in large absorption troughs \citep{SinghSARAS2017,SinghSARAS2018}. However, EDGES has recently claimed a detection of just such a signature at $78\,\textrm{MHz}$ \citep{BowmanEDGES2018}. If confirmed, the purported EDGES signal would have dramatic implications on our understanding of Cosmic Dawn. For example, the timing and narrowness of the EDGES signal suggests a more rapidly evolving star formation rate than expected \citep{MirochaFurlanetto2019}. Additionally, the amplitude of the absorption is unexpectedly large, signifying a large temperature contrast between the background radiation temperature and the neutral hydrogen spin temperature (see Equation \ref{eq:Tb_background}, which shows that large \tcm absorption signals occur when $T_s$ is much cooler than $T_\gamma$). Achieving such a large contrast would require at least one of the following revisions to our theoretical models. One possibility is the existence of a population of previously undetected high-redshift radio sources. This would increase the temperature contrast by boosting the background radiation temperature, essentially replacing $T_\gamma$ in Equation \eqref{eq:Tb_theory} with some higher temperature \citep{Feng2018,Sharma2018,EwallWice2018RadioBackground,EwallWice2019RadioBackground,Jana2019,Fialkov2019globalsig}. Alternatively, there may be exotic physics at play that allows the spin temperature to cool below the temperature expected for a gas that cools adiabatically with our Universe's expansion \citep{Falkowski2018,SlatyerWu2018,HiranoBromm2018,Barkana2018,Costa2018,Moroi2018,Berlin2018,Munoz2018a,Safarzadeh2018,Schneider2018,KovetzFuzzy2018,Kovetz2018,Clark2018,Hektor2018,Mitridate2018,Yoshiura2018,Munoz2018b,cheung2019,Chianese2019,Jia2019,Lawson2019}.

The EDGES claim is not without controversy. Discussions and debates are ongoing regarding various possible instrumental systematics \citep{BradleyPlane2018} and subtleties in the analysis methods \citep{Hills2018,EDGESreply2018,SinghSubrahmanyan2019,Jana2019}. Fortunately, other experiments such as the Probing Radio Intensity at high-$z$ from Marion experiment (PRI$^Z$M; \citealt{PRIZM2019}), the Large-aperture Experiment to Detect the Dark Age (LEDA; \citealt{Price2018}), and the Radio Experiment for the Analysis of
Cosmic Hydrogen (REACH; \citealt{EloyREACH}) are collecting data and may soon confirm or refute the EDGES claims. With the large impact of a first detection of a \tcm signal from Cosmic Dawn, these verifications will clearly be well worth the effort.

\section{Conclusion}
\label{sec:Conc}

Now is an exciting time for \tcm cosmology. The field is rapidly becoming one of experimental reality rather than theoretical promise. A large number of experiments have been built, many of which have sufficient sensitivity for a detection and characterization of the cosmological \tcm signal. Although sensitivity is not necessarily the limiting factor in these observations, having greater signal to noise also allows for a more incisive diagnosis and mitigation of systematics. In the last decade, \tcm cosmology has gone from a data-starved field to one where there is plenty of data with which to test a plethora of analysis ideas.

The primary challenges of \tcm cosmology remain the extreme sensitivity (Section \ref{sec:BasicObs}), foreground mitigation (Sections \ref{sec:fgs} and \ref{sec:FgMitigation}), and systematic control requirements (Sections \ref{sec:InstrumentSystematics} and \ref{sec:Calib}). Importantly, these challenges do not exist in isolation, and much of the problem comes from interactions between these effects. For example, in Section \ref{sec:wedge} we discussed the way in which the already-formidable foreground problem is made worse by the way in which the \emph{observed} foregrounds seen through an interferometer are less spectrally smooth than the true foreground emission on the sky. This results in a larger number of Fourier modes (``the foreground wedge") that are likely to be irretrievably contaminated by foregrounds than one would predict from intrinsic foreground properties. Thus, one sees that the data analysis techniques discussed in this paper for calibration (Section \ref{sec:skycal}), map-making (Section \ref{sec:mapmaking}), power spectrum estimation (Section \ref{sec:pspecestimation}), foreground mitigation (Section \ref{sec:FgMitigation}), and parameter estimation (Section \ref{sec:BeyondPspec}) are all inextricably linked to one another as well as one's hardware. This is true not only for experiments targeting spatial fluctuations in the \tcm line (which were emphasized for most of this paper) but also the global signal measurements discussed in Section \ref{sec:GlobalSig}. In general, \tcm experiments truly are software telescopes, where careful hardware-aware analysis pipelines are required for success.

It would be fair to say that data analysis in \tcm cosmology remains an unsolved problem, in that there is still no consensus in the community regarding the optimal way to go from raw radio telescope data to the science that is in principle enabled by measurements of the redshifted \tcm line. In many ways, however, this is beneficial, given that multiple highly sensitive experiments now exist as testbeds for the vast landscape of analysis ideas in the literature. One may thus reasonably look forward to accelerating progress in the field, with confirmation or refutation of the tentative detections of the global \tcm signal, a continued downward trend in upper limits on fluctuations in the \tcm signal at all redshifts, and eventually a set of robust detections and high-significance measurements. These will unlock the great potential of \tcm cosmology, enabling a new generation of cosmological measurements that explore our Universe in both space and time to unprecedented precision and accuracy.

\acknowledgments

The authors are delighted to acknowledge helpful discussions and/or help with figures from Marcelo Alvarez, Jo\"{e}lle-Marie Begin Miolan, Gianni Bernardi, Judd Bowman, Chris Carilli, Phil Bull, Tzu-Ching Chang, Eloy de Lera Acedo, Avinash Deshpande, Josh Dillon, Gary Hinshaw, Danny Jacobs, Nick Kern, 
Piyanat Kittiwisit, Matt Kolopanis, Henri Lamarre, Zac Martinot, Andrei Mesinger, Jordan Mirocha, Steven Murray, Michael Pagano, Nima Razavi-Ghods, Peter Sims, Harish Vedantham, and Jeff Zheng. We also gratefully thank Jeff Mangum for his editorial patience, and the anonymous referee for an astoundingly speedy yet thorough report. AL would additionally like to thank Rose Bagot, Khanh Huy Bui, Daryl Haggard, Corinne Maurice, and Kieran O'Donnell for writing group support. RS would like to thank the Berkeley Center for Cosmological Physics and the Berkeley Radio Astronomy Lab for their hospitality when this work was started. AL acknowledges support from National Science Foundation National Science Foundation Grant No. 1836019, the New Frontiers in Research Fund Exploration grant program, a Natural Sciences and Engineering Research Council of Canada (NSERC) Discovery Grant and a Discovery Launch Supplement, as well as the Canadian Institute for Advanced Research (CIFAR) Azrieli Global Scholars program.

\appendix

\section{Drift scan observations}
\label{sec:DriftScans}

Many upcoming \tcm experiments are designed to perform drift scan observations, where they simply point to a fixed location relative to the ground and use the Earth's rotation to build up sky coverage. This is in contrast to more traditional tracking observations where the telescope has steerable antennas which are pointed to a fixed location on the sky during the observation. In this section we analyse the behaviour of two telescopes differing only in whether they track, or drift scan to show how these two different modes of observation are conceptually quite distinct. We will also connect the $m$-mode formalism outlined in \secref{sec:mmodes} with the more traditional $uv$-plane formalism for radio interferometry.

To make this comparison we will use the flat sky approximation; however, rather than modelling the sky as a plane around some fixed point, we will approximate it as a cylinder around some fixed declination. In this limit, the sky is flat in the sense there is no curvature we need to worry about, but it is periodic in right ascension. This approximation is reasonable provided the primary beam of our telescope is sensitive to a limited range of declinations such that we measure the sky on a thin strip. We will also assume that the primary beam of the telescope is limited in right ascension, such that we can use small angle approximations.

In the equations below we will define $\delta_\text{dec}$ as the central declination of the strip we are observing, which has a width $\Delta$. $\theta_x$ and $\theta_y$ are the cartesian coordinates within the observed strip in the east-west and north-south directions, with the origin of $\theta_x$ coinciding with the origin of the right ascension axis, and the origin of $\theta_y$ being at the centre of the strip, i.e. declination $\delta_\text{dec}$. The telescope is located at a latitude of $\delta_\text{lat}$. We write the local sidereal time (as an angle) as $\phi$, such that at $\phi = 0$ the meridian is $\theta_x = 0$. The drift-scan telescope is permanently pointed at the meridian, such that it is centred at right ascension $\alpha = \phi$. The tracking telescope is pointed at a fixed right ascension which we choose to be the origin of our cylindrical coordinate system $\theta_x = 0$ (to make this a general point we just need to apply an offet to $\phi$).

We will use the vector $\vu$ as the Fourier conjugate to the angular position in our strip. To avoid confusion we will write the baseline vector in the ground frame as $\vb$, which (as is convention) will be expressed a physical length (and not in wavelengths).

To proceed let us write the standard equation for an unpolarized visibility as an integral over our cylindrical surface
\begin{align}
    \label{eq:vis_2d_approx}
V(\vb, \phi) & = \int A_p(\vrhat; \phi) \, e^{-2\pi i \vb(\phi) \cdot [\vrhat - \vrhat_0] / \lambda} \, T(\vrhat) d\Omega \nonumber \\
& \approx \int_0^{2\pi} \!\! d\theta_x \int_{-\Delta/2}^{\Delta/2} \!\! d\theta_y A_p^\prime(\theta_x, \theta_y; \phi) \, e^{-2\pi i \vb(\phi) \cdot [\vrhat(\theta_x, \theta_y) - \vrhat_0] / \lambda} \: T(\theta_x, \theta_y) \; .
\end{align}
For convenience we have defined $A_p^\prime(\theta_x, \theta_y; \phi) \equiv \cos{(\delta_\text{dec} + \theta_y)} A_p(\vrhat; \phi)$, which incorporates the declination dependent term from the measure $d\Omega$. We have also included a phase-centring term that references the complex phase to centre of the field we are pointing at which we call $\vrhat_0$. For the drift-scan telescope this will be at the centre of the observed strip, on the meridian (i.e. $\theta_x = \phi$, $\theta_y = 0$), for the tracking telescope it will be at the origin ($\theta_x = \theta_y = 0$).

Let us define the transfer function
\begin{equation}
    B(\theta_x, \theta_y; \vb, \phi) = A_p^\prime(\theta_x, \theta_y; \phi) e^{-2\pi i \vb(\phi) \cdot (\vrhat(\theta_x, \theta_y) - \vrhat_0) / \lambda} \; ,
\end{equation}
and write $B$ and $T$ in terms of their Fourier series:
\begin{align}
    T(\vtheta) &= \sum_{mn} \tilde{T}(\vu_{mn}) \, e^{2\pi i \vu_{mn} \cdot \vtheta} \\
    B(\vtheta; \vb, \phi) &= \sum_{mn} \tilde{B}(\vu_{mn}; \vb, \phi) \, e^{2\pi i \vu_{mn} \cdot \vtheta}
\end{align}
where $\vtheta = (\theta_x, \theta_y)$ and $\vu_{mn} = (m / 2 \pi, n / \Delta)$. Substituting these into Equation \eqref{eq:vis_2d_approx} we find that
\begin{equation}
    \label{eq:VBdef}
    V(\vb, \phi) = 2 \pi \Delta \sum_{mn} \tilde{B}(-\vu_{mn}; \vb, \phi) \, \tilde{T}(\vu_{mn})
\end{equation}
and thus to calculate the visibilities we simply need to be able to evaluate the Fourier transform of $B(\vtheta; \vb, \phi)$

We define a group fixed coordinate system with $\vbhat_e$ (points east) and $\vbhat_n$ (points north) tangent to Earth's surface at the telescope location. In terms of the usual cartesian unit vectors these are
\begin{align}
    \vbhat_e & = -\sin{\phi} \, \vxhat + \cos{\phi} \, \vyhat \\
    \vbhat_n & = -\sin{\delta_\text{lat}} \cos{\phi} \, \vxhat  - \sin{\delta_\text{lat}} \sin{\phi} \, \vyhat + \cos\delta_\text{lat} \, \vzhat
\end{align}
and the unit vector pointing to a general position on the sky (paramterized by $\theta_x$ and $\theta_y$) is
\begin{equation}
\vrhat(\theta_x, \theta_y) = \cos{(\delta_\text{dec} + \theta_y)}\cos{\theta_x} \, \vxhat + \cos{(\delta_\text{dec} + \theta_y)} \sin{\theta_x} \, \vyhat + \sin{(\delta_\text{dec} + \theta_y)} \, \vzhat \; .
\end{equation}
To investigate the behaviour of the telescope we need to construct the product $\vb \cdot \vrhat$ in these coordinates. We can do that by breaking the baseline vector $\vb$ into east-west $b_e$ and north-south $b_n$ components as $\vb = b_e \vbhat_e + b_n \vbhat_n$. Having done that we can calculate the two components
\begin{align}
\vbhat_e \cdot \vrhat = & \cos{(\delta_\text{dec} + \theta_y)} \sin{(\theta_x - \phi)} \;  , \\
\vbhat_n \cdot \vrhat = & -\sin\delta_\text{lat} \cos{(\delta_\text{dec} + \theta_y)} \cos{(\theta_x - \phi)} + \cos\delta_\text{lat} \sin{(\delta_\text{dec} + \theta_y)} \; .
\end{align}

For both drift-scan and tracking telescopes we have made the approximation that the field of view is limited such that we don't have contributions to the integral in Equation \eqref{eq:vis_2d_approx} far from the direction we are pointing and thus we can make small angle approximations of the above terms. In both cases we can treat $\theta_y$ as a small angle, i.e. we see only a narrow strip in declination. In the drift scan case $\theta_x - \phi$ is a small angle as we don't see far from the meridian, where as for the tracking case $\theta_x$ itself is a small angle. Using this we can expand out expressions for the phase to first order.

Let us proceed with the drift-scan telescope case. Here the expansion of the phase factor is fairly simple, and we find that
\begin{equation}
    \vb \cdot [\vrhat - \vrhat_0] = b_e \cos{\delta_\text{dec}} \sin{(\theta_x - \phi)} + b_n \cos{(\delta_\text{dec} - \delta_\text{lat})} \sin{\theta_y} \; ,
\end{equation}
and we also note that for a drift-scan telescope the behaviour under Earth rotation of the primary beam is simple for $A_p^\prime(\theta_x, \theta_y; \phi) = A_p^\prime(\theta_x - \phi, \theta_y)$. As both the baseline phase component and the beam simply translate in the azimuthal direction as the Earth rotates we can take advantage of the Fourier shift theorem when evaluating the Fourier transform of the transfer function
\begin{equation}
    \label{eq:Bshift}
    \tilde{B}(\vu_{mn}; \vb, \phi) = e^{-i m \phi} \tilde{B}(\vu_{mn}; \vb, \phi = 0) \; .
\end{equation}
Using this we only need to evaluate the transform at $\phi = 0$ which can be done as
\begin{align}
    B(\vu_{mn}; \vb, \phi = 0) & = \frac{1}{2\pi\Delta} \int_0^{2\pi}\!\! \int_{-\Delta/2}^{\Delta/2} \!\! A_p^\prime(\vtheta) \: e^{-2\pi i [\vb \cdot \vrhat(\theta) / \lambda - \vb\cdot\vrhat_0 / \lambda + \vu_{mn}\cdot\theta]} \, d\theta_x d\theta_y \\
    & \approx \frac{1}{2\pi\Delta} \int_0^{2\pi}\!\! \int_{-\Delta/2}^{\Delta/2} \!\! A_p^\prime(\theta_x, \theta_y) \: e^{-2\pi i [(b_e \cos{\delta_\text{dec}} / \lambda + m / 2\pi) \theta_x + (b_n \cos{(\delta_\text{dec} - \delta_\text{lat})} / \lambda + n / \Delta) \theta_y]} \, d\theta_x d\theta_y \\
    & \approx \tilde{A_p}(\vu_{mn} + \vu_\text{drift}(\vb, \delta_\text{dec}, \delta_\text{lat})) \label{eq:BAapprox}
\end{align}
where to get to the second line we needed to expand small angles in $\sin\theta \approx \theta$ and in the final line we implicitly introduced the Fourier transform of the beam function $\tilde{A_p}(\vu_{mn})$ and have defined
\begin{equation}
\vu_\text{drift}(\vb, \delta_\text{dec}, \delta_\text{lat}) =
\begin{pmatrix}
    b_e \cos{\delta_\text{dec}} \\
    b_n \cos{(\delta_\text{dec} - \delta_\text{lat})}
\end{pmatrix} \; .
\end{equation}
Remembering that $\tilde{A_p}$ is a discrete quantity, the final line is only exact when $2 \pi u \cos{\delta_\text{dec}}$ and $\Delta v \cos{(\delta_\text{dec} - \delta_\text{lat})}$ are integers\footnote{This limitation can be removed in the azimuthal direction by performing the transform without making the small angle approximation. This yields a convolution with a Bessel function $J_m(2 \pi u \cos{\delta_\text{dec}})$, which for large $m$ large approximate the delta function shift generated by the small angle approximation. Unfortunately this does not work in the polar direction as it is not strictly periodic.}.

To account for the time direction, let us take the Fourier transform of the visibilities $V(\vb, \phi)$ in the $\phi$ direction. Using Equations \eqref{eq:VBdef}, \eqref{eq:Bshift}, and \eqref{eq:BAapprox} we come to our final result for the drift-scan telescope
\begin{align}
    V_m(\vb) & = \frac{1}{2\pi} \int V(\vb, \phi) \: e^{-i m \phi} d\phi \\
    & = \sum_{n} \tilde{A_p}(\vu_\text{drift}(\vb, \delta_\text{dec}, \delta_\text{lat}) - \vu_{mn}) \: T(\vu_{mn}) \; .
\end{align}
This is the flat sky version of the $m$-mode formalism outlined in \secref{sec:mmodes}. For the drift-scan telescope we can easily see that the modes probed by the telescope are fixed, i.e. the primary beam footprint $\tilde{A_p}$ is shifted by an amount which depends only on the baseline separation ($\vb$) and the telescope and field location ($\delta_\text{lat}$ and $\delta_\text{dec}$). However, we only convolve over the north-south direction (the $n$-mode), meaning that the observed $m$'s map directly to the $m$'s on the sky.

The result for the tracking telescope is more complex. The phase factor after expansion to first order in small parameters is
\begin{align}
    \notag
    \vb\cdot\ls \vrhat - \vrhat_0\rs = & \lp b_e \cos{\phi} \cos{\delta_\text{dec}} - b_n \sin{\delta_\text{lat}} \cos{\delta_\text{dec}} \sin{\phi} \rp \sin{\theta_x} \\ & + (-b_e \sin{\delta_\text{dec}} \sin{\phi} + b_n \sin{\delta_\text{lat}} \sin{\delta_\text{dec}} \cos{\phi} + b_n \cos{\delta_\text{lat}} \cos{\delta_\text{dec}}) \sin{\theta_y} \; .
\end{align}
Similar to the drift-scan telescope let us define an effective vector
\begin{equation}
    \vu_\text{tracking}(\vb, \phi, \delta_\text{dec}, \delta_\text{lat}) = \begin{pmatrix}b_e \cos{\phi} \cos{\delta_\text{dec}} - b_n \sin{\delta_\text{lat}} \cos{\delta_\text{dec}} \sin{\phi}\\
    -b_e \sin{\delta_\text{dec}} \sin{\phi} + b_n \sin{\delta_\text{lat}} \sin{\delta_\text{dec}} \cos{\phi} + b_n \cos{\delta_\text{lat}} \cos{\delta_\text{dec}}\end{pmatrix}
\end{equation}
such that
\begin{equation}
    \vb \cdot [\vrhat - \vrhat_0] = \vu_\text{tracking}(\vb, \phi, \delta_\text{dec}, \delta_\text{lat}) \cdot \vtheta
\end{equation}
at first order in small angles. Note that unlike the drift-scan case, this vector is a function of time (i.e. $\phi$), and also that $\vu_\text{drift}(\vbhat, \delta_\text{dec}, \delta_\text{lat}) = \vu_\text{tracking}(\vbhat, \phi = 0, \delta_\text{dec}, \delta_\text{lat})$.
\begin{equation}
    V(\vb, \phi) = \sum_{mn} \tilde{A_p}(\vu_\text{tracking}(\vb, \phi, \delta_\text{dec}, \delta_\text{lat}) - \vu_{mn}) \: T(\vu_{mn}) \; .
\end{equation}
This does not have any simple reduction across the time domain in the same way that the drift-scan telescope does, as the shift vector is a function of time. We can interpret the observations as sampling part of the $uv$-plane where the position moves as an ellipse through the plane as function of the Earth's rotation. For more details (including the full three-dimensional treatment) see \cite{TMS2017}.

One important conclusion from this is that for a drift-scan telescope, the coverage in the $uv$-plane must be \emph{instantaneously} sufficient, observing for extended periods of time never changes the overall area of Fourier modes that we have access. However observing the changing of the visibilites in time gives us access to modes \emph{within} the footprint of the primary beam, and if we observe for a full sidereal day down to each individual azimuthal $m$-mode. This is stark contrast to the behaviour of a tracking telescope where Earth rotation synthesis can dramatically improve the overall footprint of modes we can measure, but we have no access to higher resolution Fourier modes than the primary beam.

\begin{figure}
    \begin{center}
        \includegraphics[width=0.8\textwidth]{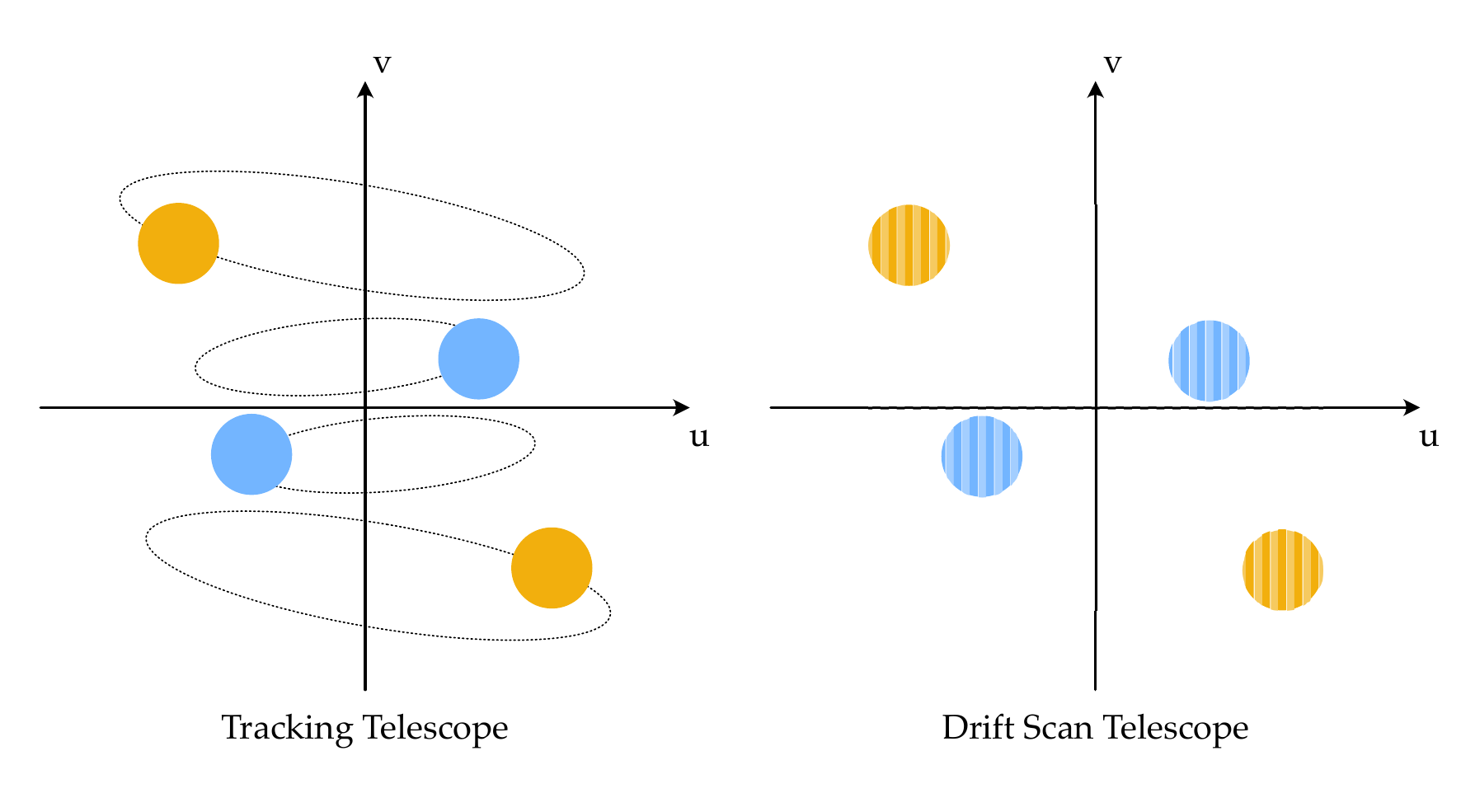}
    \end{center}
    \caption{For a tracking telescope the part of the $uv$-plane that we measure as the Earth rotates follows an ellipse, probing different Fourier modes of the sky (left panel or Figure \ref{fig:uv_tracks}). Through this we fill in the gaps in the $uv$-plane, building up a larger set of measured modes. In contrast, a drift-scan telescope measures the same set of Fourier modes (right panel), but its observing strategy means that it can directly distinguish Fourier modes in the $u$ direction within the footprint of the primary beam.}
\end{figure}

\section{Explicit Quadratic Estimator Examples}
\label{sec:OQEexamples}
In this Appendix, we construct some explicit, concrete examples of the powerful and general---but abstract---quadratic estimator formalism presented in Section \ref{sec:pspecestimation}. Many of our results will be approximate, for the pedagogical sake of analytical tractability.

\subsection{Measuring $P(\mathbf{k})$ from a three-dimensional configuration-space map}
\label{sec:Toy3DPk}
Suppose our input data consists of a three-dimensional map in configuration space, which is something that one might obtain from one of the map-making algorithms described in Section \ref{sec:mapmaking}. To be precise, let the $i$th component of our data vector $\vx$ be given by
\begin{equation}
\label{eq:pixelization}
x_i = \int \! d^3 r\, \psi_i (\vr) T(\vr)  = \int \! \frac{d^3 k}{(2\pi)^3}\, \widetilde{\psi}^*_i (\mathbf{k}) \widetilde{T} (\mathbf{k})
\end{equation}
where $T(\vr)$ is the (continuous) temperature field and $\psi_i (\vr)$ is the footprint of the $i$th voxel of our map. In the second equality, we made use of Parseval's theorem to rewrite the integral in Fourier space. As a concrete example, we might imagine a set of uniform cuboid voxels, where
\begin{equation}
\label{eq:Voxels}
\psi_i (\vr) =
\begin{cases}
1, \quad\textrm{if }\vr\textrm{ is within }\pm \Delta \vr / 2\textrm{ of }\vr_i\\
0, \quad\textrm{otherwise,}
\end{cases}
\end{equation}
with $\Delta \vr \equiv (\Delta x, \Delta y, \Delta z)$ denoting the extent of the voxels.

Recall from Section \ref{sec:pspecestimation} that the key to establishing a link between the data and the power spectrum that we wish to estimate is the covariance matrix $\mC \equiv \langle \vx \vx^\dagger \rangle$ and its response with respect to changes to a generic bin (say, the $\alpha$th bin) of the power spectrum, $\mQ^\alpha \equiv \partial \mC / \partial p^\alpha$. We can obtain the covariance matrix by inserting Equation \eqref{eq:pixelization} into the definition of the covariance. The result is that the $(i,j)$th element of the covariance is given by
\begin{equation}
\label{eq:CovIntegral}
C_{ij} = \int \! \frac{d^3 k_a}{(2\pi)^3} \frac{d^3 k_b}{(2\pi)^3} \widetilde{\psi}^*_i (\mathbf{k}_a) \widetilde{\psi}_j (\mathbf{k}_b) \langle \widetilde{T} (\mathbf{k}_a) \widetilde{T}^* (\mathbf{k}_b) \rangle = C^0_{ij} + \int \! \frac{d^3 k}{(2\pi)^3} \widetilde{\psi}^*_i (\mathbf{k}) \widetilde{\psi}_j (\mathbf{k}) P(\vk),
\end{equation}
where in the second equality we decomposed $\widetilde{T}$ into a sum of contributions from the cosmological signal $\widetilde{T}_\textrm{sig}$ and other contaminants or noise, $\widetilde{T}_\textrm{other}$. This allowed us to write $\langle \widetilde{T} (\mathbf{k}_a) \widetilde{T}^* (\mathbf{k}_b) \rangle$ as
\begin{equation}
\langle \widetilde{T}_\textrm{sig} (\mathbf{k}_a) \widetilde{T}^*_\textrm{sig} (\mathbf{k}_b) \rangle + \langle \widetilde{T}_\textrm{sig} (\mathbf{k}_a) \widetilde{T}^*_\textrm{other} (\mathbf{k}_b) \rangle + \langle \widetilde{T}_\textrm{other} (\mathbf{k}_a) \widetilde{T}^*_\textrm{sig} (\mathbf{k}_b) \rangle+ \langle \widetilde{T}_\textrm{other} (\mathbf{k}_a) \widetilde{T}^*_\textrm{other} (\mathbf{k}_b) \rangle,
\end{equation}
which reduces to $  (2\pi)^3 \delta^D (\vk_a - \vk_b) P(\vk_a) + \langle \widetilde{T}_\textrm{other} (\mathbf{k}_a) \widetilde{T}^*_\textrm{other} (\mathbf{k}_b) \rangle $ because the cosmological signal is on average uncorrelated with the ``other" signals and the signal-signal term is related to the power spectrum $P(\vk)$ via Equation \eqref{eq:PspecDef}. Inserting this into our integral above and defining the integral containing $\langle \widetilde{T}_\textrm{other} (\mathbf{k}_a) \widetilde{T}^*_\textrm{other} (\mathbf{k}_b) \rangle$ to be $C^0_{ij}$ then gives the right hand side of Equation \eqref{eq:CovIntegral}.

Having obtained the covariance matrix, our next step is to differentiate with respect to the bandpower $p^\alpha$. This requires discretizing our power spectrum. The most straightforward way to do this is to take our power spectrum to be piecewise constant, such that within the $\alpha$th $\vk$-space voxel, the value of the power spectrum is $p^\alpha$. Under this assumption, we have
\begin{equation}
\label{eq:CintermsofP}
C_{ij} = C^0_{ij} + \sum_\alpha p^\alpha \int_{V_{k_\alpha}} \! \frac{d^3 k}{(2\pi)^3} \widetilde{\psi}^*_i (\mathbf{k}) \widetilde{\psi}_j (\mathbf{k}) ,
\end{equation}
where the bandpower has come out of the integral because of our piecewise constant assumption, and $V_{k_\alpha}$ signifies the region of Fourier space covered by the $\alpha$th $\vk$-space voxel. The derivative is now straightforward, yielding
\begin{equation}
\label{eq:Qintermsofpsi}
Q^\alpha_{ij} \equiv \frac{\partial C_{ij} }{ \partial p^\alpha} = \int_{V_{k_\alpha}} \! \frac{d^3 k}{(2\pi)^3} \widetilde{\psi}^*_i (\mathbf{k}) \widetilde{\psi}_j (\mathbf{k}).
\end{equation}

We now have all the pieces that are required to construct the optimal quadratic estimator of the power spectrum. Recall from Equation \eqref{eq:palphaunnormed} that an optimal estimator for the $\alpha$th bandpower is given by $\hat{p}^\alpha \propto \vx^\dagger  \mC^{-1} \mQ^\alpha \mC^{-1} \vx $. Inserting our expression for $\mQ^\alpha$ into this and simplifying the result gives
\begin{equation}
\hat{p}^\alpha \propto \sum_{ij} (\mC^{-1} \vx)_i \left[ \int_{V_{k_\alpha}} \! \frac{d^3 k}{(2\pi)^3} \widetilde{\psi}^*_i (\mathbf{k}) \widetilde{\psi}_j (\mathbf{k}) \right] (\mC^{-1} \vx)_j =\int_{V_{k_\alpha}} \! \frac{d^3 k}{(2\pi)^3} \Big{|} \sum_i \widetilde{\psi}_i (\mathbf{k}) (\mC^{-1} \vx)_i \Big{|}^2.
\end{equation}
To gain some intuition for this, consider the voxelization of our data that we defined in Equation \eqref{eq:Voxels}. Taking the Fourier transform to get $\widetilde{\psi}_i (\vk)$ gives
\begin{equation}
\widetilde{\psi}_i (\vk) = (\Delta x \Delta y \Delta z) e^{ -i \vk \cdot \vr_i } \sinc \left( \frac{k_x \Delta x}{2} \right) \sinc \left( \frac{k_y \Delta y}{2} \right) \sinc \left( \frac{k_z \Delta z}{2} \right) \approx (\Delta x \Delta y \Delta z) e^{ -i \vk \cdot \vr_i },
\end{equation}
where the last approximation can be made if the resolution of the survey is very fine compared to the lengthscales of interest, such that $\vk \cdot \Delta \vr \ll 1$ and the sinc terms are approximately unity. With this approximation, we have
\begin{equation}
\hat{p}^\alpha \propto \int_{V_{k_\alpha}} \! \frac{d^3 k}{(2\pi)^3} \Big{|} \sum_i  e^{ -i \vk \cdot \vr_i }  (\mC^{-1} \vx)_i \Big{|}^2.
\end{equation}
In words, this says that to optimally estimate the power spectrum one should first apply an inverse covariance weighting to the data (this is what makes the optimal quadratic estimator an \emph{optimal} estimator, as shown in Section \ref{sec:pspecestimation}). Following that, notice that the sum in our expression is precisely a discrete Fourier transform of the data. Once the Fourier transform is taken, the results are squared before they are simply summed (integrated) over the Fourier space volume $V_{k_\alpha}$ of the Fourier space voxel of interest. The normalization of this power spectrum estimate can be computed using the expressions given in Section \ref{sec:pspecestimation}. One sees that
the optimal quadratic estimator formalism essentially yields the same thing as our guess for how to compute the power spectrum in Equation \eqref{eq:PkContinuous}, with two differences: the data is inverse covariance weighted in order to achieve the best possible error bars in the final result, and the quadratic estimator formalism allows one to generalize to more complicated situations where the pixelization might be complicated.

\subsection{Measuring $P(k_\perp, k_\parallel)$ or $P(k)$}

As discussed in Section \ref{sec:ImageVsStat}, it is often appropriate to report a binned version of $P(\vk)$. For diagnosing systematic effects, it is helpful to bin our Fourier space into modes parallel and perpendicular to the line of sight, given the differences in how parallel and perpendicular information is obtained in \tcm cosmology. For a final cosmological result, all directions should be statistically equivalent (ignoring redshift-space distortions), so binning $\vk$ down into a scalar $k$ is appropriate.

Within the quadratic estimator formalism, obtaining $P(k_\perp, k_\parallel)$ or $P(k)$ is trivial. Take $P(k)$ as an example. Forming such a power spectrum requires binning together all $\vk$ with the same magnitude $k$ (i.e., those that fall on a particular spherical shell in Fourier space). If we simply define $V_{k_\alpha}$ to be the Fourier space region falling on the $\alpha$th shell, all our expressions from the previous section follow. The same is true for $P(k_\perp, k_\parallel)$ with appropriately adapted definitions of the bandpower bins.

\subsection{Measuring $P(\mathbf{k})$ from a Fourier space map}

As we remarked in Section \ref{sec:pspecestimation}, the quadratic estimator formalism allows the input data to be expressed in any basis. To illustrate this, suppose our data is in Fourier space. This can be accommodated in our formalism by setting $\psi_i (\vr) = \phi(\vr) \exp (- i \vk_i \cdot \vr)$, where $\phi(\vr)$ is $1$ inside our survey and $0$ outside. With this choice of $\psi_i$, Equation \eqref{eq:pixelization} implies that our data vector $\vx$ is in Fourier space.

Taking the Fourier transform gives $\widetilde{\psi}_i = \widetilde{\phi} ( \vk - \vk_i)$. In turn, this gives
\begin{equation}
Q^\alpha_{ij}  = \int_{V_{k_\alpha}} \! \frac{d^3 k}{(2\pi)^3} \widetilde{\psi}^*_i (\mathbf{k}) \widetilde{\psi}_j (\mathbf{k}) = \int_{V_{k_\alpha}} \! \frac{d^3 k}{(2\pi)^3} \widetilde{\phi} ^*( \vk - \vk_i) \widetilde{\phi} ( \vk - \vk_j) \appropto
\begin{cases}
| \widetilde{\phi} (\vk_i - \vk_j) |^2& \textrm{if }\overline{\vk}\textrm{ is in }V_{k_\alpha} \\
0 & \textrm{otherwise},
\end{cases}
\end{equation}
where $\overline{\vk}$ is $(\vk_i + \vk_j) / 2$, and we have assumed that the survey volume is large, so that $\phi(\vr)$ has a large footprint while $\widetilde{\phi}$ is sharply peaked. With this assumption, $Q^\alpha_{ij}$ is approximately zero unless $\vk_i \approx \vk_j$. If we assume that $V_{k_\alpha}$ is relatively small (i.e., we have discretized our Fourier space reasonably finely) and denote the location of the Fourier space volume by $\vk_\alpha$, we come to the conclusion that $\vk_\alpha \approx \vk_i \approx \vk_j$ for $Q^\alpha_{ij}$ to be nonzero. In other words, we have $Q_{ij}^\alpha \approx \delta_{ij} \delta_{i\alpha}$, which means our quadratic estimator is
\begin{equation}
\hat{p}^\alpha \propto \vx^\dagger  \mC^{-1} \mQ^\alpha \mC^{-1} \vx  = \big{|} (\mC^{-1} \vx)_\alpha \big{|}^2.
\end{equation}
In words, this says that if our data is already in Fourier space, power spectrum estimation is straightforward: simply inverse covariance weight the data, and square the component corresponding to the wavevector that we are interested in. While this may seem an obvious result, note that it was the result of assuming a large survey volume. In practice, this approximation will never be exact. Fortunately, the full matrix formulation of the quadratic estimator formalism allows finite survey volumes to be treated properly.

\subsection{Measuring $C_\ell (\nu)$ from an angular sky map}

The quadratic estimator formalism can be used to estimate any quadratic statistic of the data. Here, we demonstrate that by using it to estimate the angular power spectrum $C_\ell$ that was defined in Equation \eqref{eq:AngPspec}.

For convenience, suppose that our data comes in the form of a sky map at a particular frequency $\nu$, expressed in angular coordinates. Our pixelization is then given by
\begin{equation}
x_i = \int \! d\Omega \,\psi_i (\hat{\vn}) T(\hat{\vn}) = \sum_{\ell m} w^{i*}_{\ell m} a_{\ell m},
\end{equation}
where in the second equality we expressed the sky in terms of its spherical harmonic expansion $a_{\ell m} \equiv \int \! d\Omega \,Y_{\ell m}^*(\hat{\vn}) T(\hat{\vn})$, where $Y_{\ell m}(\hat{\vn})$ is the spherical harmonic function with quantum numbers $\ell$ and $m$. We similarly defined $w^i_{\ell m} \equiv \int \! d\Omega \,Y_{\ell m}^*(\hat{\vn}) \psi_i(\hat{\vn})$ and assumed that $\psi_i (\hat{\vn})$ is real. An appropriate choice for $\psi_i$ might be
\begin{equation}
\label{eq:Pixels}
\psi_i (\hat{\vn}) =
\begin{cases}
1, \quad\textrm{if }\hat{\vn}\textrm{ is within }\Delta \Omega \textrm{ of }\hat{\vn}_i\\
0, \quad\textrm{otherwise,}
\end{cases}
\end{equation}
where $\Delta \Omega$ is the angular area of a single pixel. Following an analogous set of steps as we did for our rectilinear power spectrum, we obtain
\begin{equation}
C_{ij} = C_{ij}^0 + \sum_{\ell m} w^{i*}_{\ell m} w^{j}_{\ell m} C_\ell
\end{equation}
and
\begin{equation}
\label{eq:QsphH}
Q^\ell_{ij} \equiv \frac{\partial C_{ij}}{\partial C_\ell } = \sum_m w^{i*}_{\ell m} w^{j}_{\ell m}.
\end{equation}
Note the unfortunate near-clash of notations, where the symbol $C$ is used both for the covariance matrix and the angular power spectrum. The former will always be written in boldface or with two indices; the latter contains just a single index $\ell$. We also note that Equation \eqref{eq:QsphH} represents a very specific choice for the binning of our bandpowers, with each spherical harmonic $\ell$ mode being its own bin. While we made this choice here for the sake of pedagogy, there are often very good reasons to use thicker bins that encompass multiple $\ell$ modes. For example, incomplete sky coverage will cause different $\ell$ modes to be coupled to one another, making it reasonable to average them together \citep{Alonso2019PseudoCl}. Alternatively, such couplings can be computed using the window function and error covariance formulae presented in Section \ref{sec:pspecestimation}, which can then be incorporated into one's comparisons between data and theory.

Continuing with constructing our power spectrum estimator, we have
\begin{equation}
\hat{C}_\ell \propto \vx^\dagger  \mC^{-1} \mQ^\ell \mC^{-1} \vx  = \sum_m \bigg{|} \sum_i w_{\ell m}^i(\mC^{-1} \vx)_i \bigg{|}^2 = \sum_m \bigg{|} \int \! d\Omega \,Y_{\ell m}^*(\hat{\vn})  d(\hat{\vn}) \bigg{|}^2,
\end{equation}
where we have eschewed the symbol $\hat{p}_\alpha$ in favour of the more explicit $\hat{C}_\ell$ and have defined $d(\hat{\vn}) \equiv \sum_i \psi_i(\hat{\vn})(\mC^{-1} \vx)_i$. Given the form of Equation \eqref{eq:Pixels}, this is essentially an inverse covariance-weighted map of the sky that is pixelized but remapped into a continuous function (or as close to continuous as possible given the discrete nature of real data). After this, our expression for $\hat{C}_\ell$ instructs us to take the spherical harmonic transform, square, and sum over $m$. With a proper normalization, this treatment agrees with Equation \eqref{eq:CellGuess} up to the inverse covariance weighting.

Also of interest in \tcm cosmology (and intensity mapping in general) is the cross power spectrum between two frequencies, $C_\ell (\nu, \nu^\prime)$, as defined in Equation \eqref{eq:Cellnunuprime}. Formally, one can go through the same process as we did above, but with all the data from different frequencies stacked into one long data vector. Working through the algebra then gives a similar expression for $\hat{C}_\ell$, except with a cross multiplication between spherical harmonic coefficients from different frequencies rather than the squaring of coefficients from a single frequency.

\subsection{Measuring a power spectrum from visibilities}

As a final example in this Appendix, we consider how one might estimate a power spectrum directly from the visibility measured by a single baseline of an interferometer. We begin by writing the $i$th frequency channel of the visibility as
\begin{equation}
V_i \approx \int  \! d\nu d^2 \theta\, \gamma_i (\nu) A_p(\boldsymbol \theta, \nu) T(\boldsymbol \theta, \nu) \exp \left( -i \frac{2 \pi \nu }{c} \mathbf{b} \cdot \boldsymbol \theta\right) = \int d^3 r \left[ \frac{H_0 \nu_{21} E(z) \gamma_i (\nu) }{c (1+z)^2 D_c^2 (z)}A_p(\boldsymbol \theta, \nu) e^{ -i 2 \pi \nu \mathbf{b} \cdot \boldsymbol \theta/c}  \right] T(\vr),
\end{equation}
where $\gamma_i (\nu)$ describes the profile of the $i$th frequency channel. In the first equality we made the narrow-field, flat-sky approximation, and in the second equality we rewrote our visibility in terms of comoving cosmological coordinates, with the tacit understanding that $\boldsymbol \theta$, $\nu$, and $z$ are all implicit functions of $\vr$. Storing the visibilities from our different frequency channels in one data vector $\vx$, we have $x_i \equiv V_i$, and comparing this expression to Equation \eqref{eq:pixelization}, we see that the measurement equation of an interferometer can be viewed as a generalized pixelization of the sky. The term within the square brackets is then simply a complicated $\psi_i (\vr)$ function. Taking this and inserting it into all the subsequent expressions like Equations \eqref{eq:CintermsofP} and \eqref{eq:Qintermsofpsi} yields a quadratic estimator of the power spectrum that operates directly on the visibilities. If the frequency dependence of everything but $\gamma_i (\nu)$ is neglected, the result is the delay-spectrum estimator discussed in Section \ref{sec:delayspec}. Including the frequency dependence in the complex exponential term incorporates the phenomenology of the foreground wedge (Section \ref{sec:wedge}) into the estimator. This is worked out in detail in \citet{Liu:2014a}.

    \bibliography{review,cyl}

\end{document}